\documentclass[a4paper,11pt]{article}
\usepackage{heppub}
\pdfoutput=1
\usepackage{xspace}
\usepackage{placeins}
\usepackage{wasysym}
\usepackage{slashed}
\usepackage{empheq}

\newcommand{\pL}{p_L}
\newcommand{\lpazimuthalsymmetry}{L^{(0)}(\slash\!\!\!\!\varphi)}

\newcommand{\abs}[1]{\lvert#1\rvert}
\newcommand{\Abs}[1]{\bigl\lvert#1\bigr\rvert}
\newcommand{\ord}[1]{\mathcal{O}(#1)}

\newcommand{\ORd}[1]{\mathcal{O}\Bigl(#1\Bigr)}
\newcommand{\mae}[3]{\langle#1\lvert#2\rvert#3\rangle}

\newcommand{\df}{\mathrm{d}}
\newcommand{\img}{\mathrm{i}}

\newcommand{\eps}{\epsilon}
\newcommand{\la}{\lambda}
\newcommand{\w}{\omega}

\newcommand{\fb}{\,\mathrm{fb}}

\newcommand{\GeV}{\,\mathrm{GeV}}
\newcommand{\TeV}{\,\mathrm{TeV}}

\newcommand{\nn}{\nonumber}

\newcommand{\bn}{{\bar n}}
\newcommand{\bq}{{\bar q}}
\newcommand{\bC}{{\bar C}}

\newcommand{\bt}{{\vec b}_T}
\newcommand{\qt}{{\vec q}_T}

\newcommand{\cM}{\mathcal{M}}
\newcommand{\cO}{\mathcal{O}}
\newcommand{\cP}{\mathcal{P}}
\newcommand{\Tau}{\mathcal{T}}

\newcommand{\tf}{\tilde{f}}

\newcommand{\tB}{\tilde{B}}

\newcommand{\tS}{\tilde{S}}
\newcommand{\tgamma}{\tilde{\gamma}}

\newcommand{\as}{\alpha_s}
\newcommand{\aem}{\alpha_\mathrm{em}}

\newcommand{\etaMax}{\eta_\mathrm{max}}
\newcommand{\pTmin}{p_T^\mathrm{min}}
\newcommand{\qTcut}{q_T^\mathrm{cut}}
\newcommand{\dY}{{\Delta y}}
\newcommand{\pTlep}{p_T^\ell}
\newcommand{\Ecm}{E_\mathrm{cm}}
\newcommand{\id}{\mathbf{1}}
\newcommand{\lqcd}{\Lambda_\mathrm{QCD}}
\newcommand{\MSbar}{$\overline{\text{MS}}$\xspace}
\newcommand{\epol}{\varepsilon}
\newcommand{\mtozero}{\,\stackrel{m_{a,b}\to 0}{\longrightarrow}\,}

\newcommand{\zero}{{(0)}}
\newcommand{\one}{{(1)}}
\newcommand{\two}{{(2)}}
\newcommand{\cusp}{\mathrm{cusp}}
\newcommand{\cut}{\mathrm{cut}}

\newcommand{\incl}{\mathrm{incl}}
\newcommand{\fr}{\mathrm{fr}}
\newcommand{\lab}{\mathrm{lab}}
\newcommand{\lep}{\mathrm{lep}}
\newcommand{\match}{\mathrm{match}}

\newcommand{\res}{\mathrm{res}}
\newcommand{\run}{\mathrm{run}}
\newcommand{\sub}{\mathrm{sub}}
\newcommand{\total}{\mathrm{total}}
\newcommand{\vary}{\mathrm{vary}}

\newcommand{\CS}{\mathrm{CS}}
\newcommand{\GJ}{\mathrm{GJ}}
\newcommand{\GJbar}{\overline{\mathrm{GJ}}}
\newcommand{\FO}{\mathrm{FO}}
\newcommand{\I}{\mathrm{I}}
\newcommand{\II}{\mathrm{II}}
\newcommand{\SO}{\mathrm{SO}}

\newcommand*\widefbox[1]{\fbox{\hspace{1em}#1\hspace{1em}}}

\newcommand{\scetlib}{{\tt SCETlib}}
\newcommand{\mcfm}{{\tt MCFM}}

\newcommand{\WidthTwoSubfigs}{0.48\textwidth}
\newcommand{\WidthThreeSubfigs}{0.33\textwidth}

\allowdisplaybreaks[3]

\title{\boldmath Drell-Yan $q_T$ Resummation of Fiducial Power Corrections at N$^3$LL}

\author[a]{Markus A.~Ebert,\hspace{-0.2ex}}
\emailAdd{ebert@mit.edu}

\author[b]{Johannes K.~L.~Michel,\hspace{-0.2ex}}
\emailAdd{johannes.michel@desy.de}

\author[a]{Iain W.~Stewart,\hspace{-0.2ex}}
\emailAdd{iains@mit.edu}

\author[b]{and Frank J.~Tackmann}
\emailAdd{frank.tackmann@desy.de}

\affiliation[a]{Center for Theoretical Physics, Massachusetts Institute of Technology, Cambridge, MA 02139, USA}
\affiliation[b]{Theory Group, Deutsches Elektronen-Synchrotron (DESY), D-22607 Hamburg, Germany}

\abstract{%
We consider Drell-Yan production $pp\to V^* X \to L X$ at small
$q_T \ll Q$, where $q_T$ and $Q$ are
the total transverse momentum and invariant mass of the leptonic final state $L$.
Experimental measurements require fiducial cuts on $L$,
which in general introduce enhanced, linear power corrections in $q_T/Q$.
We show that they can be unambiguously predicted from factorization,
and resummed to the same order as the leading-power contribution.
For the fiducial $q_T$ spectrum, they constitute the complete linear
power corrections.
We thus obtain predictions for the fiducial $q_T$ spectrum to N$^3$LL and
next-to-leading-power in $q_T/Q$.
Matching to full NNLO ($\alpha_s^2$), we find that the linear power
corrections are indeed the dominant ones, and once included by factorization,
the remaining fixed-order corrections become almost negligible below $q_T \lesssim 40\GeV$.
We also discuss the implications for more complicated observables,
and provide predictions for the fiducial $\phi^*$ spectrum at N$^3$LL$+$NNLO.
We find excellent agreement with ATLAS and CMS measurements of $q_T$ and $\phi^*$.
We also consider the $p_T^\ell$ spectrum. We show that it develops
leptonic power corrections in $q_T/(Q - 2p_T^\ell)$, which diverge near the Jacobian
peak $p_T^\ell \sim Q/2$ and must be kept to all powers to obtain a meaningful result there.
Doing so, we obtain for the first time an analytically resummed result for the
$p_T^\ell$ spectrum around the Jacobian peak at N$^3$LL$+$NNLO.
Our method is based on performing a complete tensor decomposition for hadronic
and leptonic tensors. We show that in practice this is equivalent to
often-used recoil prescriptions, for which our results now provide rigorous, formal
justification.
Our tensor decomposition yields nine Lorentz-scalar hadronic structure functions, which
for $Z/\gamma^* \to \ell\ell$ or $W\to\ell\nu$ directly map onto
the commonly used angular coefficients, but also holds for arbitrary leptonic final states.
In particular, for suitably defined Born-projected leptons it still yields a LO-like
angular decomposition even when including QED final-state radiation.
Finally, we also discuss the application to $q_T$ subtractions. Including the
unambiguously predicted fiducial power corrections significantly improves their
performance, and in particular makes them applicable near kinematic edges
where they otherwise break down due to large leptonic power corrections.
}
\date{June 19, 2020}

\preprint{\vbox{%
\hbox{DESY 20-016}
\hbox{MIT--CTP 5205}}
}

\begin{document}

\maketitle

%%%%%%%%%%%%%%%%%%%%%%%%%%%%%%%%%%%%%%%%%%%%%%%%%%%%%%%%%%%%%%%%%%%%%%%%%%%%%%%%
\section{Introduction}
\label{sec:intro}
%%%%%%%%%%%%%%%%%%%%%%%%%%%%%%%%%%%%%%%%%%%%%%%%%%%%%%%%%%%%%%%%%%%%%%%%%%%%%%%%

The neutral and charged Drell-Yan processes, $pp\to Z/\gamma^*\to \ell\ell$ and $pp\to W\to \ell\nu$,
are important benchmark processes at the LHC. We are interested in the kinematic region
where the vector boson is produced with small or moderate transverse momentum $q_T$,
which contains the bulk of the total cross section. In this region, differential distributions
can be measured to sub-percent precision~\cite{%
Aad:2012wfa, Aad:2014xaa, Aad:2015uau, Aad:2015auj, Aad:2016izn, Aaboud:2017ffb, Aaboud:2017svj, Aad:2019wmn,
Chatrchyan:2011wt, Khachatryan:2015oaa, Khachatryan:2015paa, Khachatryan:2016nbe, Sirunyan:2017igm, Sirunyan:2018swq, Sirunyan:2019bzr},
allowing for high-precision tests of the electroweak sector of the SM,
including the precise measurement of the $W$ boson mass~\cite{Aaboud:2017svj}
and the weak mixing angle~\cite{Aad:2015uau, Sirunyan:2018swq}.

The Drell-Yan process can also be considered an important
benchmark process on the theoretical side and continues to be an important development
ground for theoretical predictions at hadron colliders.
Inclusive and fully-differential cross sections are known at NNLO~\cite{%
Altarelli:1979ub,
Hamberg:1990np, vanNeerven:1991gh, Harlander:2002wh,
Anastasiou:2003yy, Anastasiou:2003ds, Melnikov:2006di, Melnikov:2006kv, Catani:2009sm, Gavin:2010az, Gavin:2012sy}
and also combined with parton showers~\cite{Hoeche:2014aia, Karlberg:2014qua, Alioli:2015toa}.
Partial results are also available beyond NNLO~\cite{%
Moch:2005ky, Ravindran:2006bu, Ahmed:2014cla, Li:2014bfa, Ahmed:2014uya, Catani:2014uta, Lustermans:2019cau},
and the first N$^3$LO result for the total cross section of $pp\to \gamma^*\to\ell\ell$
was obtained recently in \refcite{Duhr:2020seh}.
The NLO electroweak corrections have also been calculated~\cite{%
Baur:1997wa, Baur:1998kt, Baur:2001ze, Dittmaier:2001ay, Baur:2004ig, Zykunov:2005tc,
Arbuzov:2005dd, Arbuzov:2007db, CarloniCalame:2006zq, CarloniCalame:2007cd, Dittmaier:2009cr},
as well as the
mixed NNLO QCD$+$QED and QCD$+$electroweak corrections in the limit where
production and decay are factorized~\cite{%
Dittmaier:2014qza, Dittmaier:2015rxo, deFlorian:2018wcj, Delto:2019ewv, Bonciani:2019nuy, Cieri:2020ikq, Buccioni:2020cfi}.

For small transverse momentum $q_T \ll Q$, where $Q$ is the invariant mass of the
color-singlet final state, the differential cross section admits an expansion in $q_T/Q$
%%%
\begin{align} \label{eq:xsection_expansion}
\frac{\df\sigma}{\df q_T^2}
&= \,\, \frac{\df\sigma^\zero}{\df q_T^2}
\,\,\,+\quad
\frac{\df\sigma^\one}{\df q_T^2}
\,\,\,+\quad
\frac{\df\sigma^\two}{\df q_T^2}
\,\,\,+\quad \cdots
\nn \\
&\,\sim\,\, \frac{1}{q_T^2} \biggl[
   1 \,\,\, +\,\, \cO\Bigl(\frac{q_T}{Q}\Bigr)
   \,\,+\,\, \cO\Bigl(\frac{q_T^2}{Q^2}\Bigr)
   \,+\quad  \cdots\,\, \biggr]
\,.\end{align}
%%%
The dominant term scales as $\df\sigma^\zero/\df q_T^2 \sim 1/q_T^2$
and is referred to as the leading-power (LP) contribution.
The additional terms $\df\sigma^{(n)}$ are suppressed by $(q_T/Q)^n$ relative to $\df\sigma^\zero$,
and are referred to as power corrections or subleading-power contributions.

At small $q_T$, the fixed-order expansion contains logarithmically enhanced
terms\linebreak[4] $\as^n \ln^m(q_T/Q)$ caused by soft and collinear emissions.
These series of logarithms need to be resummed to all orders in perturbation theory to
obtain precise and reliable perturbative predictions.
For the LP term, this resummation is possible thanks to the
$q_T$-dependent (TMD) factorization theorem for $\df\sigma^\zero$, originally derived
in \refscite{Collins:1981uk, Collins:1981va, Collins:1984kg},
with several equivalent formulations~%
\cite{Collins:1350496, Becher:2010tm, GarciaEchevarria:2011rb, Chiu:2012ir, Collins:2012uy, Li:2016axz}
based on different regularization methods.
A large variety of approaches for the resummation exist~%
\cite{Balazs:1995nz, Balazs:1997xd, Ellis:1997sc, Bozzi:2005wk, Bozzi:2010xn, Banfi:2012du,
Echevarria:2015uaa, Neill:2015roa, Catani:2015vma, Camarda:2019zyx,
Collins:2016hqq, Monni:2016ktx, Ebert:2016gcn, Kang:2017cjk, Coradeschi:2017zzw, Lustermans:2019plv}
and by now have reached N$^3$LL precision~%
\cite{Bizon:2017rah, Chen:2018pzu, Bizon:2018foh, Becher:2019bnm, Bizon:2019zgf, Scimemi:2019cmh,
Bacchetta:2019sam, Kallweit:2020gva},
the inclusion of quark-mass effects~\cite{Pietrulewicz:2017gxc}
and of QED corrections~\cite{Cieri:2018sfk, Bacchetta:2018dcq, Billis:2019evv}.

The power corrections $\df\sigma^{(n)}$ in \eq{xsection_expansion}
are classified by their relative $(q_T/Q)^n$ suppression,
and we refer to $\df\sigma^\one$ as the next-to-leading power (NLP) term,
$\df\sigma^\two$ as NNLP etc.
Due to their suppression, they are less relevant at small $q_T \ll Q$,
and are included by matching to the full fixed-order calculations, which amounts
to numerically extracting the complete set of power-suppressed terms at a given
fixed order in $\alpha_s$. They are in principle known to $\ord{\as^3}$ from
the NNLO $V+1$-parton calculations~\cite{Boughezal:2015dva, Boughezal:2015dra, Ridder:2015dxa, Caola:2015wna, Boughezal:2015ded, Gehrmann-DeRidder:2016jns, Gauld:2017tww, Gehrmann-DeRidder:2017mvr}.

Nevertheless, the subleading-power terms also contain logarithms $\as^n \ln^m(q_T/Q)$,
and so in principle should be resummed as well to maintain their power suppression relative
to the resummed LP term.
Hence, given the high precision reached at LP, it is important to
investigate the resummation of the subleading-power corrections to avoid them
limiting the theoretical precision.
First progress towards this direction has been made in \refcite{Ebert:2018gsn},
where the power corrections were explicitly calculated at NLO, and
in \refcite{Moult:2019vou}, where the resummation at subleading power
in a related, simpler context was studied.
In \refcite{Ebert:2018gsn}, it was explicitly shown that the linear NLP corrections
for the inclusive $q_T$ spectrum are absent, i.e.\ $\df\sigma^\one = 0$,
consistent with earlier numerical observations, see e.g.~\refcite{Grazzini:2016ctr}.
On the other hand, in \refcite{Ebert:2019zkb}, it was shown explicitly that linear
corrections do generically arise once fiducial cuts on the final-state leptons
are applied.%
\footnote{In case of isolation cuts, the power corrections can be even further
enhanced~\cite{Ebert:2019zkb}.}

In this work, we consider the generic Drell-Yan process $pp \to VX \to LX$, with
the intermediate vector boson decaying to the ``leptonic'' (color-singlet) final
state $L$. We study the origin and resummation of power corrections that arise
from applying fiducial cuts or performing measurements on $L$, which we will
refer to as \emph{fiducial power corrections}. While our primary application
will be to $Z/\gamma^*\to\ell\ell$ and $W\to\ell\nu$, most of our general
analysis, which is carried out in \sec{theory}, will be for generic $L$. Our
analysis and general results also immediately apply to the simpler case of an
intermediate color-singlet scalar, such as Higgs production, though we will not
consider this case explicitly here.

We encounter two classes of fiducial power corrections in our analysis:
%%%
\begin{enumerate}
\item
\emph{Linear fiducial} power corrections in $q_T/Q$ arise
when azimuthal symmetry is preserved by the leptonic measurement at leading power,
but is broken at $\ord{q_T/Q}$.
For such measurements, the linear fiducial power corrections constitute the
\emph{complete} NLP corrections $\df\sigma^\one$, and can be unambiguously
predicted from factorization, and resummed to the same logarithmic order as the
LP term $\df\sigma^\zero$.

The prototypical example is the $q_T$ spectrum
in the presence of fiducial cuts on $L$, which generically
break azimuthal symmetry and induce linear power corrections.
It also applies to other more complicated $q_T$-like observables,
that resolve the recoil of the leptonic final state and vanish at Born level,
e.g.~the $\phi^*$ observable or the scalar $p_T$-imbalance $p_{T1}^\ell - p_{T2}^\ell$.

\item
\emph{Leptonic fiducial} power corrections in $q_T/\pL$
arise when the leptonic measurement is sensitive to
the edge of Born phase space, with $p_L$ corresponding to the distance to the Born edge.
In the bulk of
the leptonic phase space $\pL \sim Q$, and the discussion
in point 1) applies.
As $p_L$ gets smaller, the leptonic power corrections become
enhanced,
and for $q_T \sim \pL$ they become $\ord{1}$ and must be
retained exactly to all powers to obtain the \emph{actual} LP result.

The prototypical example is the lepton $p_T^\ell$ spectrum close to the Jacobian peak
$p_T^\ell = Q/2$, with $p_L = Q - 2p_T^\ell$.
Close to the Jacobian peak $q_T \sim p_L \ll Q$, fixed-order
predictions are not reliable, which is a well-known effect.
The resummation  at strict LP is also not
sufficient as it neglects the $\ord{1}$ corrections in $q_T/p_L$.
Hence, in this limit the resummation including all leptonic power corrections is required.
\end{enumerate}

The inclusion of the fiducial power corrections in the $q_T$ factorization is
derived in \sec{theory}. As we will see, the fiducial power corrections are a
property of the leptonic decay and are independent of the underlying production
of the decaying vector boson. This allows one to include them in the
factorization theorem by treating the leptonic vector-boson decay exactly in
$q_T$ and consequently makes it possible to resum them at the same level of
precision as the singular cross section $\df\sigma^\zero$. In particular, this
yields a resummation of the NLP terms $\df\sigma^\one$ to N$^3$LL.

Our derivation in \sec{theory} is general and independent of the specific method
to perform the actual resummation, and of whether $q_T$ is treated as a
perturbative scale or not.
It is based on performing a Lorentz decomposition of the hadronic
and leptonic tensors, which encode the production and decay of the intermediate
vector boson.
The basic idea of Lorentz-decomposing the hadronic tensor is
of course not new and has been used before, typically to analyze
the angular dependence for lepton pair production,
see e.g.~\cite{Lam:1978pu, Mirkes:1992hu, Boer:2006eq, Arnold:2008kf, Gauld:2017tww}.
Here, we use it for both hadronic and leptonic tensors to discuss the power
counting at small $q_T$. The tensor decomposition is discussed in
\sec{tensor_decomposition}. It is constructed in a fully
Lorentz-covariant way based on minimal requirements on symmetry and to make the
small-$q_T$ limit maximally transparent, which leads to a direct equivalence with
the Collins-Soper (CS) frame.

Our tensor decomposition holds for any leptonic final state $L$.
In \sec{leptonic_decomposition}, we show that for the specific cases of
$Z/\gamma^*\to\ell\ell$ and $W\to\ell\nu$ it directly maps onto the angular
decomposition of the fully differential cross section in terms of CS angles.
In \sec{L_beyond_2body}, we discuss that Born leptons have a well-defined theoretical interpretation
as a Born projection of the full leptonic final state, and that in this case
an analogous angular decomposition in terms of generalized angular coefficients
also holds for generic $L$, in particular when including QED final-state
radiation. This implies that the use of so-defined Born leptons is
theoretically preferred over other lepton definitions in this context.

Our main power-counting analysis of both linear and leptonic fiducial corrections
and their inclusion in the factorization is given in \sec{factorization_fiducial_power_corrections}.
Some of the more technical details, such as the required power-counting of the
hadronic tensor, are discussed in
\sec{uniqueness_linear_power_corrections} using soft-collinear effective theory
(SCET)~\cite{Bauer:2000ew, Bauer:2000yr, Bauer:2001ct, Bauer:2001yt, Bauer:2002nz},
which provides a systematic expansion of QCD in $q_T/Q\sim\lambda$.
Our analysis does not rely on the precise formalism to factorize $\df\sigma^\zero$,
and thus provides formal justification for existing approaches
in the literature that include the exact lepton kinematics in the factorized
cross section~%
\cite{Balazs:1995nz, Balazs:1997xd, Ellis:1997sc, Catani:2015vma, Camarda:2019zyx,
Scimemi:2019cmh, Bacchetta:2019sam}, as discussed in \sec{literature}.

In \sec{resummation_lp}, we summarize our specific $q_T$ resummation setup,
implemented in the \texttt{C++}\ library \scetlib\ \cite{scetlib},
which we use to obtain numerical results for all factorized cross sections
at fixed order and including resummation up to N$^3$LL.
Some additional details on the numerical inputs and computational setup can
be found in \sec{numerics_setup}.
In \sec{numerics}, we discuss and illustrate in more detail the different sources
of fiducial power corrections and the mechanism for their resummation.
We consider three concrete examples, the $q_T$
spectrum with fiducial cuts (\sec{numerics_qT}),
the $\pTlep$ distribution near the Jacobian peak (\sec{numerics_pTlep}),
and the $\phi^*$ distribution (\sec{numerics_phistar}).
In all cases, we validate numerically that the fiducial power corrections
are indeed captured by the $q_T$ factorization, that their resummation
significantly improves their perturbative stability, and that the size of
remaining fixed-order power corrections is significantly reduced.
In addition, we provide for the first time the resummed $\pTlep$ spectrum
at N$^3$LL$+$NNLO accuracy.

In \sec{qT_subtraction}, we discuss the
immediate implications of our findings for the $q_T$ subtraction method for
fixed-order calculations~\cite{Catani:2007vq},
which is briefly reviewed in \sec{qT_subtraction_review}.
By including the fiducial power corrections predicted from $q_T$ factorization
in the subtractions, their numerical performance improves tremendously.
In the presence of fiducial cuts, it reduces the size of missing power corrections
by an order of magnitude or more.
Moreover, it makes the subtractions applicable also near the edges of Born
phase space, where it otherwise breaks down due to uncontrolled leptonic power
corrections.
In \sec{qT_subtraction_pTlep}, we demonstrate this explicitly for the example
of the $\pTlep$ spectrum in the vicinity of the Jacobian peak $\pTlep \sim Q/2$.
In \sec{qT_subtraction_DY}, we discuss the example of Drell-Yan production with
symmetric lepton cuts, for which large corrections due to a sensitivity to small $q_T$
have been observed before~\cite{Frixione:1997ks, Grazzini:2017mhc}.
In fact, some of our numerical results in \secs{numerics}{data_comparison}
rely on $q_T$ subtractions with fiducial power corrections to obtain stable results.

In \sec{data_comparison} we compare our resummed predictions at N$^3$LL$+$NNLO
for the fiducial $q_T$ and $\phi^*$ distributions in $pp \to Z/\gamma^* \to \ell^+ \ell^-$
with measurements by ATLAS~\cite{Aad:2015auj} and CMS~\cite{Sirunyan:2019bzr}.
We compare the results both with the fiducial power corrections at
fixed order as well as resummed, illustrating the improvement from resumming
the fiducial power corrections and the fact that this significantly reduces the
impact of the remaining fixed-order matching corrections.

To summarize, our general analysis is given in \sec{theory},
with the general setup and definitions in \sec{factorizing_production_and_decay},
a review of the factorization for the inclusive $q_T$ spectrum in \sec{factorization_inclusive},
the hadronic tensor decomposition in \sec{tensor_decomposition},
the leptonic tensor and relation to angular coefficients in \sec{leptonic_decomposition},
the main power-counting analysis in \sec{factorization_fiducial_power_corrections},
some of the more technical details in \sec{uniqueness_linear_power_corrections},
and the relation to other approaches in \sec{literature}.
In \sec{resummation_lp}, we summarize our $q_T$ resummation setup.
In \sec{numerics}, we provide a detailed analysis of fiducial power corrections
for the fiducial $q_T$ spectrum, the $\pTlep$ distribution, and the $\phi^*$ observable.
\Sec{qT_subtraction} discusses the applications to $q_T$ subtractions.
In \sec{data_comparison}, we compare our resummed N$^3$LL$+$NNLO results for
$q_T$ and $\phi^*$ to ATLAS and CMS measurements. We conclude in \sec{conclusions}.
In \app{DY_ingredients}, we collect the hard functions and leptonic tensors
required for Drell-Yan. Some additional data comparisons are provided in
\app{data_comparison}.

%%%%%%%%%%%%%%%%%%%%%%%%%%%%%%%%%%%%%%%%%%%%%%%%%%%%%%%%%%%%%%%%%%%%%%%%%%%%%%%%
\section{Theory}
\label{sec:theory}
%%%%%%%%%%%%%%%%%%%%%%%%%%%%%%%%%%%%%%%%%%%%%%%%%%%%%%%%%%%%%%%%%%%%%%%%%%%%%%%%

%===============================================================================
\subsection{Factorizing production and decay}
\label{sec:factorizing_production_and_decay}
%===============================================================================

We consider the production of a single (generically off-shell)
electroweak vector boson $V$ in unpolarized proton-proton collisions
and its subsequent decay into a set of colorless particles,
which we refer to as the ``leptonic'' final state $L$.
Throughout this paper we will work at leading order in the electroweak interactions.
At this order, the matrix element for this process factorizes as
%%%
\begin{align} \label{eq:factorization_matrix_element}
\cM(pp \to V + X \to L + X)
= \cM^{\mu}_{V\to L} \, \mae{X}{J_{V\mu}}{pp}
\,,\end{align}
%%%
where $\cM_{V\to L}^\mu$ is the amplitude for $V$ to propagate and decay into the
leptonic final state $L$,
$X$ is any additional hadronic radiation in the final state,
and $J_V^\mu$ is the electroweak $q\bar q$ current that couples to $V$,
including electroweak charges and couplings.
The polarizations of the hadronic current and of the propagating vector boson
are encoded in the Lorentz index of $J_V^\mu$ and $\cM_{V\to L}^\mu$.
The currents for $V = \gamma$ and $V = Z$ read
%%%
\begin{align} \label{eq:hadronic_current_Z_gamma}
J^\mu_{\gamma} = |e| \sum_f Q_f \, \bar{q}_f \gamma^\mu q_f
\,, \qquad
J^\mu_{Z} = - |e| \sum_f \bar{q}_f \gamma^\mu \bigl( v_f - a_f \gamma_5 \bigr) q_f
\,,\end{align}
%%%
where the sum runs over quark flavors $f = \{u, d, s, c, b, t\}$,
and $Q_f$ is the electromagnetic charge in units of $\abs{e}$.
The vector and axial couplings of flavor $f$ to the $Z$ boson are
%%%
\begin{align} \label{eq:Z_vector_axial_coupling}
v_f = \frac{T_3^f - 2Q_f \sin^2 \theta_w}{\sin (2\theta_w)}
\,, \qquad
a_f = \frac{T_3^f}{\sin (2\theta_w)}
\,,\end{align}
%%%
where $T_3^{u,d} = \pm 1/2$ is the weak isospin, and $\theta_w$ is the weak mixing angle.
For $V = W^\pm$, the currents are given by
%%%
\begin{align} \label{eq:hadronic_current_W}
J^\mu_{W^+} = -\frac{|e|}{\sqrt{2}\sin \theta_w} \sum_{f,f'} V_{f\!f'} \, \bar{q}_f \gamma^\mu
\frac{1 - \gamma_5}{2} q_{f'}
\,, \qquad
J^\mu_{W^-} = \bigl( J^\mu_{W^+} \bigr)^\dagger
\,,\end{align}
%%%
where the sum runs over $f = \{u, c, t\}$ and $f' = \{d, s, b\}$,
and $V_{f\!f'}$ is the corresponding CKM-matrix element.

The differential cross section for $pp \to VX \to LX$
in the lab frame, which we take to be the hadronic center-of-mass frame,
is given by the square of \eq{factorization_matrix_element},
integrated over phase space, and factorizes as
%%%
\begin{align} \label{eq:factorization_leptonic_hadronic}
\frac{\df\sigma(\Theta)}{\df^4 q \, \df \cO}
&= \frac{1}{2\Ecm^2} \sum_{V,V'} L_{VV'\,\mu\nu}(q, \cO, \Theta) \, W_{VV'}^{\mu\nu}(q, P_a, P_b)
\nn \\
&\equiv \frac{1}{2\Ecm^2} L_{\mu\nu}(q, \cO, \Theta) \, W^{\mu\nu}(q, P_a, P_b)
\,.\end{align}
%%%
Here, $q$ is the total momentum of the leptonic final state $L$ (i.e., the momentum
carried by $V$ or $V'$).
Parametrizing it in terms of the total leptonic invariant mass $Q = \sqrt{q^2}$ and
the rapidity $Y$ and transverse momentum $\qt$ defined in the lab frame and with respect to the
beam axis along the $z$ direction, we have
%%%
\begin{equation} \label{eq:qmu_lab}
q^\mu = (m_T \cosh Y, \qt, m_T \sinh Y)
\,,\quad
m_T = \sqrt{Q^2 + q_T^2}
\,,\qquad
\df^4 q = \frac{1}{2}\df Q^2 \, \df Y \, \df^2 \qt
\,.\end{equation}
%%%
Importantly, in addition to $q$, the cross section depends on a set of differential
observables $\cO$ measured on $L$, as well as a set of fiducial cuts or angular projections
$\Theta$ applied to $L$.
The sum over $V,V'$ in \eq{factorization_leptonic_hadronic}
runs over all intermediate vector bosons that contribute to the observed final state. In particular,
it encodes the interference of $V = \gamma$ with $V' = Z$ for neutral-current Drell-Yan.
In the following, we suppress the dependence on $V,V'$, as in the second line
of \eq{factorization_leptonic_hadronic}, unless it is of some relevance.

The hadronic tensor $W_{VV'}^{\mu\nu}$ describes the QCD dynamics of the proton-proton collision.
It is integrated over any additional hadronic radiation $X$
and independent of the measurement and cuts performed on $L$,
%%%
\begin{align} \label{eq:def_hadronic_tensor}
W_{VV'}^{\mu\nu}(q, P_a, P_b)
= \sum_X \mae{pp}{J_V^{\dagger\mu}}{X} \mae{X}{J_{V'}^\nu}{pp} \, \delta^4(P_a + P_b - q - p_X)
\,,\end{align}
%%%
where the matrix elements of $J_V^\mu$ are implicitly averaged over proton spins.
In addition to $q$, it depends on the incoming proton momenta $P_{a,b}$.
In the lab frame (and neglecting proton masses),
%%%
\begin{equation} \label{eq:Pab_massless}
P_a^\mu = \frac{\Ecm}{2}(1, 0, 0, 1)
\,,\qquad
P_b^\mu = \frac{\Ecm}{2}(1, 0, 0, -1)
\,,\end{equation}
%%%
where $\Ecm^2 \equiv (P_a + P_b)^2$ is the hadronic center-of-mass energy.

The leptonic tensor $L_{VV'}^{\mu\nu}$ describes the propagation and decay of
the intermediate vector boson,
%%%
\begin{align} \label{eq:def_leptonic_tensor}
L_{VV'}^{\mu\nu}(\Phi_L) &= \cM_{V\to L}^{*\mu}(\Phi_L) \cM_{V' \to L}^\nu(\Phi_L)
\,, \nn \\
L_{VV'}^{\mu\nu}(q, \cO, \Theta) &= \int\df\Phi_L(q) \, L_{VV'}^{\mu\nu}(\Phi_L) \,
\delta[\cO - \hat\cO(q,\Phi_L)] \, \hat \Theta(q,\Phi_L)
\,.\end{align}
%%%
In addition to $q$ and the polarization of $V$ encoded in the Lorentz indices,
it depends on the measurement and cuts acting on the leptonic phase space point $\Phi_L$.
The leptonic phase-space measure with total momentum $q$ is defined as
%%%
\begin{align} \label{eq:phiL}
\df\Phi_L(q) &= \biggl[\prod_{i \in L} \frac{\df^4 p_i}{(2\pi)^3}\, \theta(p_i^0)\delta(p_i^2 - m_i^2) \biggr]
\, (2\pi)^4 \delta^4\Bigl(q - \sum_{i \in L} p_i\Bigr)
\,.\end{align}
%%%

It is straightforward to extend this setup to the collision of generic hadrons $h_{a}(P_a)$ and $h_b(P_b)$ with nonzero, possibly distinct masses $m_a$ and $m_b$.
This is relevant for treating proton or ion mass corrections in $pp\to XL$, $pA \to XL$, or $AA'\to XL$, where $A$ and $A'$ are ions with these atomic numbers.
In this case, the differential cross section in the lab frame becomes
%%%
\begin{align} \label{eq:factorization_leptonic_hadronic_massive_hadrons}
\frac{\df\sigma(\Theta)}{\df^4 q \, \df \cO}
&= \frac{1}{2E_a \, 2E_b (v_a + v_b)} L_{\mu\nu}(q, \cO, \Theta) \, W^{\mu\nu}(q, P_a, P_b)
\,,\end{align}
%%%
where the incoming hadron momenta in the lab frame are given by
%%%
\begin{equation} \label{eq:Pab_general}
P_a^\mu = E_a(1, 0, 0, v_a)
\,,
\qquad
P_b^\mu = E_b(1, 0, 0, -v_b)
\,,\qquad
v_{a,b} = \frac{1}{E_{a,b}} \sqrt{E_{a,b}^2 - m_{a,b}^2}
\,,\end{equation}
%%%
i.e., $E_{a,b}$ and $v_{a,b}$ are the lab-frame beam energies and
beam velocities. Here, we also allow $E_a v_a \neq E_b v_b$, so the lab frame
does not necessarily have to coincide with the hadronic center-of-mass frame.
The leptonic tensor in \eq{factorization_leptonic_hadronic_massive_hadrons} is
unchanged and given by \eq{def_leptonic_tensor}.
The hadronic tensor is given by replacing the proton states $\lvert pp\rangle$ by
hadron states $\lvert h_a h_b\rangle$ in \eq{def_hadronic_tensor}.
For notational simplicity in the following we will always write the flux factor
as $1/(2\Ecm^2)$ with the obvious replacement as in \eq{factorization_leptonic_hadronic_massive_hadrons} for the massive case understood.

Lorentz invariance dictates that any Lorentz-scalar functions of $q$, $P_a$,
$P_b$ can only depend on Lorentz invariants formed out of $q$, $P_a$, $P_b$.
There are six independent invariants, three of which contain nontrivial
kinematic information, which we choose as (recall $m_T = \sqrt{Q^2 + q_T^2}$)
%%%
\begin{alignat}{9} \label{eq:kin_invariants}
q^2 &= Q^2
\,,\nn\\
s_{aq} &\equiv 2 q \cdot P_a &&= 2 E_a\, m_T\, (\cosh Y - v_a \sinh Y)
&&\mtozero \Ecm\, m_T\, e^{-Y}
\,,\nn\\
s_{bq} &\equiv 2 q \cdot P_b &&= 2 E_b\, m_T\, (\cosh Y + v_b \sinh Y)
&&\mtozero \Ecm\, m_T\, e^{+Y}
\,.\end{alignat}
%%%
In the second step we plugged in \eqs{qmu_lab}{Pab_general} to write $s_{aq}$
and $s_{bq}$ in terms of lab-frame quantities, and in the last step we took the
limit of massless, center-of-mass collisions, which corresponds to taking
$v_{a,b}\to 1$ and $E_{a,b}\to \Ecm/2$. It is clear that these are in one-to-one
correspondence to the three kinematic variables $Q$, $Y$, and $q_T^2 =
\qt^{\,2}$. The three remaining invariants only encode the beam parameters,
%%%
\begin{equation} \label{eq:trivial_invariants}
P_{a,b}^2 = m_{a,b}^2 \mtozero 0
\,,\qquad
s_{ab} \equiv 2 P_a \cdot P_b = 2 E_a E_b (1 + v_a v_b) \mtozero \Ecm^2
\,.\end{equation}
%%%
The conservation of the vector current in QCD, $\partial_\mu J^\mu_{\gamma} = 0$,
implies
%%%
\begin{equation}
q_\mu W_{\gamma\gamma}^{\mu \nu} = q_\nu W_{\gamma\gamma}^{\mu \nu} = 0
\,.\end{equation}
%%%
The same relation for $V = Z$ does not automatically follow from gauge
invariance, because the axial-vector current is not conserved in QCD due to
finite quark masses and because of the Adler-Bell-Jackiw axial
anomaly~\cite{Adler:1969gk, Bell:1969ts, Adler:1969er}. In the unbroken
electroweak theory, the axial anomaly cancels in all gauge currents thanks to
the anomaly cancellation in the SM. Since the anomaly coefficient is mass
independent, it also does not contribute to the divergence of $J_Z^\mu$ after
electroweak symmetry breaking, namely it still cancels between up-type and
down-type quarks due to their opposite $T_3^{u,d} = \pm 1/2$. However, the nonzero quark
masses now explicitly break the axial-vector current conservation. Therefore, we
have the non-conservation relation%
\footnote{As discussed, this relation is \emph{not} anomalous. It also holds after
suitable renormalization that preserves the
non-renormalization of the axial anomaly~\cite{Adler:1969er}, see e.g.\
\refscite{Larin:1993tq, Bernreuther:2005rw} for a detailed discussion.}
%%%
\begin{equation}
-\img\partial_\mu J_Z^\mu = \abs{e} \sum_f a_f\, 2m_f\, \bar q_f \gamma_5 q_f
\,.\end{equation}
%%%
In practice, neglecting all but the top-quark masses, we thus have the chiral Ward
identity
%%%
\begin{equation} \label{eq:JZ_noncons}
q_\mu \mae{X}{J_Z^\mu}{pp} = \abs{e}a_t\, 2m_t\, \mae{X}{\bar t \gamma_5 t}{pp}
\,.\end{equation}
%%%
At the partonic level, the leading contribution to this relation (without explicit
top quarks in the final state) is the gluon-fusion top-quark triangle diagram.
To isolate these non-conserved contributions, we write the hadronic matrix element as
%%%
\begin{align} \label{eq:four_vectors_duh}
\Bigl[\mae{X}{J_V^\mu}{pp} - \frac{q^\mu q_\nu }{q^2} \mae{X}{J_V^\nu}{pp} \Bigr]_{\rm cons}
  + \frac{q^\mu q_\nu }{q^2} \mae{X}{J_V^\nu}{pp}
\,,\end{align}
%%%
where the first term is ``conserved'' by construction, i.e., it vanishes when
contracted with $q_\mu$, while the second term $\sim q^\mu$ contains the
non-conserved pieces in \eq{JZ_noncons}. Similarly, we can write the hadronic
tensor as
%%%
\begin{align} \label{eq:qmuWmunu_is_zero}
W^{\mu\nu} &= W^{\mu\nu}_{\rm cons} + (\text{terms} \propto q^\mu \text{ or } q^\nu)
\,, \nn \\
q_\mu W^{\mu\nu}_{\rm cons} &= q_\nu W^{\mu\nu}_{\rm cons} = 0
\,,\end{align}
%%%
where the conserved part $W_{\rm cons}^{\mu\nu}$ arises from squaring the
conserved parts of the currents.

In practice, the non-conserved pieces rarely matter for various reasons: First,
for a real, on-shell massive vector boson with physical polarization
$\varepsilon$, they vanish due to $q\cdot \varepsilon = 0$. As a result, for an
off-shell vector boson near the resonance, they are suppressed by $1 -
q^2/m_V^2$. This is easy to see in unitary gauge, where all Goldstone bosons
have been eaten up and the vector-boson propagator is proportional to $g^{\mu\nu} - q^\mu
q^\nu/m_V^2$. (In 't Hooft-Feynman gauge, the second term is generated by the exchange of
Goldstone bosons.) Second, we can repeat the analogous discussion on the
leptonic decay side, and split the leptonic tensor into conserved parts, $q_\mu
L^{\mu\nu}_{\rm cons} = q_\nu L^{\mu\nu}_{\rm cons} = 0$, and non-conserved
parts. The non-conserved parts of $W^{\mu\nu}$ are $\propto q^{\mu,\nu}$, and
thus they only survive when contracted with the non-conserved parts of the
leptonic tensor. However, considering leptonic decays (i.e., with the
intermediate vector boson coupling to a leptonic current) the non-conserved
leptonic parts are proportional to the lepton masses and can thus be neglected.%
\footnote{%
A notable exception is associated Higgs production, which has a
$gg\to Z^\ast \to ZH$ contribution.
As a consequence of Yang's theorem, the $ggZ$ vertex vanishes if all three
bosons are real and on shell. Therefore, for real, on-shell gluons, the effective
$gg\to Z$ contribution via a top-quark triangle is purely $\propto q^\mu$ and thus
the $gg\to Z^\ast \to ZH$ process proceeds entirely via the non-conserved parts in \eq{JZ_noncons}.
Starting at $\ord{\as^2}$, one or both gluons are off shell, and the $ggZ$
vertex also contributes to the conserved parts, and therefore also to the
Drell-Yan process $Z\to\ell\ell$~\cite{Dicus:1985wx, Hamberg:1990np}.}
Therefore, for simplicity, we will ignore the non-conserved contributions for the
most part, though we emphasize that they do not pose any additional conceptual
problems and could be straightforwardly included in our analysis.

%===============================================================================
\subsection{Factorization for the inclusive \texorpdfstring{$q_T$}{qT} spectrum}
\label{sec:factorization_inclusive}
%===============================================================================

If the measurement on $L$ is inclusive, i.e., if we integrate over $\cO$ and set
$\hat{\Theta}(q, \Phi_L) = 1$ in the leptonic tensor in \eq{def_leptonic_tensor},
it reduces to
%%%
\begin{align} \label{eq:leptonic_tensor_inclusive}
L^{\mu\nu}(q) = \int\df\Phi_L(q) \, L^{\mu\nu}(\Phi_L)
= \Bigl( \frac{q^\mu q^\nu}{q^2} - g^{\mu\nu} \Bigr) L(q^2)
\,.\end{align}
%%%
In the second equality we used leptonic current conservation and
the fact that after the integration $L^{\mu\nu}(q)$ can only depend on $q$.
Explicit expressions for the scalar coefficients $L(q^2)$ in case of Drell-Yan
are given in \app{leptonic_tensors}.
Inserting \eq{leptonic_tensor_inclusive} into \eq{factorization_leptonic_hadronic} yields
%%%
\begin{align} \label{eq:inclusive_xsec}
\frac{\df \sigma}{\df^4 q}
= \frac{1}{2\Ecm^2}\, L(q^2) \, W_\mathrm{incl}(q^2, s_{aq}, s_{bq})
\,,\end{align}
%%%
where all QCD dynamics are encoded in the Lorentz-scalar inclusive hadronic
structure function
%%%
\begin{align} \label{eq:Wincl_def}
W_\mathrm{incl}(q^2, s_{aq}, s_{bq}) \equiv \Bigl( \frac{q^\mu q^\nu}{q^2} - g^{\mu\nu} \Bigr) W_{\mu\nu}(q, P_a, P_b)
\,.\end{align}
%%%
By Lorentz invariance, $W_\mathrm{incl}$ can only depend on the three kinematic
invariants $q^2$, $s_{aq}$, $s_{bq}$, which are in one-to-one correspondence to
the three kinematic variables $Q$, $Y$, $q_T^2 = \qt^{\,2}$, see
\eq{kin_invariants}. (In addition, $W_\mathrm{incl}$ depends on the beam
invariants in \eq{trivial_invariants}, which we keep implicit.) In particular,
since $L(q^2)$ only depends on $q^2$, the entire dependence on $Y$ and $q_T^2$
in \eq{inclusive_xsec} is carried by $W_\mathrm{incl}$, and there is no
dependence on the direction of $\qt$. Hence, $W_\incl$ encodes the
\emph{inclusive} (without fiducial cuts) $q_T$ distribution for fixed $Q$, $Y$.

We are interested in the region of small transverse momentum $q_T \ll Q$.
In this limit, $W_\mathrm{incl}$ satisfies the factorization theorem
\cite{Collins:1981uk, Collins:1981va, Collins:1984kg, Collins:1350496, Becher:2010tm, GarciaEchevarria:2011rb, Chiu:2012ir, Li:2016axz}
%%%
\begin{align} \label{eq:tmd_factorization_incl}
W_\mathrm{incl}
&= \sum_{a,b} H_{ab}(Q^2, \mu) \, [B_a B_b S](Q^2, x_a, x_b, \qt, \mu) \,
\Bigl[ 1 + \ORd{\frac{q_T^2}{Q^2}, \frac{\lqcd^2}{Q^2}} \Bigr]
\,.\end{align}
As indicated, $W_{\rm incl}$ receives power corrections in $(q_T/Q)^2$ and $(\lqcd/Q)^2$,
but remains valid in the nonperturbative regime $q_T \sim \lqcd$.

In \eq{tmd_factorization_incl}, $H_{ab}(Q^2,\mu)$ denotes the hard function,
which encodes the production of the vector boson in the underlying
hard interaction $ab \to V$. The \MSbar scheme result for $H_{ab}$ can
be obtained either by matching QCD onto SCET or as the IR-finite part of the
corresponding form factor using dimensional regularization to regulate IR divergences.
Explicit expressions for different vector bosons are given in \app{hard_functions}.
In practice, the leptonic prefactor $L(q^2)$ is often included in the hard function in the inclusive case.

The second factor in \eq{tmd_factorization_incl} encodes physics at the low scale $\mu \sim q_T$,
and can be written in several equivalent forms,
%%%
\begin{subequations} \label{eq:tmd_factorization}
\begin{align} \label{eq:tmd_factorization_bbs}
& [B_a B_b S](Q^2, x_a, x_b, \qt, \mu)
\nn\\
&\qquad
\equiv \int \! \df^2 \vec{k}_a \, \df^2 \vec{k}_b \, \df^2 \vec{k}_s \,
\delta^2(\qt - \vec{k}_a - \vec{k}_b - \vec{k}_s)
\nn\\ & \qquad\qquad \times
   B_a(x_a, \vec{k}_a, \mu, \nu/\w_a) \, B_b(x_b, \vec{k}_b, \mu, \nu/\w_b)
   S(\vec{k}_s, \mu, \nu)
\\ \label{eq:tmd_factorization_bbs_FT}
&\qquad
\equiv \int\!\!\frac{\df^2\bt}{(2\pi)^2} \, e^{\img \bt \cdot \qt} \,
   \tB_a(x_a, b_T, \mu, \nu/\w_a) \, \tB_b(x_b, b_T, \mu, \nu/\w_b)
   \tS(b_T, \mu, \nu)
\\ \label{eq:tmd_factorization_ff}
&\qquad
\equiv \int\!\!\frac{\df^2\bt}{(2\pi)^2} \, e^{\img \bt \cdot \qt} \,
   \tf_a(x_a, b_T, \mu, \zeta_a) \, \tf_b(x_b, b_T, \mu, \zeta_b)
\,.\end{align}
\end{subequations}
%%%
In \eq{tmd_factorization_bbs}, the beam functions $B_i(x, \vec{k}_T, \mu, \nu/\w)$ describe
the extraction of an unpolarized parton $i$ with longitudinal momentum fraction $x$
and transverse momentum $\vec{k}_T$ from an unpolarized proton, the soft function
$S(\vec{k}_T, \mu, \nu)$ encodes soft radiation with total transverse momentum $\vec{k}_T$,
and $\delta^2(\qt-\cdots)$ encodes momentum conservation in the transverse plane.
\Eq{tmd_factorization_bbs_FT} shows the equivalent result in Fourier space,
where $\tB_i$ and $\tS$ are the Fourier conjugates of $B_i$ and $S$.
Equivalently, one can write this as shown in \eq{tmd_factorization_ff},
where the transverse-momentum dependent beam and soft functions have been combined
into transverse-momentum dependent PDFs (TMDPDFs)
%%%
\begin{align} \label{eq:TMDPDF}
\tf_i(x, b_T, \mu, \zeta) = \tB_i(x,b_T,\mu,\nu/\sqrt{\zeta}) \sqrt{\tS(b_T,\mu,\nu)}
\,.\end{align}
%%%

A key feature of both transverse-momentum dependent beam functions and TMDPDFs
is their explicit dependence on the energy of the colliding parton, encoded
either in its lightcone component $\w$ or in the Collins-Soper scale $\zeta$,
where
%%%
\begin{align} \label{eq:def_x_i_w_i}
x_{a,b} = \frac{\w_{a,b}}{P_{a,b}^{\mp}}
\,, \qquad
\w_{a,b} = Q
\,, \qquad
\zeta_{a,b} \propto \w_{a,b}^2
\,, \qquad
(\w_a \w_b)^2 = \zeta_a \zeta_b = Q^4
\,,\end{align}
%%%
where the lightcone components of the hadron momenta are given by
%%%
\begin{alignat}{9} \label{eq:hadron_lc_momenta_massive_hadrons}
P_a^- &= E_a (1 + v_a) e^{-Y} &&\mtozero \Ecm e^{-Y}
\,, \qquad &
P_a^+ &= \frac{m_a^2}{P_a^-} &&\mtozero 0
\,, \qquad
P_{a\perp}^\mu = 0
\,, \nn \\
P_b^+ &= E_b (1 + v_b) e^{+Y} &&\mtozero \Ecm e^{+Y}
\,, \qquad &
P_b^- &= \frac{m_b^2}{P_b^+} &&\mtozero 0
\,, \qquad
P_{b\perp}^\mu = 0
\,,\end{alignat}
%%%
and we also indicated the massless, center-of-mass limit.
Accounting for the mass dependence of $P_a^-$ and $P_b^+$ implicit in the velocities $v_{a,b} \leq 1$
captures kinematic hadron-mass corrections to the factorization theorem
in \eq{tmd_factorization_incl}.
The factors of $e^{\pm Y}$ in $P_{a,b}^\pm$ come from our lightcone
conventions, see \eq{def_lc_coordinates}, which imply that in the lab frame
$p^\pm = e^{\pm Y}(p^0 \mp p^z)$.

The $\nu/\omega$ dependence of the beam functions or the $\zeta$ dependence of the TMDPDFs
is a remnant of so-called rapidity divergences~\cite{Collins:1981uk,Collins:1992tv,Collins:2008ht,Becher:2010tm,GarciaEchevarria:2011rb,Chiu:2011qc,Chiu:2012ir}.
Their regularization and renormalization induces an additional scale
in the individual beam and soft functions in \eq{tmd_factorization_bbs}, here denoted as $\nu$,
analogously to the appearance of the \MSbar scale $\mu$ from renormalizing UV divergences.
Importantly, the $\nu$ dependence cancels between the beam and soft functions,
such that \eq{tmd_factorization} is independent of $\nu$. This fact allows one
to combine beam and soft functions into $\nu$-independent TMDPDFs as shown in
\eq{TMDPDF}, where the Collins-Soper scale $\zeta$ is the remnant of the rapidity divergences.

In principle, the beam and soft functions (or TMDPDFs) are nonperturbative objects,
and thus allow for a rigorous field-theoretic treatment of the $\qt$ spectrum
in the nonperturbative regime $q_T \sim \lqcd$.
For perturbative $|\vec k_T| \gg \lqcd$,
the beam functions (or TMDPDFs) can be matched perturbatively onto collinear PDFs~\cite{Collins:1981uw,Collins:1984kg},
while the soft function is perturbatively calculable.
The required perturbative results are known at N$^3$LO~\cite{Catani:2011kr,Catani:2012qa,Gehrmann:2014yya,Luebbert:2016itl,Echevarria:2015byo,Echevarria:2016scs,Li:2016ctv,Luo:2019hmp,Luo:2019szz, Ebert:2020yqt}.
In the perturbative regime, \eqs{tmd_factorization_incl}{tmd_factorization}
allow one to resum large logarithms $\ln(q_T/Q)$ arising to all orders in $\as$.
In \sec{resummation_lp} we review this procedure and describe the specific
resummation setup used for the numerical results in this paper.

We note that there are various approaches in the literature on how to perform
this resummation. While they all aim to describe the same inclusive hadronic
tensor $W_{\rm incl}$ and must ultimately all be based on the factorization
theorem in \eq{tmd_factorization_incl}, they can differ in practice, e.g.,
due to differences in the rapidity regularization scheme, the different
equivalent forms of \eq{tmd_factorization}, different mathematical methods of
performing the actual resummation, and different choices for the precise form of
the logarithms that are being resummed.
Crucially, all our results in this section 2 are general and
hold \emph{independently} of how precisely the resummation is carried out,
and thus immediately apply to all formulations in the literature.%
\footnote{%
This of course only holds to the extent that an approach itself does not induce new
power corrections.}
This is because they only rely on general arguments, such as Lorentz invariance
and power counting, and the general structure of the hadronic and leptonic
tensors.

The factorization theorem for the inclusive $q_T$ spectrum in \eq{tmd_factorization_incl}
receives corrections that are suppressed by powers of $q_T/Q$ relative to the leading term.
As indicated in \eq{tmd_factorization_incl}, the leading corrections scale as $(q_T/Q)^2$,
while linear power corrections are absent.
This can be understood intuitively from the azimuthal symmetry of $W_\incl$,
i.e., the fact that it only depends on the
Lorentz invariants in \eq{kin_invariants}, which in turn only depend on $q_T^2$.
The absence of linear power corrections in $W_\incl$ has
been verified explicitly by analytic $\ord{\as}$ calculations at
next-to-leading power~\cite{Ebert:2018gsn}.
More formally, an argument for their absence to all orders in the inclusive
case is presented in \sec{uniqueness_linear_power_corrections}.
In the remainder of this section, we discuss how \eq{tmd_factorization_incl} is extended
to the case where the decay products are resolved
and, notably, linear power corrections arise.

%===============================================================================
\subsection{Hadronic tensor decomposition}
\label{sec:tensor_decomposition}
%===============================================================================

We now return to the generic, fiducial cross section in
\eq{factorization_leptonic_hadronic_massive_hadrons},
and bring it into a form suitable for factorization at small $q_T$.
The manipulations of this section are exact in $q_T$,
i.e., we do not yet expand in $q_T \ll Q$.
The key idea is to decompose the hadronic tensor $W^{\mu\nu}(q, P_a, P_b)$
into Lorentz-scalar projections with respect to four orthogonal unit four-vectors
that are constructed from the four-vectors
$P_{a,b}^\mu$ and $q^\mu$ and their invariants, and by imposing reasonable
symmetry constraints.

For the decomposition to be complete, we should pick one timelike vector $t^\mu$ and three spacelike vectors $x^\mu, y^\mu, z^\mu$,
%%%
\begin{align}
t^2 &= 1
\,, \qquad
x^2 = y^2 = z^2 = -1
\,.\end{align}
%%%
Motivated by \eq{qmuWmunu_is_zero}, we take the timelike vector to be
%%%
\begin{align}
t^\mu = \frac{q^\mu}{\sqrt{q^2}}
\,,\end{align}
%%%
such that the conserved and non-conserved parts of $W^{\mu\nu}$ will get
projected onto orthogonal components.
The spacelike vectors must be given by linear combinations of $P_{a,b}^\mu$ and $q^\mu$.
It will prove convenient to take $z^\mu$ to lie in the plane spanned by $P_a^\mu$ and $P_b^\mu$,
%%%
\begin{align} \label{eq:z_basic_assumption}
z^\mu = \lambda_a P_a^\mu + \lambda_b P_b^\mu
\,,\end{align}
%%%
where $\lambda_a$ and $\lambda_b$ are scalar functions of the kinematic invariants.
Imposing $t \cdot z = 0$ and $z^2 = -1$ then uniquely fixes $z^\mu$ to
%%%
\begin{align} \label{eq:z_CS_frame_independent_form}
z^\mu
= \frac{s_{bq}\, P_a^\mu - s_{aq}\, P_b^\mu}
 {(s_{ab} s_{aq} s_{bq} - m_b^2 s_{aq}^2 - m_a^2 s_{bq}^2)^{1/2}}
\,,\end{align}
%%%
up to a conventional overall sign. 
The $s_{ij}$ are all positive definite, as can be seen from their explicit
expressions in \eqs{kin_invariants}{trivial_invariants},
and
$s_{ab} s_{aq} s_{bq} - m_b^2 s_{aq}^2 - m_a^2  s_{bq}^2 = [2 E_a E_b (v_a + v_b)\, m_T]^2 > 0$,
so $z^\mu$ is real.
Interchanging $P_a \leftrightarrow P_b$, \eq{z_CS_frame_independent_form} satisfies $z^\mu \mapsto - z^\mu$.
The choice for the remaining $x^\mu$ and $y^\mu$ is degenerate in principle.
To reflect the fact that interchanging the initial-state hadrons
is equivalent to a $180^\circ$ rotation about an axis in the transverse plane,
we require $x^\mu$ to be invariant under $P_a \leftrightarrow P_b$
and $y^\mu$ to only change sign. All together we then have
%%%
\begin{align} \label{eq:xy_symmetry_constraints}
P_a \leftrightarrow P_b
\,: \qquad
x^\mu \mapsto +x^\mu
\,, \qquad
y^\mu \mapsto -y^\mu
 \,, \qquad
 z^\mu \mapsto - z^\mu
\,.\end{align}
%%%

We can write $x^\mu$ as a linear combination of $q^\mu$ and $P_{a,b}^\mu$,
%%%
\begin{align} \label{eq:xy_CS_frame_independent_form}
x^\mu
= \frac{c_x}{\sqrt{q^2}} \bigl( q^\mu - \kappa_a P_a^\mu - \kappa_b P_b^\mu \bigr)
\,,\end{align}
%%%
where we chose the $q^\mu$ coefficient to be positive to fix the overall sign of $x^\mu$.
Imposing $t\cdot x = z\cdot x = 0$ and $x^2 = -1$, we find
for the scalar coefficients and normalization factor
%%%
\begin{align}  \label{eq:RaRbcx}
\kappa_a &= \frac{q^2 (s_{ab}  s_{bq} - 2 m_b^2 s_{aq})}{s_{ab} s_{aq} s_{bq} - m_a^2 s_{bq}^2 - m_b^2 s_{aq}^2}
= \frac{Q^2}{m_T E_a}\,\frac{v_b \cosh Y + \sinh Y}{v_a + v_b}
\,, \nn \\
\kappa_b &= \frac{q^2 (s_{ab} s_{aq} - 2 m_a^2 s_{bq})}{s_{ab} s_{aq} s_{bq} - m_a^2 s_{bq}^2 - m_b^2 s_{aq}^2}
= \frac{Q^2}{m_T E_b}\,\frac{v_a \cosh Y - \sinh Y}{v_a + v_b}
\,,  \nn\\
c_x^2 &= \frac{s_{ab} s_{aq} s_{bq} - m_b^2 s_{aq}^2 -m_a^2 s_{bq}^2}
{s_{ab} s_{aq} s_{bq} - q^2 s_{ab}^2 - m_b^2 s_{aq}^2 -m_a^2 s_{bq}^2+4 m_a^2 m_b^2 q^2}
  = 1 + \frac{Q^2}{q_T^2} = \frac{m_T^2}{q_T^2}
\,. \end{align}
%%%
Finally, $y^\mu$ is chosen to complete a righthanded coordinate system
%%%
\begin{equation}
y^\mu = \eps^{\mu\nu\rho\sigma} t_\nu x_\rho z_\sigma
\,,\end{equation}
%%%
where we use the convention $\eps^{0123} = +1$.
For completeness, the results for the unit vectors in the massless limit are
%%%
\begin{align}
t^\mu &= \frac{q^\mu}{\sqrt{q^2}}
\,,\qquad
x^\mu
= \frac{s_{aq}s_{bq} \, q^\mu - s_{bq} q^2 \, P_a^\mu - s_{aq} q^2 \, P_b^\mu}{[ s_{aq} s_{bq} q^2 (s_{aq}s_{bq} - s_{ab} q^2)]^{1/2}}
\nn \\
z^\mu
&= \frac{s_{bq}\, P_a^\mu - s_{aq}\, P_b^\mu}
 {(s_{ab} s_{aq} s_{bq})^{1/2}}
\,,\qquad
y^\mu = \eps^{\mu\nu\rho\sigma} t_\nu x_\rho z_\sigma
\,.\end{align}
%%%

%~~~~~~~~~~~~~~~~~~~~~~~~~~~~~~~~~~~~~~~~~~~~~~~~~~~~~~~~~~~~~~~~~~~~~~~~~~~~~~~
\subsubsection{Reference frame interpretation}
\label{sec:reference_frame}
%~~~~~~~~~~~~~~~~~~~~~~~~~~~~~~~~~~~~~~~~~~~~~~~~~~~~~~~~~~~~~~~~~~~~~~~~~~~~~~~

The four-vectors $t^\mu,x^\mu,y^\mu,z^\mu$ are orthogonal and normalized, and thus
uniquely define a reference frame, namely the frame in which they have components
$t^\mu = (1, 0, 0, 0)$,
$x^\mu = (0, 1, 0, 0)$, $y^\mu = (0, 0, 1, 0)$, and $z^\mu = (0, 0, 0, 1)$.
Since $t^\mu = q^\mu/\sqrt{q^2}$, this frame is automatically a frame where the
vector boson is at rest, i.e., where $q^\mu = (\sqrt{q^2}, 0, 0, 0)$.
A goal of this section is to show that this frame turns out to be
the well-known Collins-Soper (CS) frame~\cite{Collins:1977iv}.
We will also find and discuss some subtleties in the massive case due to the fact that different CS-frame definitions that are equivalent in the massless case are no
longer equivalent in the massive case.

Let us first remind the reader that the vector-boson rest frame is not unique in itself
because different rest frames can still differ by spatial rotations, i.e., by their
orientation of the $x,y,z$-axes. There are many ways to perform a sequence of
pure boosts to go from a given frame, say the lab frame, to the rest frame, and
the difference between them precisely corresponds to an overall spatial rotation
in the rest frame. Hence, a unique way to define a specific vector-boson rest
frame is to specify the precise boost sequence to go from the lab frame to the rest frame.
We will discuss how to rotate between different rest frames in \sec{frame_choice} below.

Intuitively, the CS frame is defined such that its $z$-axis points into the
same direction as in the lab frame and its $x$-axis points in the direction of $\qt$ in the lab
frame. In terms of boosts from the lab frame, the CS frame is defined by performing
two boosts (see \fig{boosts}):
%%%
\begin{enumerate}
\item
A longitudinal boost by $Y$ in the beam direction (taken to be
the $z$-axis) that takes us to the \emph{leptonic frame}
in which $\tilde Y = 0$ and $\tilde q^z = 0$. Here and in the following, the
tilde denotes the same physical quantity but evaluated in the leptonic frame.
\item
A transverse boost in the direction of $\qt$ (taken to be the $x$-axis) with
boost parameters
%%%
\begin{align} \label{eq:def_eps_gamma}
\beta\gamma = \eps = \frac{q_T}{Q}
\,, \qquad
\gamma = \sqrt{1 + \eps^2} = \frac{m_T}{Q}
\,,\end{align}
%%%
which takes us from the leptonic frame to the rest frame.
\end{enumerate}
%%%
Under these boosts a generic four-vector $p^\mu$ transforms as
%%%
\begin{align} \label{eq:p}
p^\mu
&= (p^0, p^x, p^y, p^z)_\lab
\nn \\
&= (p^0 \cosh Y - p^z \sinh Y, p^x, p^y, p^z \cosh Y - p^0 \sinh Y)_\lep
\equiv (\tilde p^0, \tilde p^x, \tilde p^y, \tilde p^z)_\lep
\nn \\
&= (\gamma \tilde p^0 - \eps \tilde p^x, \gamma \tilde p^x - \eps \tilde p^0, \tilde p^y , \tilde p^z)_\CS
\,,\end{align}
%%%
where we explicitly indicated by a subscript in which frame the component-form
is given, with $p^{0,x,y,z}$ always denoting the lab-frame components and
$\tilde p^{0,x,y,z}$ always denoting the leptonic-frame components.
To illustrate the boosts, applying them to $q^\mu$ itself, we obtain
%%%
\begin{align} \label{eq:qmu_boost_to_CS}
q^\mu
&= Q(\gamma \cosh Y, \eps, 0, \gamma \sinh Y)_\lab
= Q(\gamma, \eps, 0, 0)_\lep
= Q(1, 0, 0, 0)_\CS
\,.\end{align}
%%%
Hence, we indeed arrive in the vector-boson rest frame, which is of course how
\eq{def_eps_gamma} was chosen in the first place.

We can now use this definition of the CS frame to make contact with our unit
vectors $t^\mu, x^\mu, y^\mu, z^\mu$. To do so, we perform the same exercise
for them, i.e., evaluate them in the lab frame and then boost them to the CS
frame. For $t^\mu = q^\mu/Q$, this would just repeat \eq{qmu_boost_to_CS}.
For $z^\mu$, evaluating its general covariant expression in
\eq{z_CS_frame_independent_form} in the lab frame and applying the two boosts
to the CS frame, we obtain
%%%
\begin{align} \label{eq:z}
z^\mu
&= \lambda_a P_a^\mu + \lambda_b P_b^\mu
\nn \\
&= \frac{1}{v_a + v_b}
\biggl(\frac{\cosh Y + v_b \sinh Y}{E_a}\, P_a^\mu - \frac{\cosh Y - v_a \sinh Y}{E_b}\, P_b^\mu \biggr)
\nn \\
&= (\sinh Y, 0, 0, \cosh Y)_\lab
= (0, 0, 0, 1)_\lep
= (0, 0, 0, 1)_\CS
\,.\end{align}
%%%
Similarly, starting from the expression for $x^\mu$ in \eq{xy_CS_frame_independent_form},
we obtain
%%%
\begin{align} \label{eq:x}
x^\mu
&= \frac{c_x}{\sqrt{q^2}} \bigl(q^\mu - \kappa_a P_a^\mu - \kappa_b P_b^\mu \bigr)
\nn \\
&= \frac{\gamma}{\eps} \frac{q^\mu}{Q} - \frac{1}{\eps(v_a + v_b)}\Bigl(\frac{v_b \cosh Y + \sinh Y}{E_a}\, P_a^\mu + \frac{v_a \cosh Y - \sinh Y}{E_b}\, P_b^\mu \Bigr)
\nn \\
&= (\eps \cosh Y, \gamma, 0, \eps \sinh Y)_\lab
= (\eps, \gamma, 0, 0)_\lep
= (0, 1, 0, 0)_\CS
\,.\end{align}
%%%
This shows explicitly that the frame defined by $t^\mu, x^\mu, y^\mu, z^\mu$ is
equivalent to the CS frame (in its boost definition), and that this equivalence also
holds in the general massive case.
It is quite pleasing to see that the CS frame naturally appears in a
covariant way only by imposing \eq{z_basic_assumption} and the symmetry constraints in \eq{xy_symmetry_constraints}.

Another definition of the CS frame~\cite{Collins:1977iv},
which is also often used in practice,
is to consider $\vec P_a$ and $\vec P_b$ in the vector-boson rest frame, and to
define the $z$-axis to bisect the angle between $\vec P_a$ and $-\vec P_b$,
while the $x$-axis is chosen to lie in the plane defined by $\vec P_a$ and $\vec P_b$.
Denoting individual components in the CS frame (defined via the above boosts)
by hats, we have
%%%
\begin{align} \label{eq:Pab_CS}
P_a^\mu
&= E_a (1, 0, 0, +v_a)_\lab
 = (\hat P_a^0, \hat P_a^x, 0, \hat P_a^z)_\CS
\,,\nn\\
P_b^\mu
&= E_b (1, 0, 0, -v_b)_\lab
 = (\hat P_b^0, \hat P_b^x, 0, \hat P_b^z)_\CS
\,,\end{align}
%%%
where explicit expressions for the components can be straightforwardly obtained
from \eq{p}. The angles $\gamma_{a,b}$ between $\vec P_{a,b}$ and the $z$-axis
(see \fig{frames} right) are given by
%%%
\begin{align} \label{eq:bisector_2}
\tan\gamma_a &= +\frac{\hat P_a^x}{\hat P_a^z}
= \eps \frac{v_a \sinh Y - \cosh Y}{v_a \cosh Y - \sinh Y}
\,, \nn \\
\tan\gamma_b &= -\frac{\hat P_b^x}{\hat P_b^z}
= -\eps \frac{v_b \sinh Y + \cosh Y}{v_b \cosh Y + \sinh Y}
\,.\end{align}
%%%
The bisector criterion amounts to requiring these two angles to be equal, i.e.,
%%%
\begin{align} \label{eq:bisector_res}
\tan\gamma_a - \tan\gamma_b
= \eps\, \frac{(v_a-v_b) \cosh(2Y) - (1- v_a v_b) \sinh(2Y)}{(v_a \cosh Y - \sinh Y)(v_b \cosh Y + \sinh Y)}
\stackrel{!}{=} 0
\,.\end{align}
%%%
This can only be satisfied for generic $Y$ if and only if $v_a = v_b = 1$,
i.e., both hadrons are massless. This means the bisector definition of the CS
frame is only equivalent to the above boost definition for massless hadrons,
for which both definitions where originally introduced in \refcite{Collins:1977iv},
while for nonzero hadron masses the two definitions are \emph{no longer equivalent}.%
\footnote{%
In some of the literature, the equivalence of the two definitions for the massive
case seems to be incorrectly assumed.
For example, in \refcite{Arnold:2008kf} expressions for the proton momenta
in the CS frame are given that would suggest the equivalence to also hold in the
massive case, but can be easily seen
to contradict the explicit expression for the Lorentz boost.}
The key advantage of our construction of $t^\mu$, $x^\mu$, $y^\mu$, $z^\mu$ and
the corresponding boost definition of the CS frame is that they are symmetric
under interchanging the beams (see \eq{xy_symmetry_constraints}) and furthermore
are manifestly independent of the beam parameters, i.e., they only depend on
$q^\mu$ without reference to the beam momenta beyond the beam direction itself.
In the rest of the paper, we will always use this definition, unless stated otherwise.

%===============================================================================
\subsubsection{Helicity decomposition}
%===============================================================================

Using $t^\mu, x^\mu, y^\mu, z^\mu$,
we can define polarization vectors for the vector boson in a fully
covariant way as
%%%
\begin{align} \label{eq:def_polarization_vectors}
\epol^\mu_\pm = \frac{1}{\sqrt{2}}\bigl( x^\mu \mp \img y^\mu \bigr)
\,, \qquad
\epol^\mu_0 = z^\mu
\,,\end{align}
%%%
which correspond to positive/negative helicity and longitudinal polarization with
respect to $z^\mu$.
Using these, we project the hadronic tensor onto the entries of a helicity density matrix~\cite{Mirkes:1992hu},
%%%
\begin{align} \label{eq:def_W_sigmasigma}
W_{\lambda \lambda'}(q, P_a, P_b)
\equiv \epol^\mu_\lambda \epol^{*\nu}_{\lambda'} \, W_{\mu\nu}(q, P_a, P_b)
\qquad\text{with}\qquad
\lambda  = \{+,-,0\}
\,.\end{align}
%%%
Since the $\epol^\mu_{\pm,0}$ span the space orthogonal to $t^\mu = q^\mu/Q$,
this decomposition fully captures the conserved part of the hadronic tensor,
see \eq{qmuWmunu_is_zero}. (To also account for the non-conserved parts, we would
just have to include the fourth time-like polarization $t^\mu$.)

From its definition in \eq{def_hadronic_tensor}, it is clear that $W^{\mu\nu}$
is hermitian, $W^{*\mu\nu} = W^{\nu\mu}$, so its symmetric (antisymmetric) components
are purely real (imaginary).
Therefore, the nine helicity components $W_{\lambda \lambda'}$
are fully specified by a total of nine real-valued, Lorentz-scalar hadronic structure
functions.
We will use the following linear combinations:
%%%
\begin{alignat}{5} \label{eq:def_W_i}
W_{-1} &= W_{++} + W_{--}
&&= ( x_\mu x_\nu + y_\mu y_\nu )\, W^{\mu\nu}
\,, \nn \\
W_0 &= 2W_{00}
&&= 2\, z_\mu z_\nu \, W^{\mu\nu}
\,, \nn \\
W_1 &= -\frac{1}{\sqrt{2}} \bigl( W_{+0} + W_{0+} + W_{-0} + W_{0-} \bigr)
&&= - ( x_\mu z_\nu + x_\nu z_\mu )\, W^{\mu\nu}
\,, \nn \\
W_2 &= -2\bigl( W_{+-} + W_{-+} \bigr)
&&= 2\, ( y_\mu y_\nu - x_\mu x_\nu )\, W^{\mu\nu}
\,, \nn \\
W_3 &= -\sqrt{2} \bigl( W_{+0} + W_{0+} - W_{-0} - W_{0-} \bigr)
&&= 2\img\, ( y_\mu z_\nu - y_\nu z_\mu )\, W^{\mu\nu}
\,, \nn \\
W_4 &= 2 \bigl( W_{++} - W_{--} \bigr)
&&= 2\img\, (x_\mu y_\nu - x_\nu y_\mu )\, W^{\mu\nu}
\,, \nn \\
W_5 &= -\img \bigl( W_{+-} - W_{-+} \bigr)
&&= - ( x_\mu y_\nu + x_\nu y_\mu )\, W^{\mu\nu}
\,, \nn \\
W_6 &= -\frac{\img}{\sqrt{2}} \bigl( W_{+0} - W_{0+} - W_{-0} + W_{0-} \bigr)
&&= - ( y_\mu z_\nu + y_\nu z_\mu )\, W^{\mu\nu}
\,, \nn \\
W_7 &= -\img \sqrt{2} \bigl( W_{+0} - W_{0+} + W_{-0} - W_{0-} \bigr)
&& = -2\img\, ( x_\mu z_\nu - x_\nu z_\mu )\, W^{\mu\nu}
\,.\end{alignat}
%%%
The reason for the somewhat odd numbering and normalization will become apparent
shortly. In the second equality, we have given the projections in terms of
$x^\mu$, $y^\mu$, $z^\mu$, corresponding to linear vector-boson polarizations.
The inclusive structure function from \eq{Wincl_def} is given by
%%%
\begin{equation}
W_\incl \equiv W_{++} + W_{--} + W_{00} = W_{-1} + \frac{1}{2}W_0
\,.\end{equation}
%%%

Since the projections of $W^{\mu\nu}$ that define the $W_i$ are orthogonal, we
can easily invert them and write $W^{\mu\nu}$ in terms of the $W_i$ as
%%%
\begin{align} \label{eq:def_W_i_inverse}
W^{\mu\nu} = \sum_{i=-1}^7 P_i^{\mu\nu} W_i
\qquad\text{($+$ terms $\propto q^\mu$ or $q^\nu$)}
\,,\end{align}
%%%
where the $P_i^{\mu\nu}$ are the same projections as in \eq{def_W_i} up to a
trivial difference in normalization, for example,
%%%
\begin{alignat}{2}
P_{-1}^{\mu\nu}
&= \frac{1}{2}\, \bigl(\epol^{*\mu}_+ \epol^\nu_+ + \epol^{*\mu}_- \epol^\nu_- \bigr)
&&= \frac{1}{2}\, ( x^\mu x^\nu + y^\mu y^\nu )
\,,\nn\\
P_0^{\mu\nu}
&= \frac{1}{2}\, \epol^{*\mu}_0 \epol^\nu_0
&&= \frac{1}{2}\, z^\mu z^\nu
\,,\nn\\
P_4^{\mu\nu}
&= \frac{1}{4}\, \bigl( \epol^{*\mu}_+ \epol^\nu_+ - \epol^{*\mu}_- \epol^\nu_- \bigr)
&&= \frac{1}{4\img}\, ( x^\mu y^\nu - x^\nu y^\mu )
\,.\end{alignat}
%%%
Contracting the leptonic tensor $L_{\mu\nu}$ with $W^{\mu\nu}$ decomposed as in \eq{def_W_i_inverse},
we have
%%%
\begin{equation}
L_{\mu\nu} W^{\mu\nu} = \sum_i L_{\mu\nu} P_i^{\mu\nu} W_i \equiv \sum_i L_i W_i
\,,\end{equation}
%%%
with the corresponding leptonic structure functions defined as
%%%
\begin{equation} \label{eq:def_L_i_integrated}
L_i(q, \cO,\Theta)
= \int \! \df\Phi_L(q) \, P_i^{\mu\nu} L_{\mu\nu}(\Phi_L) \, \delta[\cO - \hat\cO(q,\Phi_L)] \, \hat \Theta(q,\Phi_L)
\,.\end{equation}
%%%

The cross section in \eq{factorization_leptonic_hadronic_massive_hadrons}
in terms of the scalar structure functions now becomes
%%%
\begin{align} \label{eq:fully_differential_generic_param}
\frac{\df\sigma(\Theta)}{\df^4 q \, \df \cO}
&\equiv \frac{1}{2\Ecm^2} \sum_i L_i(q, \cO, \Theta) \,
W_i(q^2, s_{aq}, s_{bq})
\,,\end{align}
%%%
which generalizes the inclusive cross section in \eq{inclusive_xsec} to
arbitrary leptonic observables and fiducial cuts. As for $W_\incl$ before,
Lorentz invariance implies that the hadronic structure functions $W_i$ only
depend on the three kinematic invariants $q^2$, $s_{aq}$, $s_{bq}$, or
equivalently the three kinematic variables $Q^2$, $Y$, $q_T^2$, see
\eq{kin_invariants}. In particular, they do not depend on the orientation of
$\qt$. Since the $x^\mu$, $y^\mu$, $z^\mu$ reduce to the spatial coordinate axes
in the CS frame, the structure functions correspond to the individual tensor
components of the hadronic tensor $\hat W^{\mu\nu}$ evaluated in the CS frame,
e.g., $W_{-1} = \hat W^{xx} + \hat W^{yy}$, $W_0 = 2 W^{zz}$, etc. We will refer to
\eqs{def_W_i}{def_W_i_inverse} as the \emph{CS tensor decomposition}.

We note that one may also decompose the hadronic tensor in terms of Lorentz
structures directly formed out of $g^{\mu\nu} - q^\mu q^\nu/q^2$ and its
contractions with $P_{a,b}^\mu$, see e.g.~\refscite{Lam:1978pu, Boer:2006eq,
Arnold:2008kf, Gauld:2017tww}. This automatically ensures that the projectors are
covariant combinations of $q^\mu$ and $P_{a,b}^\mu$ and that the corresponding
coefficients are Lorentz-scalar functions. This is usually not manifest when one
considers the individual tensor components in the CS frame (or any other rest
frame). However, as we have seen, the CS-frame components are reproduced by the
CS tensor decomposition in a manifestly covariant manner as the Lorentz-scalar
structure functions $W_i$ that only depend on Lorentz invariants. Hence, there
is no formal preference for either decomposition and the two are related by a
straightforward change of basis. We will see in the following sections that the
physics at small $q_T \ll Q$ becomes particularly transparent when using the CS
tensor decomposition.

%===============================================================================
\subsection{Leptonic decomposition and relation to angular coefficients}
\label{sec:leptonic_decomposition}
%===============================================================================

\begin{figure*}
\centering
\includegraphics[width=\WidthTwoSubfigs]{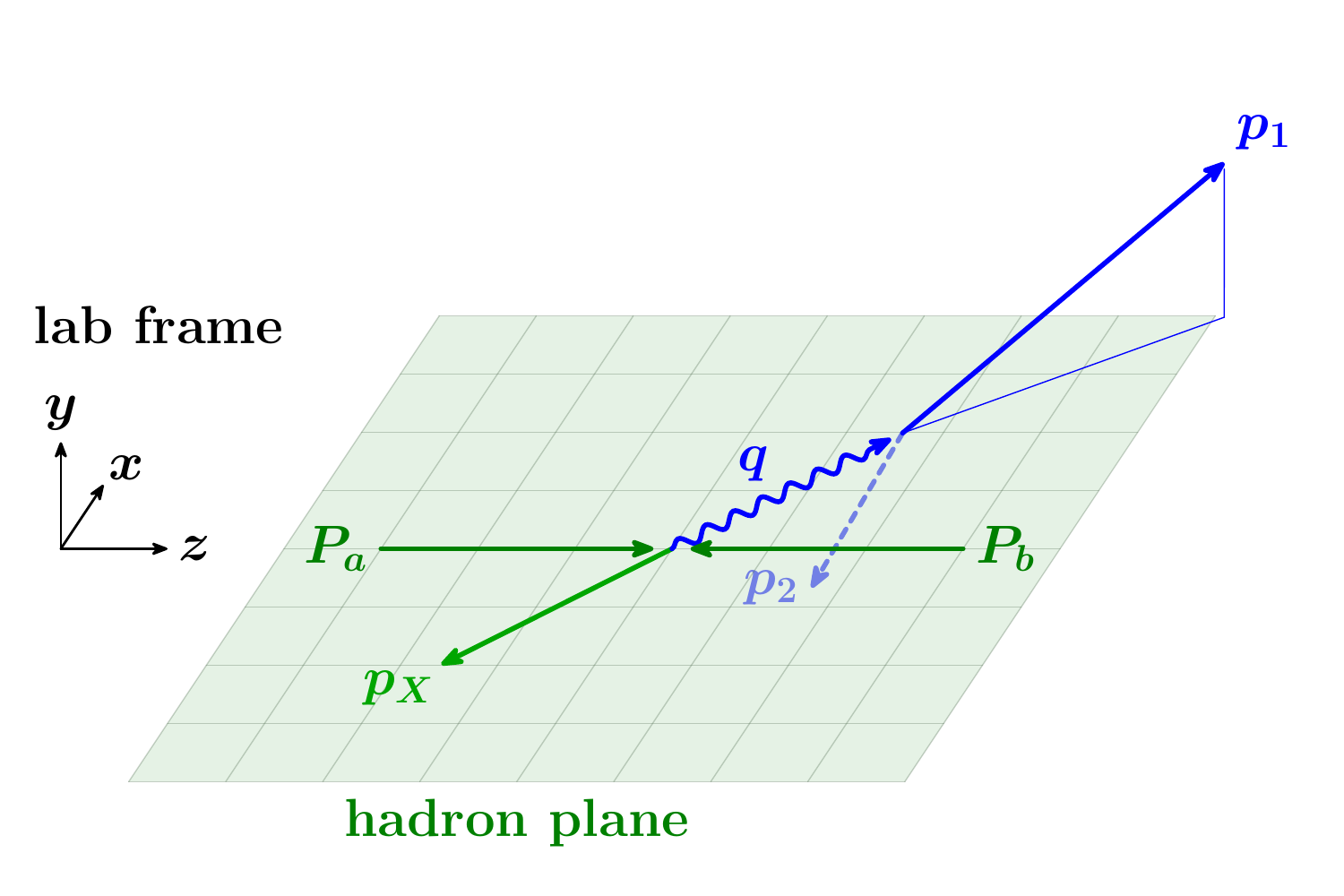}%
\includegraphics[width=\WidthTwoSubfigs]{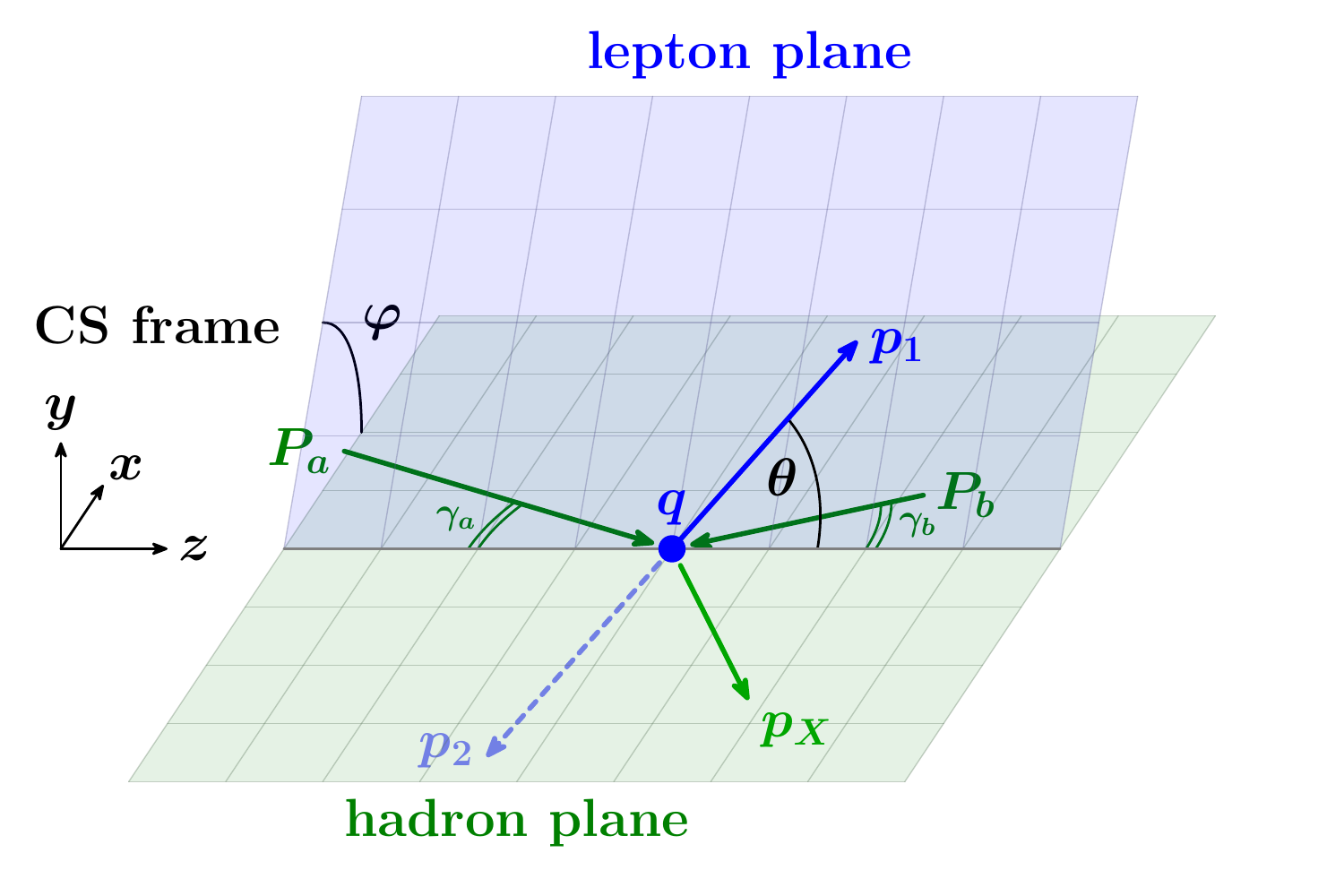}%
\caption{Kinematics in the lab frame (left) and the Collins-Soper frame (right).
In the lab frame, the incoming hadron momenta are head-to-head (assuming
the lab frame and hadronic center-of-mass frame coincide), while
the vector boson has nonvanishing three-momentum $\vec q$.
The scattering $p(P_a)\, p(P_b) \to V(q)\, X(p_X)$ defines the hadron plane (green).
In the CS frame (right), the vector boson is at rest.
The leptons are produced back to back in the lepton plane (blue).
The magnitudes of the hadron momenta in general differ for $Y \neq 0$,
but their angles $\gamma_{a,b}$ with respect to the $z$ axis (indicated by the double arcs)
become equal in the limit of vanishing hadron masses.
The Collins-Soper angles $\theta$ and $\varphi$ are defined as indicated.}
\label{fig:frames}
\end{figure*}

In this subsection, we discuss the leptonic decay in more detail. For the most
part, we specifically consider the leading-order Drell-Yan decays
%%%
\begin{alignat}{3} \label{eq:L_2body}
Z/\gamma^\ast(q) &\to \ell^-(p_1) \, \ell^+(p_2)
\,,\qquad &
Z &\to \nu(p_1)\, \bar\nu(p_2)
\,, \nn \\
\, \quad
W^+(q) &\to \nu_\ell(p_1) \, \ell^+(p_2)
\,, \quad &
W^-(q) &\to \ell^-(p_1) \, \bar{\nu}_\ell(p_2)
\,,\end{alignat}
%%%
neglecting lepton masses, $m_{1,2} = 0$, and summing over lepton polarizations.
These are the primary application we are eventually interested in.
The kinematics of the process in the lab and CS frames are illustrated in the left
and right panels of \fig{frames}.

In \sec{L_beyond_2body}, we discuss the extension to more complicated
leptonic final states, e.g.\ including QED final-state radiation, which
is important at the precision of current Drell-Yan measurements. In particular,
there we show to what extent the LO discussion carries over for measurements that
are performed in terms of suitably defined Born leptons.

%~~~~~~~~~~~~~~~~~~~~~~~~~~~~~~~~~~~~~~~~~~~~~~~~~~~~~~~~~~~~~~~~~~~~~~~~~~~~~~~
\subsubsection{Definition of CS angles}
%~~~~~~~~~~~~~~~~~~~~~~~~~~~~~~~~~~~~~~~~~~~~~~~~~~~~~~~~~~~~~~~~~~~~~~~~~~~~~~~

It is convenient to introduce spherical coordinates $(\cos\theta, \varphi)$ in the CS
frame, in terms of which we can parametrize $p_{1,2}$, as illustrated in the
right panel of \fig{frames}, as
%%%
\begin{align} \label{eq:p1_p2_CS_angles_CS_frame}
p_{1,2}^\mu = \frac{Q}{2} \Bigl( t^\mu \pm x^\mu \sin \theta \cos \varphi \pm y^\mu \sin \theta \sin \varphi \pm z^\mu \cos \theta \Bigr)
\,.\end{align}
%%%
The angles $\theta, \varphi$ are known as Collins-Soper angles.%
\footnote{%
To be precise, here we have defined the CS angles by $\theta \equiv\theta_1$ and
$\varphi\equiv \varphi_1$, where $(\theta_1, \varphi_1)$ are the spherical
coordinates of $p_1$. Since at LO in QED $p_1$ and $p_2$ are back-to-back,
the spherical coordinates for $p_2$ are then $(\pi-\theta, \pi+\varphi)$.
}
From \eq{p1_p2_CS_angles_CS_frame}, one can easily derive their explicit
expressions in terms of lab-frame quantities $E_{1,2}$, $p_{1,2}^{x,y,z}$,
%%%
\begin{align} \label{eq:CS_angles_lab_frame}
\cos \theta
&= \frac{1}{Q m_T} \bigl[ (E_1 + p_1^z)(E_2 - p_2^z) - (E_1 - p_1^z)(E_2 + p_2^z) \bigr]
\,, \nn \\
\cos \varphi
&= \frac{1}{\sin \theta} \frac{p_{1T}^2 - p_{2T}^2}{q_T m_T}
\,, \qquad
\sin \varphi
= \frac{2}{\sin \theta} \frac{p_1^y p_2^x - p_2^y p_1^x}{q_T Q}
\,,\end{align}
%%%

Note that we have arbitrarily chosen the positive orientation of the $z$ axis by
having hadron $a$ move in the $z$ direction in the lab frame. As a result, the
negatively charged lepton moves into the same
rest-frame hemisphere as hadron $a$ for $\cos \theta > 0$.
In experimental measurements at the LHC, where the choice of $a$ and $b$ is arbitrary,
hadron $b$ is often taken to be the one closer to the vector boson in rapidity
to ensure that angular distributions do not average out when integrating over rapidity,
see e.g.\ \refscite{Aad:2015uau, Khachatryan:2015paa, Aad:2016izn, Sirunyan:2018swq}.
The resulting angles $\theta^*$ and $\varphi^*$, which are often also referred to as Collins-Soper angles, are then related to \eq{CS_angles_lab_frame} by
%%%
\begin{align} \label{eq:the_other_CS_angles}
\cos \theta^* = \frac{Y}{\abs{Y}} \cos \theta
\,, \qquad
\varphi^* = \frac{Y}{\abs{Y}} \varphi
\,.\end{align}
%%%
On the other hand, \eq{CS_angles_lab_frame} does not depend on the chosen
orientations of the $x$ and $y$ axes in the lab frame as long as they form a
right-handed coordinate system.

The advantage of \eq{p1_p2_CS_angles_CS_frame}, or equivalently the boost
definition to define the CS frame, is that it stays true regardless of whether
hadron masses are included or neglected, and thus also any relations like
\eq{CS_angles_lab_frame} that are derived from it are independent of any beam
parameters. On the other hand, with the bisector construction including hadron
masses, \eq{CS_angles_lab_frame} no longer holds, see also the discussion in
\sec{reference_frame}.

%~~~~~~~~~~~~~~~~~~~~~~~~~~~~~~~~~~~~~~~~~~~~~~~~~~~~~~~~~~~~~~~~~~~~~~~~~~~~~~~
\subsubsection{Leptonic decay parametrization by angles}
%~~~~~~~~~~~~~~~~~~~~~~~~~~~~~~~~~~~~~~~~~~~~~~~~~~~~~~~~~~~~~~~~~~~~~~~~~~~~~~~

The fully-differential leptonic tensor for the $1\to 2$ Drell-Yan decays in
\eq{L_2body} at tree level has the form
%%%
\begin{align} \label{eq:Lmunu_1to2}
L^{\mu\nu}(p_1, p_2)
&= \frac{24\pi}{q^2} \Bigl[ L_{+}(q^2) \bigl( p_1^\mu p_2^\nu + p_1^\nu p_2^\mu - g^{\mu\nu} p_1 \cdot p_2 \bigr)
+ \img L_{-}(q^2) \, \eps^{\mu\nu\rho\sigma} p_{1\rho} p_{2\sigma} \Bigr]
\,.\end{align}
%%%
Only the contribution proportional to $L_+$ ($L_-$)
survives the contraction with the symmetric (antisymmetric) $P_i^{\mu\nu}$
corresponding to the parity-even (parity-odd) hadronic structure functions
$W_{-1,0,1,2,5,6}$ ($W_{3,4,7}$).
The normalization is chosen such that $L_+(q^2) = L(q^2)$ agrees with the inclusive
coefficient in \eq{leptonic_tensor_inclusive}, and such that
$L_-(q^2) = L_+(q^2)$ for $W$ decays, where parity is maximally violated.
Explicit expressions for the $L_\pm(q^2)$ are given in \app{leptonic_tensors}.

It is convenient to parametrize the $2$-body decay phase space
using the CS angles $\theta, \varphi$, in terms of which
the phase-space measure is isotropic,
%%%
\begin{align}
\df \Phi_L(q) = \frac{\df \cos \theta \, \df \varphi}{32\pi^2}
\,.\end{align}
%%%
Applying this parametrization to \eq{def_L_i_integrated}, we find
%%%
\begin{align} \label{eq:L_i_CS_angles}
L_i(q, \cO, \Theta) &= \int_{-1}^1 \! \df \cos \theta \, \int_0^{2\pi} \df \varphi \, L_i(q^2, \theta, \varphi) \, \delta[\cO - \hat\cO(q, \theta, \varphi)] \, \hat \Theta(q, \theta, \varphi)
\,, \\ \nn
L_i(q^2, \theta, \varphi) &= \frac{3}{16\pi} L_{\pm(i)}(q^2)\, g_i(\theta, \varphi)
\qquad\text{with}\qquad
\pm\!(i) =
\begin{cases}
   + \,, \qquad i \in \{-1, 0, 1, 2, 5, 6\} \,, \\
   - \,, \qquad i \in \{3, 4, 7\}
\,,\end{cases}
\end{align}
%%%
where the angular dependence arises from contracting $P_i^{\mu\nu}$ with the Lorentz
structures in \eq{Lmunu_1to2}, and is encoded in nine (real combinations of)
spherical harmonics
%%%
\begin{alignat}{7} \label{eq:def_g_i}
g_{-1}(\theta, \varphi) &= 1 + \cos^2 \theta
\,, \qquad &
g_2(\theta, \varphi) &= \tfrac{1}{2} \sin^2 \theta \cos(2\varphi)
\,, \qquad &
g_5(\theta, \varphi) &= \sin^2 \theta \sin(2\varphi)
\,, \nn \\
g_0(\theta, \varphi) &= 1 - \cos^2 \theta
\,, \qquad &
g_3(\theta, \varphi) &= \sin \theta \cos \varphi
\,, \qquad &
g_6(\theta, \varphi) &= \sin(2\theta) \sin \varphi
\,, \nn \\
g_1(\theta, \varphi) &= \sin(2\theta) \cos \varphi
\,, \qquad &
g_4(\theta, \varphi) &= \cos \theta
\,, \qquad &
g_7(\theta, \varphi) &= \sin \theta \sin \varphi
\,.\end{alignat}
%%%
Putting everything together, we obtain
%%%
\begin{align} \label{eq:fully_differential_CS_angles}
\frac{\df\sigma(\Theta)}{\df^4 q \, \df \cO}
&= \int_{-1}^1 \! \df \cos \theta \, \int_0^{2\pi} \df \varphi \,
\frac{\df\sigma}{\df^4 q \, \df\cos\theta\, \df\varphi}
\, \delta[\cO - \hat\cO(q, \theta, \varphi)] \,
\hat \Theta(q, \theta, \varphi)
\,, \nn \\
\frac{\df\sigma}{\df^4 q \, \df\cos\theta\,\df\varphi}
&=  \frac{1}{2\Ecm^2} \sum_{i = -1}^7 L_i(q^2, \theta, \varphi)\,W_i(q^2, s_{aq}, s_{bq})
\equiv
\frac{3}{16\pi} \sum_{i = -1}^7 \frac{\df \sigma_i}{\df^4 q}\, g_i(\theta, \varphi)
\,,\end{align}
%%%
where in the last step we defined the so-called helicity cross sections
%%%
\begin{align} \label{eq:def_sigma_i}
\frac{\df \sigma_i}{\df^4 q} = \frac{1}{2\Ecm^2} L_{\pm(i)}(q^2) \,
W_i(q^2, s_{aq}, s_{bq})
\,.\end{align}
%%%
Integrating over $\cO$ and setting $\hat \Theta = 1$,
we recover the inclusive cross section in \eq{inclusive_xsec},
%%%
\begin{align}
\frac{\df \sigma}{\df^4 q} = \frac{\df \sigma_{-1}}{\df^4 q} + \frac{1}{2} \frac{\df \sigma_{0}}{\df^4 q}
\,, \qquad
W_\incl = W_{-1} + \frac{W_0}{2}
\,.\end{align}
%%%

%~~~~~~~~~~~~~~~~~~~~~~~~~~~~~~~~~~~~~~~~~~~~~~~~~~~~~~~~~~~~~~~~~~~~~~~~~~~~~~~
\subsubsection{Relation to angular coefficients}
\label{sec:angular_coefficients}
%~~~~~~~~~~~~~~~~~~~~~~~~~~~~~~~~~~~~~~~~~~~~~~~~~~~~~~~~~~~~~~~~~~~~~~~~~~~~~~~

From \eq{fully_differential_CS_angles}, we can write the fully-differential cross
section in the CS angles as
%%%
\begin{align} \label{eq:xsec_Ai}
\frac{\df \sigma}{\df^4 q \, \df \cos \theta \, \df \varphi}
= \frac{3}{16\pi} \frac{\df \sigma}{\df^4 q} \biggl[ 1 + \cos^2 \theta + \frac{A_0}{2} \bigl( 1 - 3\cos^2 \theta \bigr) + \sum_{i = 1}^7 A_i \, g_i(\theta, \varphi) \biggr]
\,,\end{align}
%%%
where the \emph{angular coefficients} $A_i$ are given in terms of the helicity
cross sections or the hadronic structure functions as
%%%
\begin{alignat}{3} \label{eq:Ai_def}
A_i &= \frac{\df \sigma_i}{\df \sigma_{-1} + \frac{1}{2}\df \sigma_0}
&= \frac{L_{\pm(i)}(q^2)\, W_i(q^2, s_{aq}, s_{bq})}{L_+(q^2)\, (W_{-1 }+ \frac{1}{2} W_0\bigr)(q^2, s_{aq}, s_{bq})}
\,.\end{alignat}
%%%
We deliberately chose the numbering and normalization of the $W_i$ in \eq{def_W_i}
to match the often used form of the cross section in \eq{xsec_Ai}.
The only exception is the inclusive cross section, which is split into orthogonal
contributions from $W_0$ and $W_{-1}$.
For the same reason, we refrained from normalizing the spherical harmonics in \eq{def_g_i}.
We remind the reader that both numerator and denominator in
\eq{Ai_def} in general involve a sum over the
intermediate vector bosons, $L_\pm W_i \equiv \sum_{V,V'} L_{\pm VV'} W_{i\,VV'}$
so for neutral-current Drell-Yan ($V = Z,\gamma$),
the parity-even leptonic prefactors $L_+(q^2) \equiv L_{+VV'}(q^2)$
do not in general cancel in the ratio in \eq{Ai_def}.

A priori, \eq{fully_differential_CS_angles} or \eq{xsec_Ai} simply
provide a convenient way to parametrize the fully-differential Drell-Yan cross section
for massless $2$-body decays.
For this purpose, it is irrelevant whether or not the CS angles can be
reconstructed experimentally. Similarly, the choice of the CS tensor decomposition
is a priori arbitrary, and we could have used another decomposition. Of course,
the combination of using the CS tensor decomposition for the hadronic tensor
together with using the CS angles to parametrize the leptonic tensor is what
leads to the simple angular dependence in \eq{fully_differential_CS_angles}.
If we were to choose a different tensor decomposition $W_i'$, we would also
choose polar coordinates $\cos\theta'$, $\varphi'$ with respect to its corresponding
rest frame, and arrive at \eqs{xsec_Ai}{Ai_def} in terms of $\cos\theta'$,
$\varphi'$, $A_i'$, and $W_i'$.
On the other hand, when $\cos \theta$ and $\varphi$ are explicitly measured, or
when \eq{xsec_Ai} is used as a template to measure the $A_i$, it obviously does
matter with respect to which frame they are defined.
It is also straightforward to relate the $W_i$ or $A_i$ for different frames, see \sec{uniqueness_linear_power_corrections} below.

%~~~~~~~~~~~~~~~~~~~~~~~~~~~~~~~~~~~~~~~~~~~~~~~~~~~~~~~~~~~~~~~~~~~~~~~~~~~~~~~
\subsubsection{Extension to more complicated leptonic final states}
\label{sec:L_beyond_2body}
%~~~~~~~~~~~~~~~~~~~~~~~~~~~~~~~~~~~~~~~~~~~~~~~~~~~~~~~~~~~~~~~~~~~~~~~~~~~~~~~

Up to now, our discussion in this subsection assumed the leading-order dilepton
final states in \eq{L_2body}, and so in particular \eq{xsec_Ai} is derived in
this limit. For a generic leptonic final state $L$, e.g.\ when including QED
final-state radiation (FSR)
or for more complicated electroweak decays like $V^*\to VH$ or $V^* \to V_1
V_2$, there is a priori no reason that the $L_i$ are proportional to
spherical harmonics $g_i(\theta, \varphi)$ any longer, in which case
one cannot use \eq{xsec_Ai} to define the $A_i$ beyond this LO.

On the other hand, as we saw in \eq{Ai_def}, the $A_i$ are in one-to-one
correspondence with the underlying hadronic structure functions $W_i$. The $W_i$
are by construction independent of $L$ (apart from its total momentum $q^\mu$)
and thus well-defined for arbitrary $L$. The physical reason for the appearance
of nine independent structures in both cases is exactly the same, namely the
spin-1 nature of the intermediate vector boson (and the fact that we ignore the
non-conserved parts). Hence, the cross section in the CS tensor decomposition in
\eq{fully_differential_generic_param} should be considered as the generalization
of the LO angular decomposition in \eq{xsec_Ai} to arbitrary leptonic final
states and measurements. One could also use \eq{Ai_def} as the all-order
definition of the $A_i$ in terms of the $W_i$ and conventional LO weak couplings
and propagators included in the $L_\pm(q^2)$. One could then easily rewrite
\eq{fully_differential_generic_param} in terms of the so-defined $A_i$
multiplied by generic leptonic coefficients $L_i(q, \cO, \Theta)$, which in the
simplest case reduce to $L_\pm(q^2) g_i(\theta, \varphi)$ as in
\eq{L_i_CS_angles}, but in general can also be more complicated.
Although at that point, it is easier and perhaps less confusing to directly work in terms of
the $W_i$ and \eq{fully_differential_generic_param} as it is.

Nevertheless, in the context of Drell-Yan measurements, the LO
relation in \eq{xsec_Ai} is very useful in practice because the $g_i$ are
orthogonal spherical harmonics. This allows one to directly measure the $A_i$
(or $W_i$) by performing a fit to the angular dependence of the $(\theta,\varphi)$
distribution or by projecting out different terms by taking suitably
weighted angular integrals of it~\cite{Khachatryan:2015paa, Aad:2016izn}.
This procedure has received some
criticism, since it seemingly relies on a LO QED interpretation of the
angular dependence, while QED final-state radiation can be relevant at the level
of precision reached by Drell-Yan measurements. In fact, even the definition of
the CS angles $(\theta, \varphi)$ themselves becomes nonobvious, because with
additional QED radiation in the final state, the lepton momenta generically
no longer add to the full vector-boson momentum $q^\mu$.
Instead, we now have
%%%
\begin{equation} \label{eq:L_mom_cons}
q^\mu = p_1^\mu + p_2^\mu + k^\mu
\,,\end{equation}
%%%
where $p^\mu_{1,2}$ are the measured lepton momenta, which depend on the lepton
definition, and $k^\mu$ is the remaining momentum not included in the definitions
of $p_{1,2}$.
We stress that here we are \emph{not} concerned with the experimental methods to
reconstruct and calibrate the leptons or to recover photon radiation. The
``measured'' lepton momenta $p_{1,2}^\mu$ refer to the \emph{truth-level} lepton
definition to which the raw reconstructed momenta are corrected or unfolded.
This truth-level definition must be theoretically well-defined to have a
meaningful measurement that can be compared to theoretical calculations, and one
can consider the question whether certain truth-level definitions are theoretically
preferred or not.%
\footnote{%
On the other hand, whether a specific
truth-level definition receives more or less
associated experimental uncertainties is a separate, experimental question, to
which we have nothing to say here.
While these two questions are not entirely unrelated,
they should nevertheless be kept well separated.
We thank Daniel Froidevaux for discussions on this issue.
}

Obviously, the $(\theta_1, \varphi_1)$ angles describing the
orientation of $p_1$ now depend on the lepton definition and
also on whether they are defined in the full \emph{vector-boson} rest frame
(where $q^\mu$ or equivalently the full $L$ is at rest)
or the \emph{dilepton} rest frame (where only $p_1^\mu + p_2^\mu$ is at rest).
Especially in the latter case, there is no guarantee (in fact it seems quite
unlikely) that the angular distribution
in $(\theta_1, \varphi_1)$ will still admit a decomposition in terms of the
nine spherical harmonics $g_i(\theta_1, \varphi_1)$.

For ``bare'' leptons, $p_{1,2}^\mu$ are defined without including any FSR
photons. This means infrared QED singularities are regulated by the lepton
mass leading to potentially large logarithms of the lepton mass. This effect is
reduced by defining ``dressed'' leptons, which include all photons radiated
within a cone of some size around the leptons, and hence can be theoretically
thought of as QED ``lepton jets''. With either definition, the remaining
momentum $k^\mu$ in \eq{L_mom_cons} is nonzero and so the dilepton and
vector-boson rest frames are no longer equivalent.

Another option is to include all $k^\mu$ into $p_{1,2}^\mu$, i.e., the lepton
momenta are (partially) defined by the condition $q^\mu = p_1^\mu + p_2^\mu$.
This is basically what ``Born'' leptons are. Their full definition corresponds
to defining an IR-safe projection of the full leptonic final state $L$ onto a
Born-like $2$-body final state. In principle there are many ways to do so, but
as long as the Born projection is well defined so are the Born leptons.

To illustrate this, let us consider an explicit example: We start by defining the
leptonic thrust axis $\vec n_L$ of the full leptonic final state $L$ in its rest
frame. The thrust axis $\vec n_L$ is defined in the usual way as the axis $\vec
n$, with $\vec n^2 = 1$, that minimizes
%%%
\begin{equation}
\vec n_L:\qquad
\min_{\vec n} \sum_{i \in L} (E_i - |\vec n \cdot \vec p_i|)
= Q - \max_{\vec n} \sum_i |\vec n \cdot \vec p_i|
\,,\end{equation}
%%%
where the sum runs over all particles in $L$, including in particular
all QED FSR, and $E_i$, $\vec p_i$ are defined
in the rest frame of $L$. The overall positive (negative) orientation of $\vec
n_L$ can be fixed by convention, e.g., to point into the hemisphere that
contains the lepton (antilepton).%
\footnote{%
In practice, one would use a flavor-aware minimization or clustering procedure
to exclude minima or solutions for $\vec n_L$ for which lepton and antilepton
are clustered into the same hemisphere.
}
Imposing the condition $p_1^\mu + p_2^\mu = q^\mu$, requiring massless on-shell momenta,
$p_{1,2}^2 = 0$, and using $\vec n_L$ to define the direction of $\vec p_1 = -
\vec p_2$, then uniquely determines (recall $Q = \sqrt{q^2}$)
%%%
\begin{equation} \label{eq:p12_generic_lep_def}
p_{1,2}^\mu = \frac{Q}{2}\,(1, \pm \vec n_L)_\CS \equiv \frac{Q}{2}(t^\mu \pm n_L^\mu)
\,.\end{equation}
%%%
In the second step we defined the unit vectors
%%%
\begin{equation}
t^\mu = \frac{q^\mu}{Q} = (1, \vec 0)_\CS
\,,\qquad
n_L^\mu = (0, \vec n_L)_\CS
\,,\quad
n_L^2 = -1
\,,\qquad
t \cdot n_L = 0
\,,\end{equation}
%%%
where $t^\mu$ is the same as before, and $n_L^\mu$ describes the
overall orientation of $L$.

\begin{figure*}
\centering
\includegraphics[width=\WidthTwoSubfigs]{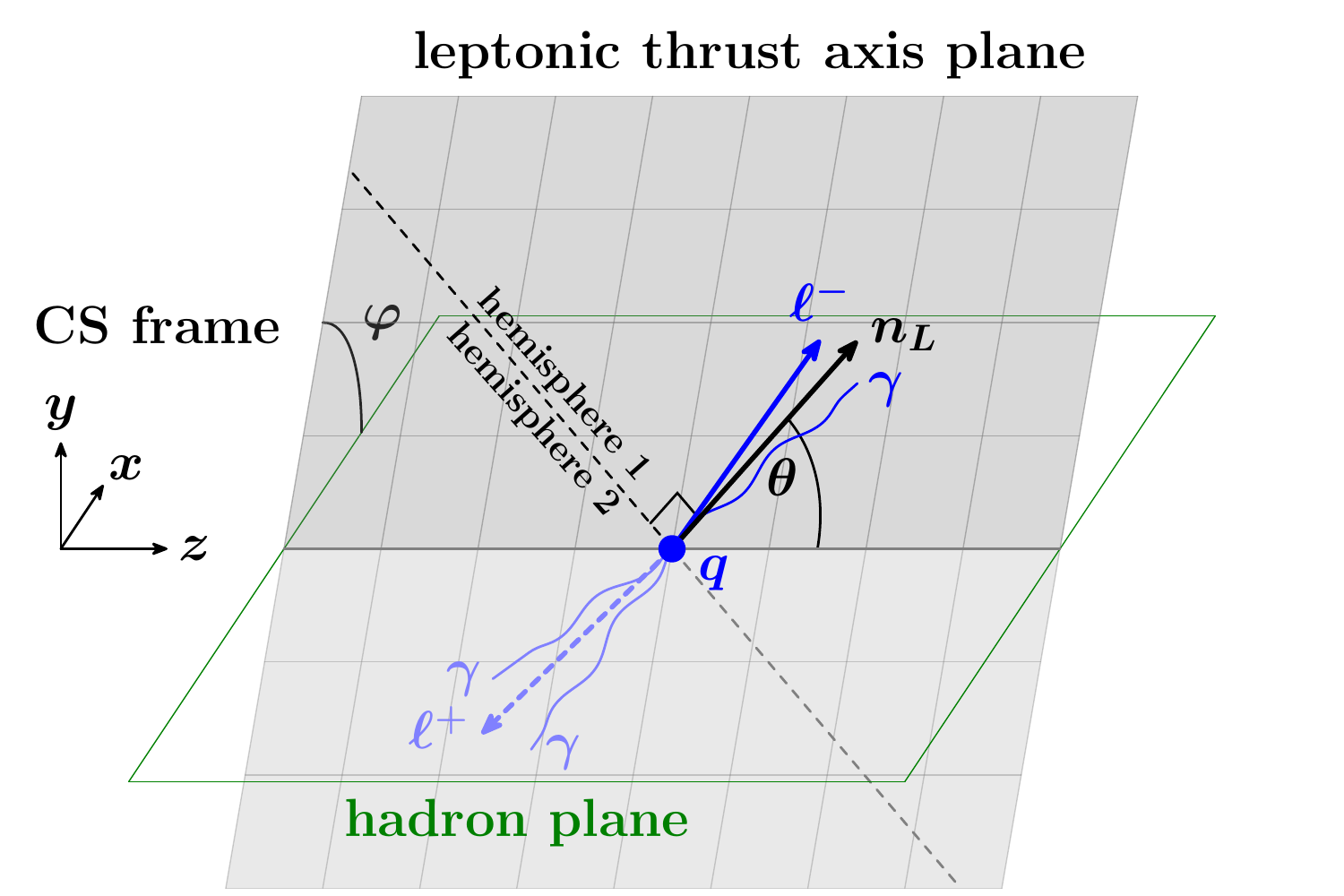}%
\caption{%
Definition of the Collins-Soper angles $\theta, \varphi$ for a generic leptonic
final state including FSR (blue) in terms of the leptonic thrust axis $n_L$.
The leptonic thrust axis plane in the CS frame is spanned by the $z$ axis and
$n_L$ and generalizes the lepton plane in \fig{frames}.
Beyond LO in QED, the decay products do no longer have to lie in this plane.
The hadronic scattering in the hadron plane (green) is as in \fig{frames}
and omitted for clarity.
The hemisphere boundary (dashed line) is perpendicular to the
thrust axis and separates the emissions into hemisphere~1 (clustered with the lepton)
and hemisphere~2 (clustered with the antilepton).
}
\label{fig:cs_angles_born_leptons}
\end{figure*}

More generally, we can also carry out the construction in two steps, first constructing
$q^\mu = P_1^\mu + P_2^\mu$ with massive $P_{1,2}^2\ne 0$ and then projecting them
onto massless $p_{1,2}$. Here, we first cluster all emissions with either the lepton or antilepton based on whose hemisphere they are in, which yields the massive hemisphere momenta $P_{1,2}^\mu$,
%%%
\begin{align}
P_{1,2}^\mu
&= \frac{Q}{2}\bigl[x_{1,2}\, t^\mu \pm \lambda(q^2, P_1^2, P_2^2)\, n_L^\mu \bigr]
\,, \nn \\
x_{1,2} &= 1 + \frac{P_{1,2}^2 - P_{2,1}^2}{q^2}
\,, \qquad
\lambda(q^2, P_1^2, P_2^2)
= \frac{1}{q^2}\sqrt{(q^2 - P_1^2 - P_2^2)^2 - 4P_1^2 P_2^2}
\,.\end{align}
%%%
Next, we project $P_{1,2}^\mu$ onto massless momenta $p_{1,2}^\mu$ by preserving
the three-momentum direction, $\vec p_1 / \abs{\vec p_1} = \vec P_1 / \abs{\vec
P_1} = \vec n_L$, and the total energy, $p_1^0 + p_2^0 = P_1^0 + P_2^0 = Q$,
which yields \eq{p12_generic_lep_def}. The spherical coordinates $(\theta_L,
\varphi_L)$ of $\vec n_L$ in the CS frame now provide a unique, all-order
definition of the CS angles $(\theta, \varphi) \equiv (\theta_L, \varphi_L)$,
i.e.,
%%%
\begin{equation}
n_L^\mu = x^\mu \sin \theta \cos \varphi + y^\mu \sin \theta \sin \varphi + z^\mu \cos \theta
\,.\end{equation}
%%%
This is the generalization of \eq{p1_p2_CS_angles_CS_frame}, where the CS angles
now describe the overall orientation of $L$ in the CS frame,
as illustrated in \fig{cs_angles_born_leptons}.

It is easy to see that the above definitions are IR safe and reduce to the
respective LO definitions. In principle, any other IR-safe way of clustering the
emissions into $P_{1,2}^\mu$ is possible. Other ways to project them onto
massless $p_{1,2}^\mu$ are also possible, as long as the projection is IR safe
and preserves the total leptonic momentum $q^\mu = P_1^\mu + P_2^\mu = p_1^\mu +
p_2^\mu$. In practice, defining the projection by keeping the orientation fixed
is the most natural and also the easiest, as it avoids any confusion about which
particular direction is used to define the CS angles.

The advantage of Born leptons is that they \emph{do admit} an analogous
LO-like angular decomposition as we will now show. More generally, it is
sufficient to restrict to leptonic measurements that can be written in terms of
$P_{1,2}^\mu$,
%%%
\begin{align}
\hat \cO(q, \Phi_L)
&\equiv \hat \cO(q, P_1, P_2) = \hat \cO(q, \theta, \varphi, P_1^2, P_2^2)
\,, \nn \\
\hat \Theta(q, \Phi_L) &\equiv \hat\Theta(q, P_1, P_2) = \hat\Theta(q, \theta, \varphi, P_1^2, P_2^2)
\,.\end{align}
%%%
For such measurements we can write the general leptonic tensor
in \eq{def_leptonic_tensor} as
%%%
\begin{align} \label{eq:Lmunu_P1P2}
L^{\mu\nu}(q, \cO,\Theta)
&= \int\!\frac{\df^4 P_1}{(2\pi)^3}\, \frac{\df^4 P_2}{(2\pi)^3}\,
   (2\pi)^4\delta^4(q - P_1 - P_2)\, F^{\mu\nu}(P_1, P_2)
\nn \\ & \qquad \times
\delta[\cO - \hat\cO(q, P_1, P_2)]\,\hat\Theta(q, P_1, P_2)
\,,\end{align}
%%%
where $F^{\mu\nu}(P_1, P_2)$ is the projection of the full leptonic decay
$L^{\mu\nu}(\Phi_L)$ onto the massive $2$-body $(P_1, P_2)$ phase space,
%%%
\begin{align}
F^{\mu\nu}(P_1, P_2)
&= (2\pi)^2 \int\!\df\Phi_L(P_1 + P_2)\,L^{\mu\nu}(\Phi_L)\,
   \delta^4[P_1 - \hat P(\Phi_L)]
\,,\end{align}
%%%
where $\hat P^\mu(\Phi_L)$ implements the clustering of $\Phi_L$ into $P_1^\mu$,
and $P_2^\mu$ is implicitly defined via $q^\mu = P_1^\mu + P_2^\mu$.
For the LO decays in \eq{L_2body}, we have $\Phi_L = (p_1, p_2)$
and $\hat P(\Phi_L) = p_1$ such that
%%%
\begin{equation}
F_{\rm LO}^{\mu\nu}(P_1, P_2) = L^{\mu\nu}(P_1, P_2)\,\delta(P_1^2)\,\delta(P_2^2)
\,,\end{equation}
%%%
with $L^{\mu\nu}(p_1, p_2)$ given by the LO result in \eq{Lmunu_1to2}.

The key point is that $F^{\mu\nu}(P_1, P_2)$ is defined in a Lorentz-covariant
way, and therefore obeys the following Lorentz decomposition (ignoring as before
the non-conserved parts)
%%%
\begin{align} \label{eq:Lmunu_Fi_def}
F^{\mu\nu}(P_1, P_2)
&= 12\pi\Bigl[
   (t^\mu t^\nu - g^{\mu\nu} - n_L^\mu n_L^\nu)\, F_+
   + (t^\mu t^\nu - g^{\mu\nu})\, F_0
   + \img \eps^{\mu\nu\rho\sigma} n_{L\rho} t_\sigma\, F_-
   \Bigr]
\nn \\ & \qquad
\text{($+$ terms $\propto q^\mu$ or $q^\nu$)}
\,,\end{align}
%%%
where $F_{\pm,0} \equiv F_{\pm,0}(q^2, P_1^2, P_2^2)$ are Lorentz-scalar functions that can
only depend on three independent invariants formed out of $P_{1,2}$, which we
chose as $q^2 = (P_1 + P_2)^2$ and $P_{1,2}^2$. The decomposition in
\eq{Lmunu_Fi_def} is chosen so that at LO, comparing to \eq{Lmunu_1to2}, we have
%%%
\begin{align}
F_\pm(q^2, P_1^2, P_2^2)
&= L_\pm(q^2)\, \delta(P_1^2)\delta(P_2^2) + \ord{\aem}
\,, \qquad
F_0(q^2, P_1^2, P_2^2) = \ord{\aem}
\,.\end{align}
%%%
The leptonic structure functions are now obtained as defined in \eq{def_L_i_integrated},
by contracting \eq{Lmunu_Fi_def} with the projectors $P_i^{\mu\nu}$ and performing the phase-space integrals in \eq{Lmunu_P1P2},
%%%
\begin{align} \label{eq:Li_CS_angles_bornproj}
L_i(q, \cO, \Theta)
&= \int\!\df\cos\theta\,\df\varphi\, \df P_1^2\, \df P_2^2\,\lambda(q^2, P_1^2, P_2^2)\,
L_i(q^2, \theta, \varphi, P_1^2, P_2^2)
\nn \\ & \qquad \times
\delta\bigl[\cO - \hat\cO(q, \theta, \varphi, P_1^2, P_2^2)\bigr]
\, \hat\Theta(q, \theta, \varphi, P_1^2, P_2^2)
\,,\end{align}
%%%
with the underlying leptonic structure functions given by
%%%
\begin{align} \label{eq:Li_bornproj_Fi}
L_{-1}(q^2, \theta, \varphi, P_1^2, P_2^2)
&= \frac{3}{16\pi} \Bigl[
F_+(q^2, P_1^2, P_2^2)\, g_{-1}(\theta, \varphi) + 2 F_0(q^2, P_1^2, P_2^2) \Bigr]
\,, \nn \\
L_0(q^2, \theta, \varphi, P_1^2, P_2^2)
&= \frac{3}{16\pi}
\Bigl[F_+(q^2, P_1^2, P_2^2)\, g_{0}(\theta, \varphi) + F_0(q^2, P_1^2, P_2^2) \Bigr]
\,, \nn \\
L_{1,2,5,6}(q^2, \theta, \varphi, P_1^2, P_2^2)
&= \frac{3}{16\pi}
F_+(q^2, P_1^2, P_2^2)\, g_{1,2,5,6}(\theta,\varphi)
\,, \nn \\
L_{3,4,7}(q^2, \theta, \varphi, P_1^2, P_2^2)
&= \frac{3}{16\pi}
F_-(q^2, P_1^2, P_2^2)\, g_{3,4,7}(\theta,\varphi)
\,.\end{align}
%%%
\Eqs{Li_CS_angles_bornproj}{Li_bornproj_Fi} are the generalization of
\eq{L_i_CS_angles} to an arbitrary Born-projected leptonic final state $L$.
The $(\theta, \varphi)$ dependence is still completely described by the same
$g_i(\theta, \varphi)$ in \eq{def_g_i}. The $L_i$ for $i \geq 1$ are still given
by their own respective $g_i$ times a common leptonic form factor $F_+$ for $i =
1,2,5,6$ and $F_-$ for $i = 3,4,7$. On the other hand, the angular dependence of
$L_{-1}$ and $L_0$ now gets mixed up by $F_0$, which enters with a flat
$(\theta, \varphi)$ dependence corresponding to $g_{-1} + g_0 = 2$.

If the measurements are defined in terms of massless Born leptons, then they are
also independent of $P_{1,2}^2$, such that the $P_{1,2}^2$ integrals in
\eq{Li_CS_angles_bornproj} can be performed to give $L_i(q^2, \theta, \varphi)$
that are given by the same expressions as in \eq{Li_bornproj_Fi} but in terms of
corresponding integrated
%%%
\begin{equation}
F_{\pm,0}(q^2)
= \int\! \df P_1^2\, \df P_2^2\,\lambda(q^2, P_1^2, P_2^2)\, F_{\pm,0}(q^2, P_1^2, P_2^2)
\,.\end{equation}
%%%

Removing all leptonic measurements, the inclusive $q_T$ spectrum is now given by
%%%
\begin{equation} \label{eq:incl_xsec_bornproj}
\frac{\df \sigma}{\df^4 q}
= \frac{1}{2\Ecm^2}\Bigl[F_+(q^2) + \frac{3}{2} F_0(q^2)\Bigr] W_\incl(q^2, s_{aq}, s_{bq})
\,,\qquad
W_\incl = W_{-1} + \frac{W_0}{2}
\,,\end{equation}
%%%
i.e., in terms of the \emph{same} inclusive hadronic structure function multiplied
by a generalized inclusive leptonic function.
We remind the reader that also here there is always an implicit sum over
intermediate vector bosons, $F_{\pm,0} W_i \equiv \sum_{V,V'} F_{\pm,0\, VV'} W_{i\, VV'}$.
The cross section differential in the CS angles becomes
%%%
\begin{align} \label{eq:xsec_Ai_bornproj}
\frac{\df \sigma}{\df^4 q \, \df \cos \theta \, \df \varphi}
&= \frac{1}{2\Ecm^2}\frac{3}{16\pi} \biggl[
   2F_0\, W_\incl + \sum_{i = -1}^7 F_{\pm(i)} W_i \, g_i(\theta, \varphi)
\biggr]
\\ \nn
&= \frac{1}{2\Ecm^2}\frac{3}{16\pi} \biggl[
   \Bigl(F_+\! + \frac{3}{2} F_0\Bigr)W_\incl (1 + \cos^2 \theta)
   + \frac{F_+ W_0 + F_0\, W_\incl}{2}\, \bigl( 1 - 3\cos^2 \theta \bigr)
\nn \\ & \qquad\qquad\qquad
   + \sum_{i = 1}^7 F_{\pm(i)} W_i \, g_i(\theta, \varphi)
\biggr]
\nn \\
&\equiv \frac{\df \sigma}{\df^4 q} \biggl[
   1 + \cos^2 \theta
   + \frac{\tilde A_0}{2}\, \bigl( 1 - 3\cos^2 \theta \bigr)
   + \sum_{i = 1}^7 \tilde A_i \, g_i(\theta, \varphi)
\biggr]
\nn\,,\end{align}
%%%
where we suppressed the arguments of the structure functions for brevity, and
the analogous expression also holds differential in $P_{1,2}^2$.
To make contact with \eq{xsec_Ai}, in the second step we split the flat
contribution from $F_0$ as
$(3/2)(1 + \cos^2\theta) + (1/2)(1 - 3\cos^2\theta) = 2$,
and in the last step we factored out the
inclusive cross section in \eq{incl_xsec_bornproj}, denoting the resulting
normalized coefficients of the angular dependence as $\tilde A_i$,
%%%
\begin{equation} \label{eq:tildeAi_def}
\tilde A_0 = \frac{F_+\, W_0 + F_0\, W_\incl}{\bigl(F_+ + \tfrac{3}{2} F_0\bigr)\, W_\incl}
\,, \qquad
\tilde A_{i \geq 1} = \frac{F_{\pm(i)} W_i}{\bigl(F_+ + \tfrac{3}{2} F_0\bigr)\, W_\incl}
\,.\end{equation}
%%%
These are the generalization of the $A_i$ in \eq{Ai_def} for an arbitrary
Born-projected final state. They implicitly depend on the specific Born
projection used because the CS angles $(\theta, \varphi)$ implicitly depend on it.
The $\tilde A_i$ are the angular coefficients that are measured by decomposing or
projecting the $(\theta, \varphi)$ dependence defined in terms of
Born-projected leptons. It would be interesting to precisely identify the
underlying Born projection that is effectively used in the
measurements~\cite{Khachatryan:2015paa, Aad:2016izn}.

Generically, the QED corrections to $F_+$, $F_-$, and $F_0$ will differ and
thus not cancel in \eq{tildeAi_def}. In other words, even though
Born-projected leptons admit a well-defined LO-like angular decomposition as
shown in \eq{xsec_Ai_bornproj}, the resulting $\tilde A_i$ in \eq{tildeAi_def}
still differ by QED FSR corrections from the LO $A_i$ in \eq{Ai_def}. These
corrections should be of generic $\ord{\aem}$ size, i.e., neither enhanced by
soft or collinear photon emissions nor suppressed near the $Z$ pole. In the
limit of an on-shell $Z$ boson, they would produce the QED corrections to the
$Z$ decay rate to leptons.
For $i \geq 1$, the hadronic contributions to $A_i$ and $\tilde A_i$ are the same.
As we will discuss below, $W_0$ is suppressed by $\ord{q_T^2/Q^2}$ relative to
$W_\incl$ at small $q_T$, such that at LO in QED $A_0$ vanishes like $q_T^2$ for
$q_T\to 0$. Interestingly, $\tilde A_0$ receives an additional contribution
$F_0\, W_\incl$, and therefore it no longer vanishes for $q_T \to 0$ but goes to
a calculable $\ord{\aem}$ constant.

\Refcite{Buonocore:2019puv} considered QED radiation off massive final-state
leptons, and found linear power corrections even in the inclusive case. Since
their massive leptons correspond to bare leptons, this is not entirely
surprising. It would be interesting to identify the precise source of linear
power corrections, i.e, whether the bare leptons induce linear corrections in
the leptonic tensor itself, or populate additional leptonic structure functions
that come with linearly suppressed hadronic structure functions, or both.

Finally, while most of the above discussion was phrased in terms of QED FSR
corrections to Drell-Yan, it applies to an arbitrary Born-projected final state
$L$. For example, keeping the $P_{1,2}^2$ dependence, it applies to
Drell-Yan-like electroweak diboson production $V^*\to VH$ or $V^* \to V_1 V_2$
if one remains inclusive over the decays of the final-state bosons.

%===============================================================================
\subsection{Factorization for fiducial power corrections}
\label{sec:factorization_fiducial_power_corrections}
%===============================================================================

We now investigate the structure of power corrections in the limit $q_T \ll Q$
in the presence of measurements on the leptonic final state.
To expand in $q_T\ll Q$, we introduce a formal power-counting parameter
%%%
\begin{align}
\la \sim q_T/Q
\,.\end{align}
%%%
The leptonic measurements $\hat\cO$ and $\hat \Theta$ in \eq{def_L_i_integrated}
are functions of the total four-momentum $q$ of the final state,
and admit an expansion in $\la$ as
%%%
\begin{align} \label{eq:def_observable_power_expansion}
\hat\cO(q, \Phi_L) &= \hat\cO^{(0)}(q, \Phi_L) \bigl[ 1 + \ord{\la} \bigr]
\,, \nn \\
\hat \Theta(q, \Phi_L) &= \hat \Theta^{(0)}(q, \Phi_L) \bigl[ 1 + \ord{\la} \bigr]
\,.\end{align}
%%%
We refer to the corrections in $\lambda$ in these expansions as \emph{fiducial power corrections}.
For observables that exist at Born level, e.g., cuts on the lepton momenta,
the leading-power (LP) observables $\hat\cO^{(0)}$ and $\hat \Theta^{(0)}$ are
simply obtained by taking the Born limit $q_T \to 0$.
For $q_T$-like resolution variables like $\phi^*$ that scale like $q_T$ itself and
vanish at Born level, $\hat\cO^{(0)}$ or $\hat \Theta^{(0)}$ are given by the
leading, nontrivial contribution in the $q_T\to 0$ limit.

%~~~~~~~~~~~~~~~~~~~~~~~~~~~~~~~~~~~~~~~~~~~~~~~~~~~~~~~~~~~~~~~~~~~~~~~~~~~~~~~
\subsubsection{Linear fiducial power corrections}
\label{sec:linear_fiducial_pc}
%~~~~~~~~~~~~~~~~~~~~~~~~~~~~~~~~~~~~~~~~~~~~~~~~~~~~~~~~~~~~~~~~~~~~~~~~~~~~~~~

We first assume that the leptonic measurement does not induce any additional
nontrivial dynamic scale $\pL$, such that the power expansion in
\eq{def_observable_power_expansion} is genuinely an expansion in $q_T/Q$.
We can then focus on the \emph{linear} $\ord{\lambda}$ fiducial power corrections.

Let us consider leptonic measurements that are \emph{azimuthally symmetric at
leading power}, which we will indicate by $\lpazimuthalsymmetry$ and define more
precisely in a moment.
We will show that for such measurements the only linear $\ord{\la}$ power corrections that
arise are due to the linear fiducial power corrections in \eq{def_observable_power_expansion}.
As a result, the $\ord{\la}$ power corrections can be uniquely predicted
and resummed in terms of leading-power hadronic structure functions.

For measurements that can be parameterized in terms of CS angles $\theta$, $\varphi$,
which includes our default Drell-Yan cases in \eq{L_2body}, azimuthal symmetry
means that they do not depend on $\varphi$. Azimuthal symmetry at leading power
then simply means that the LP measurements $\hat\cO^{(0)}(q, \theta)$ and
$\hat \Theta^{(0)}(q, \theta)$ are $\varphi$ independent, which implies that
they average out against $\cos(n\varphi)$ and $\sin(n\varphi)$,
%%%
\begin{align}
\lpazimuthalsymmetry \,: \qquad
&\int_0^{2\pi} \df \varphi \, e^{\img n \varphi} \, \delta[\cO - \hat\cO^{(0)}(q, \theta)]\,
\hat \Theta^{(0)}(q, \theta) = 0
\,, \qquad (n \geq 1)
\,.\end{align}
%%%
In particular, the integration against all spherical harmonics $g_i(\theta, \varphi)$
in \eqs{L_i_CS_angles}{def_g_i} vanishes, except for $i = -1, 0, 4$, which do not
depend on $\varphi$.
More generally, we \emph{define} a generic leptonic measurement as azimuthally symmetric
if it only contributes to $L_{-1, 0, 4}$, such that azimuthal symmetry at leading power
is defined by
%%%
\begin{align} \label{eq:def_lp_azimuthal_symmetry}
\lpazimuthalsymmetry \,: \qquad
L_i^\zero(q, \cO, \Theta) = 0
\,,\qquad
i \neq -1, 0, 4
\,.\end{align}
%%%
Note that this definition is also natural from the point of view of the CS tensor
decomposition in \eq{def_W_i}.
Azimuthal symmetry corresponds to symmetry under rotations of the $x$ and $y$ axes
around the $z$ axis. The projections for $i = -1, 0, 4$ are precisely
those that are invariant under azimuthal rotations (corresponding to the norm
and cross product, or are independent of $x$ and $y$), which is the physical
reason why their corresponding $g_i(\theta, \varphi)$ do not depend on $\varphi$.

A primary example is a fiducial cut on the lepton transverse
momenta $p_{T1,2} \geq \pTmin$, for which we have
%%%
\begin{align} \label{eq:pTlep_cut_lp}
\hat \Theta^{(0)}(Q, Y, \qt, \theta)
= \theta\Bigl(\pTmin \leq \frac{Q}{2} \sin \theta \Bigr)
= \hat \Theta(Q, Y, \qt = 0, \theta, \varphi)
\,.\end{align}
%%%
In words, the leptons are exactly back-to-back at leading power, and whether
they pass the cut only depends on their rest-frame energy $Q/2$ and scattering
angle $\theta$. We discuss this case as well as a cut on the lepton rapidity
in more detail in \sec{numerics_qT}.
On the other hand, angular asymmetries that are designed to project out the
$\cos \varphi$ or $\cos (2\varphi)$ dependence in the angular distribution
(by construction) do not qualify under \eq{def_lp_azimuthal_symmetry}.

Power expanding the leptonic structure functions, which includes the power
expansion of the measurement, we have
%%%
\begin{align} \label{eq:def_L_i_power_expansion}
L_i(q, \cO, \Theta) &= L_i^{(0)}(q, \cO, \Theta) + L_i^{(1)}(q, \cO, \Theta) + \dotsb
\,,\end{align}
%%%
where with the assumption in \eq{def_lp_azimuthal_symmetry} only
$L_i^{(0)}$ with $i = -1, 0, 4$ are nonzero. The $L_i^{(1)}$ contain the linear
fiducial power corrections. They can be, and in general are, nonzero for other
$i$, as our azimuthal symmetry assumption only concerns the leading-power
$L_i^\zero$.

\begin{table}
\renewcommand{\arraystretch}{1.2}
\centering
\begin{tabular}{c|c|c|c|c}
\hline\hline
$W_i$    & Scaling             & $\pm(i)$ & $g_i(\theta, \varphi)$                 & $L_i^{(0)}$ \\
\hline
$W_{-1}$ & $\sim\la^0$         & $+$ & $1 + \cos^2 \theta$                         & \checkmark \\
$W_0$    & $\sim\la^2$         & $+$ & $1 - \cos^2 \theta$                         & \checkmark \\
$W_1$    & $\sim\la^1$         & $+$ & $\sin(2\theta) \cos \varphi$                & $-$ \\
$W_2$    & $\sim\la^0$         & $+$ & $\tfrac{1}{2} \sin^2 \theta \cos(2\varphi)$ & $-$ \\
\hline
$W_3$    & $\sim\la^{\geq 1}$  & $-$ & $\sin \theta \cos \varphi$                  & $-$ \\
$W_4$    & $\sim\la^0$         & $-$ & $\cos \theta$                               & \checkmark \\
\hline
$W_5$    & $\sim\la^0$         & $+$ & $\sin^2 \theta \sin(2\varphi)$              & $-$ \\
$W_6$    & $\sim\la^{\geq 1}$  & $+$ & $\sin(2\theta) \sin \varphi$                & $-$ \\
\hline
$W_7$    & $\sim\la^{\geq 1}$  & $-$ & $\sin \theta \sin \varphi$                  & $-$ \\
\hline\hline
\end{tabular}%
\caption{Scaling of the hadronic structure functions $W_i$ in the CS tensor
decomposition in \eq{def_W_i} in the limit $\la \sim q_T/Q \ll 1$ relative to
the leading $W_{-1,4} \sim 1/q_T^2$.
In this table we generically count $\lqcd \sim q_T$.
In several cases we only derive bounds on the scaling,
where $\sim \la^{\geq m}$ means the $W_i$ is suppressed by at least $\la^m$.
We group the structure functions by parity
and whether they arise from the dispersive ($i = -1 \dots 4$) or absorptive
parts ($i = 5 \dots 7$) of the production amplitude~\cite{Hagiwara:1984hi}.
The second-to-last column shows the corresponding angular dependence
$g_i(\theta, \varphi)$ on the Collins-Soper angles for $2$-body decays,
and in the last column we indicate whether there is a nonvanishing
LP leptonic tensor $L_i^{(0)}$
for observables that are azimuthally symmetric at Born level.
}
\label{tab:power_counting_W_i}
\end{table}

We also need to power-count the hadronic structure functions $W_i$,
%%%
\begin{align}
W_i = \sum_{m=0}^\infty W_i^{(m)}
\,, \qquad
W_i^{(m)} \sim \frac{\la^m}{q_T^2}
\,.\end{align}
%%%
The $\la$ scaling of the first nonzero contributions $W_i^{(m)}$
relative to the leading-power $W_{-1,4}^{(0)} \sim 1/q_T^2$ is summarized in \tab{power_counting_W_i},
and is derived more carefully using SCET in \sec{uniqueness_linear_power_corrections}.
From \tab{power_counting_W_i}, we see that the only nonvanishing LP structure
functions $W_{i}^{(0)}$ are for $i = -1, 2, 4, 5$.
The physical reason is that at LP,
angular momentum conservation works the same way as at tree level,
i.e., as in the collision of two massless partons with $p_a + p_b = q$, $p_{Ta} = p_{Tb} = q_T = 0$. In this limit, the CS frame coincides with the leptonic frame,
and the longitudinal polarization vector is given by
%%%
\begin{align}
\epol^\mu_0 = \frac{p_a^\mu - p_b^\mu}{Q}
\qquad \text{(tree level)}
\,.\end{align}
%%%
It is easy to see that projections of the tree-level partonic matrix element onto $\epol^\mu_0$ vanish,
%%%
\begin{align} \label{eq:long_pol_vanishes_tree_level}
\epol^*_{0\mu} \mae{0}{J_{\gamma}^\mu}{q(p_a, s_a) \bar{q}(p_b, s_b)}
\propto \bar{v}^{s_b}(p_b) (p\!\!\!\slash_a - p\!\!\!\slash_b) u^{s_a}(p_a) = 0
\end{align}
%%%
for any polarization $s_a$ ($s_b$) of the quark (antiquark),
and similarly for the axial-vector current.
It follows that structure functions $W_i$ that involve contractions with $\epol^\mu_0$ vanish at tree level.
We will see in \sec{uniqueness_linear_power_corrections}
that to all orders, each contraction with $\epol^\mu_0$ is in fact penalized by at least one power of $\la$.

\begin{figure*}
\centering
\includegraphics[width=\WidthThreeSubfigs]{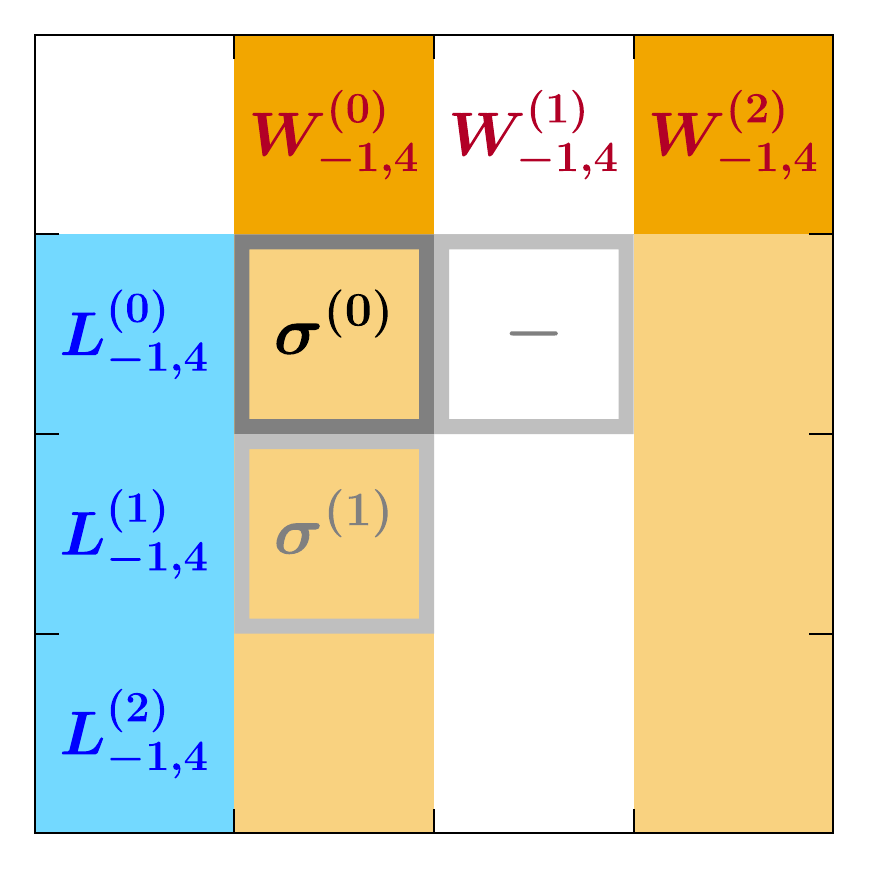}%
\includegraphics[width=\WidthThreeSubfigs]{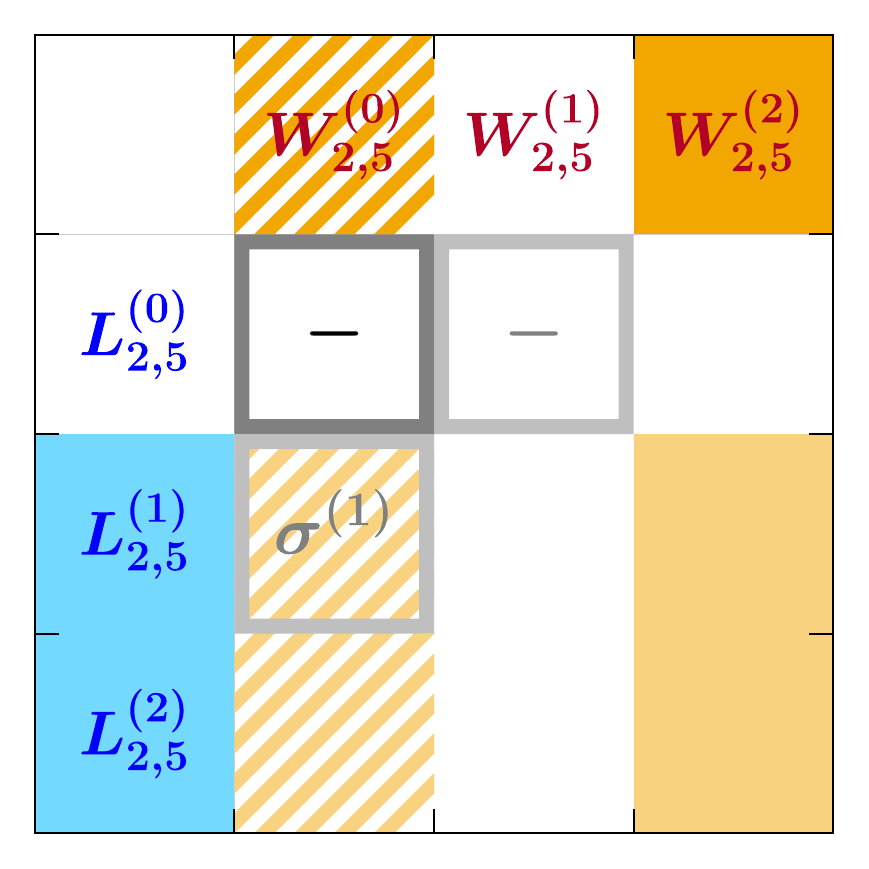}%
\\
\includegraphics[width=\WidthThreeSubfigs]{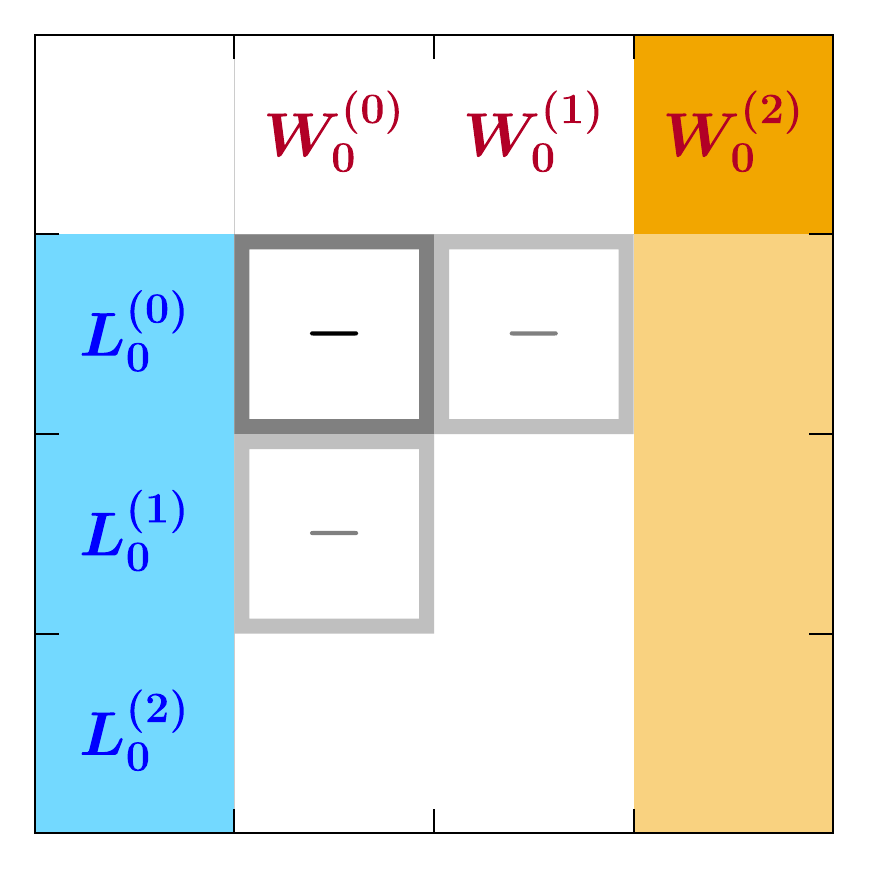}%
\includegraphics[width=\WidthThreeSubfigs]{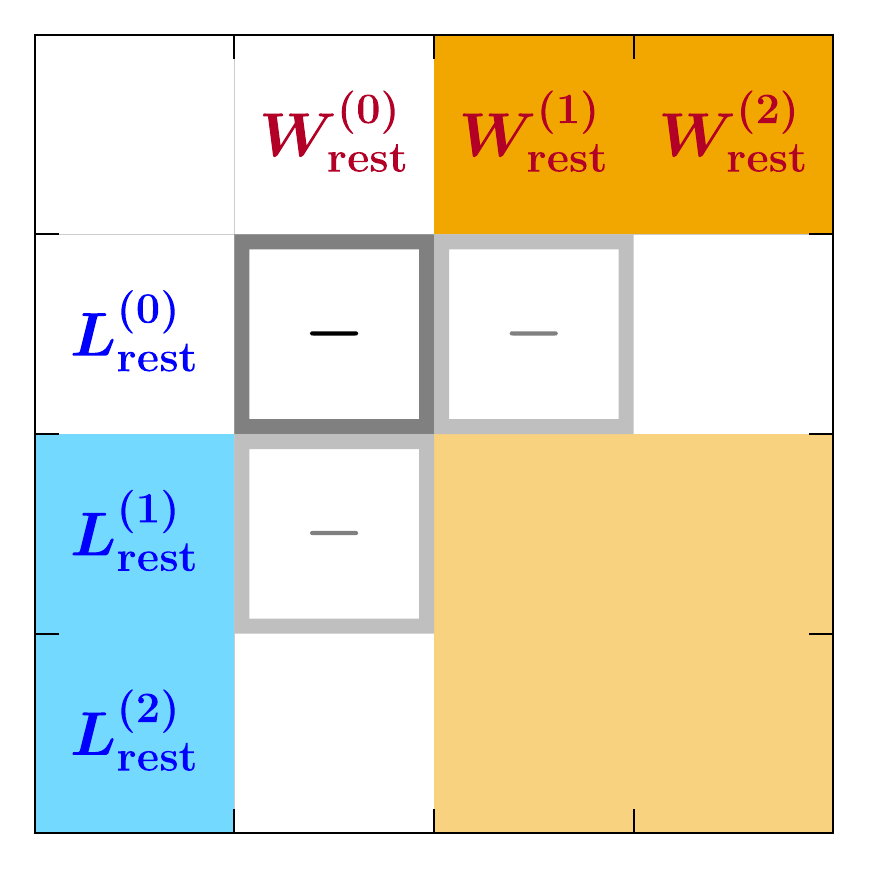}%
\caption{Power counting of hadronic structure functions and their leptonic
counterparts for $\la \sim q_T/Q \ll 1$, assuming azimuthal symmetry at leading power.
Nonzero contributions to the hadronic and leptonic tensors are indicated by
solid orange and blue filling, respectively.
Nonzero contributions to the cross section are indicated by solid light orange filling.
Hatched filling for $W_{2,5}^\zero$ indicates that this contribution does not
match onto leading-twist collinear PDFs and is suppressed for $\lqcd \ll q_T$.
Gray boxes indicate contributions to the LP and linear NLP cross sections $\sigma^{(0)}$
and $\sigma^{(1)}$. The latter solely arise through the leptonic tensor as
$L_i^\one W_i^\zero$ while all contributions of the form $L_i^\zero W_i^\one$ vanish.
Dashes indicate that a contribution vanishes.
In the bottom right panel, ``rest'' refers to structure functions $i = 1, 3, 6, 7$.
}
\label{fig:power_counting}
\end{figure*}

Suppressing the arguments of $L_i$ and $W_i$, the strict LP cross section is given by
%%%
\begin{align} \label{eq:xsec_stric_lp}
\lpazimuthalsymmetry \,: \qquad
\frac{\df\sigma^{(0)}(\Theta)}{\df^4 q \, \df \cO}
&= \frac{1}{2\Ecm^2} \sum_{i = -1, 4} L_i^{(0)} \, W_i^{(0)}
\,.\end{align}
%%%
The $i = 2, 5$ contribution does not survive
because $L_{2,5}^{(0)} = 0$ due to \eq{def_lp_azimuthal_symmetry},
and the nonzero $L_0^{(0)}$ does not contribute because $W_0^\zero = 0$.

Next, the linear $\ord{\la}$ power corrections to the cross section are given by
%%%
\begin{align} \label{eq:factorization_sigma1_preliminary}
\frac{\df\sigma^{(1)}(\Theta)}{\df^4 q \, \df \cO}
&= \frac{1}{2\Ecm^2} \sum_{i = -1}^7 \bigl[ L_{i}^{(1)} \, W_{i}^{(0)} + L_i^{(0)} \, W_i^{(1)} \bigr]
\,.\end{align}
%%%
In the first term, only $i = -1, 2, 4, 5$ contribute to the sum due to
\tab{power_counting_W_i}. For the second term, assuming LP azimuthal symmetry,
only $i = -1, 0, 4$ contribute.
From \tab{power_counting_W_i}, $W_0 \sim \ord{\la^2}$, and as we will argue in
\sec{uniqueness_linear_power_corrections}, all power
corrections to $W_{-1,4}$ are quadratic in $\lambda$, so we have
%%%
\begin{align} \label{eq:W_m1_4_linear_vanish}
W_{-1}^{(1)} = 0
\,, \qquad
W_{0}^{(1)} = 0
\,, \qquad
W_{4}^{(1)} = 0
\,.\end{align}
%%%
For $W_{-1}$ and $W_0$, this statement is equivalent to the absence of linear power corrections
in the inclusive cross section $\propto W_{-1} + \frac{1}{2}W_0$. For $W_4$, it is
equivalent to the absence of linear power corrections in the inclusive forward-backward
asymmetry.
Hence, the second
term in \eq{factorization_sigma1_preliminary} vanishes, and we arrive at
%%%
\begin{align} \label{eq:factorization_sigma1_final_result}
\lpazimuthalsymmetry \,: \qquad
\frac{\df\sigma^{(1)}(\Theta)}{\df^4 q \, \df \cO}
&= \frac{1}{2\Ecm^2} \sum_{i = -1, 2, 4, 5} L_{i}^{(1)} \, W_{i}^{(0)}
\,.\end{align}
%%%
We have thus shown that for leptonic measurements that are azimuthally symmetric at
leading power, \emph{all linear} power corrections uniquely arise from \emph{linear
fiducial} power corrections $L_i^\one$ multiplying the leading-power hadronic
structure functions $W_{-1,2,4,5}^\zero$.
The power-counting logic leading to \eq{factorization_sigma1_final_result} is
illustrated in \fig{power_counting}.

%~~~~~~~~~~~~~~~~~~~~~~~~~~~~~~~~~~~~~~~~~~~~~~~~~~~~~~~~~~~~~~~~~~~~~~~~~~~~~~~
\subsubsection{Leptonic fiducial power corrections}
\label{sec:leptonic_fiducial_pc}
%~~~~~~~~~~~~~~~~~~~~~~~~~~~~~~~~~~~~~~~~~~~~~~~~~~~~~~~~~~~~~~~~~~~~~~~~~~~~~~~

We now turn to leptonic fiducial power corrections that arise from the presence
of an additional, physical scale $\pL$ induced by the leptonic
measurement. In this case, the power expansion of the measurements a priori
receives power corrections in both $q_T/Q$ and $q_T/\pL$,
%%%
\begin{align}
\hat\cO(q, \Phi_L) &= \hat\cO^{(0)}(q, \Phi_L) \Bigl[ 1 + \ORd{\frac{q_T}{Q}, \frac{q_T}{\pL}} \Bigr]
\,, \nn \\
\hat \Theta(q, \Phi_L) &= \hat \Theta^{(0)}(q, \Phi_L) \Bigl[ 1 + \ORd{\frac{q_T}{Q}, \frac{q_T}{\pL}} \Bigr]
\,.\end{align}
%%%
The case of linear fiducial power corrections discussed above corresponds to
$p_L \sim Q$. We refer to the $q_T/p_L$ corrections as \emph{leptonic} fiducial
power corrections.
For $q_T \ll p_L \ll Q$, they become enhanced compared to the $q_T/Q$ corrections
and for $q_T \sim p_L$ they become $\ord{1}$ and cause the naive expansion in
$q_T$ to break down.

Generically this happens when the leptonic measurement is close to an edge
of Born phase space that is sensitive to additional radiation, such that
a nonzero $q_T$ opens up new phase space beyond the Born edge, with $p_L \sim q_T$
parametrizing the distance from the Born edge.
We will demonstrate this effect in detail in \sec{numerics_pTlep} for the
important example of the $\pTlep$ spectrum near the Jacobian peak $\pTlep \sim
Q/2$, in which case $p_L = Q - 2\pTlep$.

To expand in such regions, it is necessary to count both
%%%
\begin{equation} \label{eq:pL_pc}
\frac{q_T}{Q} \sim \frac{p_L}{Q} \sim \lambda
\,,\end{equation}
%%%
which explicitly avoids expanding in $q_T/p_L \sim \ord{1}$ and thereby retains
all leptonic power corrections exactly to all powers. Expanding the leptonic
measurements in this limit
%%%
\begin{align} \label{eq:pL_LP_meas}
\hat\cO(q, \Phi_L) &= \hat\cO^{(0)}(q, \Phi_L; q_T/p_L) \Bigl[ 1 + \ORd{\frac{q_T}{Q}, \frac{p_L}{Q}} \Bigr]
\,, \nn \\
\hat \Theta(q, \Phi_L) &= \hat \Theta^{(0)}(q, \Phi_L; q_T/p_L) \Bigl[ 1 + \ORd{\frac{q_T}{Q}, \frac{p_L}{Q}} \Bigr]
\,,\end{align}
%%%
where with a slight abuse of notation the superscript $\zero$ now refers to the
leading-power term in $\lambda$ with the modified power counting in \eq{pL_pc},
and the $q_T/p_L$ argument is meant to remind us that we have not expanded in
this ratio.

The cross section at leading power in $\lambda$ in this limit is given by
%%%
\begin{align} \label{eq:xsec_strict_lp_leptonic_pcs}
\frac{\df\sigma(\Theta)}{\df^4 q \, \df \cO}
&= \frac{1}{2\Ecm^2} \sum_{i = -1, 2, 4, 5} L_i^{(0)}(q_T/p_L) \, W_i^{(0)}
\Bigl[ 1 + \ORd{\frac{q_T}{Q}, \frac{p_L}{Q}} \Bigr]
\,,\end{align}
%%%
where the $L_i^\zero(q_T/p_L)$ arise from the modified LP leptonic measurements
in \eq{pL_LP_meas}. The hadronic tensor does not know anything about $p_L$, and
so its power expansion is unaffected by \eq{pL_pc}. However, as we are now
keeping some terms in the leptonic measurement that we would otherwise drop in
the strict $q_T\to 0$ limit, the azimuthal symmetry we might have in the strict
$q_T\to 0$ limit is typically lost now, and so we do not require it. As a
result, also the $W_{2,5}$ contribute at LP in \eq{xsec_strict_lp_leptonic_pcs}.
Furthermore, there are now generically linear power corrections to
\eq{xsec_strict_lp_leptonic_pcs} from both $L_i^\one$ and $W_i^\one$.

%~~~~~~~~~~~~~~~~~~~~~~~~~~~~~~~~~~~~~~~~~~~~~~~~~~~~~~~~~~~~~~~~~~~~~~~~~~~~~~~
\subsubsection{Generic fiducial power corrections}
\label{sec:generic_fiducial_pc}
%~~~~~~~~~~~~~~~~~~~~~~~~~~~~~~~~~~~~~~~~~~~~~~~~~~~~~~~~~~~~~~~~~~~~~~~~~~~~~~~

Since \eq{xsec_strict_lp_leptonic_pcs} relies on counting $p_L/Q\sim\lambda$ it
is not valid for $p_L \sim Q$. Hence, to cover the full leptonic phase space, we
have to satisfy two competing conditions from the different regions. For $q_T
\ll p_L \sim Q$ we must not expand in $p_L/Q$, while for $q_T \sim p_L \ll Q$ we
must count $q_T \sim p_L$ to avoid uncontrolled power corrections in $q_T/p_L$.
The natural way to satisfy both requirements is to expand the leptonic
measurements neither in $q_T$ nor $p_L$ and thus keep the exact leptonic tensor,
%%%
\begin{empheq}[box=\widefbox]{equation} \label{eq:def_sigma0pL}
\frac{\df\sigma^{(0+L)}(\Theta)}{\df^4 q \, \df \cO}
\equiv \frac{1}{2\Ecm^2} \sum_{i = -1, 2, 4, 5} L_i(q, \cO, \Theta) \,
W_i^{(0)}(q^2, s_{aq}, s_{bq})
\,.\end{empheq}
%%%
Of course, we still need to expand the hadronic tensor in $q_T/Q\sim\lambda$,
and all four LP hadronic structure functions in principle contribute.
For $q_T \ll p_L \sim Q$, \eq{def_sigma0pL} obviously captures the linear
power corrections as in \eq{factorization_sigma1_final_result}, while
for $q_T \sim p_L \ll Q$ it captures as required all leptonic power corrections
as in \eq{xsec_strict_lp_leptonic_pcs}. In the following, we always use the
notation $\df\sigma^{(0+L)}$ to denote the inclusion of the exact leptonic
tensor as in \eq{def_sigma0pL}.

\Eq{def_sigma0pL} is our final master formula. By treating the leptonic tensor
exactly, it in fact incorporates \emph{all} fiducial power corrections that
multiply the leading-power hadronic structure functions. The leptonic tensor
does not produce small-$q_T$ logarithms, which solely arise from the hadronic
tensor. Therefore, \eq{def_sigma0pL} automatically resums all logarithms in
fiducial power corrections to the same order as the resummation is included for
the hadronic tensor. All further power corrections to $\df \sigma^{(0+L)}$ are
obtained by working to subleading power in the hadronic structure functions, and
arise purely from subleading-power QCD dynamics.

One might argue that we could have immediately kept the leptonic tensor exact
from the start, just because there is no reason or benefit to expanding it, and
so one should not. On the other hand, one might argue that by doing so one keeps
a seemingly arbitrary set of power corrections in the cross section, and there
is a priori no guarantee that doing so would make things better and not worse,
and so one should expand the leptonic tensor in order to have a consistent power
expansion for the cross section at each order in the power expansion. Both
arguments are found in the literature.

Our analysis provides several formal justifications for keeping the exact
leptonic tensor. First, for the common case of leptonic measurements that are
azimuthally symmetric at Born level (and generic $p_L \sim Q$), it uniquely
predicts all linear $\ord{\la}$ next-to-leading power corrections in the cross
section. In other words, in this limit the retained power corrections are not
arbitrary but provide an unambiguous, systematic improvement in the power
expansion of the cross section, including their logarithmic resummation.
Second, it retains all leptonic power corrections, which as argued is mandatory
to correctly obtain the actual \emph{leading-power} result for $q_T \sim p_L$.
Third, the actual regions and relevant scales $p_L$ depend on the leptonic
measurement and identifying them can be quite involved. Keeping the exact
leptonic tensor is by far the simplest (and perhaps only sensible) way to
guarantee that all such possible regions are correctly treated. Finally, it
ensures that any in-between regions $q_T \ll p_L \ll Q$ are smoothly covered.

The LP hadronic structure functions entering in \eq{def_sigma0pL} are given by
%%%
\begin{align} \label{eq:tmd_factorization_m1_2_4_5}
W_i^{(0)}
&= \sum_{a,b} H_{i\,ab}(Q^2, \mu) \,
\begin{cases}
[B_a B_b S](Q^2, x_a, x_b, \qt, \mu)
\,, & i = -1, 4
\\
[h_{1a}^{\perp} h_{1b}^{\perp}](Q^2, x_a, x_b, \qt, \mu)
\,, & i = 2, 5
\,.\end{cases}
\end{align}
%%%
The cases $i = -1, 4$ are a straightforward generalization of the standard
inclusive factorization theorem in
\eqs{tmd_factorization_incl}{tmd_factorization}, where $B_{a,b}$ and $S$ are the
same beam and soft functions, and only the hard functions $H_{i\,ab}$ depend on
the projection $i$. They are collected in \app{hard_functions}.

The $W_{2,5}^{(0)}$ contribution, which corresponds to the $\cos(2\varphi)$ and
$\sin(2\varphi)$ angular modulations of the cross section, are proportional to a
(weighted) convolution of two Boer-Mulders functions $h_{1}^{\perp}$ in the
transverse plane~\cite{Boer:1999mm, Boer:2002ju, Wang:2018naw}, where
$h_{1a}^{\perp}$ measures the net transverse polarization of flavor $a$,
longitudinal momentum fraction $x$, and given transverse momentum $\vec{k}_T$
within an unpolarized proton~\cite{Boer:1997nt}. It does not match onto leading
twist-2 collinear PDFs, i.e., for $\lqcd \ll q_T$, each $h_1^{\perp}$ is
suppressed by at least one power of $\lqcd/q_T$~\cite{Bacchetta:2008xw} relative
to the leading-power beam functions $B_a$ in $W_{-1,4}$, which do match onto
leading twist-2 PDFs. The matching of $h_1^{\perp}$ onto subleading twist-3 PDFs
was carried out in \refcite{Scimemi:2018mmi}. On the other hand, the first
contribution to $W_{2,5}$ that does match onto leading twist-2 PDFs is
suppressed by $q_T^2/Q^2$ relative to $W_{-1,4}$~\cite{Boer:2006eq}. For these
reasons, we will neglect the $i = 2, 5$ contributions in our numerical results.
However, it should be stressed that for $q_T \sim \lqcd$ they do become formally
leading-power contributions.

Perturbatively, \eqs{def_sigma0pL}{tmd_factorization_m1_2_4_5} allow us
to resum fiducial power corrections to the same order
to which the LP hadronic structure functions are known.
We stress that due to the different sum over flavors with different weights $H_{i\,ab}$,
the resummation effects do not in general cancel in the ratio $W_4^{(0)}/W_{-1}^{(0)}$.
This is relevant when computing the angular coefficient $A_4$,
corresponding to the forward-backward asymmetry, at small $q_T$.

%===============================================================================
\subsection{Uniqueness of linear power corrections}
\label{sec:uniqueness_linear_power_corrections}
%===============================================================================

There are several loose ends in the nontechnical discussion of the previous subsection
that we now tie up to establish that \eq{factorization_sigma1_final_result}
uniquely and unambiguously captures all linear power corrections for
LP-azimuthally symmetric observables.
%%%
\begin{enumerate}
\item
In \sec{pc_Wi}, we derive the power counting of the hadronic structure functions in \tab{power_counting_W_i}.
\item
In \sec{Ola_vanishes}, we argue that power corrections to $W_{-1,4}^{(0)}$ are quadratic,
such that \eq{W_m1_4_linear_vanish} holds.
\item
In \sec{frame_choice}, we show explicitly that the linear power corrections in
\eq{factorization_sigma1_final_result} are unique, i.e., switching to a
different basis induces only quadratic power corrections.
\end{enumerate}
%%%

%~~~~~~~~~~~~~~~~~~~~~~~~~~~~~~~~~~~~~~~~~~~~~~~~~~~~~~~~~~~~~~~~~~~~~~~~~~~~~~~
\subsubsection{Power counting hadronic structure functions}
\label{sec:pc_Wi}
%~~~~~~~~~~~~~~~~~~~~~~~~~~~~~~~~~~~~~~~~~~~~~~~~~~~~~~~~~~~~~~~~~~~~~~~~~~~~~~~

To derive the $\lambda$ scaling of the $W_i$ in \tab{power_counting_W_i},
we use SCET~\cite{Bauer:2000ew, Bauer:2000yr, Bauer:2001ct, Bauer:2001yt, Bauer:2002nz},
which is the effective theory of QCD in the limit $\la \ll 1$ and
provides a systematic expansion of QCD in powers of $\lambda$.
In SCET, each collinear sector is parametrized by two lightlike reference
vectors $n_i^\mu$ and $\bn_i^\mu$, where $n_i \cdot \bn_i = 2$.
In our case, the relevant collinear sectors describe the incoming hadrons,
the colliding partons and collinear radiation off of them.
We choose the reference vectors along the beam axis in the leptonic frame as
%%%
\begin{align} \label{eq:def_n_a_b}
n_a^\mu = (1, 0, 0, 1)_\lep
\,,\qquad
n_b^\mu = (1, 0, 0, -1)_\lep
\,, \qquad
\bn_a^\mu = n_b^\mu
\,, \qquad
\bn_b^\mu = n_a^\mu
\,,\end{align}
%%%
in terms of which any four-vector $p^\mu$ can be decomposed as
%%%
\begin{align} \label{eq:def_lc_coordinates}
p^\mu = (n_b \cdot p)\, \frac{n_a^\mu}{2} + (n_a \cdot p)\, \frac{n_b^\mu}{2} + p_\perp^\mu
\equiv p^- \frac{n_a^\mu}{2}  + p^+ \frac{n_b^\mu}{2}  + p_\perp^\mu
\,.\end{align}
%%%
With this choice, the space perpendicular to $n_{a,b}$ coincides with the
transverse plane. It is useful to define the transverse metric and antisymmetric tensor,
%%%
\begin{align}
g_\perp^{\mu\nu} = g^{\mu\nu} - \frac{n_a^\mu n_b^\nu + n_a^\nu n_b^\mu}{2}
\,, \qquad
\eps_\perp^{\mu\nu} = \frac{1}{2} {\eps^{\mu\nu}}_{\rho\sigma} \, n_a^\rho n_b^\sigma
\,.\end{align}
%%%
In addition, we need to distinguish a direction in the transverse plane,
which we take to be
%%%
\begin{align} \label{eq:def_n_perp}
n_\perp^\mu = \frac{q_\perp^\mu}{(-q_\perp^2)^{1/2}} = (0, 1, 0, 0)_\lep
\,, \qquad
n_\perp^2 = -1
\,, \qquad
n_\perp \cdot n_a = n_\perp \cdot n_b = 0
\,,\end{align}
%%%
where $q_\perp^\mu \equiv g^{\mu\nu}_\perp q_\nu$,
and we remind the reader that we aligned the $x$ axis in the leptonic frame with
the transverse component of $q^\mu$.

To discuss the power counting of the hadronic structure functions in SCET,
we first write $t^\mu$, $x^\mu$, $z^\mu$ in terms of $n_{a,b}^\mu$ and $n_\perp^\mu$.
From their explicit expressions in the leptonic frame in \eqs{z}{x}, we have
%%%
\begin{alignat}{3} \label{eq:CS_frame_na_nb_nperp}
t^\mu &= \gamma \frac{n_a^\mu + n_b^\mu}{2} + \eps \, n^\mu_\perp
\,, \qquad &
y^\mu &= \eps_\perp^{\mu\nu} n_{\perp\nu}
\,, \nn \\
x^\mu &= \eps \frac{n_a^\mu + n_b^\mu}{2} + \gamma \, n_\perp^\mu
\,, \qquad &
z^\mu &= \frac{n_a^\mu - n_b^\mu}{2}
\,,\end{alignat}
%%%
where as before $\eps = q_T / Q \sim \la$ and $\gamma = \sqrt{1 + \eps^2} = 1 + \ord{\la^2}$.
It is straightforward to expand \eq{CS_frame_na_nb_nperp} in $\la$,
%%%
\begin{alignat}{9} \label{eq:CS_frame_power_series}
t^\mu &= \frac{n_a^\mu + n_b^\mu}{2} + \eps \, n_\perp^\mu + \ord{\la^2}
\,, \qquad &
y^\mu &= \eps_\perp^{\mu\nu} n_{\perp \nu}
\,, \nn \\
x^\mu &= n_\perp^\mu + \eps \frac{n_a^\mu + n_b^\mu}{2} + \ord{\la^2}
\,, \qquad &
z^\mu &= \frac{n_a^\mu - n_b^\mu}{2}
\,.\end{alignat}
%%%
Note that the relations for $y^\mu$ and $z^\mu$ are exact and do not receive power corrections, which is a direct consequence of the symmetry we imposed on $z^\mu$.
The simple form of \eq{CS_frame_power_series} motivates our choice of $n_{a,b}$ in the leptonic frame in \eq{def_n_a_b}.
If we had chosen $n_{a,b}$ as $(1, 0, 0, \pm 1)_\lab$ in the lab frame instead,
there would be additional factors of $e^{\pm Y}$ in \eq{CS_frame_power_series}.

The power counting of the hadronic structure functions is determined by the
order in $\lambda$ at which contractions of \eq{CS_frame_power_series} with the hadronic
current are populated when expanding the hadronic currents $J_V^\mu$
in \eqs{hadronic_current_Z_gamma}{hadronic_current_W} in an explicit power
expansion in $\lambda$ in terms of the corresponding SCET currents. Schematically,
%%%
\begin{align} \label{eq:hard_matching}
J_V^\mu = \sum_{k \geq 0} J_V^{(k)\mu} = \sum_{k \geq 0} \sum_m C^{(k)\mu}_m O^{(k)}_m
\,,\end{align}
%%%
where the $\ord{\lambda^k}$ current $J_V^{(k)\mu}$ is a combination of
$\ord{1}$ matching coefficients $C^{(k)\mu}_m$
and SCET hard-scattering operators $O^{(k)}_m$
that have an explicit power suppression of $\ord{\la^k}$ relative to the leading $k = 0$
term.

At leading power, $\ord{\lambda^0}$, the hard matching takes the form~\cite{Stewart:2009yx}
%%%
\begin{align} \label{eq:hard_matching_lp}
J_V^{(0)\mu}(x) &= \sum_{n_1, n_2} \int \! \df \w_1 \, \df \w_2 \, e^{-\img(\w_1 n_1 \cdot x + \w_2 n_2 \cdot x)/2}
\nn \\ & \qquad \times
\sum_{q,q'} C^{(0)\mu\,\alpha \beta}_{V\, q\bar q'}(n_1, n_2; \w_1, \w_2) \,
O^{(0) \alpha \beta}_{q\bar q'}(n_1, n_2; \w_1, \w_2; x)
\,,\end{align}
%%%
where $x$ is the current's spacetime position,
$\alpha, \beta$ are spinor indices, and the second sum runs
over the flavor labels $q, q'$ carried by the SCET operator,
which in general are distinct from the flavors coupling directly to the vector boson.
The $\w_{1,2}$ and $n_{1,2}$ are (at this point arbitrary) large label momenta and directions.
The leading-power hard-scattering operator $O^{(0)\alpha \beta}_{q\bar q'}$ reads
%%%
\begin{align} \label{eq:Oqq_def}
O^{(0)\alpha \beta}_{q\bar q'}(n_1, n_2; \w_1, \w_2; x) = \bar \chi^{\alpha}_{q\,n_1, -\w_1}(x) \, \chi^{\beta}_{q'\,n_2, \w_2}(x)
\,,\end{align}
%%%
where $\chi_{q\,n, \w}(x)$ is an $n$-collinear field with total label momentum $\w$,
with color indices implicit. It is defined as
%%%
\begin{equation} \label{eq:chin_def}
\chi_{q\,n,\omega}(x)
= \bigl[\delta(\omega - \bn \cdot \cP)\, W_n^\dagger(x) \, \xi_{q\,n}(x) \bigr]
\,,\end{equation}
%%%
where $\xi_{q\,n}$ is an $n$-collinear quark field with flavor $q$,
and $W_n$ is a Wilson line constructed from $n$-collinear gluons, such that the product
$W_n^\dagger \xi_n$ is invariant under $n$-collinear gauge transformations. The
$\delta$ function picks out the total large label momentum $\w$. The sign conventions for
$\omega$ in \eqs{Oqq_def}{chin_def} are chosen such $\w_{1,2} >0 $ for incoming
particles~\cite{Stewart:2009yx}.

In principle, \eq{hard_matching_lp} also receives a contribution from a
corresponding leading-power $O^\zero_{gg}$ operator, whose matching coefficient is
proportional to $q^\mu$~\cite{Stewart:2009yx}. It precisely captures the
non-conserved part of the current, see \eq{JZ_noncons} and the discussion below it.
Since it does not contribute to Drell-Yan for massless leptons, we do not
consider it here.

When evaluating proton matrix elements of \eq{hard_matching_lp},
momentum conservation requires $n_{2,1} = n_{a,b}$ and $\w_{2,1} = \w_{a,b}$
in the case where parton $a$ is a quark.
Making use of these identifications and the fact that $\w_a \w_b = q^2$,
the hard matching coefficients for $V=\gamma,Z,W$ are given by~\cite{Stewart:2009yx, Moult:2015aoa}
\begin{align} \label{eq:wilson_coeff_hard_matching_lp}
C^{(0)\mu\,\alpha \beta}_{\gamma\, q\bar q'}(n_b, n_a; \w_b, \w_a)
&= \delta_{qq'} |e| (\gamma_\perp^\mu)^{\alpha \beta}
\Bigl[  Q_q\, C_q(q^2) + \sum_{f} Q_f \, C_{v f}(q^2) \Bigr]
%%%
\,,\nn\\
%%%
C^{(0)\mu\,\alpha \beta}_{Z\, q\bar q'}(n_b, n_a; \w_b, \w_a)
&= \delta_{qq'} (-|e|) \biggl\{
\bigl[ \gamma_\perp^\mu(v_q - a_q \gamma_5\bigr)\bigr]^{\alpha\beta}
C_q(q^2)
\nn\\&\qquad\qquad\qquad
 + \sum_{f} \Bigl[  (\gamma_\perp^\mu)^{\alpha \beta} v_f \, C_{v f}(q^2)
                  - \bigl(\gamma_\perp^\mu \gamma_5 \bigr)^{\alpha\beta} a_f \, C_{a f}(q^2) \Bigr]\biggr\}
%%%
\,,\nn\\
%%%
C^{(0)\mu\,\alpha \beta}_{W^+ q\bar q'}(n_b, n_a; \w_b, \w_a)
&= -\frac{|e| V_{qq'}}{2\sqrt2 \sin\theta_w}\, \bigl[ \gamma_\perp^\mu(1-\gamma_5)
\bigr]^{\alpha \beta} C_q(q^2)
%%%
\,,\nn\\
%%%
C^{(0)\mu\,\alpha \beta}_{W^- q\bar q'}(n_b, n_a; \w_b, \w_a)
&= -\frac{|e| V_{q'q}^*}{2\sqrt2 \sin\theta_w}\, \bigl[ \gamma_\perp^\mu(1-\gamma_5)
\bigr]^{\alpha \beta} C_q(q^2)
\,,\end{align}
%%%
where the vector and axial-vector contributions have the same flavor-diagonal matching coefficient $C_q(q^2)$
because massless QCD preserves chirality, but in general have different singlet coefficients $C_{vf}(q^2)$ and $C_{af}(q^2)$.
The latter arise from closed quark loops coupling to the vector boson, and thus involve an electroweak coupling different from the external quark flavors.
Here, $V_{qq'}$ is the CKM-matrix element for $q \in \{u, c, t\}$ and $q' \in \{d, s, b\}$ (and we take it to vanish in all other cases).

Importantly, the spin structure of the leading-power hard matching coefficient is proportional to $\gamma_\perp^\mu = g_\perp^{\mu\nu} \gamma_\nu$,
and therefore satisfies
%%%
\begin{align} \label{eq:na_nb_C}
n_{a} \cdot C^{(0)\alpha \beta}_{V\, q\bar q'}(n_b, n_a; \w_b, \w_a)
= n_{b} \cdot C^{(0)\alpha \beta}_{V\, q\bar q'}(n_b, n_a; \w_b, \w_a) = 0
\,.\end{align}
%%%
Using \eq{CS_frame_power_series}, it is easy to see that contractions
with the longitudinal polarization vector $\epol^\mu_0 = z^\mu$ vanish to all orders at the level of the amplitude,
%%%
\begin{align} \label{eq:long_pol_vanishes_lp}
\epol^*_{0\mu} \, \mae{X}{J_V^{(0)\mu}}{pp} = 0
\,.\end{align}
%%%
This is the all-order analogue of \eq{long_pol_vanishes_tree_level} in the limit $\la \ll 1$.
It follows that projections onto $\epol^\mu_0$ in \eq{def_W_i}
are only populated by matrix elements of
the subleading-power currents $J_V^{(j)\mu}$ with $j \geq 1$ in \eq{hard_matching},
and are penalized by at least one power of $\la$.
This implies that only $W_{-1, 2, 4, 5}$, which do not involve longitudinal polarizations,
can scale as $\ord{\la^0}$,
while $W_{1, 3, 6, 7}$ are suppressed at least by $\ord{\la}$,
and $W_0 = 2W_{00}$ is suppressed by at least $\ord{\la^2}$.
This completes the derivation of \tab{power_counting_W_i}.

For $W_{-1, 2, 4, 5}$, our power-counting argument
agrees with the well-known scaling of the leading contributions given by \eq{tmd_factorization_m1_2_4_5},
while for $W_0$ and $W_1$ it reproduces the known scaling at fixed $\ord{\as}$~\cite{Boer:2006eq}.
For the remaining $W_{3,6,7}$, our argument provides a lower bound on the degree of power suppression.
To our knowledge, this is the first time that the scaling of $W_{3,6,7}$
at small $q_T$ has been explicitly considered for generic currents.

We also point out that starting from \eq{CS_frame_power_series},
it is straightforward to identify the subleading-power SCET currents
that populate a given $W_i$.
For example, $W_{1, 3, 6, 7}$ can only receive their leading contributions
from the interference
of $J_V^{(1)\mu}$ with the leading-power current $J_V^{(0)\mu}$,
while the leading contribution to $W_0$ must arise from the interference of $J_V^{(1)\mu}$ with itself due to \eq{long_pol_vanishes_lp}.
The hard-scattering operators to $\ord{\la^2}$ relevant for color-singlet
production have been constructed in \refscite{Moult:2017rpl, Feige:2017zci, Chang:2017atu} using the approach of helicity operators~\cite{Moult:2015aoa, Kolodrubetz:2016uim},
and the list of operators contributing to $J_V^{(1)\mu}$ is fairly short.
Due to the explicit power suppression from the current, it should be possible
to derive factorization theorems for these $W_i$ in the $q_T \ll Q$ limit using
SCET. This would be relevant e.g.\ to understand the degree
to which resummation effects are universal between $W_i$ and $W_{-1}$,
and hence to what extent they cancel in predictions for the angular coefficients $A_i$.
A conjecture for the factorization of $W_1$ at small $q_T$ was given in \refcite{Bacchetta:2019qkv}, and it would be interesting to analyze it using the systematic organization of subleading operators in SCET.

%~~~~~~~~~~~~~~~~~~~~~~~~~~~~~~~~~~~~~~~~~~~~~~~~~~~~~~~~~~~~~~~~~~~~~~~~~~~~~~~
\subsubsection{Vanishing \texorpdfstring{$\ord{\la}$}{O(lambda)} corrections in \texorpdfstring{$W_{-1,4}$}{Wm14}}
\label{sec:Ola_vanishes}
%~~~~~~~~~~~~~~~~~~~~~~~~~~~~~~~~~~~~~~~~~~~~~~~~~~~~~~~~~~~~~~~~~~~~~~~~~~~~~~~

We next discuss the absence of linear power corrections in $W_{-1,4}$,
cf.\ \eq{W_m1_4_linear_vanish}. The projectors defining $W_{-1,4}$ involve $x^\mu$,
which in principle receives a linear power correction, see \eq{CS_frame_power_series}.
However, this $\ord{\la}$ correction is proportional to $n_a + n_b$
and thus orthogonal to the leading-power SCET current in \eq{hard_matching_lp}
due to \eq{na_nb_C}, similar to the longitudinal polarization vector
discussed above. We therefore have up to quadratic power corrections,
%%%
\begin{align} \label{eq:W_m1_4_simplified_projections}
W_{-1} = - g_{\perp\mu\nu} W^{\mu\nu} \bigl[ 1 + \ord{\la^2} \bigr]
\,, \qquad
W_{4} = 2 \img \eps_{\perp \mu\nu} W^{\mu\nu} \bigl[ 1 + \ord{\la^2} \bigr]
\,.\end{align}
%%%
The question then reduces to why $- g_{\perp\mu\nu} W^{\mu\nu}$ and $2 \img \eps_{\perp \mu\nu} W^{\mu\nu}$
do not receive linear power corrections relative to the contribution
from the squared LP current.

It is well known that for $e^+ e^- \to \text{dijets}$ event shapes such as thrust,
the leading $\ord{\la}$ corrections vanish~\cite{Lee:2004ja, Beneke:2003pa, Freedman:2013vya, Feige:2017zci, Moult:2019mog}.
The explicit proof in \refscite{Feige:2017zci, Moult:2019mog} relies on invariance under rotations about the axis
defined by the lightlike directions that parametrize the collinear sectors for the outgoing jets.
The analogous statement here is that $- g_{\perp\mu\nu} W^{\mu\nu}$ and $2 \img \eps_{\perp \mu\nu} W^{\mu\nu}$
are indeed invariant under rotations about $n_a^\mu - n_b^\mu$.
To see that this implies the absence of linear power corrections,
we discuss the possible sources of power corrections in turn:

\begin{enumerate}
\item
Subleading hard-scattering operators were shown not to contribute
to the thrust spectrum at $\ord{\la}$ in \refcite{Feige:2017zci}, using the rotational symmetry.
While thrust is described by SCET$_\I$ and $q_T$ is a SCET$_\II$ observable,
the operator basis involving only collinear fields is identical and has manifest crossing symmetry,
so for these contributions the argument carries over to the case of Drell-Yan.

\item On the other hand, contributions involving soft fields differ between
SCET$_\I$ and SCET$_\II$, and occur both through subleading hard-scattering
operators and subleading Lagrangian insertions.  For SCET$_\I$-like event shapes
the vanishing of such terms at $\ord{\la}$ was demonstrated
in~\refscite{Feige:2017zci, Moult:2019mog}. The analysis of $\ord{\la}$ terms in
SCET$_\II$ is more difficult due to the non-locality of the theory and existence
of $\ord{\lambda^{1/2}}$ operators~\cite{Moult:2017xpp}. In \refcite{Chang:MIT5024},
subleading-power Lagrangians and hard-scattering
operators involving soft fields in SCET$_\II$ are constructed, and it is
demonstrated that soft $\ord{\la}$  contributions are absent for the
inclusive Drell-Yan small $q_T$ spectrum, including the forward-backward asymmetry.

\item
For the choice of $n_{a,b}$ in \eq{def_n_a_b},
the measurement function for $\vec{q}_T$ is the vectorial sum
of the perpendicular momenta of all particles in the hadronic final state.
Unlike the case of $e^+e^-$ event shapes, the sum factorizes into $n_a$-collinear, $n_b$-collinear,
and soft contributions without approximation,
so power corrections from the $\vec{q}_T$ measurement are absent.
Since fundamentally the hadronic structure functions
only depend on the Lorentz invariants in \eq{kin_invariants},
the measurement can be marginalized over the azimuthal angle of $\qt$
and thus preserves the rotational symmetry.
\item
A source of power corrections absent in the $e^+e^-$ case
are the Born measurements on $Q$ and $Y$ that set the arguments of the PDFs.
As has been discussed in detail in \refscite{Ebert:2018lzn, Ebert:2018gsn},
these give rise to new nonperturbative functions such as derivatives of the PDFs
at subleading power in $q_T$.
It can easily be seen from the exact result that these corrections are quadratic,
%%%
\begin{align}
\frac{q^-}{P_a^-} = \frac{\sqrt{Q^2 + q_T^2}e^Y}{\Ecm} = \frac{Qe^Y}{\Ecm} \bigl[ 1 + \ord{\la^2} \bigr] = x_a \bigl[ 1 + \ord{\la^2} \bigr]
\,.\end{align}
%%%
\end{enumerate}
%%%
Recall that $W_{\rm incl} = W_{-1} + W_0/2$ and we already showed that $W_0\sim\lambda^2$,
so the absence of linear power corrections for $W_\incl$ and $W_{-1}$ is equivalent.
We can thus conclude that $W_{-1}$ and $W_4$ do not receive linear power corrections.

%~~~~~~~~~~~~~~~~~~~~~~~~~~~~~~~~~~~~~~~~~~~~~~~~~~~~~~~~~~~~~~~~~~~~~~~~~~~~~~~
\subsubsection{Choice of tensor decomposition is \texorpdfstring{$\ord{\la^2}$}{O(lambda2)}}
\label{sec:frame_choice}
%~~~~~~~~~~~~~~~~~~~~~~~~~~~~~~~~~~~~~~~~~~~~~~~~~~~~~~~~~~~~~~~~~~~~~~~~~~~~~~~

In \sec{tensor_decomposition}, we defined a set of reference vectors $t^\mu, x^\mu,y^\mu,z^\mu$
to perform the tensor decomposition into hadronic structure functions,
which turned out to be equivalent to the CS frame (using the boost definition).
The $x^\mu,y^\mu,z^\mu$ were uniquely determined by imposing \eqs{z_basic_assumption}{xy_symmetry_constraints},
but these constraints are not technically required.
In general, we can also pick a different set $x'^\mu, y'^\mu, z'^\mu$ of
orthonormal, spacelike reference vectors.
These in turn define a vector-boson rest frame related to the CS frame by a
rotation $r \in \SO(3)$ that in general depends on $q$,
%%%
\begin{align}
\begin{pmatrix}
   x'^\mu \\
   y'^\mu \\
   z'^\mu \\
\end{pmatrix}
= r(q)
\begin{pmatrix}
   x^\mu \\
   y^\mu \\
   z^\mu \\
\end{pmatrix}
\,.\end{align}
%%%
The corresponding hadronic structure functions $W'_i$ are related to the $W_i$
by a corresponding orthogonal transformation
%%%
\begin{align}
W'_i = \sum_j R_{ij}(q) W_j
\,.\end{align}
%%%
In terms of the $W'_i$, the fully differential cross section is given by
%%%
\begin{align}
\frac{\df\sigma(\Theta)}{\df^4 q \, \df \cO}
&= \frac{1}{2\Ecm^2} \sum_i L_i(q, \cO, \Theta) \,
\sum_j R^{-1}_{ij}(q) \, W'_j(q^2, s_{aq}, s_{bq})
\,.\end{align}
%%%
Note that the parametrization of the lepton phase space used to evaluate the
$L_i$ is irrelevant here.
Of course, if corresponding angles $\theta', \varphi'$ are considered,
the corresponding spherical harmonics are related by the same rotation $R_{ij}(q)$,
such that their angular coefficients are given by ratios of the $W'_i$.

First, let us point out that there is never any frame ambiguity to the order in
the power expansion we are working in, because to the working order different
frame choices simply amount to a specific choice of basis or coordinate system,
which the final result cannot depend on. An ambiguity can only arise in
the higher-order terms that are partially retained and partially neglected,
which in general do depend on the frame choice.

%-------------------------------------------------------------------------------
\begin{figure*}
\centering
\includegraphics[width=0.6\textwidth]{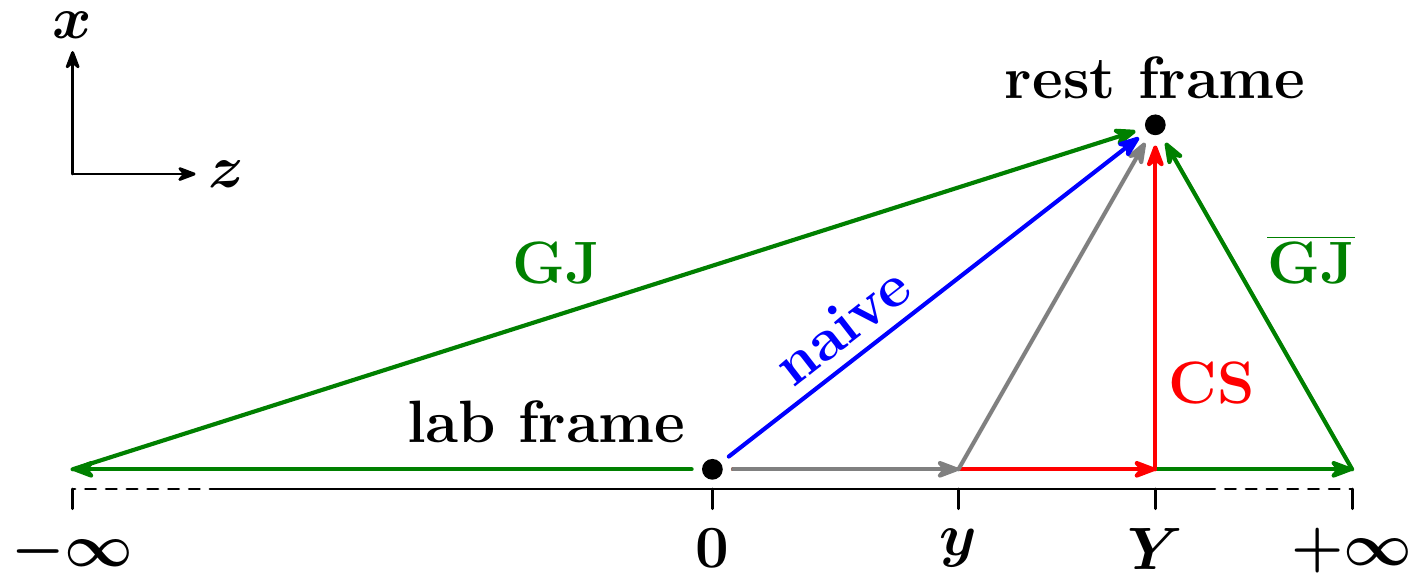}%
\vspace{-1ex}
\caption{Sequence of boosts definining different vector-boson rest frames.
In the general case (gray), we first boost by a rapidity $y$ along the lab-frame
beam axis, and then boost into the rest frame. Different choices of $y$ lead
to a relative Wigner rotation of the resulting rest frames.
Relevant special cases are $y = Y$ (Collins-Soper frame, red),
$y = 0$ (the naive rest frame, blue) and $y \to \pm \infty$,
(Gottfried-Jackson frame, green).}
\label{fig:boosts}
\end{figure*}
%-------------------------------------------------------------------------------

To remove the trivial effect from a mere basis choice at LP, we start from a
common LP rest frame at $q_T = 0$, which is the leptonic frame.
A convenient way to parametrize the different possible frames for nonzero $q_T$
is by the sequence of boosts starting from the lab frame as shown in
\fig{boosts}.
Specifically, we first boost by a rapidity $y$ along the beam direction
and then directly into the rest frame.
Some cases of interest are the ``naive'' rest frame,
obtained by performing a single direct boost from the lab frame ($y = 0$) into the rest frame,
the CS frame ($y = Y$), the Gottfried-Jackson (GJ) frame defined by the limit $y \to -\infty$,
or the $\GJbar$ frame obtained by taking $y \to +\infty$.
Another way to conceptualize these frames is through the angles
between their respective $z'^\mu$ axes and the momenta of the incoming protons.
In the CS frame, those angles become precisely equal for massless hadrons, see
\fig{frames}. In the GJ frame, the $z$ axis is aligned with the direction of $P_a$,
while the naive case falls in between for $Y > 0$.

It is easy to see that for $q_T\to 0$,
the two boosts for any frame collapse into a single boost from the lab frame to
the leptonic frame. Since for $q_T\to 0$ all frames coincide, we have
%%%
\begin{equation} \label{eq:rotated_axes_only_constraint}
r(q) = \id_{3\times 3} + \ord{\la}
\,, \qquad
R_{ij}(q) = \delta_{ij} + \ord{\la}
\,.\end{equation}
%%%
Working at LP in the hadronic tensor, the $\ord{\la}$ corrections in
\eq{rotated_axes_only_constraint} could in
principle induce an $\ord{\la}$ ambiguity in the hadronic power corrections.
We now show that this is not the case, which means that the linear power corrections
predicted by \eq{factorization_sigma1_final_result} are unique. In other words,
we have to show that the absence of additional linear corrections from the
hadronic tensor that lead to \eq{factorization_sigma1_final_result} was not just
an accident of our particular choice of tensor decomposition.

\Eq{rotated_axes_only_constraint} immediately implies that the new axes still
satisfy
%%%
\begin{align}
x'^\mu = n_\perp^\mu + \ord{\la}
\,, \qquad
y'^\mu = \eps_\perp^{\mu\nu} n_{\perp\nu} + \ord{\la}
\,, \qquad
z'^\mu = \frac{n_a^\mu - n_b^\mu}{2} + \ord{\la}
\,,\end{align}
%%%
so the scaling properties in \tab{power_counting_W_i} also hold for the $W'_j$.
Following our previous analysis, we would now discard all power-suppressed $W'_j$
structure functions, evaluate the remaining ones ($j = -1,2,4,5$) at leading power, and dress them with exact leptonic tensor components,
%%%
\begin{align} \label{eq:factorization_sigma0p1_rotated}
\frac{\df\sigma^{(0+L)}(\Theta)}{\df^4 q \, \df \cO}
&= \frac{1}{2\Ecm^2} \sum_i L_i(q, \cO, \Theta) \, \sum_{j = -1, 2, 4, 5} R^{-1}_{ij}(q) \, W'^{(0)}_j(q^2, s_{aq}, s_{bq}) \bigl[ 1 + \ord{\la^2} \bigr]
\,.\end{align}
%%%
Since all $i = -1, \ldots, 7$ are now populated by $R^{-1}_{ij}$,
one might think that the different choice of tensor decomposition
amounts to a linear power correction compared to the left-hand side
as given in \eq{def_sigma0pL},
because the $R_{ij}$ differ from unity by $\ord{\lambda}$, but as indicated
it is only of $\ord{\la^2}$.

To show that the induced difference is indeed only
$\ord{\la^2}$, first note that rotations around the $z$ axis amount to a trivial shift in $\varphi$.
This induces an $\ord{\la^2}$ difference at cross-section level
due to our assumption of azimuthal symmetry at leading power in
\eq{def_lp_azimuthal_symmetry}.
Hence, it is sufficient to consider the $\SO(2)$ subgroup of rotations around
the $y$ axis parametrized by one remaining Euler angle $\alpha$,
%%%
\begin{align}
\begin{pmatrix}
   x'^\mu \\
   z'^\mu \\
\end{pmatrix}
= \begin{pmatrix}
   c_\alpha   & s_\alpha \\
   - s_\alpha & c_\alpha \\
\end{pmatrix}
\begin{pmatrix}
   x^{\mu} \\
   z^{\mu} \\
\end{pmatrix}
\,, \qquad
y'^\mu = y^\mu
\,,\end{align}
%%%
with $c_\alpha \equiv \cos \alpha$ and $s_\alpha \equiv \sin \alpha$.
By a straightforward calculation, the $W_i$ can be shown to transform under the following representations of $\SO(2)$:
%%%
\begin{align} \label{eq:rotation_W_i_exact}
\begin{pmatrix}
   W'_{-1} \\
   W'_0 \\
   W'_1 \\
   W'_2 \\
\end{pmatrix}
&= \begin{pmatrix}
   1 - \frac{s_\alpha^2}{2}   ~& \frac{s_\alpha^2}{2} ~& - s_\alpha c_\alpha ~& \frac{s_\alpha^2}{4} \\
   s_\alpha^2 ~& 1 - s_\alpha^2 ~& 2 s_\alpha c_\alpha ~& - \frac{s_\alpha^2}{2} \\
   s_\alpha c_\alpha ~& -s_\alpha c_\alpha ~& 1 - 2 s_\alpha^2 ~& -\frac{s_\alpha c_\alpha}{2} \\
   s_\alpha^2 ~& - s_\alpha^2 ~& 2 s_\alpha c_\alpha ~& 1 - \frac{s_\alpha^2}{2}
\end{pmatrix}
\begin{pmatrix}
   W_{-1} \\
   W_0 \\
   W_1 \\
   W_2 \\
\end{pmatrix}
\,, \\[0.5em]
\begin{pmatrix}
   W'_3 \\
   W'_4 \\
\end{pmatrix}
&= \begin{pmatrix}
   c_\alpha & s_\alpha \\
   -s_\alpha & c_\alpha \\
\end{pmatrix}
\begin{pmatrix}
   W_3 \\
   W_4 \\
\end{pmatrix}
\,, \qquad
\begin{pmatrix}
   W'_5 \\
   W'_6 \\
\end{pmatrix}
= \begin{pmatrix}
   c_\alpha & s_\alpha \\
   -s_\alpha & c_\alpha \\
\end{pmatrix}
\begin{pmatrix}
   W_5 \\
   W_6 \\
\end{pmatrix}
\,, \qquad
W'_7 = W_7
\nn \,.\end{align}
%%%

It is a simple exercise in special relativity to show
that the resulting Wigner rotation of a generic frame defined by boost $y$
relative to the CS frame with $y = Y$ is
%%%
\begin{align} \label{eq:wigner_rotation}
\tan \alpha
= \frac{\eps \sinh(Y-y) \bigl[ \gamma \cosh(Y-y) - 1 \bigr]}{\gamma \sinh^2(Y-y) + \eps^2 \cosh(Y-y)}
= \eps \tanh\frac{Y-y}{2} + \ord{\eps^3}
\,,\end{align}
%%%
where $\eps = q_T/Q \sim \lambda$ and $\gamma = \sqrt{1+\eps^2}$.
For any $Y,y$, \eq{wigner_rotation} turns out to be bounded by the leading
term, $\abs{\tan \alpha} \leq \eps \abs{\tanh\frac{Y-y}{2}} \leq \eps$.
In particular for the GJ frame we have $\tan \alpha_\GJ = \eps$, with
which we recover the well-known result
for the relation between $W_{-1,0,1,2}$ in the CS and GJ frames~\cite{Boer:2006eq}.

\Eq{wigner_rotation} shows explicitly that $\alpha \sim \lambda$.
To see that \eq{factorization_sigma0p1_rotated} indeed holds up to quadratic
power corrections, we invert \eq{rotation_W_i_exact} by taking $\alpha \mapsto -\alpha$,
and expand in $\alpha\sim \la$ to find
%%%
\begin{align} \label{eq:rotation_W_i_approx}
 \begin{pmatrix} W_{-1} \\ W_0 \\ W_1 \\ W_2 \end{pmatrix}
 &= \left[\id_{4\times4} + \cO\begin{pmatrix}
   \la^2 & \la^2 & \la   &~ \la^2 \\
   \la^2 & \la^2 & \la   &~ \la^2 \\
   \la   & \la   & \la^2 &~ \la   \\
   \la^2 & \la^2 & \la   &~ \la^2 \\
   \end{pmatrix}\right]
   \begin{pmatrix} W'_{\!\!-1} \\ W'_0 \\ W'_1 \\ W'_2 \end{pmatrix}
%%%
\,,\nn\\[5pt]
%%%
 \begin{pmatrix} W_3 \\ W_4 \end{pmatrix}
 &= \left[\id_{2\times2} + \cO\begin{pmatrix}
   \la^2 & \la   \\
   \la   & \la^2 \\
   \end{pmatrix}\right]
   \begin{pmatrix} W'_3 \\ W'_4 \end{pmatrix}
%%%
\,,\nn\\[5pt]
%%%
 \begin{pmatrix} W_5 \\ W_6 \end{pmatrix}
 &= \left[\id_{2\times2} + \cO\begin{pmatrix}
   \la^2 & \la   \\
   \la   & \la^2 \\
   \end{pmatrix}\right]
   \begin{pmatrix} W'_5 \\ W'_6 \end{pmatrix}
%%%
\,, \quad W_7 = W'_7
\,.\end{align}
%%%
We see that under the rotation,
the leading structure functions mix into themselves and into each other only by an $\ord{\la^2}$ amount.
The subleading structure functions are populated precisely by an amount commensurate
with their intrinsic scaling, see \tab{power_counting_W_i}.
Combined with the scaling of the corresponding leptonic structure functions
as in \fig{power_counting},
we find that the effect of any $\ord{\la}$ rotation on the cross section
is indeed only of $\ord{\la^2}$ for LP-azimuthally symmetric observables.

It is natural to ask whether a specific frame choice should be preferred
in order to also capture an optimal set of terms at $\ord{\la^2}$.
It is well known that leading-logarithmic terms are absent in $W_1$ in the CS
frame~\cite{Boer:2006eq}.
This is natural from the point of view that the CS $z$ axis does not receive power corrections,
reducing ``spill-over'' of terms from the leading-power currents.
A reduced size of power corrections by a symmetric choice of frame has also been
found for $0$-jettiness $\Tau_0$~\cite{Moult:2016fqy, Moult:2017jsg, Ebert:2018lzn},
albeit for a somewhat different physical reason.
In \sec{numerics_phistar} we will find some numerical evidence
that the CS decomposition also reduces the size of power corrections in the
$\phi^*$ spectrum. Taken together, this suggests that the CS frame
might indeed be the optimal choice, although the size of the frame-dependent
$\ord{\la^2}$ corrections should still be assessed.

%===============================================================================
\subsection{Relation to the literature}
\label{sec:literature}
%===============================================================================

Early approaches to resummation effects on the Drell-Yan cross section
differential in the lepton kinematics~\cite{Balazs:1995nz, Balazs:1997xd, Ellis:1997sc}
typically picked the CS angles to parametrize the decay phase-space integral,
but did not discuss the ambiguity inherent in this choice or the relative size
of power corrections.
In these approaches, the CS frame primarily serves as a tool to enable generic
lepton-differential observables.
In \refcite{Boer:2006eq}, the logarithmic structure of $W_{-1,0,1,2}$ at low
$q_T$ in collinear factorization was discussed in detail, but the implications
for the structure of power corrections to e.g.\ the fiducial cross section were
not explored.

An implementation of generic lepton-differential observables in $q_T$
resummation was presented in \refcite{Catani:2015vma} and more recently in
\refcite{Camarda:2019zyx} based on a parton-level
Monte-Carlo generation of the leptonic final state. There, the choice of rest
frame used in earlier results was recast as a $q_T$-recoil prescription for how
to distribute the nonzero $q_T$ between the colliding partons in the rest frame
where the leptonic decay is evaluated. They also showed that the ambiguity in
the $q_T$-recoil prescription is in one-to-one correspondence with the ambiguity
of the rest-frame choice and that it vanishes for $q_T\to 0$. They thus argue
that the recoil effects are $\ord{q_T/Q}$ effects that cannot be unambiguously
computed through the $q_T$ resummation, but some recoil prescription is
nevertheless required for practical purposes  to satisfy transverse
momentum conservation in the parton-level generation of the leptonic decay.
(Note that similar recoil prescriptions to preserve momentum conservation are also commonly
used in parton-shower Monte Carlos, and in the event-level resummation of \refcite{Becher:2019bnm}.)

Very recently, \refscite{Scimemi:2019cmh, Bacchetta:2019sam}
used N$^3$LL perturbative baselines to fit nonperturbative models for the
rapidity anomalous dimension and TMDPDFs using fiducial $Z$ $q_T$ spectra among other data.
They also retain the exact dependence of the fiducial phase space on $q_T$,
with the analytic leptonic decay matrix element contracted against the LP hadronic tensor
$\propto g_\perp^{\mu\nu}$ (see also \refcite{Scimemi:2017etj}).%
\footnote{%
The leptonic $q_T$ dependence is extrapolated across a given $q_T$ bin in
\refcite{Bacchetta:2019sam} to simplify the $q_T$ bin integral, which should
give a good approximation of the exact bin integral, especially for small bin
width.}
This is essentially equivalent to an exact treatment of
$L_{-1}$ in our notation (up to an overall $\ord{\la^2}$ difference in the
projection itself), while $L_4$ does not contribute to the observables they
consider. They also do not provide formal arguments for the exact treatment
of the leptonic contributions.

\emph{If} the fully-differential and flavor-dependent squared leptonic decay matrix element
is evaluated and integrated over leptonic phase space accounting for the exact
dependence on the vector boson transverse momentum
and contracted with the resummed LP hadronic tensor
\emph{then} this amounts to retaining the exact $q_T$
dependence of the leptonic tensor.
This can be done explicitly at the analytic level or during Monte-Carlo generation
via some boost prescription that accounts for the exact $q_T$-dependent
recoil of the leptonic system.
To the best of our understanding this is the
case in \refscite{Catani:2015vma, Camarda:2019zyx, Becher:2019bnm, Scimemi:2019cmh,
Bacchetta:2019sam}. Our analysis thus provides formal justification for doing
so, showing that the ambiguity inherent in any implementation
meeting these criteria%
\footnote{While such an ambiguity was not discussed in
\refscite{Becher:2019bnm, Scimemi:2019cmh, Bacchetta:2019sam}, an analogous
ambiguity exists in all approaches.}
is only of $\ord{\la^2}$ and that for a large
class of common measurements (those that are azimuthally symmetric at LP) this
actually unambiguously predicts all linear power corrections along with their
resummation for single-boson production.
In addition, it is formally required in phase-space regions that
exhibit leptonic power corrections.

In \refcite{Bizon:2018foh}, fiducial lepton cuts are implemented in the resummed
cross section strictly on Born kinematics at $q_T = 0$, while fiducial power
corrections are obtained through the fixed-order matching.
Large power corrections from the fixed-order matching were observed in the fiducial
case compared to the inclusive case.
From our analysis, this is explained by the linear power corrections induced by
the fiducial cuts.

Sometimes a multiplicative fixed-order matching procedure is employed, as e.g.\
in \refscite{Bizon:2018foh, Kallweit:2020gva}, where in order to Sudakov-suppress
the fixed-order matching corrections at small $q_T$ they are multiplied by the ratio of the LP
resummed contribution to its fixed-order expansion. While this procedure is
unlikely to produce the correct Sudakov suppression for genuine hadronic power
corrections, one might ask if it correctly ``dresses'' the fiducial
power corrections with the LP resummation to achieve their resummation. For this
to be the case, the multiplicative matching at minimum has to reproduce \eq{factorization_sigma1_final_result} for the linear power corrections. Clearly,
this can only happen if the multiplicative matching only involves a single
(effective) hadronic structure function at a time and if it is performed fully
differentially in $q^2$, $Y$, and $q_T^2$. This is typically not the case.
For example, the multiplicative matching in \refscite{Bizon:2018foh, Kallweit:2020gva}
is performed at the cumulant level, and thus does not satisfy this requirement.

%%%%%%%%%%%%%%%%%%%%%%%%%%%%%%%%%%%%%%%%%%%%%%%%%%%%%%%%%%%%%%%%%%%%%%%%%%%%%%%%
\section{Resummation of leading-power hadronic structure functions}
\label{sec:resummation_lp}
%%%%%%%%%%%%%%%%%%%%%%%%%%%%%%%%%%%%%%%%%%%%%%%%%%%%%%%%%%%%%%%%%%%%%%%%%%%%%%%%

In this section, we discuss our specific resummation setup
for the leading-power hadronic structure functions $W_{-1,4}^{(0)}$ in \eq{tmd_factorization_m1_2_4_5}.
The setup follows standard procedures and is deliberately kept simple, e.g.,
by ignoring nonperturbative corrections or quark-mass effects~\cite{Pietrulewicz:2017gxc}
at small $q_T$, allowing us to focus on the effect of resumming
the fiducial power corrections in the following sections.
As discussed in the previous section, the fact that their resummation can be
obtained in terms of the leading-power hadronic structure functions, and that
this captures all linear as well as leptonic power corrections,
holds independently of how precisely the LP resummation is performed.

%===============================================================================
\subsection{Renormalization group evolution}
\label{sec:rges}
%===============================================================================

Directly resumming the logarithms of $q_T/Q$ in momentum space
is challenging due to the vectorial nature of $\vec q_T$, though
by now approaches for doing so exist~\cite{Monni:2016ktx, Ebert:2016gcn}.
As shown in \refcite{Ebert:2016gcn}, the correct solution of the
RGE system in momentum space in terms of distributions is actually equivalent to
the canonical solution in conjugate ($b_T$) space modulo different boundary
conditions. We therefore bypass this issue, as is commonly done, by solving the
RGEs in conjugate ($b_T$) space.
Using that the beam and soft functions only depend on the magnitude $b_T = |\bt|$,
\eq{tmd_factorization_bbs_FT} can be written as
%%%
\begin{align}
 [B_a B_b S](Q^2, x_a, x_b, \qt, \mu) &
 = \frac{1}{2\pi} \int_0^\infty \df b_T \, b_T J_0(b_T q_T)
   \tB_a(x_a, b_T, \mu, \nu/\w_a) \, \tB_b(x_b, b_T, \mu, \nu/\w_b)
   \nn\\&\qquad\qquad\times
   \tS(b_T, \mu, \nu)
\,,\end{align}
where $J_0(x)$ is the zeroth-order Bessel function of the first kind.

We use the framework of the rapidity renormalization group~\cite{Chiu:2012ir}
in conjunction with the exponential regulator of \refcite{Li:2016axz},
where the beam and soft functions in Fourier space obey the coupled RGEs
%%%
\begin{alignat}{2} \label{eq:RGEs}
 \frac{\df}{\df\ln\mu} \ln\tB_q(x,b_T,\mu,\nu/\omega)
 &= \tilde\gamma_B^q(\mu,\nu/\omega)
\,, \qquad
 &&\frac{\df}{\df\ln\mu} \ln\tS_q(b_T,\mu,\nu)
 = \tilde\gamma_S^q(\mu,\nu)
%%%
\,,\nn\\
%%%
 \frac{\df}{\df\ln\nu} \ln\tB_q(x,b_T,\mu,\nu/\omega)
 &= -\frac{1}{2}\tilde\gamma_\nu^q(b_T,\mu)
\,, \quad
 &&\frac{\df}{\df\ln\nu} \ln\tS_q(b_T,\mu,\nu)
 = \tilde\gamma_\nu^q(b_T,\mu)
\,,\end{alignat}
%%%
where the $\mu$ anomalous dimensions on the first line have the all-order expressions
%%%
\begin{align}
 \tilde\gamma_B^q(\mu,\nu/\omega) &
 = 2 \Gamma_\cusp^q[\as(\mu)] \ln\frac{\nu}{\omega} + \tilde\gamma_B^q[\as(\mu)]
\,,\nn\\
 \tilde\gamma_S^q(\mu,\nu) &
 = 4 \Gamma_\cusp^q[\as(\mu)] \ln\frac{\mu}{\nu} + \tilde\gamma_S^q[\as(\mu)]
\,.\end{align}
%%%
Here, $\Gamma_\cusp^q(\as)$ is the cusp anomalous dimension in the fundamental representation,
which is known to four loops~\cite{Korchemsky:1987wg, Moch:2004pa, Vogt:2004mw, Lee:2016ixa,Moch:2017uml,Lee:2019zop,Henn:2019rmi,Bruser:2019auj, Henn:2019swt, vonManteuffel:2020vjv} (see \refcite{Bruser:2019auj} for a complete list of earlier references).
At N$^3$LL we also require the QCD beta function to four loops~\cite{Tarasov:1980au, Larin:1993tp, vanRitbergen:1997va, Czakon:2004bu}.
The beam and soft noncusp anomalous dimensions $\tilde\gamma_B^q(\as)$ and $\tilde\gamma_S^q(\as)$
are related to those of the $0$-jettiness beam and soft functions through consistency
relations~\cite{Billis:2019vxg} and are known to three loops~\cite{Moch:2005id, Stewart:2010qs, Luebbert:2016itl, Bruser:2018rad}.
The all-order form of the rapidity anomalous dimension $\gamma^q_\nu$ reads
%%%
\begin{align} \label{eq:resummed_rapidity_anom_dim}
\tgamma^q_{\nu}(b_T, \mu)
&= -4\eta^q_\Gamma(\mu_0, \mu) + \tgamma^q_{\nu,\FO}(b_T, \mu_0)
\,, \qquad
\eta_\Gamma^{q}(\mu_0,\mu)
= \int_{\mu_0}^\mu \! \frac{\df \mu'}{\mu'} \, \Gamma^{q}_{\rm cusp}[\as(\mu')]
\,.\end{align}
%%%
This form follows from the commutativity of the $\mu$ and $\nu$ derivatives in \eq{RGEs}.
The fixed-order boundary term $\tgamma^q_{\nu,\FO}$ is known to three
loops~\cite{Luebbert:2016itl, Li:2016ctv, Vladimirov:2016dll}.

The RG evolution factors for $\tB_q$ and $\tS$ follow by solving the coupled
systems of equations in \eq{RGEs}.
One possible way of writing the solution is given by
%%%
\begin{align} \label{eq:RGevolution}
 \tB_q\Bigl(x,b_T,\mu,\frac{\nu}{\omega}\Bigr) &
 = \tB_q\Bigl(x,b_T,\mu_B,\frac{\nu_B}{\omega}\Bigr)
   \exp\biggl[-\frac12 \ln\frac{\nu}{\nu_B} \tgamma^q_\nu(b_T,\mu_B) \biggr]
   \exp\biggl[\int_{\mu_B}^\mu \frac{\df\mu'}{\mu'} \tgamma_B^q(\mu',\nu/\omega) \biggr]
\,,\nn\\
 \tS_q(b_T,\mu,\nu) &
 = \tS_q(b_T,\mu_S,\nu_S)
   \exp\biggl[\ln\frac{\nu}{\nu_S} \tgamma^q_\nu(b_T,\mu_S) \biggr]
   \exp\biggl[\int_{\mu_S}^\mu \frac{\df\mu'}{\mu'} \tgamma_S^q(\mu',\nu) \biggr]
\,.\end{align}
%%%
Here, we evolve first in $\nu$ and then in $\mu$ and the final scale $\mu$ is
set to the scale $\mu_H$ at which the hard function is evaluated. Any other path
in the two-dimensional $(\mu, \nu)$ space connecting
$(\mu_H, \mu_B, \nu_B, \mu_S, \nu_S)$ is viable as well,
and the path independence is ensured by exactly satisfying the RG consistency
relations between all anomalous dimensions, in particular by using
\eq{resummed_rapidity_anom_dim}.
By choosing appropriate boundary scales $\mu_{H,B,S}$ and $\nu_{B,S}$, the
hard function, as well as the beam and soft functions on the right-hand side of
\eq{RGevolution} are free of large logarithms and can be evaluated in fixed-order
perturbation theory, with all large logarithms resummed through the evolution kernels.

Different methods to numerically evaluate the evolution kernels and integrals over
anomalous dimensions entering in \eqs{resummed_rapidity_anom_dim}{RGevolution} were analyzed in
detail in \refcite{Billis:2019evv}, which found that their numerical
precision becomes relevant at N$^3$LL.
Here we use the approximate unexpanded analytic results,
which provide sufficient numerical precision for our purposes, with the explicit
expressions up to N$^3$LL given in \refcite{Billis:2019evv}.
The anomalous dimensions and fixed-order boundary conditions
required for the resummation at N$^3$LL are collected in our notation in
\refscite{Billis:2019vxg, Ebert:2017uel}.

%===============================================================================
\subsection{Canonical scales and nonperturbative prescription}
\label{sec:canonical_scales_nonp_prescription}
%===============================================================================

The canonical scales of the hard, beam, and soft functions, and the rapidity
anomalous dimension in $b_T$ space are given by
%%%
\begin{alignat}{5} \label{eq:canonical_scales_bT_space}
\mu_H &\sim Q
\,, \qquad
&\mu_B &\sim  b_0/b_T
\,, \qquad
&\mu_S &\sim b_0/b_T
\,, \qquad
&\mu_0 &\sim b_0/b_T
\,, \nn \\
&
&\nu_B &\sim Q
\,, \qquad
&\nu_S &\sim b_0/b_T
\,,\end{alignat}
%%%
where $b_0 \equiv 2 e^{-\gamma_E} \approx 1.12291$.
By evaluating the functions in the factorization theorem at their canonical scales
and evolving them to common scales $(\mu, \nu)$,
all logarithms of $\mu_B/\mu_H \sim \mu_S/\mu_H \sim \nu_S/\nu_B \sim (b_0/b_T) / Q$ are resummed.
In \refcite{Ebert:2016gcn} it was shown that the resummation in $b_T$ space with
the canonical scales in \eq{canonical_scales_bT_space}
is in fact equivalent to the exact solution of the RGE in momentum space, i.e.,
it reproduces the canonical, distributional logarithms in $q_T$,
except for the fact that one effectively uses
a shifted set of finite terms in the boundary conditions
(similar to the difference between renormalization schemes).
We exploit this  and require that for $q_T \ll Q$, \eq{canonical_scales_bT_space} is
exactly satisfied, such that the resummed $q_T$ spectrum in this region is obtained
from the numerical inverse Fourier transform of the canonical $b_T$-space result.

With the choice in \eq{canonical_scales_bT_space}, the beam and soft functions
and the rapidity anomalous dimension become sensitive to nonperturbative effects
at large $b_T \gtrsim \lqcd^{-1}$.
The Landau pole can be avoided by choosing $\mu_0$ (and $\mu_B, \mu_S$) as a function of $b_T$
such that it asymptotes to a perturbative value at large $b_T$, see e.g.~\refcite{Lustermans:2019plv} for a concrete implementation.
Alternatively, a global replacement of $b_T$ by a suitable function $b^*(b_T)$ may be performed at the level of the cross section,
where $b^*(b_T)$ itself is bounded by some perturbative value $b_\mathrm{max} \lesssim 1/\lqcd$~\cite{Collins:1981uk, Collins:1981va}.
In either case, the mismatch to the full result is then in principle captured by a
corresponding nonperturbative contribution.
Recently, it was shown that the full $\tgamma^i_\nu$ may also be determined from lattice calculations~\cite{Ebert:2018gzl, Ebert:2019okf, Vladimirov:2020ofp, Shanahan:2020zxr, Zhang:2020dbb},
and that estimates of the first subleading power in $b_T \lqcd$ can also be related to the gluon condensate~\cite{Vladimirov:2020umg}.

\begin{figure*}
\centering
\includegraphics[width=\WidthTwoSubfigs]{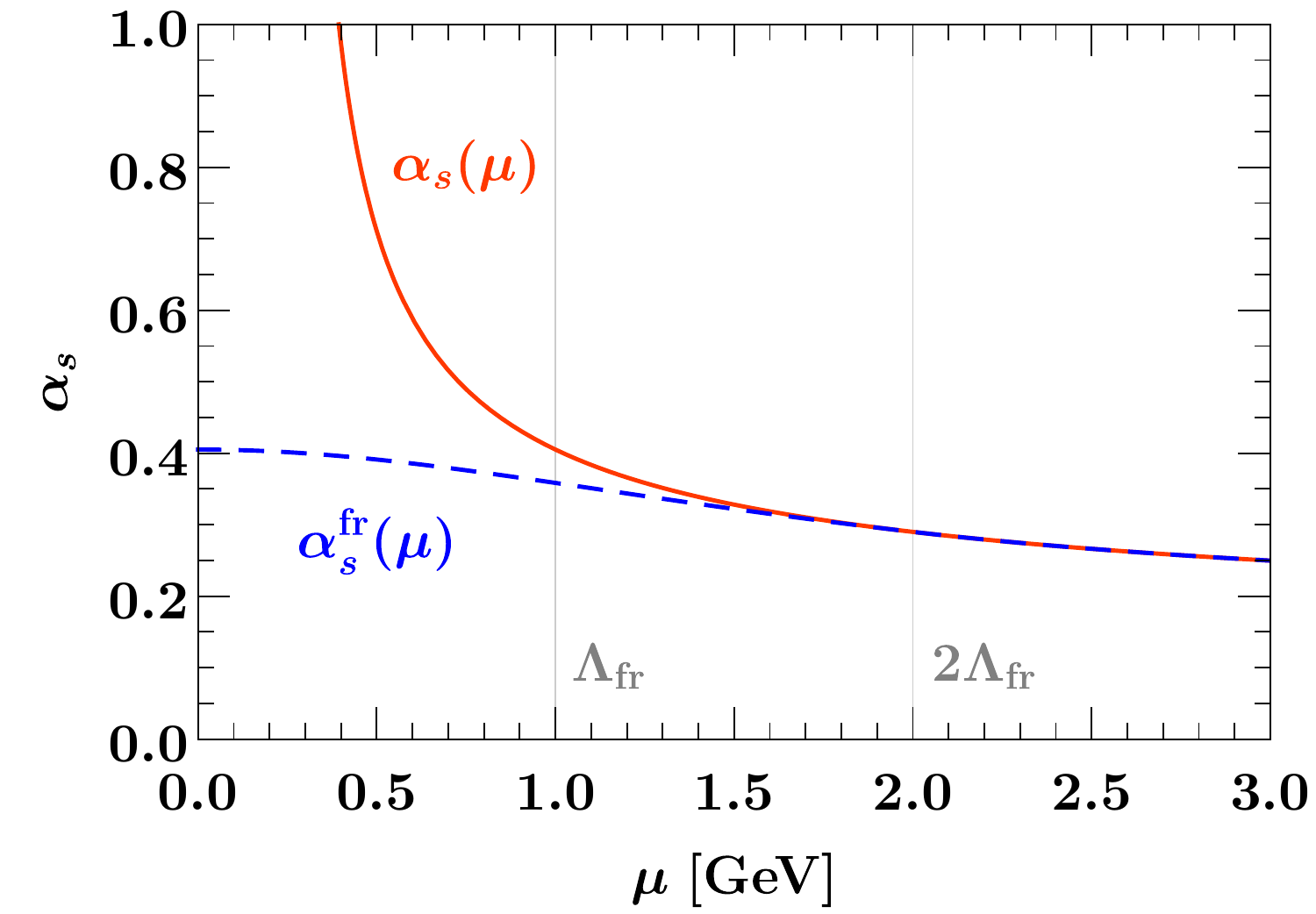}%
\hfill%
\includegraphics[width=\WidthTwoSubfigs]{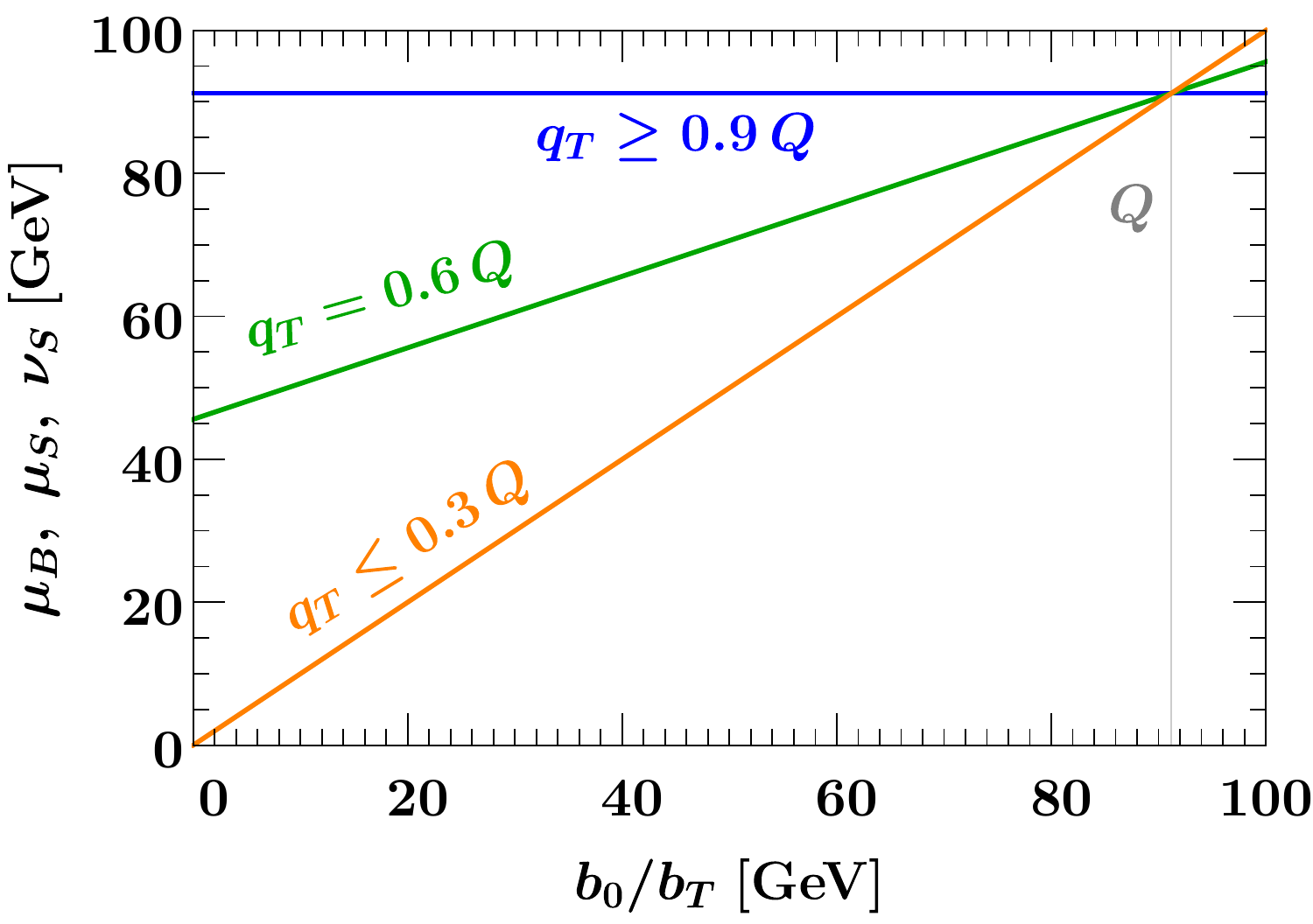}%
\caption{Left: Illustration of the freeze-out prescription used to ensure that $\as$ is evaluated at perturbative scales.
Right: Hybrid profile scales as a function of $b_0/b_T$ for representative values of $q_T$.
The thin vertical line corresponds to $b_0/b_T = Q$, where $Q = m_Z$ for illustration purposes.}
\label{fig:profiles}
\end{figure*}

Since nonperturbative effects in the region $q_T \sim \lqcd$ are not the main focus of this work,
we use a simpler prescription to ensure that $\as$ remains perturbative.
Specifically, we freeze out both the running coupling and the PDFs
entering the hadronic structure functions $W_{-1,4}^{(0)}$ at a perturbative scale
by performing the replacement
%%%
\begin{align} \label{eq:freeze_out_replacement}
\as(\mu) \mapsto \as^\fr(\mu) \equiv \as\bigl[ \mu_\fr(\mu) \bigr]
\,, \qquad
f_i(\mu) \mapsto f_i^\fr(\mu) \equiv f_i[ \mu_\fr(\mu) \bigr]
\,.\end{align}
%%%
We choose the smooth function $\mu_\fr(\mu)$ governing the freeze-out as
%%%
\begin{align} \label{eq:freeze_out_implementation}
\mu_\fr(\mu) = \begin{cases}
\Lambda_\fr + \frac{\mu^2}{4\Lambda_\fr}
& \mu \leq 2 \Lambda_\fr
\,, \\
\mu
& \mu > 2 \Lambda_\fr
\,.\end{cases}
\end{align}
%%%
In practice, we pick $\Lambda_\fr = 1 \GeV$ for our central results.
The behavior of $\as^\fr$ at low scales is illustrated in the left panel of \fig{profiles}.
This choice constitutes a (fairly crude) model for the large $b_T$ behavior of $\tgamma_\nu^i$
that is sufficient to regulate the large $b_T$ region,
and formally amounts to a power correction in $\lqcd \ll q_T$.
We similarly ignore contributions of $\ord{b_T \lqcd}$ in the beam and soft function boundary conditions,
beyond the ones encoded in our global choice of $\mu_\fr(\mu)$.
This is also consistent with neglecting the hadronic structure functions $W_{2,5}$
involving Boer-Mulders functions in the regime $\lqcd \ll q_T \ll Q$ altogether,
see the discussion below \eq{tmd_factorization_m1_2_4_5}.
We have checked that for perturbative $q_T \gtrsim 2 \GeV$,
the results from our prescription are compatible with a traditional (global)
$b^\ast$ prescription~\cite{Collins:1981uk, Collins:1981va}
within the uncertainty estimated by varying $\Lambda_\fr$ as described in \sec{estimate_pert_uncerts}.

%===============================================================================
\subsection{Fixed-order matching and profile scales}
\label{sec:fixed_order_matching_profile_scales}
%===============================================================================

We extend the description of the cross section to the fixed-order region $q_T \sim Q$
by an additive matching to the fixed-order result via profile
scales,
%%%
\begin{align} \label{eq:fixed_order_matching}
\df \sigma
= \df \sigma^{(0+L)}(\mu_\res) + \bigl[ \df \sigma^\FO(\mu_\FO) - \df \sigma^{(0+L)}(\mu_\FO) \bigr]
\,.\end{align}
%%%
Here the $\mu_\res$ argument of the first  term indicates
that we evaluate the resummed LP hadronic structure functions in $\df\sigma^{(0+L)}$
using the resummation profile scales given below.
The argument $\mu_\FO$ in the last term indicates that they are instead evaluated at common
fixed-order scales $\mu_\FO$, which can be done directly in momentum space.
The last term effectively acts as a differential subtraction term for the full
fixed-order cross section $\df\sigma^\FO$ in the second term, such that the difference in square
brackets is a nonsingular power correction.
We will refer to the outcome of \eq{fixed_order_matching} as N$^3$LL$^{(0+L)}+$NNLO$_0$
when the resummed LP hadronic structure functions
at N$^3$LL are combined with the exact leptonic tensor as discussed in \sec{factorization_fiducial_power_corrections}, and matched to the fixed $\ord{\as^2}$
NNLL$_0$ result.
When instead evaluating the leptonic tensor at strict LP, we refer to it
as N$^3$LL$^{(0)}+$NNLO$_0$, in which case the fiducial power corrections are
only included through the fixed-order matching.
The analogous notation is used at lower orders.

Approaching $q_T \sim Q$, the $q_T$ resummation must be turned off
to ensure the delicate cancellations between singular and nonsingular contributions
and to properly recover the correct fixed-order result for the spectrum, i.e., such
that the difference between the first and last terms in \eq{fixed_order_matching}
smoothly vanishes for $q_T \sim Q$.
To achieve this, we use the method of profile scales~\cite{Ligeti:2008ac, Abbate:2010xh}.
We use the hybrid profile scales constructed in \refcite{Lustermans:2019plv},
which depend on both $b_T$ and $q_T$,
and undergo a continuous deformation away from the canonical $b_T$ scales in
\eq{canonical_scales_bT_space} as a function of the target $q_T$ value, schematically,
%%%
\begin{align} \label{eq:profile_requirements}
\mu_{B,S}(q_T, b_T)
\,,
\nu_{B,S}(q_T, b_T)
\to \mu_H = \mu_\FO
\quad \text{for} \quad q_T \to Q
\,.\end{align}
%%%
We note that $\mu_0$ does not need to asymptote to $\mu_\FO$ towards large $q_T$
because its effect on the matched result is already turned off
as $\nu_S \to \nu_B$. In this limit, the first and last term in
\eq{fixed_order_matching} exactly cancel, leaving the fixed-order result $\df\sigma^\FO$.
We choose central scales as
%%%
\begin{align} \label{eq:central_scales}
&\mu_H = \nu_B = \mu_\FO
\,, \qquad
\mu_B = \mu_S = \nu_S =
\mu_\FO\, f_\run\Bigl( \frac{q_T}{Q}, \frac{b_0}{b_T\, Q} \Bigr)
\,, \qquad
\mu_0 = \frac{b_0}{b_T}
\,,\end{align}
%%%
where $f_\run$ is a hybrid profile function given by
%%%
\begin{align} \label{eq:def_f_run}
f_\run(x,y) &= 1 + g_\run(x)(y-1)
\,.\end{align}
%%%
It controls the amount of resummation by adjusting the slope of the scales in $b_T$ space
as a function of $q_T/Q$ via the function
%%%
\begin{align} \label{eq:def_g_run}
g_\run(x) &= \begin{cases}
1 & 0 < x \leq x_1 \,, \\
1 - \frac{(x-x_1)^2}{(x_2-x_1)(x_3-x_1)} & x_1 < x \leq x_2
\,, \\
\frac{(x-x_3)^2}{(x_3-x_1)(x_3-x_2)} & x_2 < x \leq x_3
\,, \\
0 & x_3 \leq x
\,.\end{cases}
\end{align}
%%%
As a result, for $q_T \leq x_1 Q$, the slope is unity yielding the canonical resummation,
while for $q_T \geq x_3 Q$, the slope vanishes so the resummation is fully turned off.
In between, the slope smoothly transitions from one to zero, which transitions the resummation from being
canonical to being turned off. This is illustrated in the right panel of \fig{profiles}.
Note that in contrast to \refcite{Lustermans:2019plv}, we do not require
a deformation away from the canonical scales to regulate the Landau pole at large $b_T$,
but instead rely on the replacement in \eq{freeze_out_replacement}.
For our central results, we use the transition points $(x_1, x_2, x_3) = (0.3, 0.6, 0.9)$.

%===============================================================================
\subsection{Estimate of perturbative uncertainties}
\label{sec:estimate_pert_uncerts}
%===============================================================================

We identify several sources of perturbative uncertainties,
which we estimate as follows.
In the limit $q_T \ll Q$, the perturbative uncertainty
is driven by the combined uncertainty from truncating the expansion
of the soft, beam, and rapidity anomalous dimensions.
To estimate them, we adopt the set of SCET$_\II$ profile scale variations
introduced in \refcite{Stewart:2013faa} for a SCET$_\II$-like jet veto.
They are given by
%%%
\begin{align}\label{eq:def_resummation_scale_variations}
\mu_S
&= \mu_\FO\, \Bigl[ f_\vary\Bigl( \frac{q_T}{Q} \Bigr) \Bigr]^{v_{\mu_S}}\,
f_\run\Bigl( \frac{q_T}{Q} , \frac{b_0}{b_T Q} \Bigr)
\,, \nn \\
\nu_S
&= \mu_\FO\, \Bigl[ f_\vary\Bigl( \frac{q_T}{Q} \Bigr) \Bigr]^{v_{\nu_S}}\,
f_\run\Bigl( \frac{q_T}{Q} , \frac{b_0}{b_T Q} \Bigr)
\,, \nn \\
\mu_B
&= \mu_\FO\, \Bigl[ f_\vary\Bigl( \frac{q_T}{Q} \Bigr) \Bigr]^{v_{\mu_B}}\,
f_\run\Bigl( \frac{q_T}{Q} , \frac{b_0}{b_T Q} \Bigr)
\,, \nn \\
\nu_B
&= \mu_\FO\, \Bigl[ f_\vary\Bigl( \frac{q_T}{Q} \Bigr) \Bigr]^{v_{\nu_B}}
\,,\end{align}
%%%
where each of the four variation exponents can be $v_i = \{+1,0,-1\}$,
and the amount of variation is governed by
%%%
\begin{align} \label{eq:def_f_vary}
f_\vary(x) = \begin{cases}
2(1 - x^2/x_3^2) & 0 \leq x < x_3/2\,, \\
1 + 2(1 - x/x_3)^2 & x_3/2 \leq x < x_3\,, \\
1 & x_3 \leq x\,,
\end{cases}
\end{align}
%%%
i.e., as $q_T \to Q$, this source of perturbative uncertainty is turned off
together with the resummation as a whole.
The central scale choice corresponds to $(v_{\mu_S}, v_{\nu_S}, v_{\mu_B}, v_{\nu_B}) = (0,0,0,0)$,
and a priori there are 80 possible different combinations of the $v_i$.
Since the arguments of the resummed logarithms are ratios of scales,
some combinations of scale variations will lead to variations of these arguments
that are larger than a factor of two, and are therefore excluded.
After dropping these combinations we are left with 36 different scale variations.
By taking their maximum envelope,
we obtain an estimate of the resummation uncertainty $\Delta_\res$.
Note that independent variations of $\mu_H$ need not be considered as part of $\Delta_\res$
because the corresponding change in the argument of the resummed logarithms is already covered by \eq{def_resummation_scale_variations}.

Second, we estimate the fixed-order perturbative uncertainty $\Delta_\FO$
from the maximum envelope of overall variations of $\mu_\FO$ by a factor of two.
These variations are inherited by all the resummation scales in \eq{central_scales},
so they leave the resummed logarithms invariant.
Third, we estimate the inherent uncertainty $\Delta_\match$ in our matching procedure \eq{fixed_order_matching}
by taking the maximum envelope of explicit variations of the transition points $x_i$,
%%%
\begin{align}
(x_1, x_2, x_3) = \bigl\{ (0.4, 0.75, 1.1), \, (0.2, 0.45, 0.7), \, (0.4, 0.55, 0.7), \, (0.2, 0.65, 1.1) \bigr\}
\,.\end{align}
%%%
Finally, we consider two independent variations of $\Lambda_\fr = \{ 0.8, 1.5\} \GeV$
away from our central choice $\Lambda_\fr = 1 \GeV$ as a rough estimate
of the uncertainty $\Delta_\Lambda$ in our nonperturbative prescription.
Combining all sources of uncertainty in quadrature, we take
%%%
\begin{align}
\Delta_\total = \sqrt{\Delta_\res^2 + \Delta_\FO^2 + \Delta_\match^2 + \Delta_\Lambda^2}
\end{align}
%%%
as an estimate of the total (perturbative) uncertainty on our results.

%%%%%%%%%%%%%%%%%%%%%%%%%%%%%%%%%%%%%%%%%%%%%%%%%%%%%%%%%%%%%%%%%%%%%%%%%%%%%%%%
\section{Resumming fiducial power corrections}
\label{sec:numerics}
%%%%%%%%%%%%%%%%%%%%%%%%%%%%%%%%%%%%%%%%%%%%%%%%%%%%%%%%%%%%%%%%%%%%%%%%%%%%%%%%

As discussed in \sec{factorization_fiducial_power_corrections}, fiducial power
corrections arise entirely from the leptonic tensors $L_i(q, \cO, \Theta)$,
and accordingly can be treated exactly in the factorization by keeping the $L_i$ exact.
In this section, we consider three applications to discuss this mechanism in more detail,
namely the $q_T$ spectrum with fiducial cuts (\sec{numerics_qT}),
the lepton transverse momentum distribution ($\pTlep$) (\sec{numerics_pTlep}),
and the $\phi^*$ observable (\sec{numerics_phistar}). In all cases, we consider
our primary examples of $Z\to \ell^+\ell^-$ or $W\to \ell\nu$.

%===============================================================================
\subsection{Numerical inputs and computational setup}
\label{sec:numerics_setup}
%===============================================================================

All our numerical results in this and the following sections
are obtained using the following setup.
We use the {\tt CT18NNLO}~\cite{Hou:2019efy} PDF set, and correspondingly use
the three-loop running to obtain the numerical value of $\as$ at any required
scale with $\as(91.1870) = 0.118$ as starting point.
We use the same PDF also at lower orders, which is consistent and allows us to
exhibit the genuine size of perturbative corrections.

For the resonant $W$ and $Z$ propagators, we work in the fixed-width pole scheme.
We use the following electroweak parameters~\cite{Tanabashi:2018oca}%
\footnote{%
The pole-scheme values are converted from the on-shell ones using
%%%
\begin{equation*}
m_V = m_V^{\rm OS}\, \bigl[1 + \bigl(\Gamma_V^{\rm OS}/m_V^{\rm OS}\bigr)^2\bigr]^{-1/2}
\,, \qquad
\Gamma_V = \Gamma_V^{\rm OS} \bigl[1 + \bigl(\Gamma_V^{\rm OS}/m_V^{\rm OS}\bigr)^2 \bigr]^{-1/2}
\,.\end{equation*}
%%%
}
%%%
\begin{alignat}{5}
m_Z &= 91.1535\GeV
\,, \qquad\,\,
\Gamma_Z = 2.4943 \GeV
\,,\nn\\
m_W &= 80.3580 \GeV
\,,\qquad
\Gamma_W = 2.0843 \GeV
\,, \\
G_F &= 1.1663787 \times 10^{-5} \GeV^{-2}
\,,\nn\\
V_{\rm CKM}
&= \begin{pmatrix}
    V_{ud} \quad V_{us} \quad V_{ub} \\
    V_{cd} \quad V_{cs} \quad V_{cb} \\
    V_{td} \quad V_{ts} \quad V_{tb} \\
   \end{pmatrix}
 = \begin{pmatrix}
    0.97446 \quad 0.22452 \quad 0.00365 \\
    0.22438 \quad 0.97359 \quad 0.04214 \\
  ~0.00896 \quad 0.04133 \quad 0.999105 \\
   \end{pmatrix}
\,.\end{alignat}
%%%
For the electroweak couplings, we use the $G_\mu$ scheme, with the values for
$\aem$ and $\sin^2\theta_w$ obtained from $m_W$, $m_Z$, and $G_F$ as
%%%
\begin{align}
\sin^2\theta_w
&= 1 - \frac{m_W^2}{m_Z^2}
 = 0.22284
\,, \nn \\
\aem \equiv \alpha_\mu
&= \frac{\sqrt{2} G_F}{\pi}\, m_W^2 \Bigl(1 - \frac{m_W^2}{m_Z^2}\Bigr)
 = \frac{1}{132.357}
\,.\end{align}
%%%

All factorized cross sections, both
at fixed order and including resummation up to N$^3$LL accuracy as described in
\sec{resummation_lp} are obtained from the \texttt{C++}\ library \scetlib\ \cite{scetlib}.
By default we use the CS tensor decomposition, and LP
cross sections including fiducial power corrections are denoted as
$\sigma^{(0+L)}$, while those at strict LP without fiducial power corrections
are denoted by $\sigma^\zero$.
We have also implemented alternative tensor decompositions using
\eq{factorization_sigma0p1_rotated}, in particular the one
that corresponds to the GJ frame, and will denote cross sections
evaluated using this choice as $\sigma^{(0+L)}_\GJ$.
\scetlib~uses the \texttt{Cuba 4.2}
library~\cite{Hahn:2004fe, Hahn:2014fua} for adaptive multi-dimensional
integration over $Q$ and $Y$, combined with $q_T$ integrals whenever they
cannot be performed analytically. To perform oscillatory Bessel integrals for
inverse Fourier transforms we use a double-exponential method for oscillatory
integrals~\cite{Takahasi1974, Ooura2001, Ooura1997}.

The integral over the leptonic phase space appearing in \eq{L_i_CS_angles},
%%%
\begin{align} \label{eq:L_i_CS_angles_again}
\int_{-1}^1 \! \df \cos \theta \, \int_0^{2\pi} \df \varphi \, g_i(\theta, \varphi) \, \delta[\cO - \hat\cO(q, \theta, \varphi)] \, \hat \Theta(q, \theta, \varphi)
\,,\end{align}
%%%
is carried out semi-analytically as follows. We focus on binned observables
$\hat\Theta(q, \theta, \varphi)$ and assume that the differential measurement $\cO$
on the decay products is being integrated over.
We assume that the measurement cuts and bins $\hat\Theta(q, \theta, \varphi)$ evaluate
to $1$ when all cuts are passed and $0$ otherwise, i.e., we take it to be a product
of $\theta$ functions. An explicit dependence on $\theta$ or $\varphi$, e.g.\ to
apply angular projections, can also easily be accommodated, but this is not needed for
the results obtained in this paper.
For given values of $q$ and $\varphi$, we first determine all intervals in $\cos\theta$
that pass the given cuts. Notably, for all observables considered here
($p_{T1,2}$, $\eta_{1,2}$, $\phi^*$, and any of their combinations),
the interval boundaries can be evaluated analytically even for nonzero $q_T$.
The integral over $\cos\theta$ over these intervals is then carried out analytically.
The remaining integral over $\varphi$ is performed by adaptive numerical integration.
In practice, the sum over hadronic structure functions $i$, see
\eq{def_sigma0pL}, can be pulled under the integral in
\eq{L_i_CS_angles_again} because the structure functions only depend on the
given value of $q$, so the decay phase-space boundaries only have to be determined once.
Typical evaluation times even for complicated phase-space volumes are in the few-millisecond range
on a single Intel\textsuperscript{\textregistered} Core\texttrademark\ \texttt{i5-7200U} CPU @ $2.50\,\mathrm{GHz}$
for a target relative numerical precision of $10^{-7}$.
The algorithm is not restricted to the leading-power structure functions, but
can also be used standalone with generic hadronic structure functions that are
provided.

Fixed-order results for $q_T$ and $\phi^*$ at LO$_1$ and NLO$_1$ for the relevant
Born$+1$-parton cross sections are obtained from
\mcfm \texttt{v8}~\cite{Campbell:1999ah, Campbell:2015qma, Boughezal:2016wmq}.
These results are used in the fixed-order matching. In addition, they are used
to obtain $q_T$ (or $\phi^*$) integrated cross sections at NLO$_0$ and NNLO$_0$
by combining them with $q_T$ (or $\phi^*$) subtractions including fiducial power
corrections supplied by \scetlib. This setup is discussed
in more detail in \sec{qT_subtraction}. We stress that the inclusion of fiducial
power corrections in the subtractions is essential to obtain numerically stable
results down to very small $q_T$ and $\phi^*$ and for $\pTlep$ near the Jacobian
peak.

%===============================================================================
\subsection[\texorpdfstring{$q_T$}{qT} spectrum with fiducial cuts]
           {$q_T$ spectrum with fiducial cuts}
\label{sec:numerics_qT}
%===============================================================================

We first discuss the impact of fiducial cuts on the Drell-Yan $q_T$ spectrum.
We consider the standard kinematic selection cuts of requiring a
minimum transverse momentum $\pTmin$ and maximum rapidity $\etaMax$
of the final-state leptons,
\begin{align} \label{eq:cuts}
 \Theta : \qquad p_{T,i} \ge \pTmin \,,\quad |\eta_i| \le \etaMax
\,,\end{align}
where $p_{T,i}$ and $\eta_i$ with $i=1,2$ are the transverse momentum
and pseudorapidity of the two leptons.

%===============================================================================
\subsubsection{Origin of power corrections}
\label{sec:qT_power_corrections}
%===============================================================================

The $\pTmin$ cut was already discussed in detail in \refcite{Ebert:2019zkb}.
Here, we briefly review the key steps and results,
and in addition discuss the rapidity cut.
A useful parametrization of the total momentum $q^\mu$
and the lepton momenta $p_{1,2}^\mu$ in the lab frame is
%%%
\begin{align} \label{eq:momenta}
 q^\mu &= \bigl( m_T \cosh Y \,,\, q_T ,\, 0 \,, m_T \sinh Y \bigr)
\,,\nn\\
 p_1^\mu &= p_{T,1} \bigl( \cosh(Y + \dY) \,,\, \cos\psi \,,\, \sin\psi \,,\, \sinh(Y + \dY) \bigr)
\,,\nn\\
 p_2^\mu &= q^\mu - p_1^\mu
\,,\end{align}
%%%
where $m_T = (Q^2 + q_T^2)^{1/2}$.
As before, we neglect the lepton masses and align the total transverse momentum
$\qt$ with the $x$-axis. We denote the azimuthal angle of the first lepton in
the lab frame as $\psi$ to distinguish it from the CS angle $\varphi$,
and write its rapidity as $y_1 = Y + \dY$. Momentum conservation determines $p_2^\mu$,
and fixes the transverse momenta and rapidities of the leptons to
%%%
\begin{alignat}{3} \label{eq:pTlep}
 p_{T,1} &= \frac{Q^2 / 2}{m_T\cosh(\dY) - q_T \cos\psi}
\,,\quad
 &&\eta_1 = Y + \dY
%%%
 \,,\nn\\
%%%
 p_{T,2} &= \sqrt{p_{T,1}^2 - 2 q_T \, p_{T,1} \cos\psi + q_T^2}
\,,\quad
 &&\eta_2 = Y + \frac12 \ln\frac{m_T - p_{T,1} e^{+\dY}}{m_T - p_{T,1} e^{-\dY}}
\,.\end{alignat}
%%%
For compactness, here we do not substitute $p_{T,1}$ in the expressions for $p_{T,2}$ and $\eta_2$.
The two-particle phase space defined in \eq{phiL} then takes the form
%%%
\begin{align} \label{eq:phi_ll}
\df\Phi_L(q) &= \frac{p_{T,1}^2}{8\pi^2Q^2}\, \df\psi \,\df\dY
\,.\end{align}
%%%
The integrated phase space with the cuts in \eq{cuts} can now be written as
%%%
\begin{align} \label{eq:L_qT_1}
\Phi_L(q,\pTmin,\etaMax)
&= \int\!\df\Phi_L(q) \,
   \theta\bigl( p_{T,1} \ge \pTmin\bigr)
   \theta\bigl( p_{T,2} \ge \pTmin\bigr)
   \theta\bigl( |\eta_1| \le \etaMax)
   \theta\bigl( |\eta_2| \le \etaMax)
\nn\\
&= \int_0^{2\pi}\df\psi \int_{-\infty}^\infty \df\dY \, \frac{p_{T,1}^2}{8\pi^2Q^2}\,
   \nn\\&\qquad\times
   \theta\bigl( \min\{p_{T,1}, p_{T,2} \} \ge \pTmin\bigr) \,
   \theta\bigl( \max\{|\eta_1|, |\eta_2|\} \le \etaMax\bigr)
\,.\end{align}
%%%
The integrand in \eq{L_qT_1} depends on $q_T$ only
through the combinations $q_T^2$ and $q_T \cos\psi$.
Thus, the expansion of the integrand in the limit $q_T \ll Q$
can only yield linear fiducial corrections if the $\psi$ integral does not average out.
This is equivalent to requiring that the cuts break azimuthal symmetry,
as otherwise the $\psi$ integral can always be trivially carried out,
such that all odd powers of $q_T \cos\psi$ integrate to zero
and only quadratic corrections in $(q_T/Q)^2$ arise.
Inclusive measurements are a special case, as without cuts
$\Phi_L(q)$ can only depend on $q^2$.

To see this mechanism explicitly for cuts on $p_T$ and $\eta$,
we expand the lepton transverse momenta and rapidities in \eq{pTlep} in $q_T/Q \sim \lambda$,
%%%
\begin{align} \label{eq:momenta_expanded}
p_{T,1} &= \frac{Q}{2 \cosh\dY} \biggl[1 + \frac{q_T}{Q} \frac{\cos\psi}{\cosh\dY}
+ \cO(q_T^2/Q^2) \biggr]
\,,\nn\\
p_{T,2} &= p_{T,1} - q_T \cos\psi + \cO(q_T^2/Q^2)
%%%
\,,\nn\\
%%%
 \eta_1 &= Y + \dY
\,,\nn\\
 \eta_2 &= Y - \dY - 2 \frac{q_T}{Q} \cos\psi \sinh\dY + \cO(q_T^2/Q^2)
\,.\end{align}
All observables in \eq{momenta_expanded} have a well-defined, nonvanishing LP limit as $q_T\to0$,
and the first correction is proportional to $q_T \cos\psi$.
Since at $q_T = 0$, $\psi$ and $\varphi$ coincide, we immediately find that
the fiducial $q_T$ spectrum obeys azimuthal symmetry at leading power, so the
discussion in \sec{linear_fiducial_pc} applies.

Naively, one might also expect that all linear fiducial corrections vanish upon
integration over $\psi$.
However, this is spoiled by the minimum and maximum in the $\theta$ functions in \eq{L_qT_1},
as can be easily seen for the $\pTmin$ cut.
For $\cos\psi > 0$, one has $p_{T,1} > p_{T,2}$, and thus the $\theta$ function in \eq{L_qT_1}
only restricts $p_{T,2}$. Vice versa, for $\cos\psi < 0$ it is $p_{T,1}$
that is constrained. This leads to two different integrands of the $\psi$ integral
in the two integration regions, leading to a nonvanishing $\psi$ integral.
This shows that the azimuthal symmetry is explicitly broken at $\ord{\lambda}$
leading to linear fiducial power corrections. However, it also shows that if
one were to only apply cuts to one of the leptons while being fully inclusive in the other,
no linear power corrections from the cuts would arise, since the $\psi$ integral
would average out when integrated against the $g_{-1,2,4,5}(\theta, \varphi)$
(using again that the difference between $\psi$ and $\varphi$ is itself of order $q_T$).

The situation is more complicated for the rapidity cut.
Determining the transition point of the maximum in the corresponding $\theta$ function
in \eq{L_qT_1}, i.e.\ the value $\psi_\mathrm{tp}$ for which $|\eta_1| = |\eta_2|$, we find that
\begin{align} \label{eq:cos_tp}
 \cos\psi_\mathrm{tp} =  \frac{Q}{2 q_T} \frac{\sinh(2Y)}{\sinh(2Y+\dY)} \times \bigl[1 + \cO(q_T^2/Q^2)\bigr]
\,.\end{align}
If $\abs{\cos\psi_\mathrm{tp}} \ge 1$, then the $\theta$ function in \eq{L_qT_1}
only restricts either $|\eta_1|$ or $|\eta_2|$ but not both for the whole $\psi$ range.
In this case, the rapidity cut does not break azimuthal symmetry.

For small but nonvanishing values of $q_T$, the $Q/q_T$ scaling
in \eq{cos_tp} can be overcome by a sufficiently small value of the vector-boson
rapidity $Y$. To be precise, \eq{cos_tp} has a solution in the physical range
$\abs{\cos\psi_\mathrm{tp}} < 1$ when
\begin{align} \label{eq:qT_tp_1}
 q_T > \frac{Q}{2} \biggl|\frac{\sinh(2Y)}{\sinh(2Y+\dY)}\biggr|
\,.\end{align}
Note that the $\eta_1$ constraint always requires that $|Y+\dY| \le \etaMax$.
Furthermore, we are only interested in the $q_T \ll Q$ limit,
which implies that \eq{qT_tp_1} only becomes important when $|Y| \ll 1$.
Hence, linear fiducial corrections will only arise in the region
\begin{align} \label{eq:qT_tp_2}
 \frac{q_T}{Q} \gtrsim \frac{q_T^\mathrm{tp}}{Q} \equiv \frac{|Y|}{\sinh(\etaMax)}
 \,,\qquad |Y| \ll 1
\,,\end{align}
while for $q_T \lesssim q_T^\mathrm{tp}$ only quadratic power corrections arise.
Note that in the region $q_T \sim q_T^\mathrm{tp} \ll Q$ the standard power counting
breaks down, as one has to simultaneously expand $|Y| \sim q_T/Q \ll 1$.
This is an example of a leptonic fiducial power correction as discussed
in \sec{leptonic_fiducial_pc}, where it is crucial to keep the lepton
phase space exact to correctly account for both small scales $q_T/Q$ and $|Y|$.
In practice, the size of this region is of $\cO(|Y|)$ and thus small by construction,
and hence its contribution to the cross section when integrated over or binned in $Y$
is suppressed as well.

\begin{figure*}
 \centering
 \includegraphics[width=\WidthTwoSubfigs]{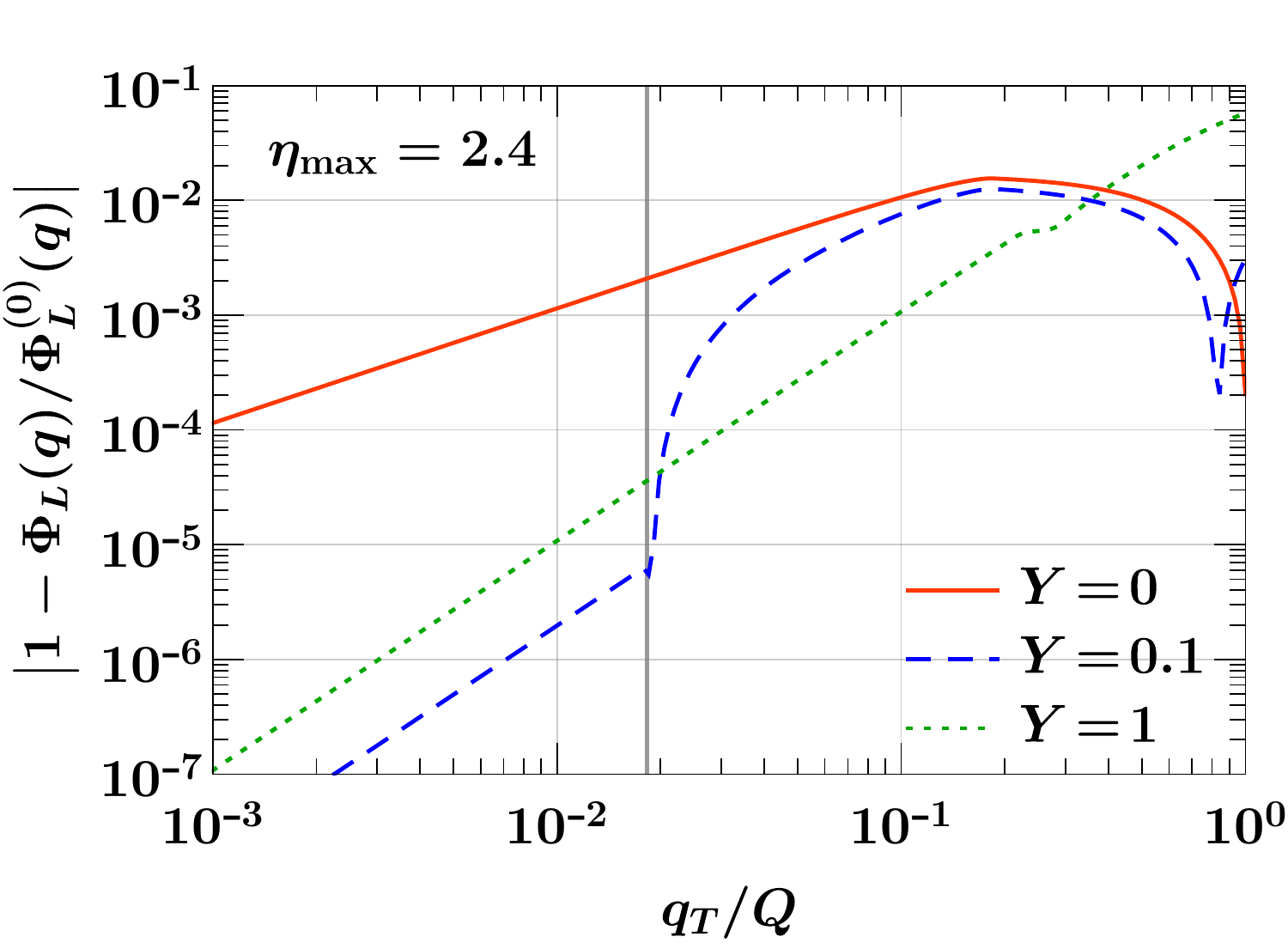}%
 \hfill%
 \includegraphics[width=\WidthTwoSubfigs]{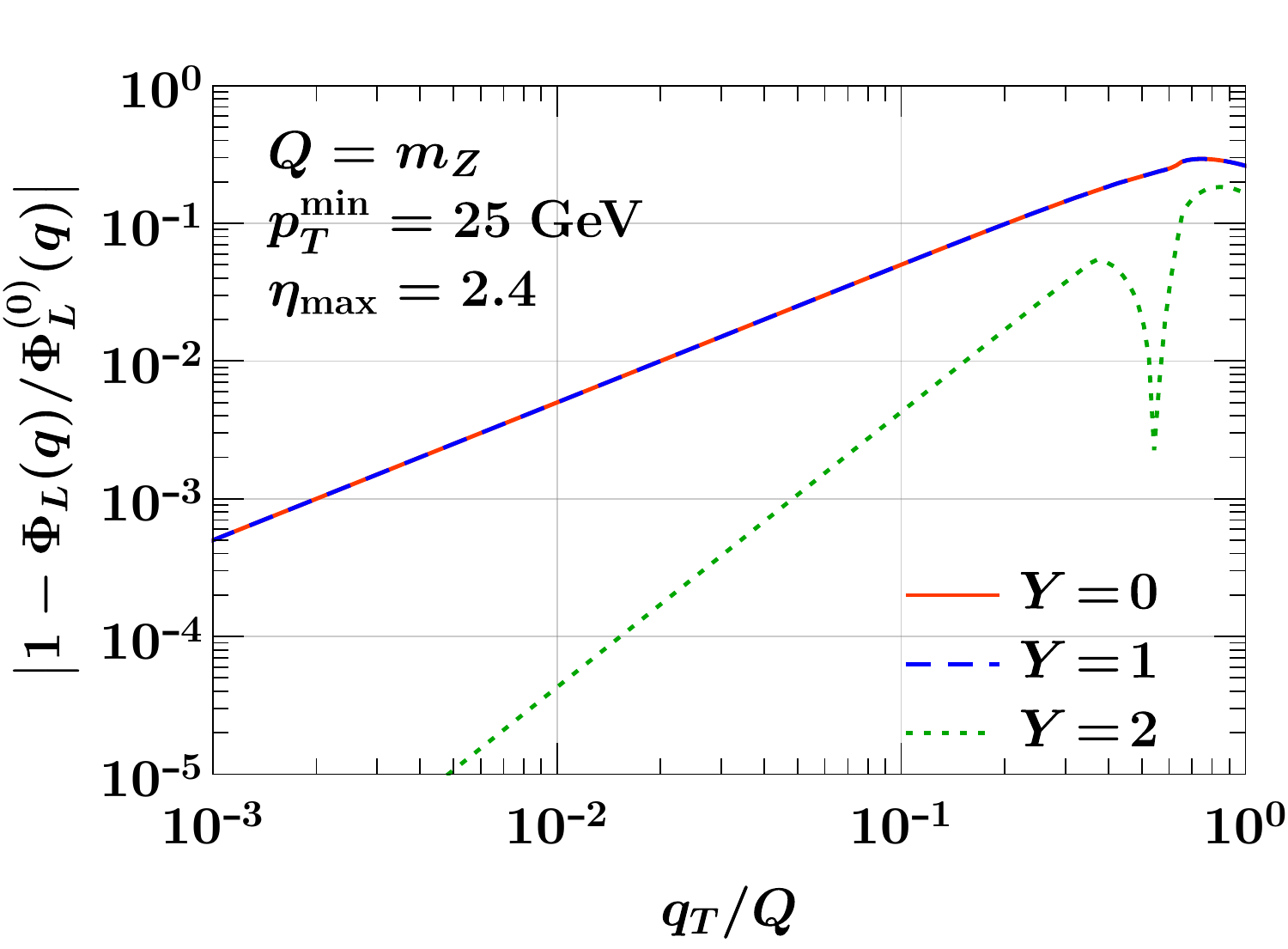}%
 \hfill%
 \caption{Relative difference of the exact dilepton phase space with fiducial cuts
 to its LP (Born) limit, for various values of the total dilepton rapidity $Y$
 for a pure rapidity cut (left) and both rapidity and $p_T$ cuts (right).
 The thick vertical line on the left shows transition point in \eq{qT_tp_1}
 for $Y=0.1$.
 }
 \label{fig:phi_scalar}
\end{figure*}

To illustrate and validate our observations, we have numerically implemented the exact
phase space with cuts in \eq{L_qT_1}. In \fig{phi_scalar}, we show the relative
difference of the dilepton phase space as a function of $q_T/Q$ for the cut
$\etaMax = 2.4$ (left) and the combined cut $\pTmin = 25~\GeV$ and $\etaMax=2.4$ (right),
for different values of the total rapidity $Y$ and fixed $Q = m_Z$.
In the left plot, one can clearly see that for $Y=0$ one has a linear power correction
for small $q_T$ up to $q_T/Q \sim 0.1$ (red solid line). For the slightly larger value $Y = 0.1$
(blue dashed line), we observe quadratic corrections up to the transition point
$q_T^\mathrm{tp}/Q \approx 0.2$, indicated by the thick vertical line,
while above this transition the correction has no simple scaling behavior.
Finally, for the relatively large value $Y=1$ one has quadratic corrections essentially
throughout the whole $q_T$ spectrum (green dotted line).
Also note that, in general, the corrections to the phase space are rather small,
as even for $Y=0.1$ they do not exceed $1\%$ for $q_T/Q < 0.1$.

In the right plot in \fig{phi_scalar}, we apply both $\etaMax$ and $\pTmin$ cuts.
For small values of $Y$, the $\pTmin$ cut is a stronger constraint
on the phase space than the $\etaMax$ cut. Hence, the two curves for $Y=0$
(red solid) and $Y=1$ (blue dashed) are equal, as the $\pTmin$ constraint
is independent of rapidity. For large $Y \sim \etaMax$, the rapidity cut
dominates over the $p_T$ cut, as illustrated for $Y=2$ (green dotted),
and one only has quadratic power corrections.

%===============================================================================
\subsubsection{Numerical results}
\label{sec:qT_results}
%===============================================================================

Having explicitly demonstrated that linear power corrections to the $q_T$
spectrum arise from fiducial cuts on the final-state leptons, we now verify
that they can indeed be captured in the factorization theorem by keeping
the lepton kinematics exact, as discussed in \sec{linear_fiducial_pc}.
In the following, we will always consider the fiducial cut
\begin{align}
 \pTlep \ge 25~\GeV \,,\quad |\eta_\ell| \le 2.4
\,,\end{align}
as employed in the CMS Drell-Yan measurement at 13 TeV \cite{Sirunyan:2019bzr}.

\begin{figure*}
 \centering
 \includegraphics[width=\WidthTwoSubfigs]{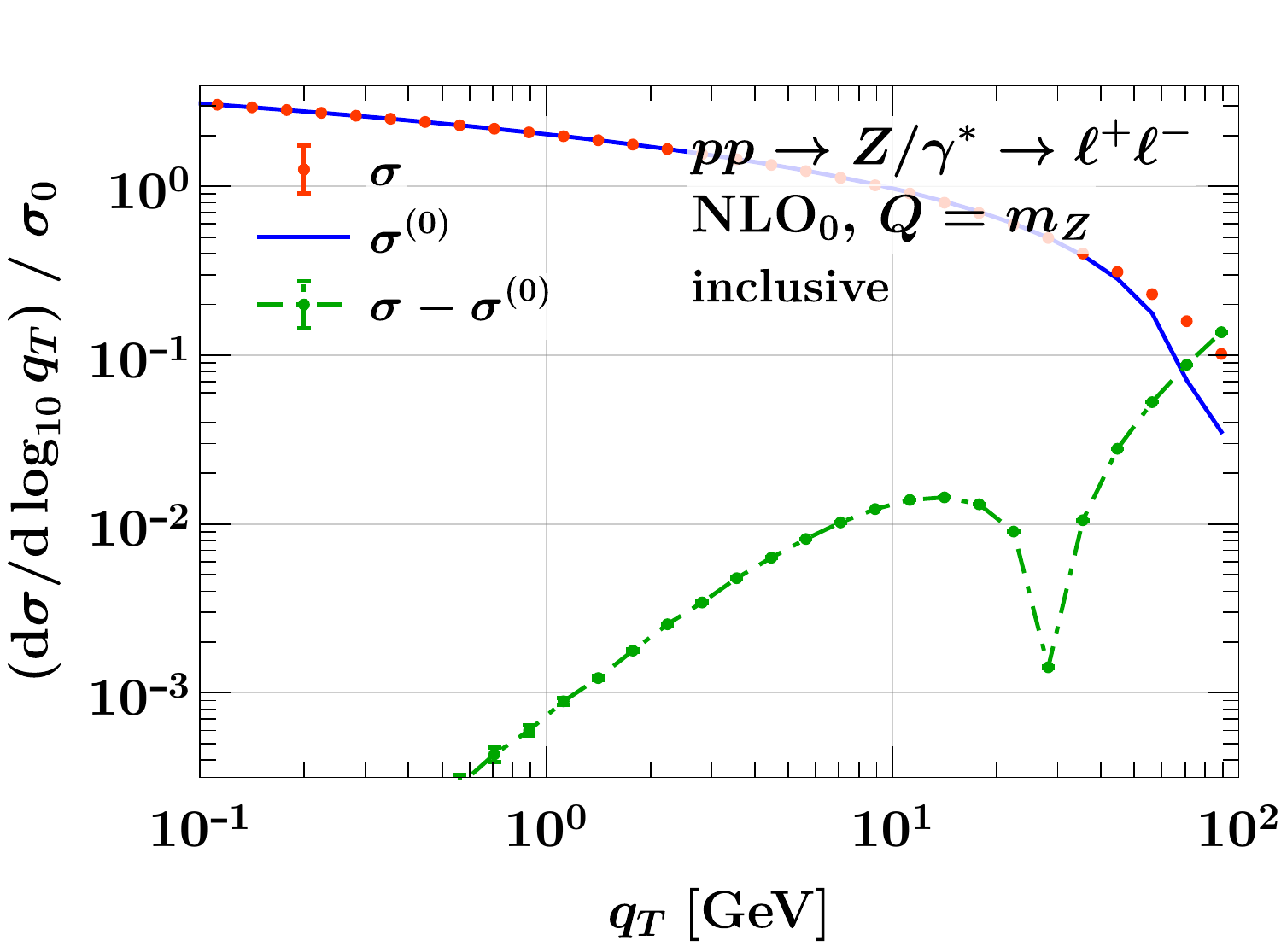}%
 \hfill%
 \includegraphics[width=\WidthTwoSubfigs]{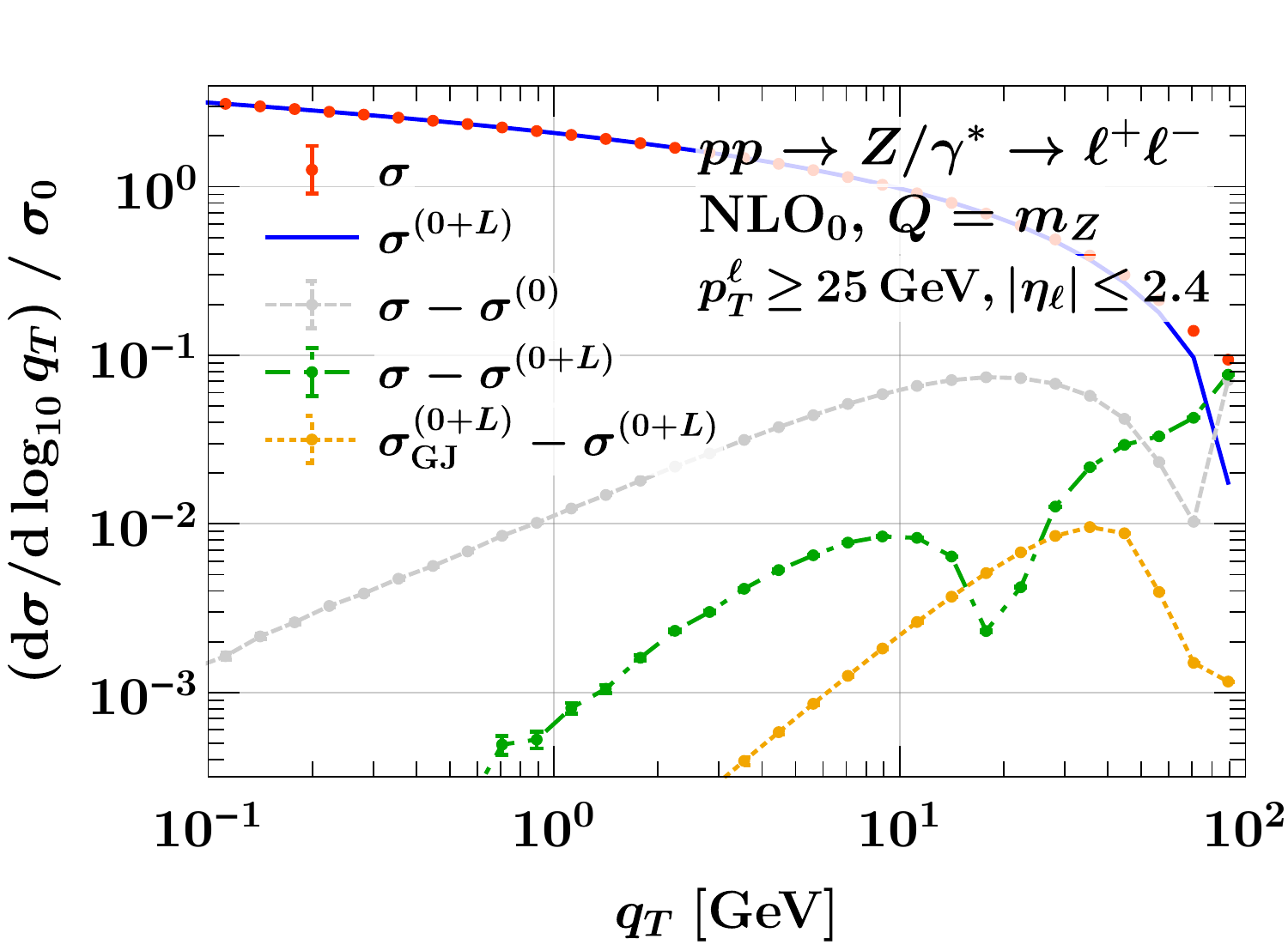}%
 \caption{Power corrections for the $q_T$ spectrum in Drell-Yan production, inclusive in the decay products (left) and with fiducial cuts (right). }
 \label{fig:Z_qT_sing_nons_13TeV}
\end{figure*}

In \fig{Z_qT_sing_nons_13TeV}, we show the $q_T$ spectrum for $Q = m_Z$ at NLO$_0$,
both without (left) and with (right) fiducial cuts.
In both figures, the red points illustrate the full NLO$_0$ result,
while the solid blue line shows the result at $\sigma^{(0)}$ (left)
and at $\sigma^{(0+L)}$ (right).
The various dotted and dashed lines show the differences between the full result and different singular limits.
In the inclusive case (left panel), there are no linear power corrections,
and thus $\sigma - \sigma^{(0)}$ (green, dashed) scales quadratically in $q_T$, as expected.
With fiducial cuts (right panel), $\sigma - \sigma^{(0)}$ (gray, dotted) clearly
suffers from linear power corrections, and as explained before, these linear corrections
can be accounted for by keeping the leptonic tensor exact.
This is illustrated by the green-dashed line, which shows the difference
$\sigma - \sigma^{(0+L)}$  between the exact and NLP result,
and only depends quadratically on $q_T$.
In particular, the size of these corrections is comparable to the quadratic corrections in the inclusive case.
The orange, dot-dashed curve shows the difference $\sigma^{(0+L)}_{\rm GJ} - \sigma^{(0+L)}$
between two choices of the tensor decomposition, corresponding to the CS frame and the GJ frame.
This difference scales quadratically in $q_T$, confirming that the ambiguity
from the choice of tensor decomposition is quadratically suppressed.
Moreover, we observe that this ambiguity is numerically much smaller than $\sigma^{(0+L)}$
itself, indicating that it may be completely negligible in practice.

\begin{figure*}
\centering
\includegraphics[width=\WidthTwoSubfigs]{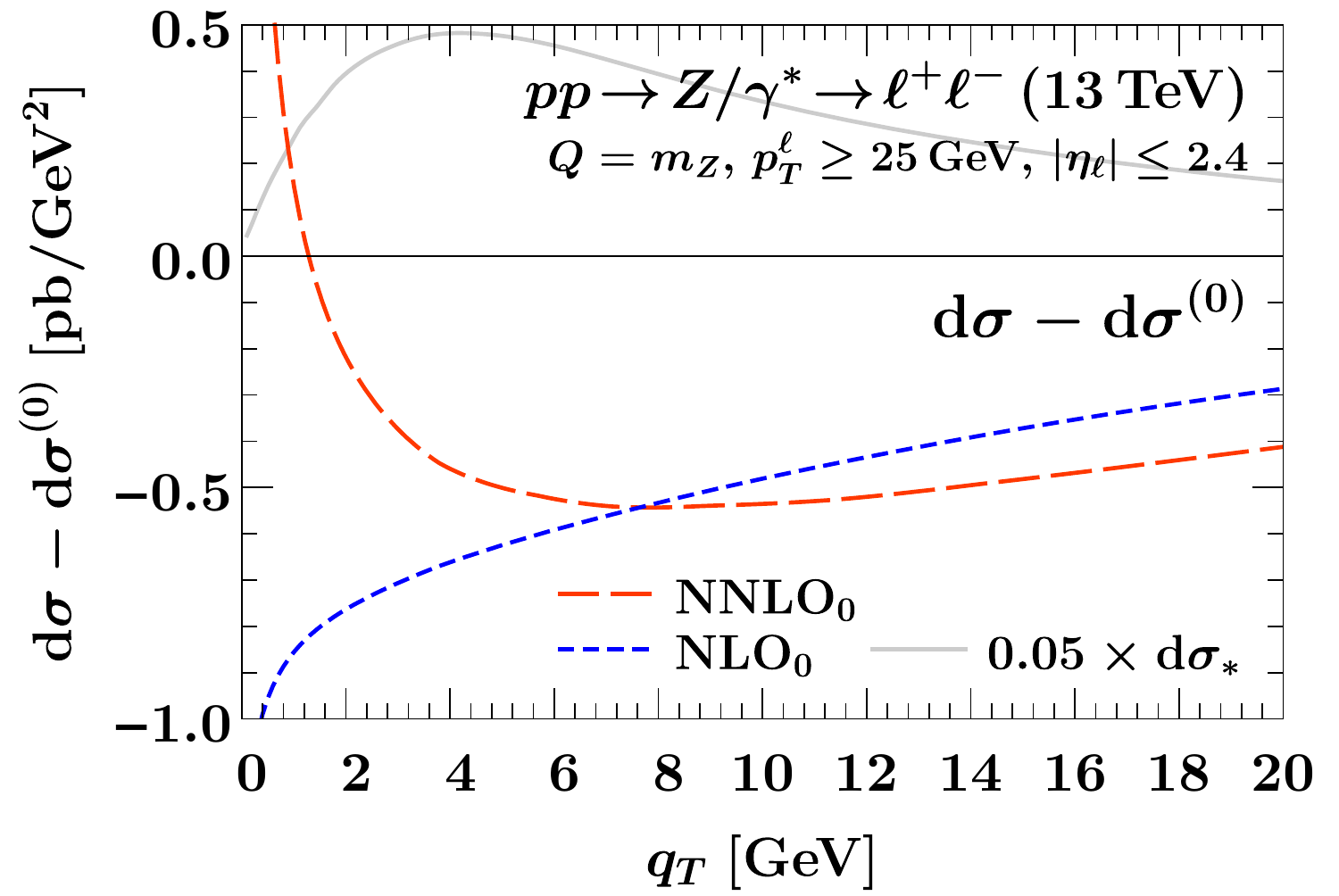}%
\hfill%
\includegraphics[width=\WidthTwoSubfigs]{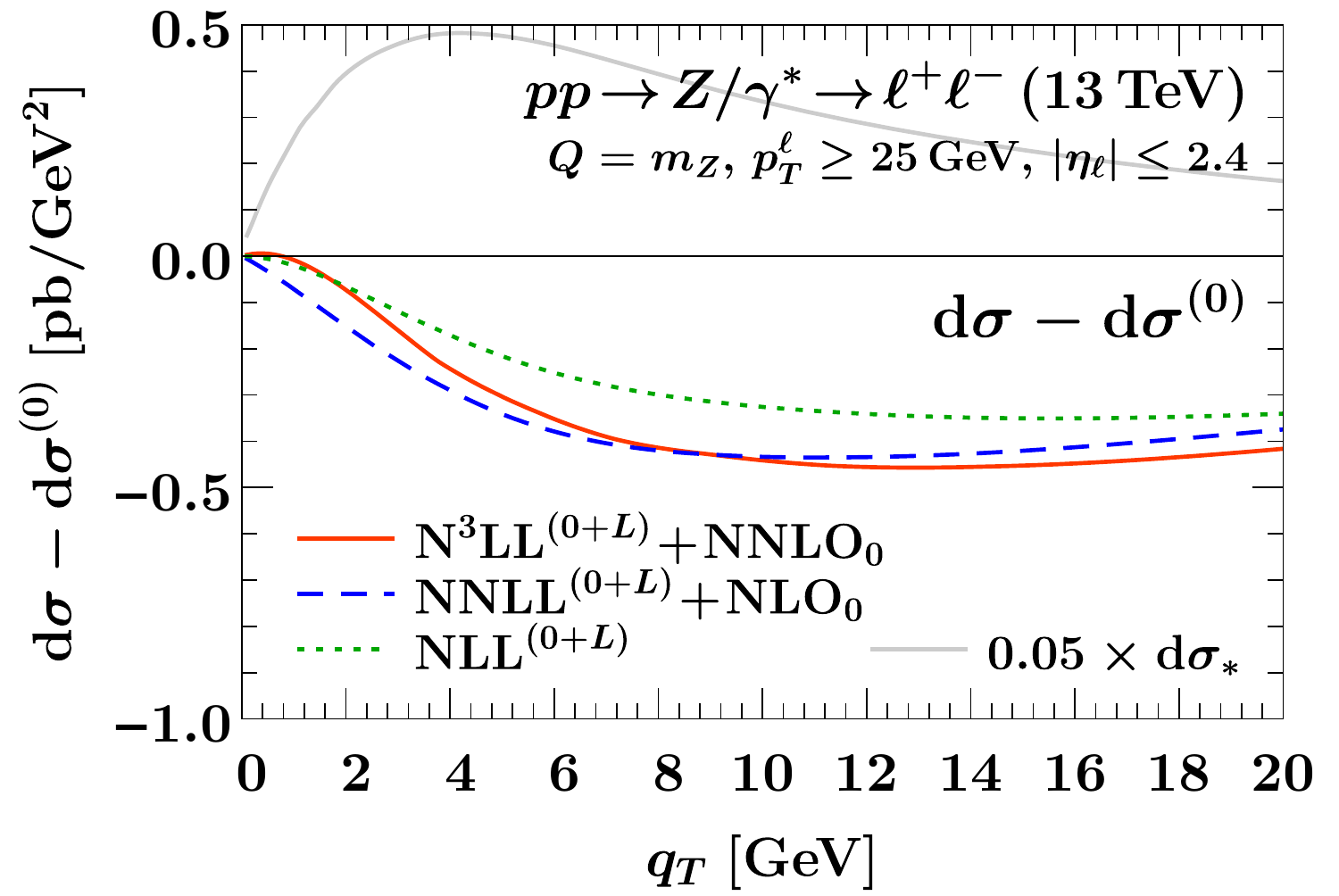}%
\\
\includegraphics[width=\WidthTwoSubfigs]{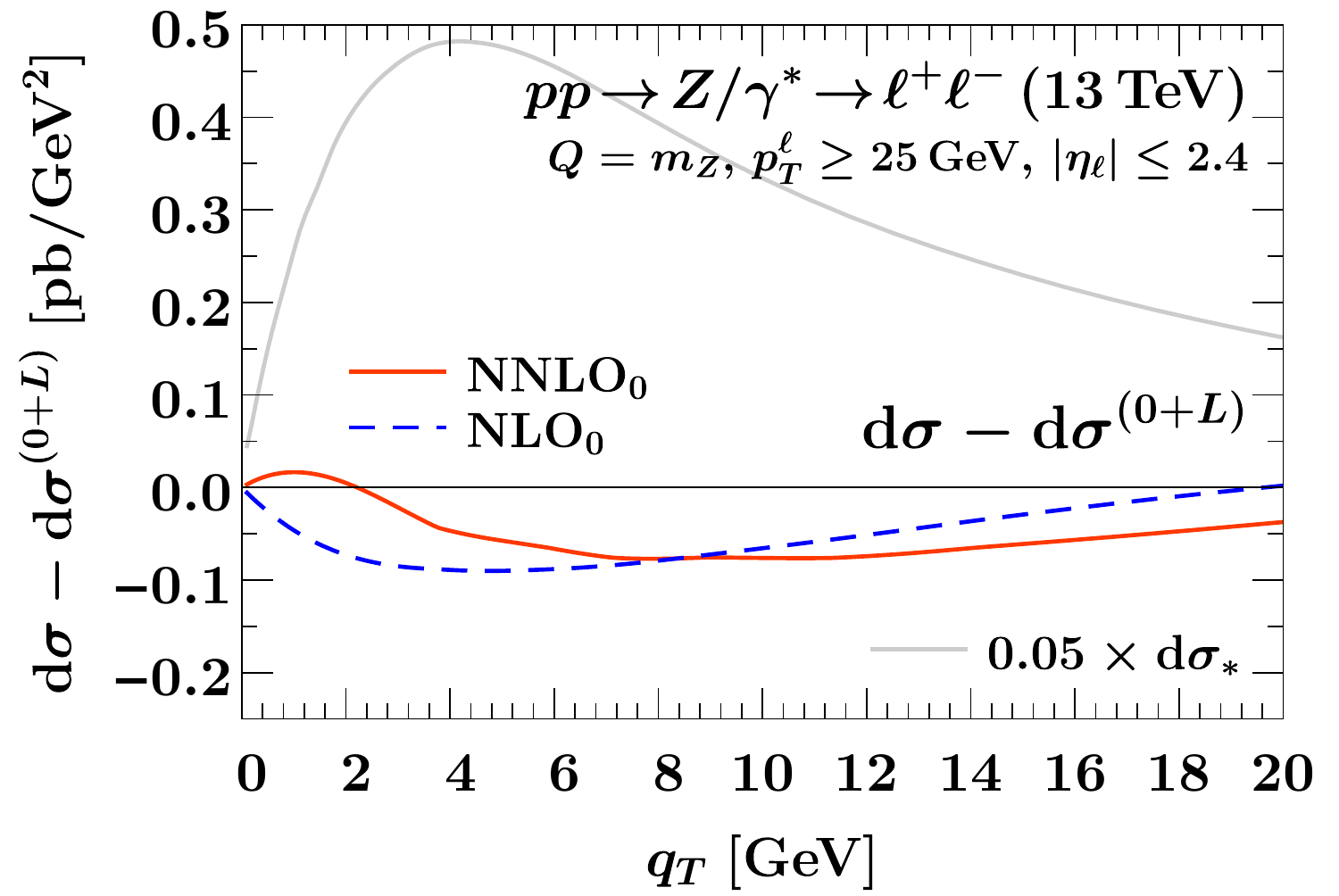}%
\caption{Breakdown of subleading-power contributions to the fiducial Drell-Yan $q_T$
on the resonance, $Q = m_Z$.
We compare the sum of all subleading power contributions,
treating the linear fiducial power corrections at fixed order (top left)
or resumming them (top right).
Our best prediction $\df \sigma_*$ for the total spectrum at N$^3$LL$^{(0+L)}+$NNLO$_0$
is indicated as a light gray line for reference, scaled down to $5\%$ of its original size.
In the bottom row we restrict to the remaining nonsingular (quadratic) power corrections,
which are finite for $q_T \to 0$ (note the difference in vertical scale).}
\label{fig:Z_qT_13TeV_illustration}
\end{figure*}

It is also interesting to study the impact of the NLP corrections on the resummed $q_T$ spectrum.
In the top-left panel of \fig{Z_qT_13TeV_illustration}, we show the difference
between the LP and the exact $q_T$ spectrum at NLO$_0$ (blue, short-dashed) and NNLO$_0$ (red, long-dashed).
For reference, the gray line shows our best prediction $\sigma_*$ at N$^3$LL$^{(0+L)}+$NNLO$_0$,
scaled down to 5$\%$ of its original size.
At both NLO$_0$ and NNLO$_0$, the power corrections diverge as $q_T\to0$
due to the overall $1/q_T$ behavior (compared to $1/q_T^2$ at LP).
The opposite signs at small $q_T$ also illustrates the poor perturbative convergence in this regime.

In the top-right panel of \fig{Z_qT_13TeV_illustration}, we show the difference
between the LP $q_T$ spectrum and the resummed and matched $q_T$ spectrum,
at NLL$^{(0+L)}$ (green, dotted), NNLL$^{(0+L)}+$NLO$_0$ (blue, dashed)
and N$^3$LL$^{(0+L)}+$NNLO$_0$ (red, solid).
Since the resummation includes the linear power corrections, the divergence as $q_T\to0$
is cured, and we observe very good perturbative convergence between the different resummed predictions.

Finally, in the bottom panel of \fig{Z_qT_13TeV_illustration} we show the difference
between the NLP and the exact $q_T$ spectrum at NLO$_0$ (blue, short-dashed) and NNLO$_0$ (red, long-dashed),
again including our best prediction $\sigma_*$ for reference.
Since all terms diverging as $1/q_T^2$ or $1/q_T$ are included in $\sigma^{(0+L)}$,
this difference is finite as $q_T \to 0$, and the overall size is much smaller
compared to the top left panel, indicating that corrections beyond the linear NLP
limit can be safely included at fixed order.

Given the large effect that the resummation of fiducial power corrections has
on the $q_T$ spectrum at small $q_T$, it would be very interesting to investigate
whether a net resummation effect on the \emph{total} fiducial cross section remains
after integration over $q_T$, as would be relevant for PDF determinations from fiducial $Z$ and $W$ rapidity spectra.
It has been argued that recoil ambiguities might prevent a first-principle calculation of these effects~\cite{Glazov:2020gza},
but since the ambiguity from the choice of tensor decomposition is fairly small,
and vanishes as soon as the resummation is fully turned off,
it is not unlikely that a surviving net effect is unambiguous and can be calculated.

%===============================================================================
\FloatBarrier
\subsection[Lepton \texorpdfstring{$p_T$}{pT} spectrum]
           {Lepton $p_T$ spectrum}
\label{sec:numerics_pTlep}
%===============================================================================

We next study the distribution in the lepton transverse momentum $\pTlep$. To
be concrete, we consider $W^\pm \to \ell^\pm \nu_\ell$, for which $\pTlep$ is an
essential observable. For simplicity, we do not consider any additional fiducial
cuts. This serves as a prototypical example of the appearance of leptonic power
corrections near a radiation-sensitive edge of Born phase space as discussed in
\sec{factorization_fiducial_power_corrections}, which in this case happens near
the Jacobian peak at $p_T^\ell \sim Q/2$.

%===============================================================================
\subsubsection{Origin of power corrections}
\label{sec:pTlep_power_corrections}
%===============================================================================

Using the parametrization of the lepton momenta in terms of CS angles
in \eq{p1_p2_CS_angles_CS_frame}, the lepton $\pTlep$ can be expressed as
%%%
\begin{align} \label{eq:pTlep_CS}
 \pTlep = \frac{Q}{2} \sqrt{(\gamma s_\theta c_\varphi + \eps)^2 + s_\theta^2 s_\varphi^2 }
 \equiv \frac{Q}{2} \kappa
\,, \qquad
\eps = \frac{q_T}{Q}
\,, \quad
\gamma = \sqrt{1 + \eps^2}
\,,\end{align}
%%%
where $s_\theta \equiv \sin\theta$, $s_\varphi \equiv \sin\varphi$,
$c_\varphi \equiv \cos\varphi$. We also introduced the variable
$\kappa = 2\pTlep/Q$ to parameterize the distance of $\pTlep$
from the Jacobian peak at $\pTlep = Q/2$, which will be useful in the following.
\Eq{pTlep_CS} can be easily solved for $s_\theta$, with physical solutions
constrained by $0 \le s_\theta \le 1$. In the following, we restrict ourselves to
the case $\kappa > \eps$, which will be the relevant region to describe large $\pTlep$.
In this case, the only physical solution for $s_\theta$ is given by
%%%
\begin{align} \label{eq:sol_st_1}
\sin(\theta_\varphi) = s(c_\varphi)
 = -\beta \gamma\, \eps c_\varphi + \sqrt{\Delta}
\,,\end{align}
where for brevity we introduced the abbreviations
\begin{align}
 \Delta = (\beta \gamma\, \eps c_\varphi)^2 + \beta (\kappa^2-\eps^2)
\,,\quad
\beta = \frac{1}{1 + (\eps c_\varphi)^2}
\,.\end{align}
The leptonic structure functions defined in \eq{L_i_CS_angles} are then given by
%%%
\begin{align} \label{eq:L_pTlep_1}
\frac{L_i(q, \pTlep)}{L_{\pm(i)}(q^2)}
&= \frac{3 \pTlep}{4\pi Q^2} \int_0^{2\pi} \df\varphi \frac{\beta}{\sqrt\Delta}\,
   \theta\Bigl(c_\varphi - \frac{\kappa - \gamma}{\eps}\Bigr)
   \int_0^\pi \df \theta \, s_\theta \, g_i(\theta, \varphi) \,
   \delta[s_\theta - s(c_\varphi)]
\,.\end{align}
%%%
Note that \eq{L_pTlep_1} still shows the full dependence on $\varphi$ and $s_\theta$,
such that one could easily reinstate fiducial cuts.

\Eq{sol_st_1} yields two solutions for $\theta_\varphi$, related by
$\theta \to \pi-\theta$. Since the $g_i(\theta,\varphi)$ for $i = 1, 4, 6$ are odd
under this transformation, \eq{L_pTlep_1} immediately vanishes for these cases.
In all other cases, the $g_i$ are even under this transformation, such that we obtain
\begin{align} \label{eq:L_pTlep_2}
 \frac{L_i(q, \pTlep)}{L_{\pm(i)}(q^2)} &
 = \frac{3\pTlep}{2\pi Q^2} \int_0^{2\pi} \df\varphi \frac{\beta}{\sqrt\Delta}\,
   \theta\Bigl(c_\varphi - \frac{\kappa - \gamma}{\eps}\Bigr)
   \frac{s(c_\varphi) \, g_i(\theta_\varphi, \varphi)}{[1 - s(c_\varphi)^2]^{1/2}}
\qquad
(i \ne 1, 4, 6)
\,,\end{align}
where $\theta_\varphi$ can be either of the two physical solutions defined by \eq{sol_st_1}.

It is now straightforward to expand in $\eps$ to study the $q_T\to0$ limit, which yields
%%%
\begin{align} \label{eq:L_pTlep_LP}
\frac{L_i(q, \pTlep)}{L_{\pm(i)}(q^2)}
&= \frac{3 \pTlep}{2\pi Q^2}\,
\frac{\theta(1 - \kappa)}{\sqrt{1-\kappa^2}} \int_0^{2\pi} \df\varphi \,
   g_i(\arcsin\kappa, \varphi) + \cO(\eps)
\nn\\
&= \frac{3 \pTlep}{Q^2} \frac{\theta(1 - \kappa)}{\sqrt{1-\kappa^2}}
 \begin{cases}
  2 - \kappa^2 \,,\qquad &i = -1 \,, \\
  \kappa^2     \,,\qquad &i = 0  \,, \\
  0 \,,&i = 1,\ldots,7 \,.
 \end{cases}
\end{align}
%%%
The constraint $\kappa \leq 1$ reflects the strict bound $\pTlep \leq Q/2$ in
the Born limit. We also recover that at LP only the $i=-1,0$ contributions survive.
The $i = 4$ contribution, which in principle can contribute at LP
and gives rise to the forward-backward asymmetry, vanishes due to the symmetry
of $\pTlep$ under $\theta \to \pi-\theta$.

Before proceeding, it is instructive to illustrate the phase space differential in $\pTlep$,
which is closely related to $L_i(q,\pTlep)$ and provides a bound on the $L_i(q,\pTlep)$
since all $g_i(\theta,\varphi)$ are bounded. It can be evaluated more easily using the
parametrization of the lepton momenta given in \eq{momenta}. After some effort, one obtains
%%%
\begin{align} \label{eq:phi_exact}
 \frac{\df\Phi_L(q)}{\df \pTlep} &
 = \int\!\df\Phi_L(q) \, \delta\bigl[\pTlep - \pTlep(\Phi_L)\bigr]
\nn \\
&= \begin{cases}
   \displaystyle\frac{1}{2\pi^2 Q} \frac{\alpha \kappa }{\sqrt{\alpha^2-\kappa^2}}\,
   K\biggl[-\frac{(1-\alpha^4)\kappa^2}{\alpha^2-\kappa^2}\biggr]
   \quad & 0 < \kappa \le \alpha
   \,,\\[2ex]
   \displaystyle\frac{1}{2\pi^2Q}  \frac{\alpha \kappa}{\sqrt{\kappa^2-\alpha^2}}\, K\biggl[\frac{\alpha^2(1-\alpha^2\kappa^2)}{\alpha^2-\kappa^2}\biggr]
   \,,\quad & \alpha < \kappa < 1/\alpha
   \,,\\[2ex]
   0  &\kappa \ge 1/\alpha
 \end{cases}
\end{align}
%%%
Here, $\alpha = \sqrt{1+\eps^2} - \eps$, and $K(x)$ is the complete
elliptic integral of the first kind. The appearance of three distinct regions
can easily be understood from \eq{L_pTlep_2}: For $\kappa > 1/\alpha$,
the $\theta$ function in \eq{L_pTlep_2} becomes incompatible with $c_\varphi \leq 1$,
while for $\kappa < \alpha$ it imposes no constraint in addition to $c_\varphi \geq -1$.

\begin{figure*}
\includegraphics[width=\WidthTwoSubfigs]{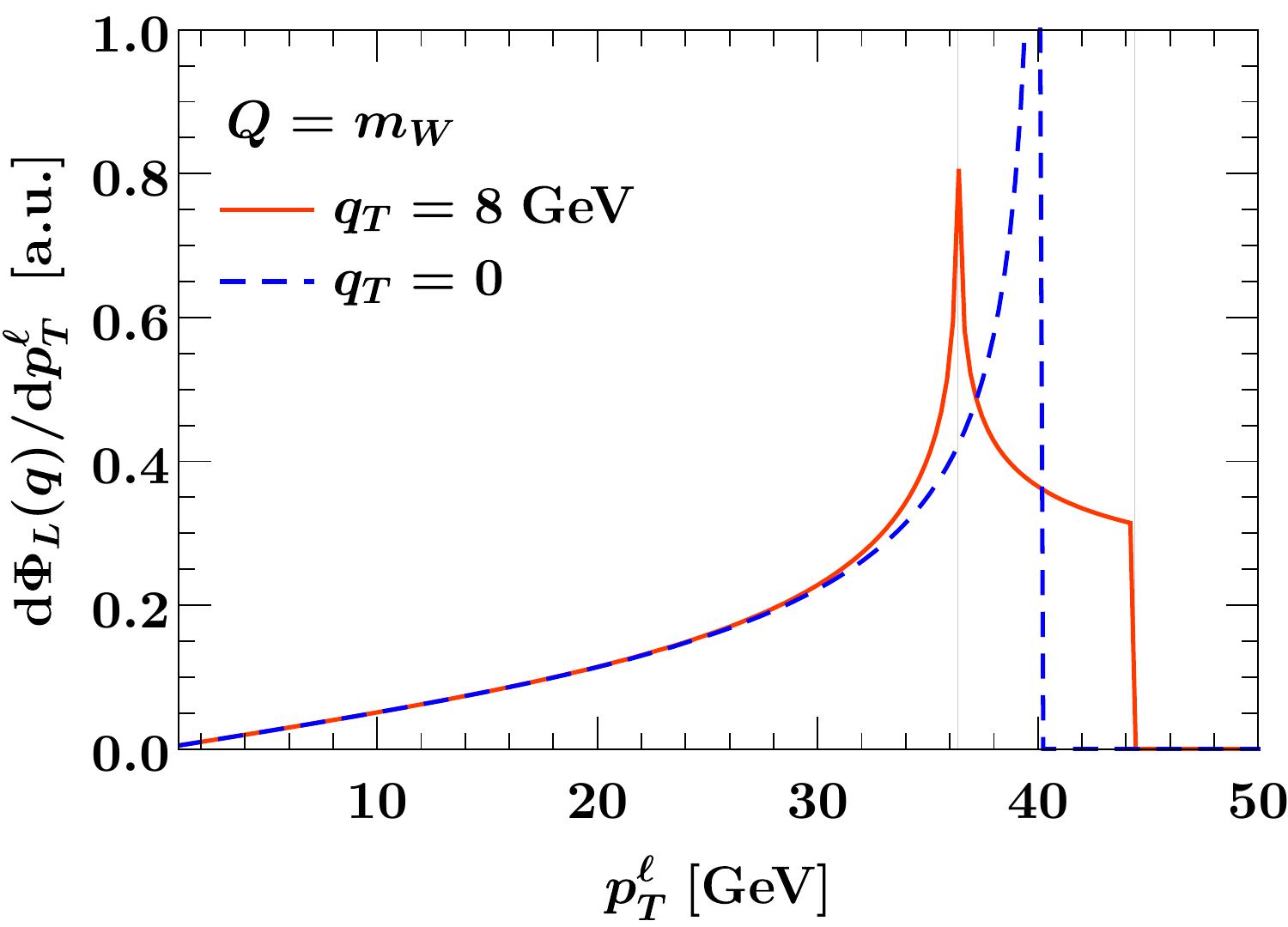}%
\hfill%
\includegraphics[width=\WidthTwoSubfigs]{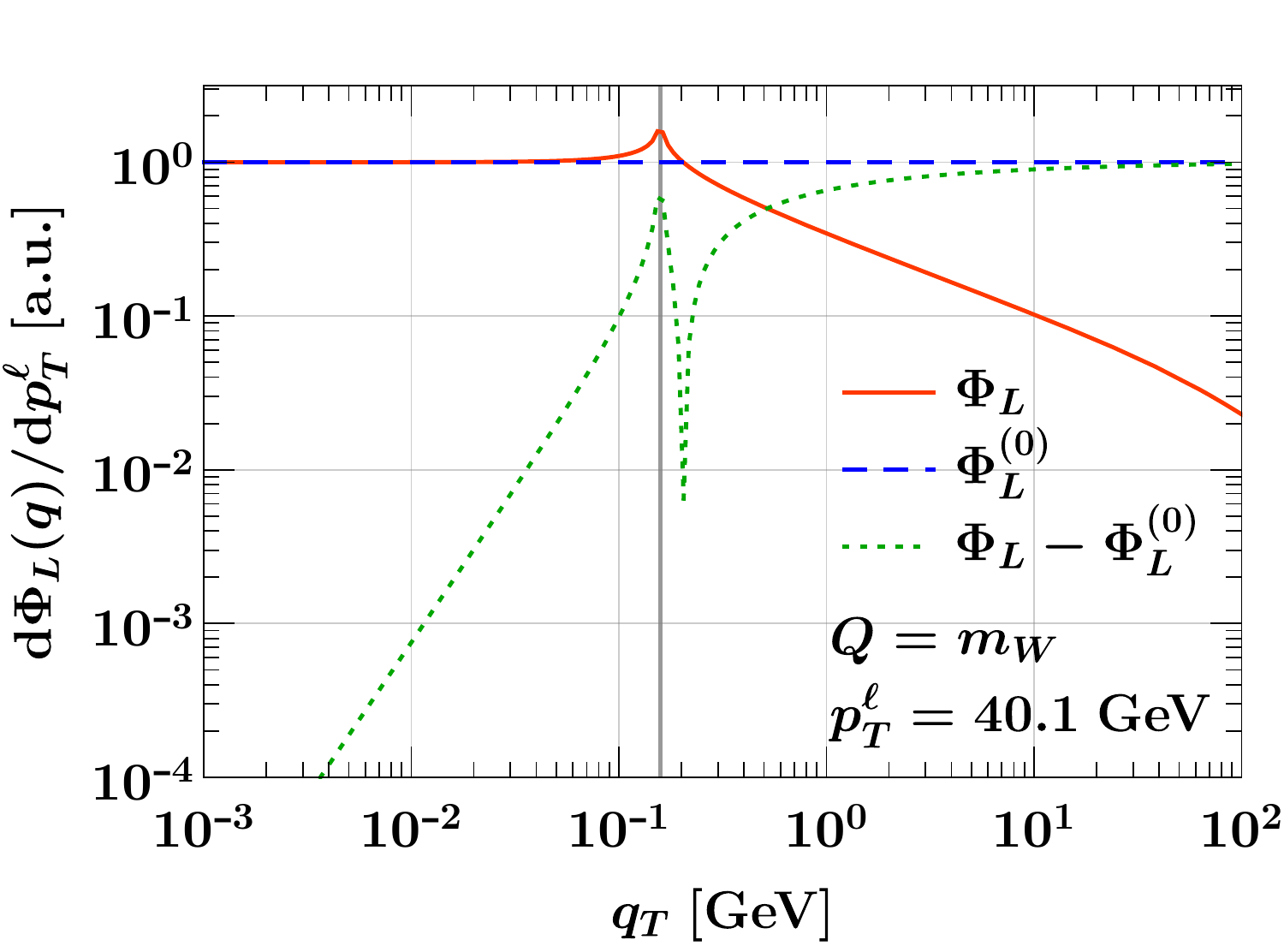}%
\caption{%
Leptonic phase space $\df\Phi_L(q)/\df \pTlep$,
as a function of $\pTlep$ for fixed $q_T$ (left)
and as a function of $q_T$ for fixed $\pTlep = 40.1\GeV$ (right).
In both cases we fix $Q=m_W=80.358\GeV$.
Left: The red solid curve shows the exact result for $q_T = 8\GeV$, and the blue
dashed curve shows the $q_T = 0$ Born limit.
Right: The red solid curve shows the exact phase space, the blue dashed line
the LP limit, and the green dotted curve their difference.
}
\label{fig:phiL}
\end{figure*}

In \fig{phiL}, we show the differential phase space in $\pTlep$
in \eq{phi_exact} for $Q=m_W$.
In the left panel, we show the exact phase space as a function of $\pTlep$
for fixed $q_T = 8\GeV$ (red solid), with the gray vertical lines indicating
the edges of the different regions in \eq{phi_exact}. In the $q_T = 0$
LP limit (blue dashed), they collapse to the kinematic Born limit $\pTlep \leq Q/2$.
In the right panel, we fix $\pTlep = 40.1\GeV$ very close
to the Born edge, and show the phase space as a function of $q_T$. The exact result
is shown by the red curve, the $q_T = 0$ LP limit by the blue dashed line, and
their difference by the green dotted line. The thick vertical line at
$q_T = Q - 2\pTlep \equiv p_L$ shows the transition to the second region of
\eq{phi_exact}. For sufficiently small $q_T \ll p_L$, we see a clear (quadratic) power
suppression, while near and above this value the power corrections become $\ord{1}$.
(The sharp dip in the green line is just an artefact of the logarithmic scale
and the green line changing its sign.)

%-------------------------------------------------------------------------------
%-------------------------------------------------------------------------------

Clearly, expanding in $\eps \sim \lambda$ is only well-defined if $\kappa \ll
1$, i.e., away from the Jacobian peak $\pTlep \ll Q/2$. This is already evident
from the divergence of \eq{L_pTlep_LP} as $\kappa\to1$. As long as $\kappa \ll
1$, $\eps$ is the only small scale in the problem, which justifies expanding in
$\eps$ and leads to at most linear fiducial power corrections.
On the other hand, close to the Jacobian peak, the distance $p_L = Q - 2\pTlep$
emerges as an additional small scale, and the naive expansion in $q_T$
is actually an expansion in $q_T/p_L$, which is only allowed for $q_T \ll p_L$ but breaks
down for $p_L \sim q_T$. To illustrate this explicitly, we can expand in the regime
$p_L \sim q_T$ by simultaneously counting both scales as small. To do so, we take
%%%
\begin{align}
p_L = Q - 2\pTlep \sim \lambda Q
\,,\qquad
1 - \kappa = \frac{p_L}{Q} \sim \lambda
\,,\end{align}
%%%
where as before $\eps = q_T/Q \sim \lambda$, such that formally
$q_T/(Q - 2\pTlep)\sim 1$. With this replacement, we can expand \eq{L_pTlep_2}
in $\lambda$,
%%%
\begin{align} \label{eq:pTlep_L_i_near_peak}
\frac{L_i(q, \pTlep)}{L_{\pm(i)}(q^2)}
&= \frac{3}{4\pi} \frac{1}{\sqrt{2Q}}
\int_0^{2\pi} \! \df \varphi \, \frac{\theta(q_T\, c_\varphi + p_L)}{\sqrt{q_T\, c_\varphi + p_L}} \, g_i\bigl(\pi/2, \varphi \bigr)
\times \bigl[ 1 + \ord{\lambda} \bigr]
\,.\end{align}
%%%
This vanishes for all $g_i$ odd in $\varphi$,
which only leaves $i = -1, 0, 2, 3$. This should be contrasted
with the naive LP result in \eq{L_pTlep_LP}, which only
receives contributions from $i=-1,0$.
The $i = 2$ contribution is proportional to the double-Boer-Mulders effect,
which we can neglect, see the discussion below \eq{tmd_factorization_m1_2_4_5}.
For $i = 3$ we have $W_3 \sim \cO(\lambda)$,
see \tab{power_counting_W_i}, which thus yields a linear power correction.
Hence, we find the interesting effect that the proximity to the Jacobian peak
induces sensitivity to new hadronic structure functions at $\ord{\la}$, which
do not contribute at $\ord{\la}$ away from the peak region.

From \eq{pTlep_L_i_near_peak} it is evident that naively expanding in $q_T$ near the
Jacobian peak would amount to expanding in $q_T/p_L$, which is not allowed.
However, \eq{pTlep_L_i_near_peak} is only valid near the peak, because by counting
$p_L/Q \sim \lambda$ we have expanded away the dependence on $\kappa = 1 + \ord{\lambda}$,
which is not allowed away from the peak.
Hence, to cover the full range of $\pTlep$, we must not expand in $p_L$, while
near the peak we must count $q_T \sim p_L$ to avoid inducing uncontrolled leptonic
power corrections in $q_T/p_L$. Clearly, the simplest way to satisfy both
requirements is to not expand at all and keep the exact result corresponding
to \eq{L_pTlep_2}.

Finally, note that the breakdown of the naive power expansion around $\pTlep = Q/2$
does not immediately affect the leptonic tensor if we only consider a fiducial
cut $\pTlep \ge \pTmin$, since we can evaluate it as
%%%
\begin{align}
\Phi_L(q,\pTmin)
&= \int_{\pTmin} \df \pTlep \frac{\df\Phi_L(q)}{\df \pTlep}
 = \frac{1}{8\pi} - \int_0^{\pTmin} \df \pTlep \frac{\df\Phi_L(q)}{\df \pTlep}
\,.\end{align}
Thus, the leptonic power corrections in this case scale as $q_T/(Q - 2\pTmin)$,
and so as long as $\pTmin \ll Q/2$, the effect of $\pTmin$ can be treated as
a linear fiducial power correction as discussed for the $q_T$ spectrum with
fiducial cuts in \sec{numerics_qT}.

%===============================================================================
\subsubsection{Numerical results}
\label{sec:pTlep_results}
%===============================================================================

%-------------------------------------------------------------------------------
\begin{figure*}
\centering
\includegraphics[width=\WidthTwoSubfigs]{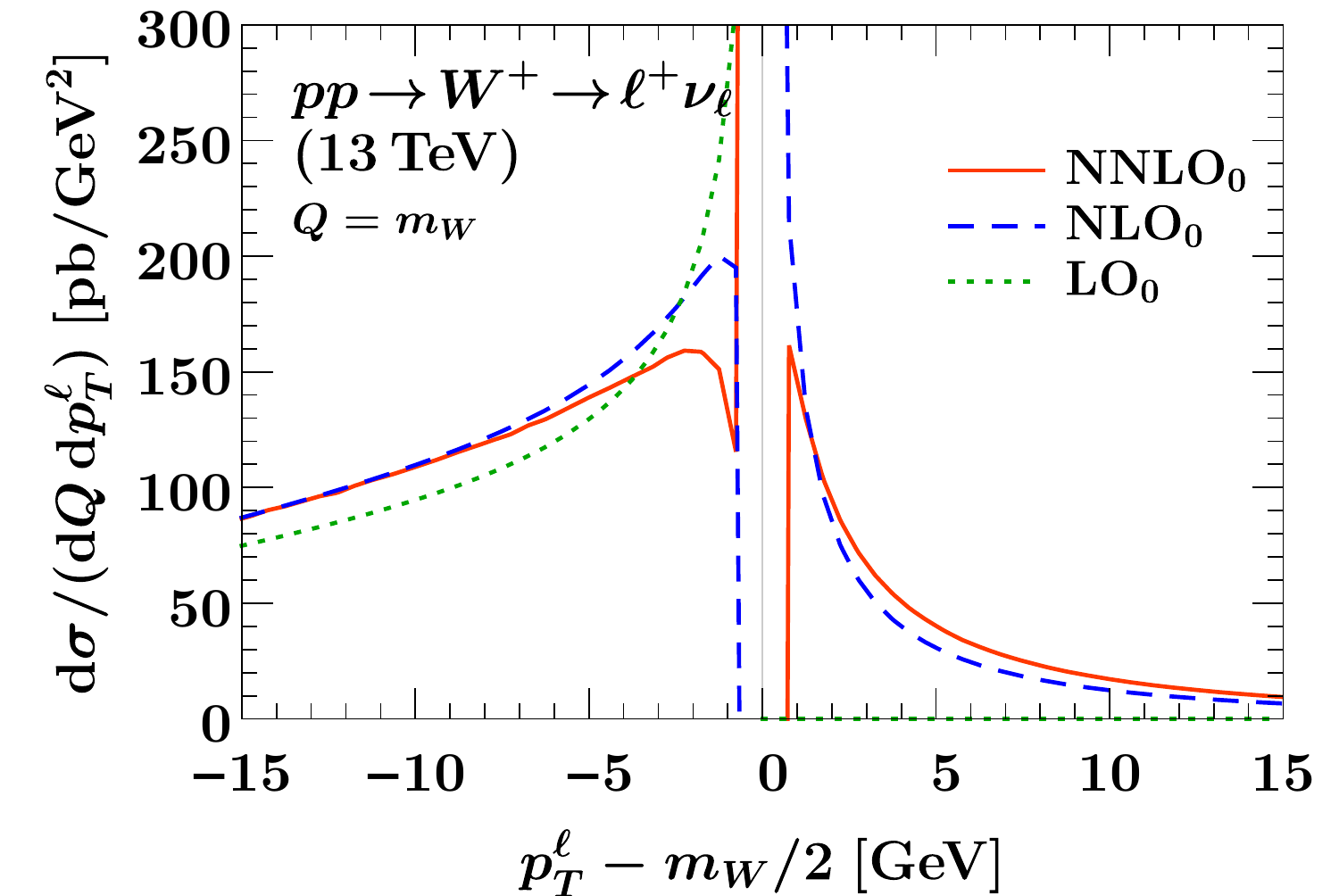}%
\hfill%
\includegraphics[width=\WidthTwoSubfigs]{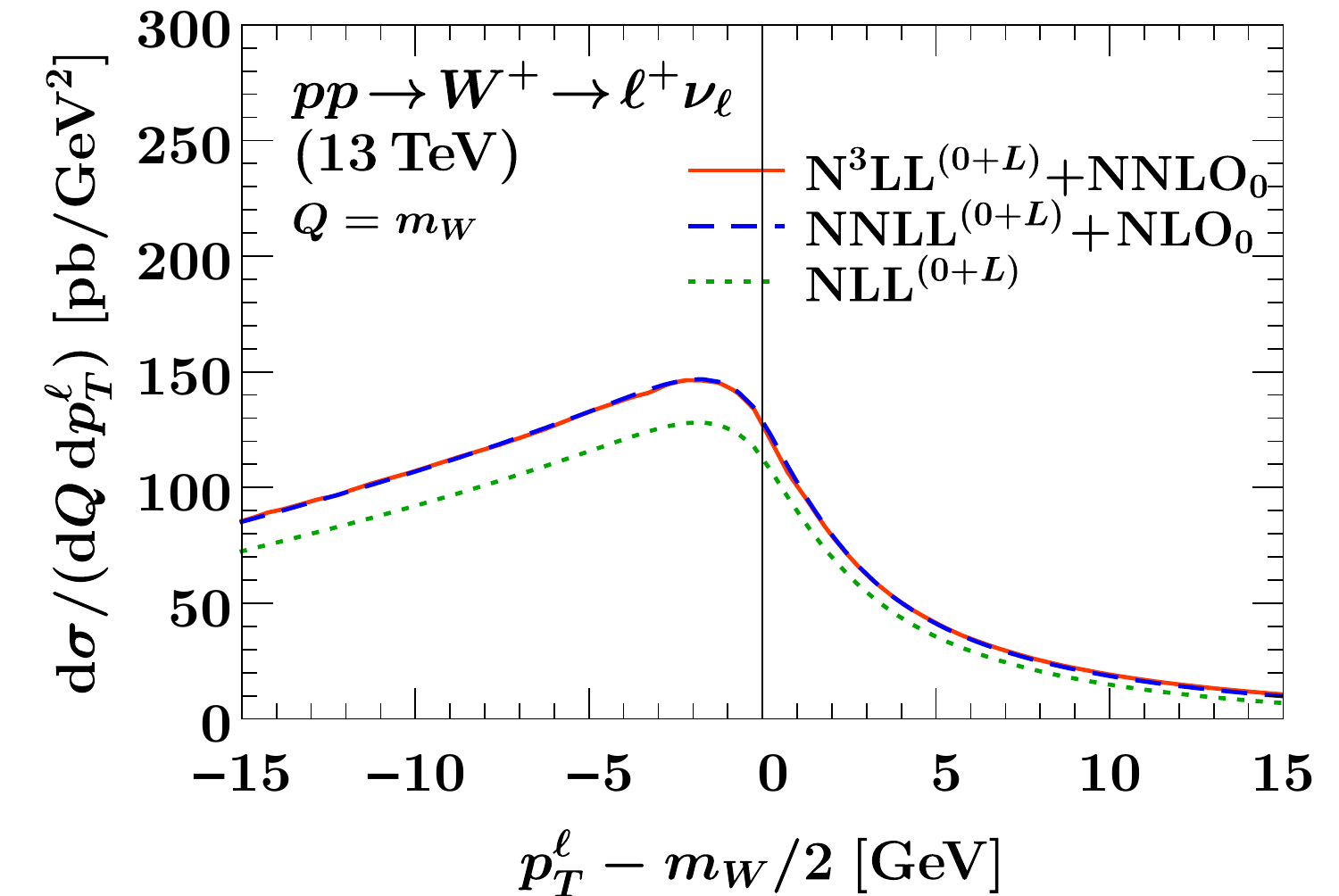}%
\caption{Lepton transverse momentum spectrum for on-resonance $W^+$ production at the LHC
at fixed order (left) and including the resummation of fiducial power corrections
to N$^3$LL (right).
The horizontal axes shows the distance to the Jacobian peak at
$\pTlep = m_W/2$.}
\label{fig:Wplus_pT2_8TeV}
\end{figure*}
%-------------------------------------------------------------------------------

There are two key insights from our analysis of the differential $\pTlep$ phase space.
First, the $\pTlep$ spectrum near the Jacobian peak is directly sensitive to
the small transverse momentum $q_T$ of the decaying vector boson.
This causes fixed-order predictions to become unreliable in this region, which
is a well-known effect.
Second, the strict $q_T\to0$ limit by itself cannot describe the $\pTlep$ spectrum
in this region, which means the strict LP $q_T$ resummation is also insufficient.
Both problems are cured simultaneously by combining the exact leptonic tensor,
which encodes the exact decay kinematics and automatically retains all leptonic
power corrections, with the $q_T$-resummed hadronic tensor,
thus allowing us to obtain physical predictions around the Jacobian peak.

We illustrate this in \fig{Wplus_pT2_8TeV} for the $\pTlep$ spectrum in
$W^+ \to \ell^+ \nu_\ell$ decays,
where we show the spectrum both at fixed order (left) and after resummation
including fiducial power corrections (right).
In both panels, the horizontal axis shows the distance of $\pTlep$ to the Jacobian peak
at $\pTlep = m_W/2$, and to avoid smearing out the peak we consider the spectrum
at a fixed point $Q=m_W$.
The fixed-order spectrum (left) is shown at LO$_0$ (green dotted), NLO$_0$ (blue dashed),
and NNLO$_0$ (red solid).
The LO$_0$ result corresponds to Born kinematics and clearly shows the kinematic
edge at $\pTlep = Q/2$. Starting at NLO$_0$, the $W$ boson can
have nonvanishing $q_T$, which opens up the phase space beyond the edge.
However, in the vicinity of the edge, the fixed-order predictions become
unstable due to the sensitivity to small $q_T$, which is clearly visible
by the diverging NLO$_0$ and NNLO$_0$ curves, and in particular by
the sign change between NLO$_0$ and NNLO$_0$ at $\pTlep \approx Q/2$.

In the right panel in \fig{Wplus_pT2_8TeV}, we show the resummed $\pTlep$
spectrum at NLL$^{(0+L)}$ (green dotted), NNLL$^{(0+L)}$+NLO$_0$ (blue dashed),
and N$^3$LL$^{(0+L)}$+NNLO$_0$ (red solid). The resummation including leptonic
power corrections cures the unphysical behaviour of the fixed-order results,
yielding a well-behaved spectrum in the full $\pTlep$ range, with a resummed
Sudakov shoulder at $\pTlep \approx m_W/2$. Note that the cross section beyond
the edge is already populated at NLL$^{(0+L)}$ without any fixed-order matching.
We stress that without including the exact leptonic tensor, the resummation
would only affect the region $\pTlep < m_W/2$, and not cure the peak region. In
fact, the results with strict LP resummation would look very similar to the
pure fixed-order results, with the N$^3$LL$^{(0)}$+NNLO$_0$ essentially
indistinguishable from the pure NNLO$_0$ result.

This is the first time that resummed N$^3$LL results for the $\pTlep$ spectrum
are presented, and we observe extremely good perturbative convergence,
with the results at NNLL$^{(0+L)}$+NLO$_0$ and N$^3$LL$^{(0+L)}$+NNLO$_0$
falling on top of each other.
We leave a more detailed phenomenological analysis of the $\pTlep$ spectrum
to future work.

%===============================================================================
\subsection[\texorpdfstring{$\phi^*$}{phi*} spectrum]
           {$\phi^*$ spectrum}
\label{sec:numerics_phistar}
%===============================================================================

The $\phi^*$ observable was first proposed in \refcite{Banfi:2010cf},
extending earlier work on the $a_T$ observable \cite{Ackerstaff:1997rc,Vesterinen:2008hx}.
Both observables are sensitive to small $q_T$, but promise better experimental resolution
than $q_T$ itself due to being based on angular measurements, which are less susceptible
to the momentum resolution of the individual lepton momenta than $q_T$ itself.

The factorization and resummation for $a_T$ was first studied in \refcite{Banfi:2009dy} at NLL,
and extended to both $a_T$ and $\phi^*$ at NNLL$+$NLO in \refscite{Banfi:2011dx, Banfi:2011dm,
Banfi:2012du}, and also in \refscite{Guzzi:2013aja} including a study of nonperturbative
contributions, and also more recently at N$^3$LL$+$NNLO in \refscite{Bizon:2017rah, Bizon:2018foh}.
None of these calculations incorporate the finite recoil of the dilepton system
in the calculation of the $\phi^*$ observable, i.e.\ the employed definition of $\phi^*$
is only an approximation of the actually measured observable, as discussed below.
The resummation with the exact definition of $\phi^*$ was considered in
\refscite{Catani:2015vma, Becher:2019bnm} at NNLL$+$NNLO and NNLL$+$NLO via
parton-level MC integration of the leptonic final state.

\Refcite{Banfi:2010cf} defines the two closely related observables
$\phi^*_\CS$ and $\phi^*_\eta$. We only consider the latter, as it is
more commonly used in experiments. It is defined as
\begin{align} \label{eq:phistar}
 \phi^* \equiv \phi^*_\eta = \tan(\phi_\mathrm{acop}/2) \, \sin\theta_\eta^*
\,,\end{align}
where the acoplanarity angle is $\phi_\mathrm{acop} = \pi - \Delta\varphi$,
with $\Delta\varphi$ being the azimuthal opening angle between the leptons in the lab frame, and
\begin{align}
 \cos\theta_\eta^* = \tanh\frac{\eta_1 - \eta_2}{2}
\,,\end{align}
where $\eta_{1,2}$ are the two lepton rapidities.

%===============================================================================
\subsubsection{Origin of power corrections}
\label{sec:phiStar_power_corrections}
%===============================================================================

Using the parametrization of the lepton momenta in the Collins-Soper frame
as given in \eq{p1_p2_CS_angles_CS_frame}, \eq{phistar} can be written as
\begin{align} \label{eq:phistar_exact}
 (\phi^*)^2 &= \frac{ 8 \kappa (\eps \sin\varphi \sin\theta)^2}{(\kappa -\eps^2+\alpha^2)^2 (\kappa + \eps^2-\alpha^2 + 2)}
\,,\nn\\
 \kappa^2 &= (\eps^2-\alpha^2)^2 + 4 \eps^2 \sin^2\theta \sin^2\varphi
 \,,\qquad
 \alpha^2 = \bigl(1 + \eps^2 \cos^2\varphi \bigr) \sin^2\theta
\,.\end{align}
Note that $\phi^*$ is boost invariant and thus independent of $Y$,
and can depend on $q_T$ only through the dimensionless ratio $\eps = q_T/Q$.
From \eq{phistar_exact}, one easily finds the special values
\begin{align} \label{eq:phistar_limits}
 \phi^*|_{\theta = \pi/2} = \eps \abs{\sin\varphi}
\,,\qquad
 \lim_{\theta \to 0, \pi} \phi^* = \infty
\,.\end{align}
The singularity arises from the case where both momenta are parallel to each other in the transverse plane, such that $\phi_{\rm acop} = \pi$ and \eq{phistar} becomes ill-defined.
Numerically, we have also tested that $\phi^*$  monotonically decreases with $|\!\sin\theta|$,
such that $\phi^*$ can be uniquely inverted on the intervals $\theta \in [0,\pi/2]$ and $\theta \in [\pi/2,\pi]$, with the two solutions trivially related by $\theta_1 = \pi-\theta_2$.

Expanding \eq{phistar_exact} in $\eps \ll 1$, one obtains the commonly employed approximation
\begin{align}  \label{eq:phistar_approx}
 \phi^{*\,(0)} = \eps \, \abs{\sin\varphi}
\,.\end{align}
The monotonicity of $\phi^*$ with $\abs{\sin\theta}$ implies that this is a lower bound to $\phi^*$,%
\footnote{This implies that a phase space generator producing events
with $q_T \ge q_T^{\rm min}$ is guaranteed to correctly describe $\phi^* \ge q_T^{\rm min}/Q$.}
but from \eq{phistar_limits} it follows that this bound is only saturated for $\theta=\pi/2$.
It is thus natural to ask whether there is a better approximation of $\phi^*$.
From \eq{phistar_exact}, it is easy to see that $\phi^*$ only vanishes
if either $\eps = 0$ or $\sin\varphi = 0$.
Expanding \eq{phistar_exact} in these limits, we find
\begin{align} \label{eq:phistar_expanded}
 (\phi^*)^2 &
 = \frac{\eps^2 \sin^2\varphi}{1+ \eps^2(1- \sin^{-2}\theta)} + \cO(\sin^4\varphi)
 \nn\\&
 = \eps^2 \sin^2\varphi \bigl[ 1 + \eps^2 \cot^2 \theta + \, \cO(\eps^4) \bigr]
\,.\end{align}
Note that the expansion in small $\eps$ in the second line can be recovered by
reexpanding the small-$\sin\varphi$ limit,
because in \eq{phistar_exact} each term in $\eps$ either multiplies $\sin\varphi$
or is enhanced relative to $\eps \sin\varphi$.
The first line in \eq{phistar_expanded} is ill-defined for $\abs{\tan\theta} < \eps$
already at the leading $\ord{\sin^2\varphi}$,
while this singularity only appears at the second order in $\eps$.
\Eq{phistar_expanded} suggests that the fundamentally small quantity to be power counted is $\eps \abs{\sin\varphi}$,
not $\eps$ itself. In particular, any given value in $\phi^*$ can in principle
receive contributions from arbitrarily large $\eps$.

\begin{figure*}
 \centering
 \includegraphics[width=\WidthTwoSubfigs]{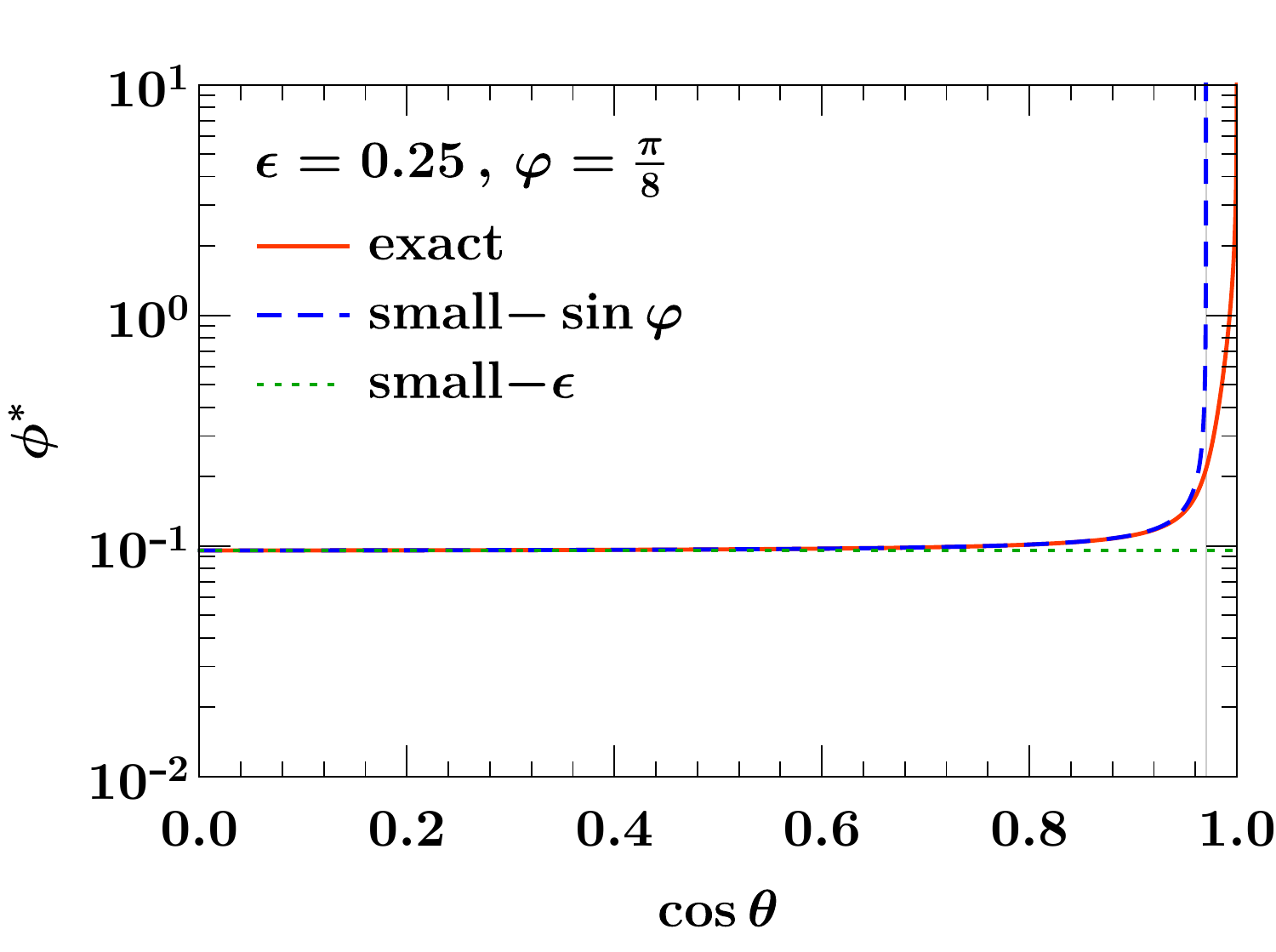}%
 \hfill%
 \includegraphics[width=\WidthTwoSubfigs]{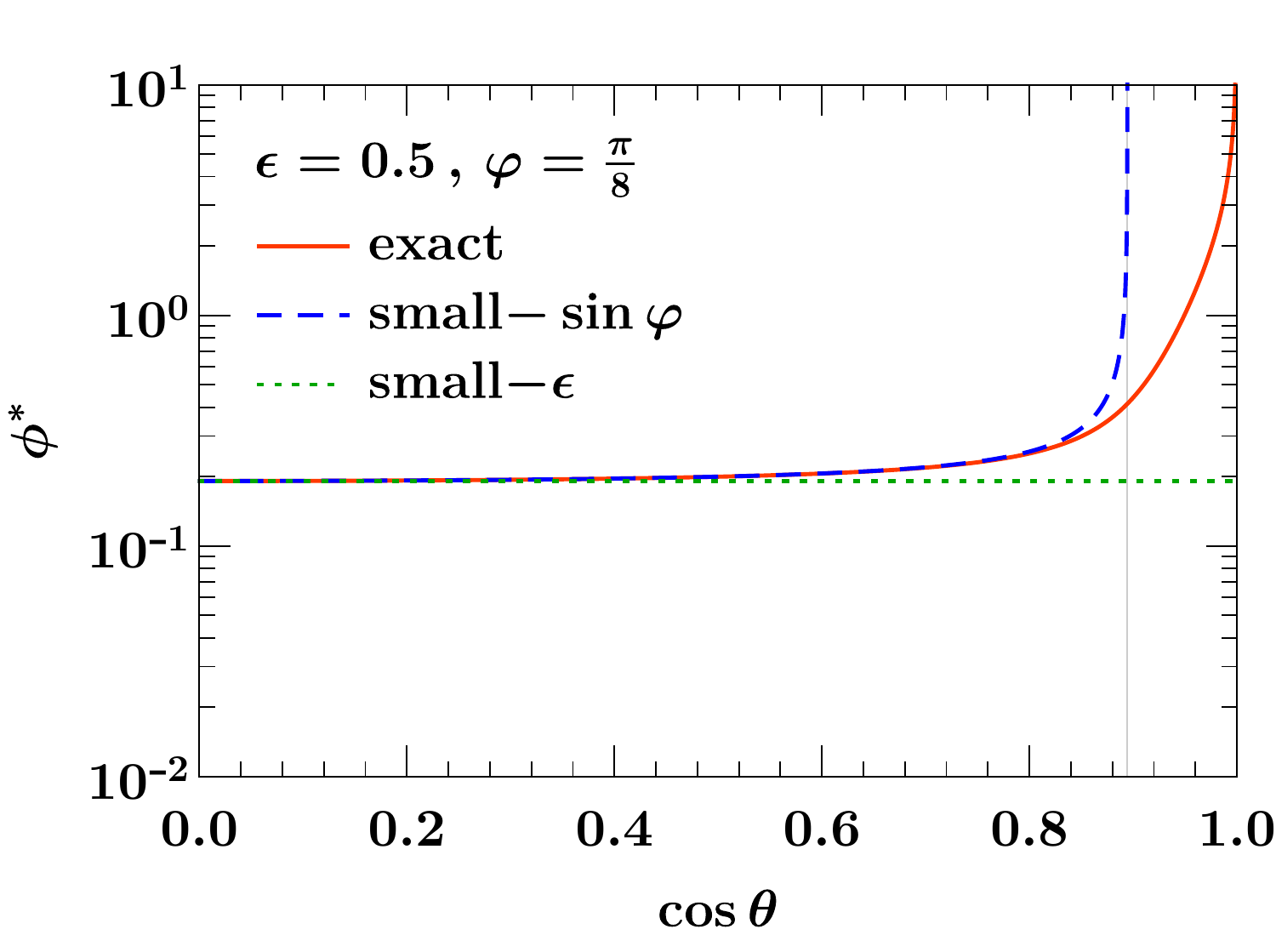}%
 \caption{Comparison of the different expansions of $\phi^*$, as given in \eq{phistar_expanded},
 as function of $\cos\theta$ for two different choices of $\eps$ and $\varphi$.
 The vertical gray line indicates the breakdown of the expansion in small $\sin\varphi$.}
 \label{fig:phistar_approximations}
\end{figure*}

In \fig{phistar_approximations}, we compare the exact result for $\phi^*$ (red solid)
to its leading expansions in small $\sin\varphi$ (blue dashed)
and small $\eps$ (green dotted). We fix $\varphi = \pi/8$ and show results
for $\eps=0.25$ (left panel) and $\eps = 0.5$ (right panel). The gray vertical line
shows the breakdown of the small-$\sin\varphi$ approximation at $\cos\theta = (1+\eps^2)^{-1/2}$.
Since $\phi^*$ is fairly insensitive to $\cos\theta$ in a rather large range of $\cos\theta$,
the small-$\eps$ expansion is a fairly good approximation in that region.
However, it quickly deteriorates for moderate to large $\cos\theta$.
In contrast, the small-$\sin\varphi$ expansions follows the exact curve remarkably well,
almost up to the point where it becomes ill-defined. At large $\cos\theta$, both approximations break down,
as $\phi^*$ is driven by the small-$\theta$ behavior $\phi^* \sim 1/\theta^2$.

We find that the region $\abs{\tan\theta} < \eps$ cannot be described
by the expansion in small-$\sin\varphi$, which breaks down,
or by the expansion small-$\eps$, which assigns an artificially
small value $\phi^{*(0)}$ in this region. However, this does not invalidate
the LP description of $\phi^*$, as this region of phase space is suppressed as $\cO(\eps^2)$,
\begin{align}
 \int\df\Phi_L(q) \, \theta(\abs{\tan\theta} < \eps)
 = \frac{1}{8\pi} \bigl(1 - \sqrt{1-\eps^2}\bigr)
 = \frac{\eps^2}{16\pi} + \cO(\eps^4)
\,.\end{align}

%-------------------------------------------------------------------------------
%-------------------------------------------------------------------------------

Another important property of $\phi^*$ is that it is not azimuthally symmetric
at LP due to its explicit dependence on $\varphi$.
To identify which $W_i$ contribute to $\phi^*$ at LP, we evaluate \eq{L_i_CS_angles}
with the approximate observable in \eq{phistar_approx}, which yields
\begin{align} \label{eq:resummed_phistar_approx}
 \frac{\df\sigma}{\df Q^2 \df Y \df\!\cos\theta \, \df\phi^{*\zero}} &
 = \frac{3 Q}{16\pi \Ecm^2} \sum_{i=-1,0,2,4} L_{\pm(i)}(Q^2) \, g_i(\theta,0)
   \nn\\&\quad\times
   \int_0^\infty \df b_T \, K_i(b_T Q \phi^{*\zero}) \tilde W_i(Q^2, Y, b_T)
\,.\end{align}
%%%
Here, the $\tilde W_i$ are the hadronic structure functions in Fourier space.
In \eq{resummed_phistar_approx}, we have already carried out the integral over $\varphi$,
which gives rise to kernels $K_i$, while we are still differential in $\cos\theta$.
One can easily incorporate any LP fiducial cuts that depend on $\theta$,
but are independent of $\varphi$ into \eq{resummed_phistar_approx},
which holds for most common cuts such as \eq{cuts}.

The nonvanishing kernels entering \eq{resummed_phistar_approx} are given by
\begin{align}
 K_{-1,0,4}(\beta) &= \cos(\beta)
 \,,\qquad
 K_2(\beta) = \cos(\beta) + 2 \beta \operatorname{si}(\beta)
\,,\end{align}
where $\operatorname{si}(\beta) = -\int_\beta^\infty \! \df t \, \sin t/t$ is the sine integral.
The kernels for $i=1,3,5,6,7$ vanish because the corresponding $g_i(\theta,\varphi)$
are odd in $\varphi$ or under $\varphi \to \varphi + \pi$.
Since $g_{-1,0,4}(\theta,\varphi)$ are independent of $\varphi$,
they give rise to the same kernel $K_{-1,0,4}$.
In contrast, $W_2$ is dressed with a different kernel $K_2$ due to
the nontrivial $\varphi$ dependence of $g_2(\theta, \varphi) \propto \cos(2\varphi)$.
In particular, for $\beta\to\infty$ one has $K_2(\beta) \approx - \cos\beta = K_{-1,0,4}(\beta)$,
and thus there is a relative phase shift of $\pi$.

\Eq{resummed_phistar_approx} is convenient,
as it effectively only reweights the (resummed) hadronic tensor
in Fourier space with $K_i(b_T Q \phi^*)$, compared to $J_0(b_T q_T)$
appearing in $q_T$ resummation. In momentum space, this is equivalent
to the fact that the spectrum for the LP $\phi^{*\zero}$ in \eq{phistar_approx}
can be obtained by reweighting the (resummed) $q_T$ distribution
with the angle to the dilepton system.
The convenient form of \eq{resummed_phistar_approx} was first noticed
in~\refscite{Banfi:2009dy,Banfi:2011dx}, where it was also noted that the
$\cos(b_T Q \phi^*)$ gives rise to a plateau in the resummed $\phi^*$ spectrum,
in contrast to the Sudakov peak encountered in $q_T$ resummation.
However, the form in \eq{resummed_phistar_approx} has the distinct disadvantage
that it does not allow one any longer to include fiducial power corrections due
to additional fiducial cuts beyond the strict LP.

The above previous works did not consider the contribution from $W_2^{(0)}$,
which involves the double-Boer-Mulders contribution, see \eq{tmd_factorization_m1_2_4_5}.
Comparing to \tab{power_counting_W_i}, $W_2$ does not contribute to azimuthally
symmetric observables at LP, and thus is not encountered in the LP $\qt$ resummation.
In practice, we expect this contribution to be rather small,
as $h_1^\perp$ only matches onto subleading twist-3 collinear PDFs,
and thus we will not consider it in our numerical study.
Nevertheless, it is interesting to note that the $\phi^*$ spectrum
may give direct access to the double Boer-Mulders effect,
and we leave a more detailed study of this for future work.
At leading twist-2, $W_2$ is suppressed as $\cO(\lambda^2)$,
such that our resummed spectrum $\sigma^{(0+L)}$ still fully captures all linear
power corrections arising from small $q_T$ with leading-twist collinear PDFs,
and this also holds when including additional fiducial cuts.

%===============================================================================
\subsubsection{Numerical results}
\label{sec:phiStar_results}
%===============================================================================

\begin{figure*}
\includegraphics[width=\WidthTwoSubfigs]{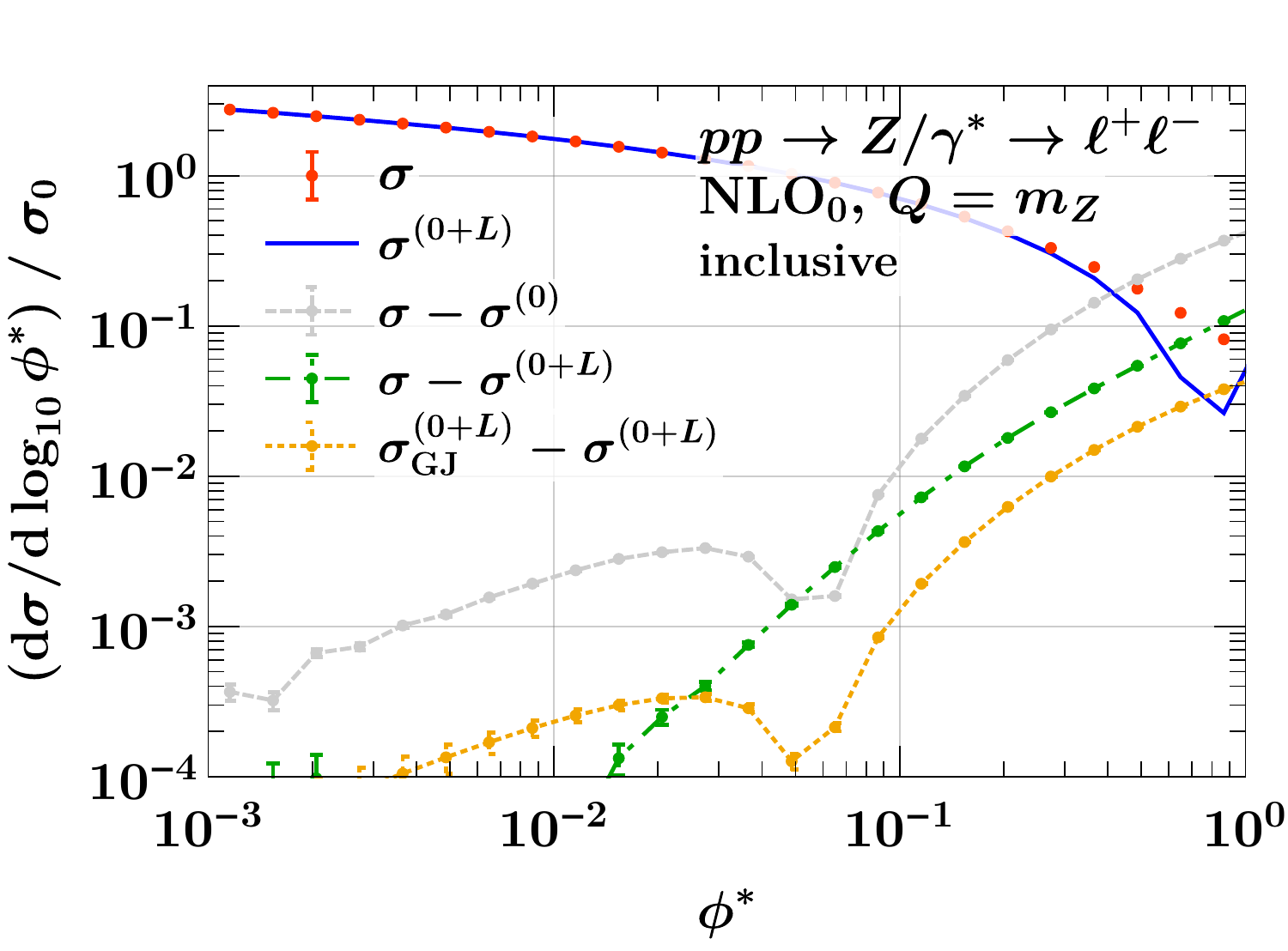}%
\hfill%
\includegraphics[width=\WidthTwoSubfigs]{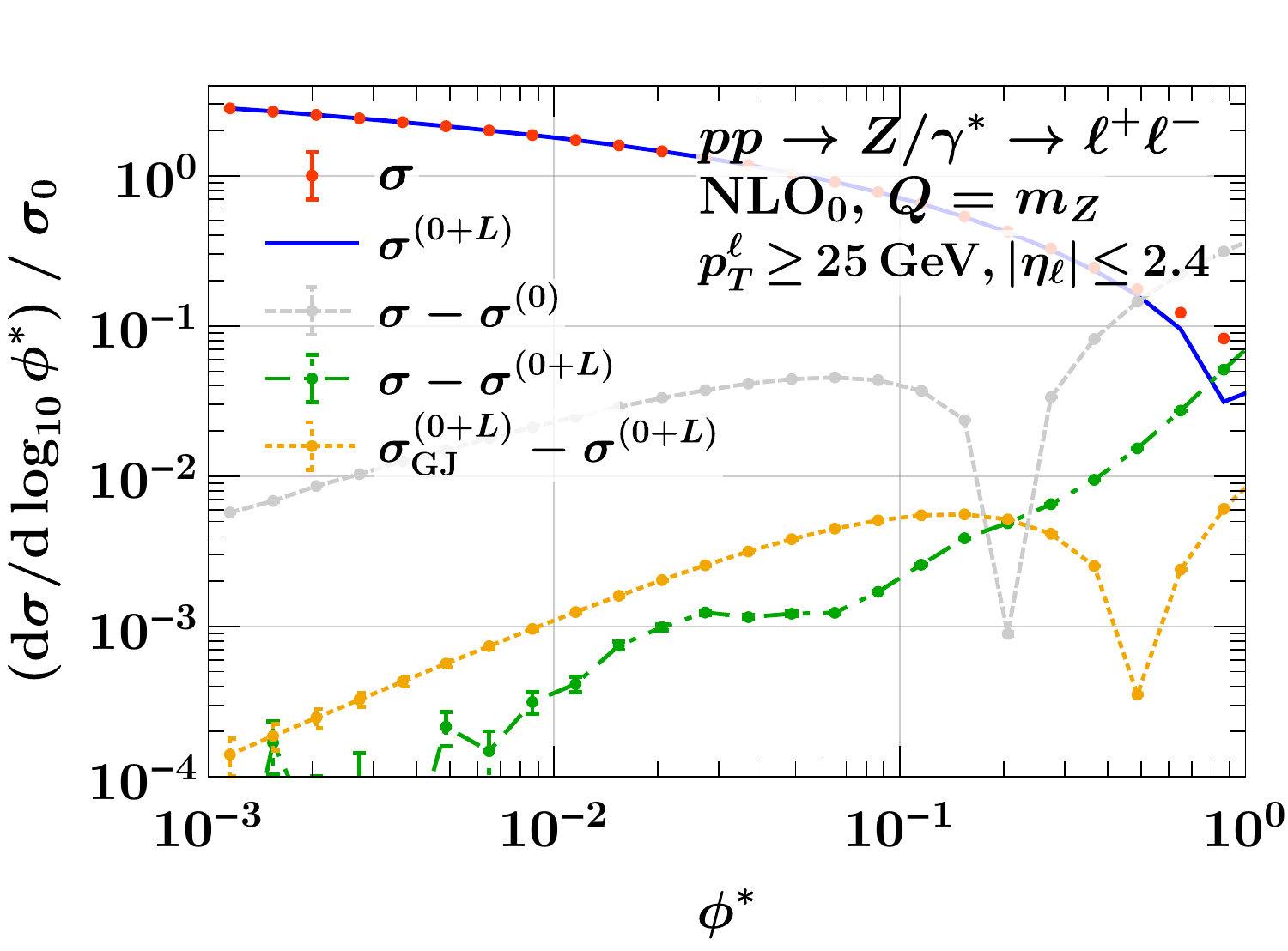}%
\hfill%
\caption{Power corrections for $\phi^*$ for inclusive (left) and fiducial (right) Drell-Yan production.}
\label{fig:Z_phistar_sing_nons_13TeV}
\end{figure*}

We now turn to the numerical study of power corrections to $\phi^*$ at one loop.
For simplicity, we work at fixed $Q = m_Z$,
and normalize all results to the tree-level cross section.
The results for inclusive and fiducial Drell-Yan are shown in the left and right
panel of \fig{Z_phistar_sing_nons_13TeV}, respectively.
In both plots, the exact results $\sigma$ are shown by the red points,
the factorized prediction including fiducial power corrections $\sigma^{(0+L)}$ by the blue line,
and their difference by the green dot-dashed curve.
This is contrasted by the gray-dashed curve which shows the difference $\sigma - \sigma^{(0)}$
between the exact and strict LP (i.e.\ employing $\phi^{*(0)}$) result.
The orange, dot-dashed curve shows the difference
$\sigma^{(0+L)}_{\rm GJ} - \sigma^{(0+L)}$
between our default CS tensor decomposition, and an alternative choice
corresponding to the GJ frame.

%-------------------------------------------------------------------------------
%-------------------------------------------------------------------------------

In the inclusive case (left panel), $\sigma^{(0+L)}$ and $\sigma^{(0)}$
only differ by whether $\phi^*$ is implemented exactly or using $\phi^{*(0)}$.
In this case, we observe large linear corrections to $\sigma^{(0)}$,
whereas corrections to $\sigma^{(0+L)}$ appear to be quadratically suppressed.
Interestingly, the $\sigma^{(0+L)}_{\rm GJ}$ seems to have linear corrections
as can be seen from the linear scaling of the difference
$\sigma^{(0+L)}_{\rm GJ} - \sigma^{(0+L)}$.
Hence, $\sigma^{(0+L)}$ for a generic frame receives linear corrections,
although they are roughly an order of magnitude suppressed compared to $\sigma^{(0)}$.

For fiducial Drell-Yan production (right panel), we observe linear power corrections
for both $\sigma^{(0)}$ and $\sigma^{(0+L)}$, which are larger than the power
corrections in the inclusive case, especially for $\sigma^\zero$.
Nevertheless, we again see see that $\sigma^{(0+L)}$ has significantly smaller
corrections than $\sigma^{(0)}$, despite having the same linear scaling in $\phi^*$.
The ambiguity between the two choices of tensor decomposition is again a linear effect,
but again at much smaller overall magnitude than the corrections beyond $\sigma^\zero$.

Overall, we find that $\phi^*$ generically receives linear power corrections.
In addition to the common fiducial corrections, $\phi^*$ is affected by corrections
from expanding the observable itself, and by the fact that even very small $\phi^*$ receives
contributions from large $q_T$, as is apparent from \eq{phistar_approx}.
Hence, \emph{a priori} there is no reason to expect that corrections to $\phi^*$
are quadratically suppressed.
Nevertheless, $\sigma^{(0+L)}$ includes all linear power corrections from
small-$q_T$, which is reflected by the corrections to $\sigma^{(0+L)}$
being significantly reduced compared to $\sigma^{(0)}$.
We also note that the choice of tensor decomposition strongly affects
how well contributions from large $q_T$ are captured.
Empirically, we find that our default choice corresponding to the CS frame
minimizes the size of power corrections, but we have not been able
to identify an underlying reason for this observation.

%-------------------------------------------------------------------------------
\begin{figure*}
\centering
\includegraphics[width=\WidthTwoSubfigs]{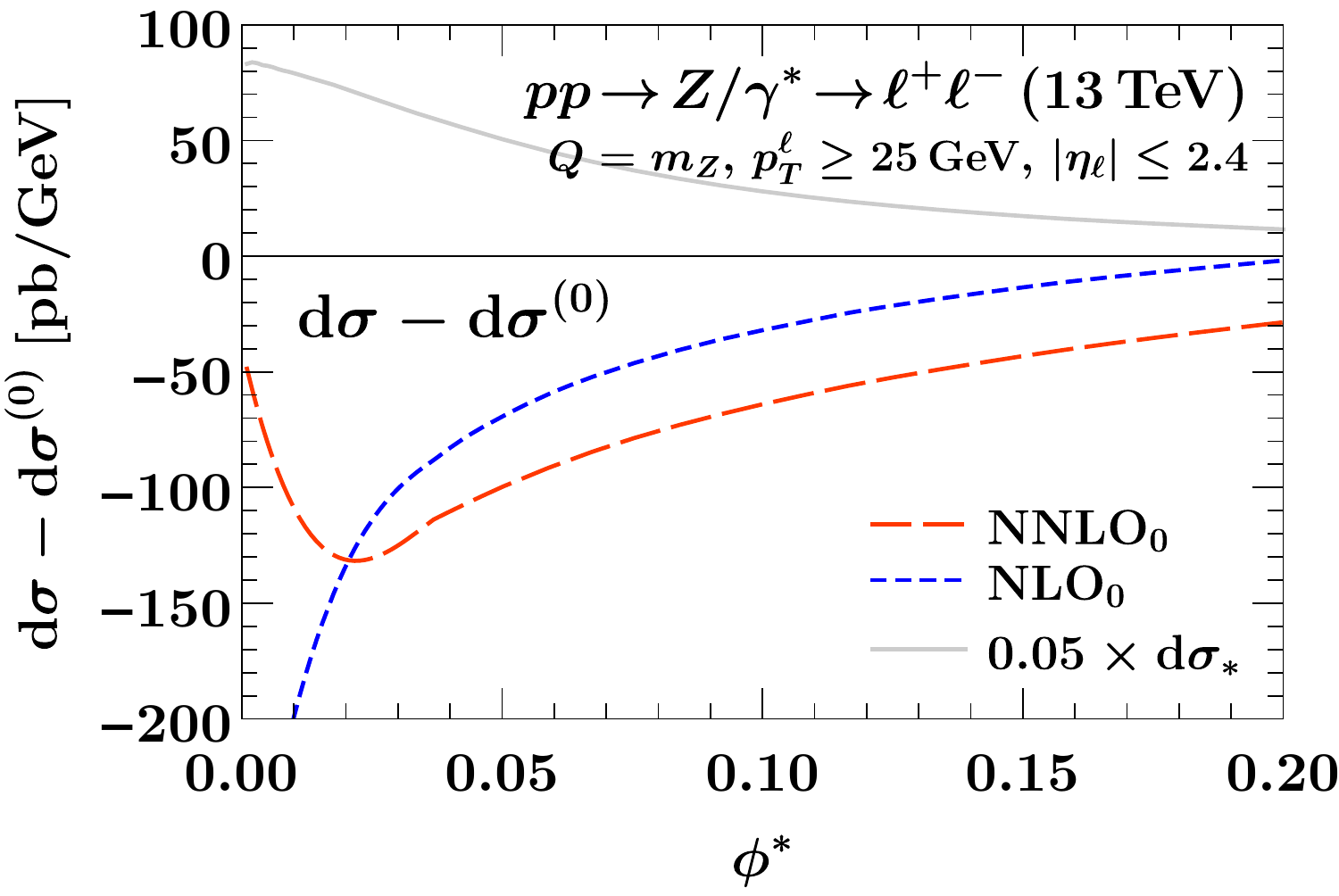}%
\hfill%
\includegraphics[width=\WidthTwoSubfigs]{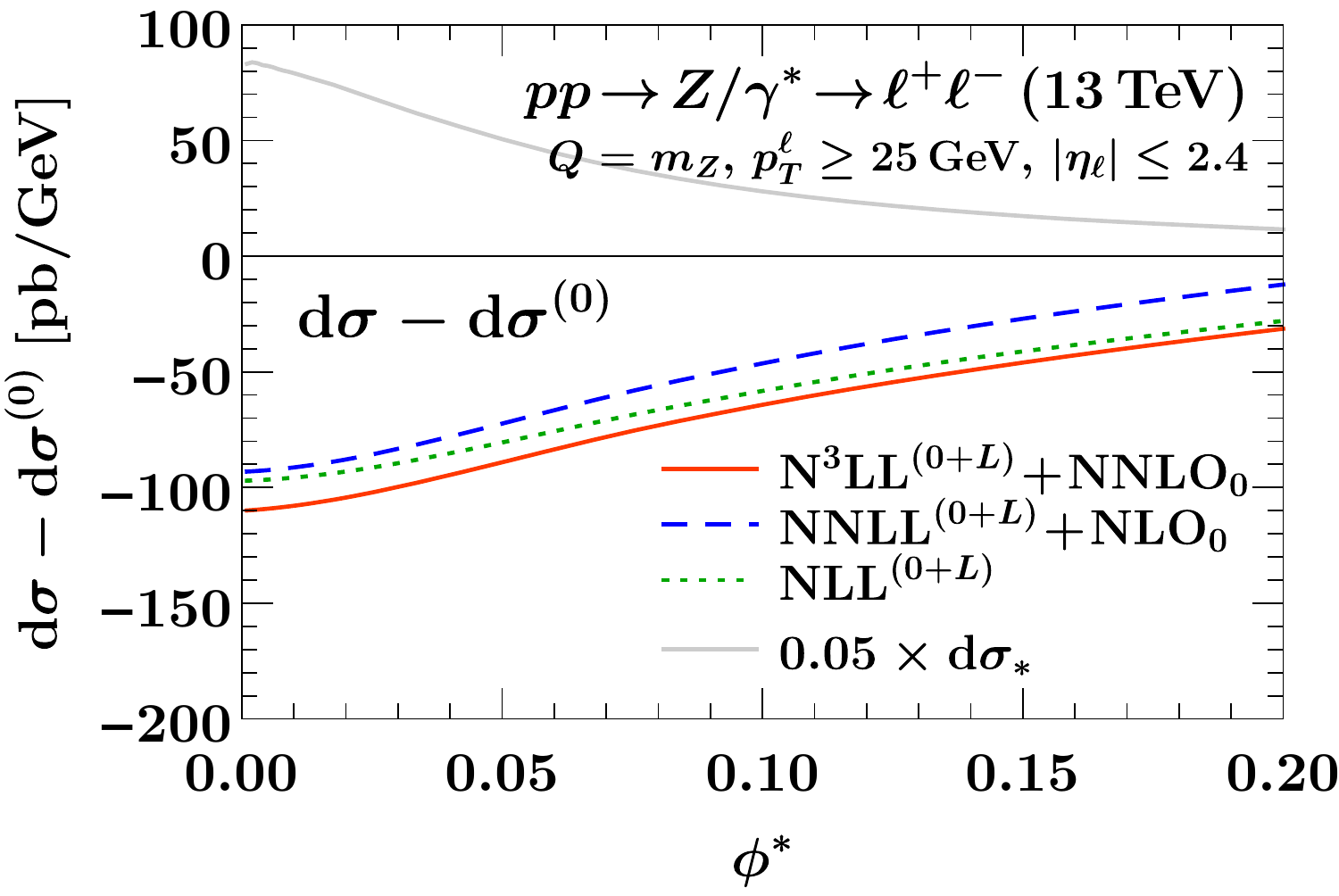}%
\\
\includegraphics[width=\WidthTwoSubfigs]{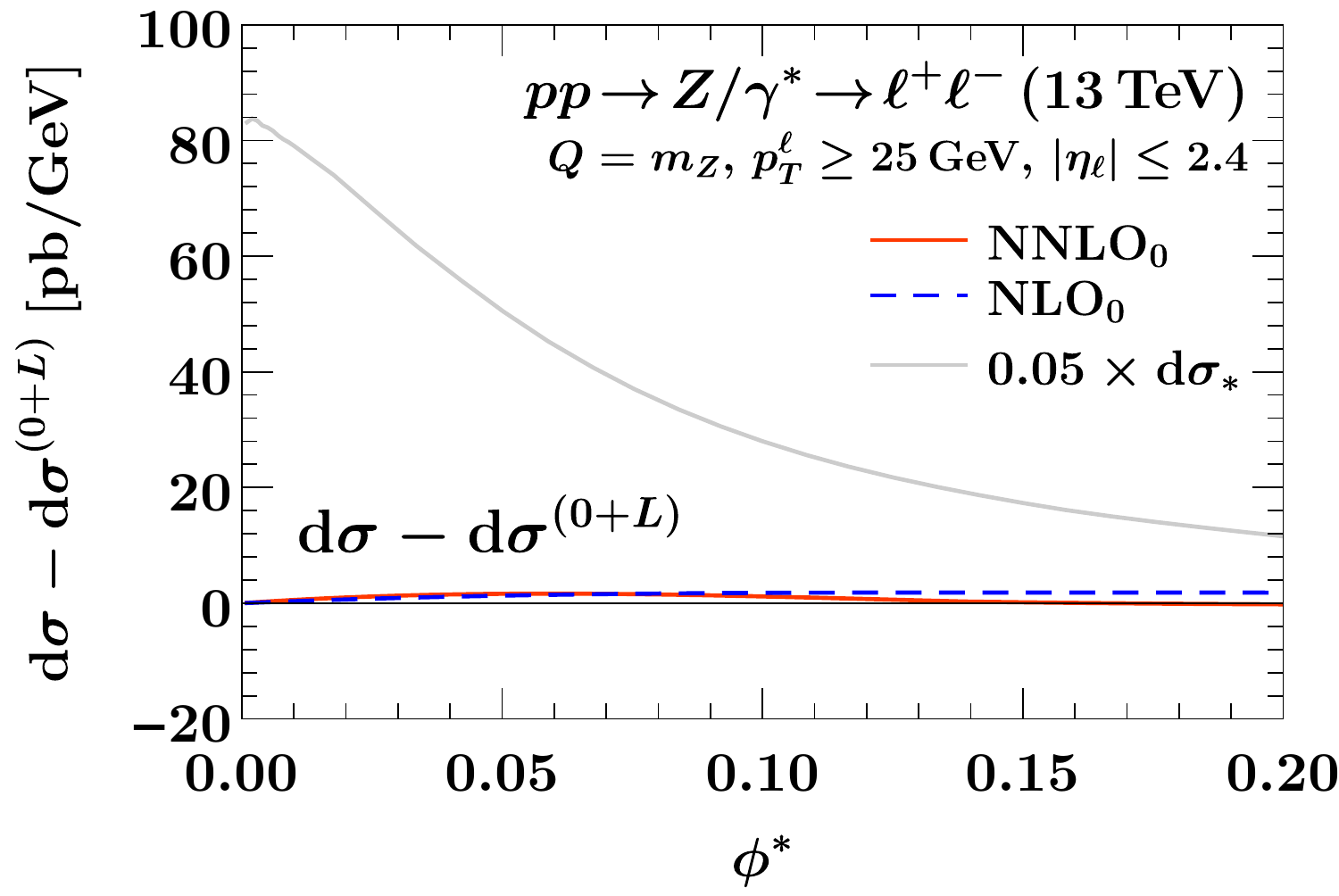}%
\caption{Subleading-power contributions to the fiducial $\phi^*$ spectrum
for Drell-Yan on the resonance, $Q = m_Z$.
We compare the sum of all subleading power contributions,
treating the linear fiducial power corrections at fixed order (top left)
or resumming them to all orders (top right).
Our best prediction $\df \sigma_*$ for the total spectrum at N$^3$LL$^{(0+L)}+$NNLO$_0$
is indicated as a light gray line for reference, scaled down to $5\%$ of its original size.
The bottom panel shows the remaining fixed-order power corrections,
which are finite for $\phi^* \to 0$ (note the difference in vertical scale).}
\label{fig:Z_phi_star_13TeV_fid_illustration}
\end{figure*}
%-------------------------------------------------------------------------------

We conclude this section by studying the impact of the fiducial power corrections on the
resummed $\phi^*$ spectrum with fiducial cuts.
In the top-left panel of \fig{Z_phi_star_13TeV_fid_illustration}, we show the difference
between the strict LP and the exact $\phi^*$ spectrum at NLO$_0$ (blue, short-dashed) and NNLO$_0$ (red, long-dashed).
For reference, the gray line shows our best prediction $\sigma_*$ at N$^3$LL$^{(L)}+$NNLO$_0$,
scaled down to 5$\%$ of its original size.
We note a large discrepancy between NLO$_0$ and NNLO$_0$ as $\phi^*\to0$,
indicating large, unresummed logarithms in the power corrections and consequently
poor perturbative convergence in this regime.

In the top-right panel of \fig{Z_phi_star_13TeV_fid_illustration}, we show the difference
between the LP $\phi^*$ spectrum and the resummed and matched $\phi^*$ spectrum,
at NLL$^{(0+L)}$ (green, dotted), NNLL$^{(0+L)}+$NLO$_0$ (blue, dashed)
and N$^3$LL$^{(0+L)}+$NNLO$_0$ (red, solid), which corresponds to the left panel
but with the linear power corrections resummed. As a result, the divergence as $\phi^*\to0$
is cured, and we observe very good perturbative convergence between the different resummed predictions.
Note that in contrast to $q_T$, see \fig{Z_qT_13TeV_illustration},
the resummed power corrections do not vanish as $\phi^*\to0$,
due to the different weighting of the Sudakov factor with $\cos(b_T Q \phi^*)$
rather than $J_0(b_T q_T)$, cf.~\eq{resummed_phistar_approx}.

Finally, in the bottom panel of \fig{Z_phi_star_13TeV_fid_illustration} we show
the remaining fixed-order power corrections after including the fiducial power
corrections in the resummation at NLO$_0$ (blue, short-dashed) and NNLO$_0$ (red solid),
again including our best prediction $\sigma_*$ for reference.
Since all terms diverging as $\phi^*\to0$ are included in $\sigma^{(0+L)}$,
the remaining power corrections are well-behaved as $\phi^*\to0$, which makes
their overall size almost negligible.

%%%%%%%%%%%%%%%%%%%%%%%%%%%%%%%%%%%%%%%%%%%%%%%%%%%%%%%%%%%%%%%%%%%%%%%%%%%%%%%%
\section{Applications in fixed-order subtractions}
\label{sec:qT_subtraction}
%%%%%%%%%%%%%%%%%%%%%%%%%%%%%%%%%%%%%%%%%%%%%%%%%%%%%%%%%%%%%%%%%%%%%%%%%%%%%%%%

The inclusion of fiducial power corrections in the $q_T$ factorization theorem
by treating the leptonic tensor exactly can be immediately applied to improve
the $q_T$ subtraction method. In fact, we have employed this in
\sec{numerics_pTlep} to obtain NNLO predictions for the $\pTlep$ spectrum. Here,
we elaborate in more detail on this, illustrating that this approach can
significantly and systematically improve the $q_T$ subtraction. In
particular, it becomes crucial in regions of phase space sensitive to leptonic power
corrections, such as $\pTlep$ close to the Jacobian peak, where the strict LP
expansion breaks down, and thereby also the subtractions based on the strict LP
limit.

%===============================================================================
\subsection{\texorpdfstring{$q_T$}{qT} subtraction with fiducial power corrections}
\label{sec:qT_subtraction_review}
%===============================================================================

The $q_T$ subtraction method has been originally proposed in \refcite{Catani:2007vq}
for the calculation of color-singlet processes at NNLO, and its application at N$^3$LO
was discussed in \refscite{Cieri:2018oms,Billis:2019vxg}.
Here, we briefly review this method, and discuss how to extend it to include
fiducial power corrections.
Here, we only discuss the simplest implementation of $q_T$ subtractions
as a phase-space slicing. As discussed in \refcite{Ebert:2019zkb}, fiducial cuts
can also be explicitly accounted for in a differential implementation of $q_T$
subtractions, which can also be further improved with the methods we discuss here,
which we leave for future work.

We denote the cross section to be calculated as $\sigma(X)$,
where $X$ summarizes all observables to be differential in, such as $Q$, $Y$, and $\pTlep$,
as well as possible fiducial cuts. The cross section can be written as
\begin{align} \label{eq:qTsub_1}
 \sigma(X) &
 = \int_0^\infty \df q_T \, \frac{\df\sigma(X)}{\df q_T}
 = \sigma(X,\qTcut) + \int_{\qTcut}^\infty \df q_T \, \frac{\df \sigma(X)}{\df q_T}
\,,\end{align}
where the cumulative cross section as a function of $\qTcut$ is defined as
\begin{align}
 \sigma(X,\qTcut) = \int_0^{\qTcut}\df q_T \, \frac{\df\sigma(X)}{\df q_T}
\,.\end{align}
The $q_T$ subtraction is typically implemented as a slicing method by adding
and subtracting a suitable subtraction term,
\begin{align} \label{eq:qTsub_2}
 \sigma(X) &
 = \sigma^\sub(X,\qTcut)
 + \int_{\qTcut}^\infty \df q_T \, \frac{\df\sigma(X)}{\df q_T}
 + \Delta\sigma(X,\qTcut)
%%%
 \,,\nn\\
%%%
 \Delta\sigma(X,\qTcut) &= \sigma(X,\qTcut) - \sigma^\sub(X,\qTcut)
\,.\end{align}
%%%
For color-singlet processes,
$q_T$ vanishes by construction in the Born limit. Hence, the integral in \eq{qTsub_2}
necessarily involves at least one resolved real emission, and can thus be calculated
from the corresponding Born$+1$-parton process at one order lower than $\sigma(X)$ itself.
The cancellation of virtual and real divergences occurs only in the limit $q_T\to0$.
By constructing $\sigma^\sub$ such that it fully describes this limit, the
cancellation of IR divergences only occurs within $\sigma^\sub$, and $\Delta\sigma$
is a power correction that vanishes as $\qTcut \to 0$, and thus can be neglected
for sufficiently small values of $\qTcut$.

To construct $\sigma^\sub$ and study the size of $\Delta\sigma$, it is useful
to expand the differential cross section and its cumulant for $q_T \ll Q$
and $\qTcut \ll Q$, respectively,
%%%
\begin{alignat}{2} \label{eq:xsec_expand}
\frac{\df \sigma(X)}{\df q_T}
&= \frac{\df\sigma^{(0)}(X)}{\df q_T} &&+ \sum_{m>0} \frac{\df\sigma^{(m)}(X)}{\df q_T}
%%%
\,, \\ \nn
%%%
\sigma(X,\qTcut)
&= \sigma^{(0)}(X,\qTcut) &&+ \sum_{m>0} \sigma^{(m)}(X,\qTcut)
\,,\end{alignat}
%%%
where the different contributions scale as
%%%
\begin{alignat}{4} \label{eq:xs_scaling}
\frac{\df\sigma^{(0)}(X)}{\df q_T} &\sim \delta(q_T) + \sum_{j \geq 0} \biggl[\frac{\ln^j(q_T/Q)}{q_T}\biggr]_+
\,,\quad &
\sigma^{(0)}(X, \qTcut) &\sim  \sum_{j \geq 0} \ln^j\frac{\qTcut}{Q}
\,,\nn\\
%%%
q_T \frac{\df\sigma^{(m)}(X)}{\df q_T} &\sim \Bigr(\frac{q_T}{Q}\Bigl)^m\,
\sum_{j\ge0} \ln^j \frac{q_T}{Q}
\,, &
 \sigma^{(m)}(X,\qTcut) &\sim \Bigl(\frac{\qTcut}{Q}\Bigr)^m\, \sum_{j\ge0} \ln^j\frac{\qTcut}{Q}
\,.\end{alignat}
%%%
The $\df\sigma^{(0)}/\df q_T$ diverges as $1/q_T$ for $q_T\to0$, and is in fact precisely
given by the factorization theorem in \eq{tmd_factorization_m1_2_4_5}. Note that it contains $\delta$ and plus distributions
to regulate this divergence, which precisely encodes the cancellation of virtual
and real IR divergences. The $\df\sigma^{(m)}/\df q_T$ with $m>0$ are integrable as $q_T\to0$,
and consequently $\sigma^{(m)}(\qTcut\to 0) \to 0$.

The subtraction term $\sigma^\sub(X, \qTcut)$ must contain all singular terms,
and has the form
%%%
\begin{align} \label{eq:sigma_sub_0}
 \sigma^\sub(X,\qTcut) &= \sigma^{(0)}(X,\qTcut) \times \bigl[ 1 + \cO(\qTcut/Q) \bigr]
\,.\end{align}
%%%
It follows that the correction term in \eq{qTsub_2} scales as
%%%
\begin{align}
 \Delta\sigma(X,\qTcut) &
 = \sigma(\qTcut) - \sigma^\sub(X,\qTcut)
 = \cO\bigl[(q_T^\cut/Q)^{m} \bigr]
\,,\end{align}
%%%
where $m$ corresponds to the first term in the sum in \eq{xsec_expand}
that is not contained in $\sigma^\sub$. So far, $q_T$ subtractions
have always been implemented in this fashion by choosing $\sigma^{\sub} = \sigma^{(0)}$.

As discussed in \sec{theory}, for inclusive processes the sum in \eq{xsec_expand}
starts with $m = 2$, i.e.\ $\Delta\sigma$ is suppressed as $\cO(q_T^2/Q^2$).
These terms were explicitly calculated in \refcite{Ebert:2018lzn}
for Higgs and Drell-Yan production differential in $Q$ and $Y$, but inclusive in their decay.
The corresponding results for inclusive Higgs and Drell-Yan production, i.e.\ integrated over $Y$,
were also calculated later in \refscite{Cieri:2019tfv, Buonocore:2019puv}.

When applying fiducial cuts, one expects corrections to be only suppressed as
$\cO(q_T/Q)$, i.e.\ $m=1$, as shown in \refcite{Ebert:2019zkb}
and discussed more generally in \secs{theory}{numerics}.%
\footnote{For more complicated cuts such as photon isolation cuts, these corrections
can be even further enhanced~\cite{Ebert:2019zkb}.}
In \sec{numerics}, we also demonstrated that more complicated observables
such as $\pTlep$ or $\phi^*$ can also have much larger corrections than expected.

In \sec{theory}, we showed that the linear power corrections in $q_T/Q$ are unique
for any observables that are azimuthally symmetric in the $q_T\to 0$ limit,
in which case they only arise from the leptonic tensor.
Since the subtraction method is inherently tied
to a calculation at leading-twist in collinear factorization,
this also holds for more complicated observables such as $\phi^*$.
This means that in many relevant cases the $\df\sigma^{(1)}/\df q_T$ term in
\eq{xsec_expand} can be unambiguously predicted and included in the subtraction term.
Hence, instead of \eq{sigma_sub_0} we use
%%%
\begin{align} \label{eq:sigma_sub_1}
 \sigma^\sub(X,\qTcut) &
 = \sigma^{(0+L)}(X,\qTcut) \times \bigl\{ 1 + \cO\bigl[(\qTcut/Q)^2\bigr] \bigr\}
\,.\end{align}
%%%
where $\sigma^{(0+L)}$ includes all fiducial power corrections using our default
CS tensor decomposition. In principle, other choices are possible as well. From
our discussion of $\phi^*$ in \sec{numerics_phistar}, we have some indication
that the CS decomposition tends to minimize the power corrections in more
complicated cases, which motivates using it as our default choice also in the
subtraction context.

%===============================================================================
\subsection{Application to the lepton \texorpdfstring{$p_T$}{pT} spectrum}
\label{sec:qT_subtraction_pTlep}
%===============================================================================

We illustrate the advantage of including the leptonic corrections in \eq{sigma_sub_1}
for the case of the $\pTlep$ spectrum. As discussed in \sec{numerics_pTlep}
(see e.g.\ \fig{phiL}), we can distinguish three regions in the spectrum
according to their sensitivity to small $q_T$. The expected dependence on $\qTcut$
and the linear subtraction terms are as follows:
\begin{enumerate}
 \item Sufficiently below the Jacobian peak, $\pTlep \lesssim (m_W - \qTcut)/2$,
       only quadratic power corrections arise, and thus we expect no significant
       impact from including the linear terms in the subtraction.
 \item Around the Jacobian peak, $|2m_W - \pTlep| \lesssim \qTcut$, the standard
       power counting breaks down, and one is sensitive to the singular behavior
       as $q_T\to0$. In this region, keeping fiducial power corrections becomes
       strictly necessary, since the strict LP expansion breaks down due to
       leptonic power corrections $q_T/(2m_W - \pTlep)$. In particular, the region
       $m_W/2 < \pTlep < (m_W + \qTcut)/2$
       requires to take these corrections into account, as otherwise
       the singular prediction is limited to $\pTlep < m_W/2$, and the region
       beyond the peak will contain leftover singular behavior.
 \item Far above the Jacobian peak, $\pTlep \gtrsim (m_W + \qTcut)/2$, there are no
       contributions from the subtraction term, and only the above-cut integral
       in \eq{qTsub_1} is relevant.
\end{enumerate}
We illustrate these effects numerically at NLO. Reference results are obtained
by evaluating $pp \to W^+ \to \ell^+ \nu_\ell$ at NLO using \mcfm, which employs
Catani-Seymour dipole subtractions~\cite{Catani:1996jh, Catani:1996vz}.
The corresponding results using $q_T$ subtraction are obtained by combining the
below-cut contribution $\df\sigma^\sub/\df\pTlep$ obtained from \scetlib,
both for $\sigma^\sub = \sigma^{(0)}$ and $\sigma^\sub = \sigma^{(0+L)}$,
with the above-cut contribution obtained by evaluating
$pp \to (W^+{\to}\ell^+ \nu_\ell)+1$ parton at LO$_1$ using \mcfm. We work in the narrow-width
approximation such that $Q = m_W$, which avoids smearing out the Jacobian peak,
as is necessary to clearly see the effects described above. To keep numerical
uncertainties from the Monte Carlo integration small, we bin the $\pTlep$ spectrum
in bins of width $1~\GeV$, with $\pTlep = m_W/2$ being aligned with a bin boundary
to cleanly separate below-peak and above-peak bins.

\begin{figure*}
\centering
\includegraphics[width=\WidthTwoSubfigs]{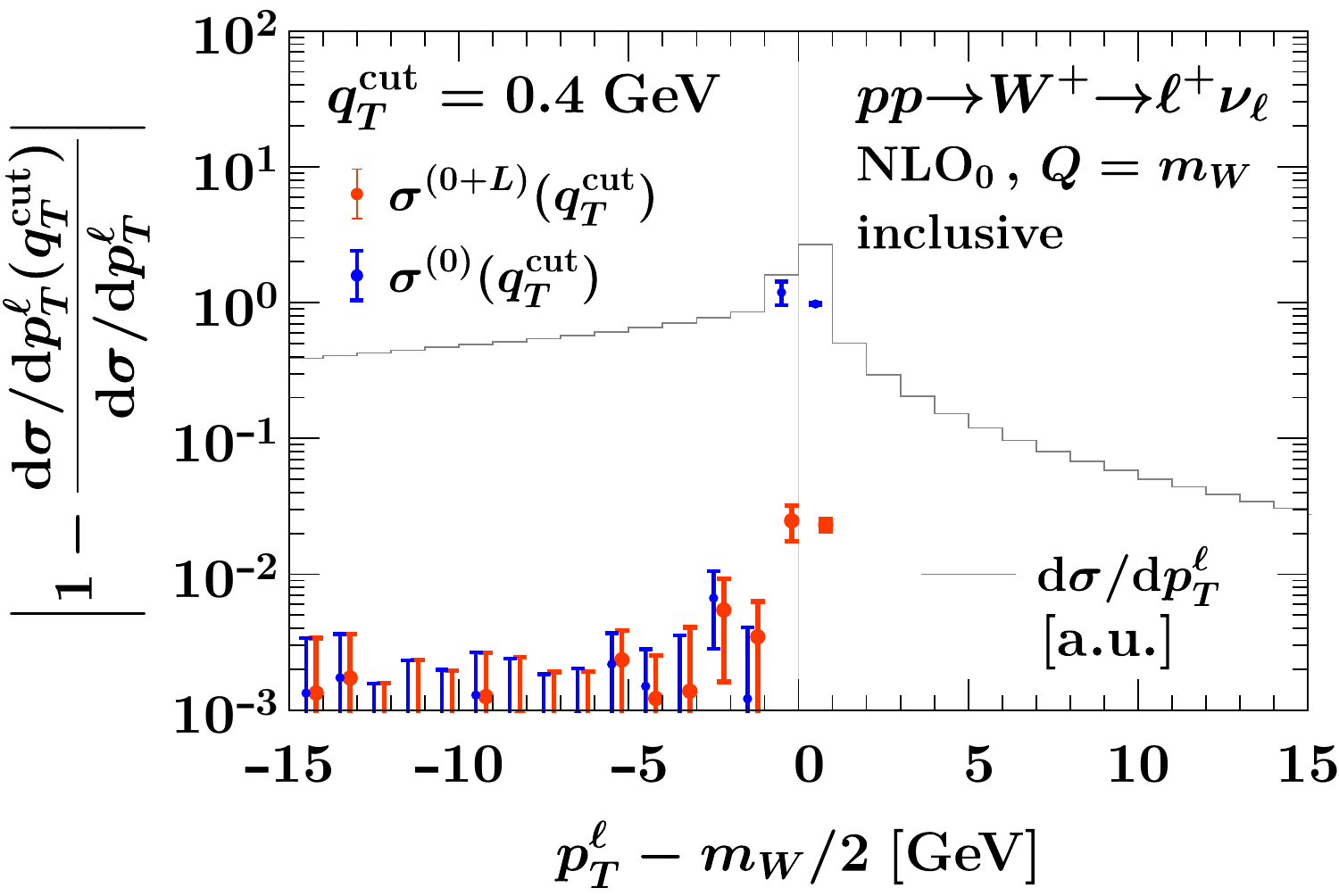}%
\hfill%
\includegraphics[width=\WidthTwoSubfigs]{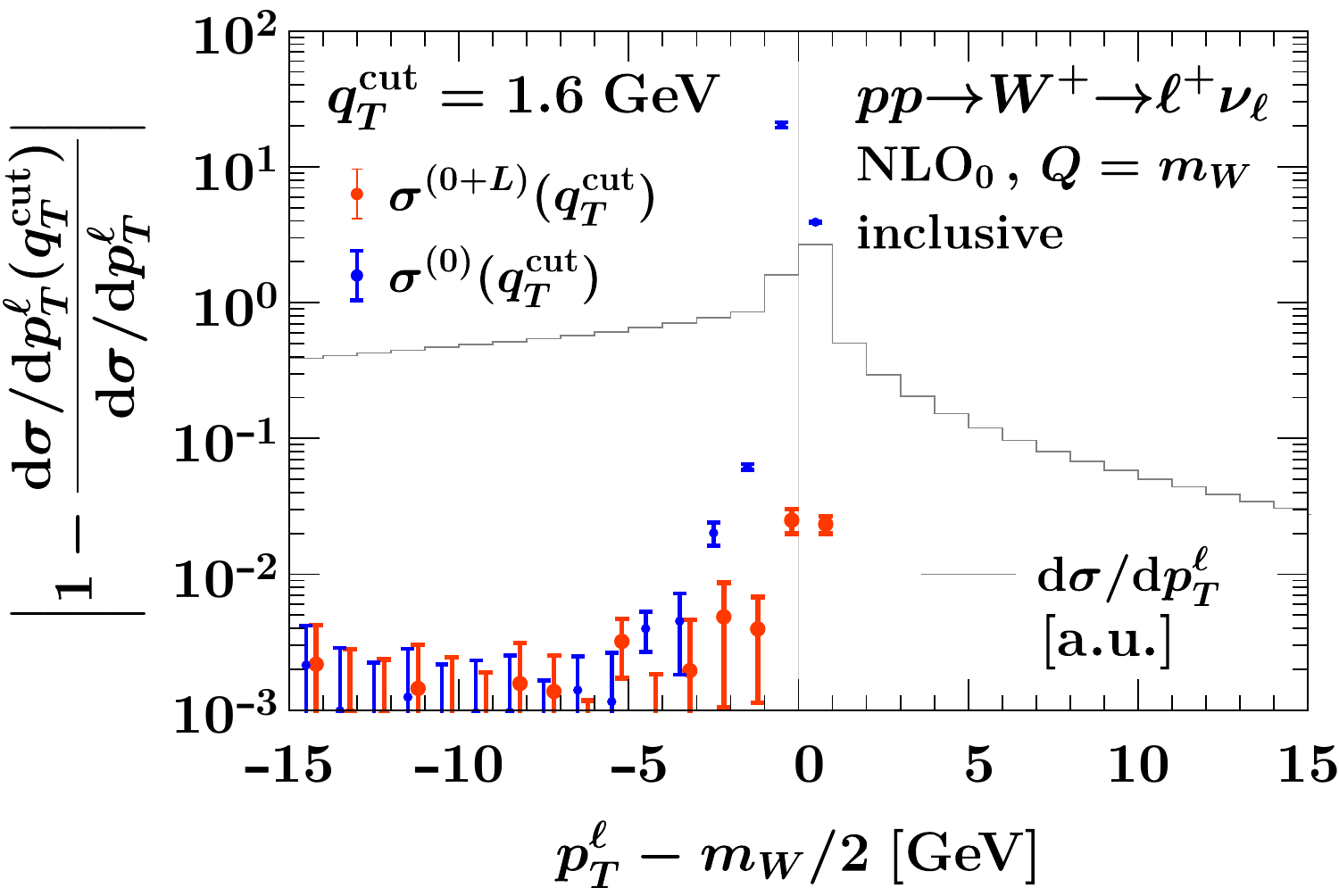}%
\\%
\includegraphics[width=\WidthTwoSubfigs]{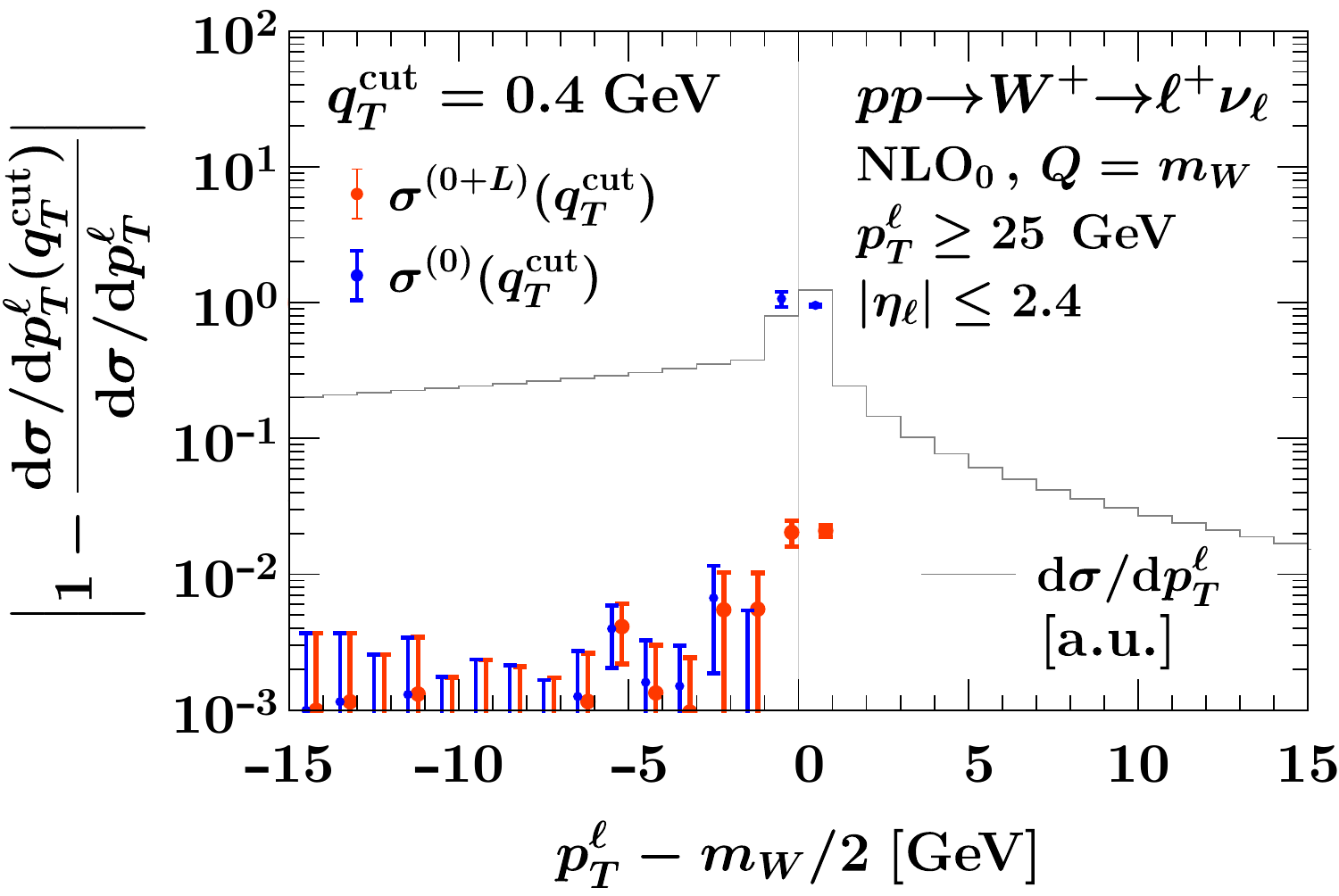}%
\hfill%
\includegraphics[width=\WidthTwoSubfigs]{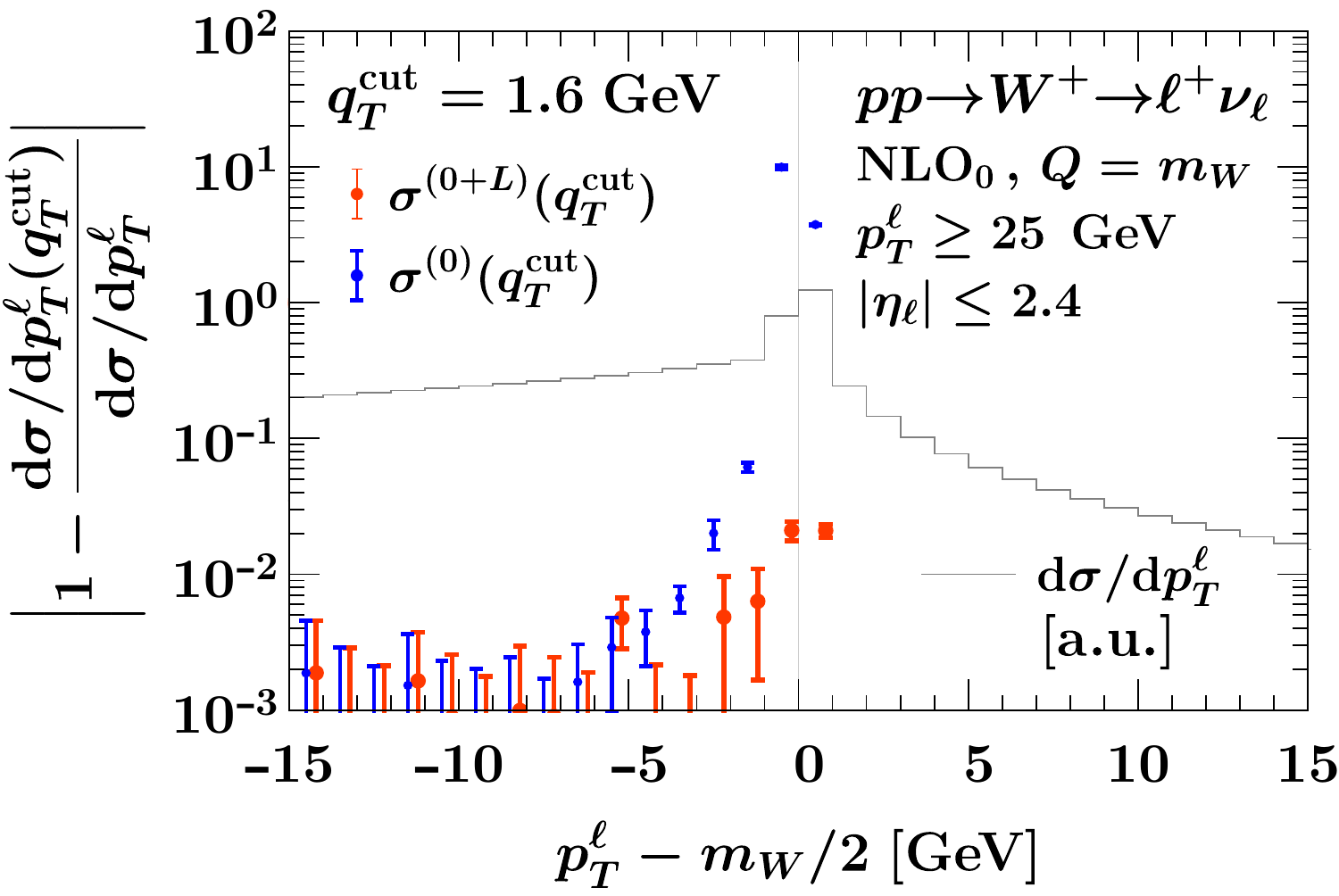}%
\caption{%
Relative difference of the $\pTlep$ spectrum obtained using $q_T$ subtraction
to the exact NLO results, for $\qTcut = 0.4~\GeV$ (left) and $\qTcut = 1.6~\GeV$ (right).
In the top row, we are inclusive in the lepton phase space, while in the
bottom row we apply the CMS cuts $\pTlep \ge 25~\GeV, |\eta_\ell| < 2.4$.
The error bands indicate uncertainties from the Monte Carlo integration.
In all plots, the gray line shows the shape of the exact NLO result in arbitrary units,
and the blue error bars are slightly shifted to right for better visibility only.
}
\label{fig:pTlep_rel_diff}
\end{figure*}

In \fig{pTlep_rel_diff}, we show the relative difference for the $\pTlep$
spectrum obtained with $q_T$ subtraction to the exact NLO result, so
smaller values imply that the $q_T$ subtractions work better with smaller
missing power corrections.
We show results for the two values $\qTcut = 0.4~\GeV$ (left) and $\qTcut = 1.6~\GeV$ (right).
In the top row, we are inclusive in the lepton, while in the bottom row we apply the cuts
$\pTlep \ge 25~\GeV, |\eta_\ell| < 2.4$.
The blue points use the strict LP subtractions, while the red points include
the fiducial power corrections in the subtraction.
The error bands indicate the MC integration uncertainties.
For reference, we also show the exact NLO spectrum in gray in arbitrary units.

%-------------------------------------------------------------------------------
%-------------------------------------------------------------------------------

Due to the rather large bin size of $1\GeV$, only the two bins adjacent to the
Jacobian peak at $\pTlep = m_W/2$ are significantly affected by the choice of
$\qTcut$. Below the first bin left of the peak, power corrections are
essentially negligible for all choices of $\qTcut$, irrespective of whether or
not the fiducial power corrections are included and independent of the cuts.
Beyond the first bin right to the peak, there is no contribution from the
subtraction term, and the spectrum is entirely given by the above-cut
contribution. However, as expected, the bins adjacent to the Jacobian peak are
significantly affected by the large leptonic power corrections. At strict LP
(blue points), these are of $\cO(80\%)$ for $\qTcut = 0.4~\GeV$, and even of
$\cO(10)$ for $\qTcut = 1.6~\GeV$, indicating the breakdown of the subtractions.
Including the fiducial power corrections (red points) automatically retains all
leptonic power corrections and as a result the power corrections are of the
order of a few percent, leading to stable predictions even very close to the
peak. The results with and without addition fiducial cuts are very similar, as
expected by the absence of linear fiducial power corrections when being
inclusive in the  neutrino.

\begin{figure*}
\includegraphics[width=\WidthTwoSubfigs]{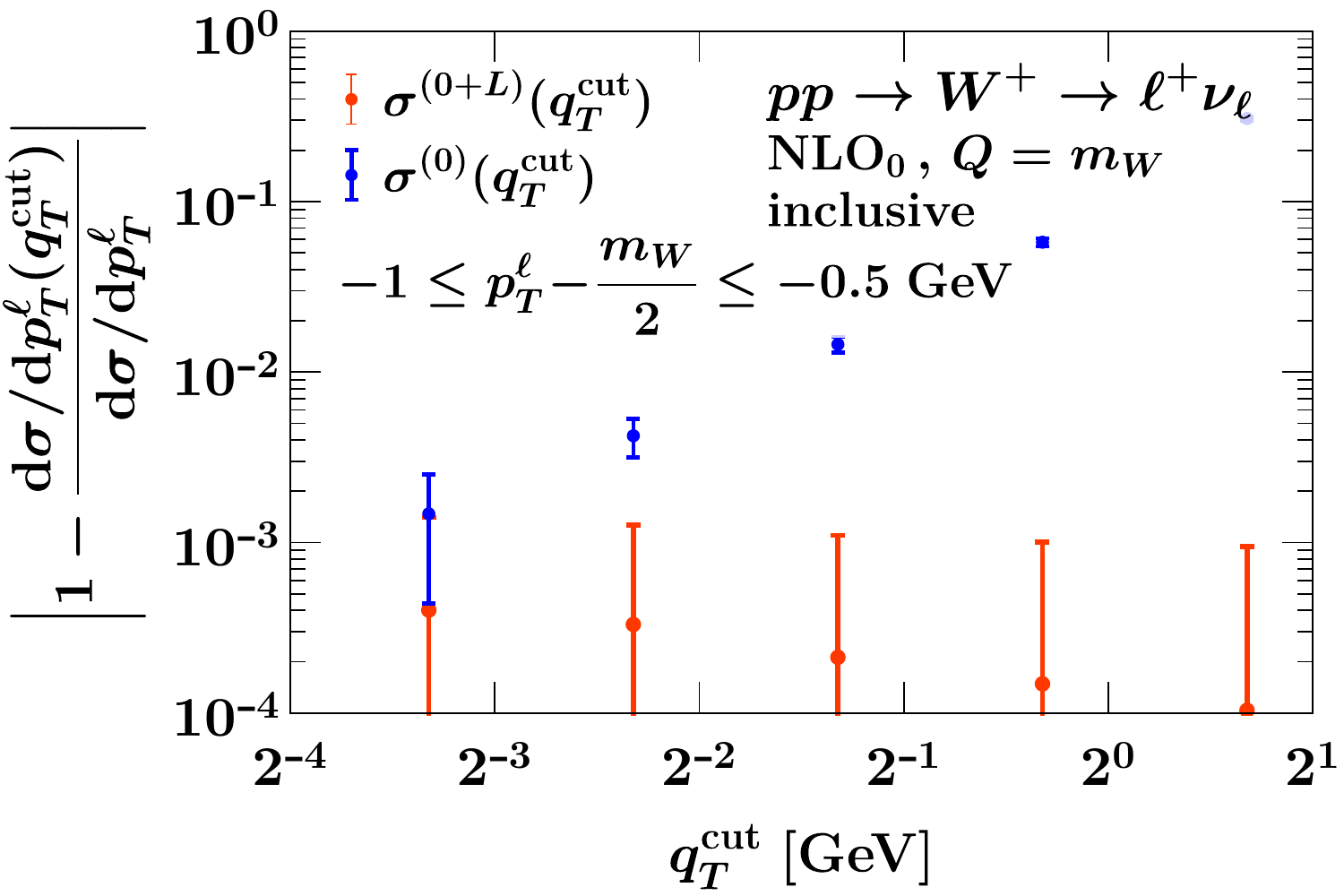}%
\hfill%
\includegraphics[width=\WidthTwoSubfigs]{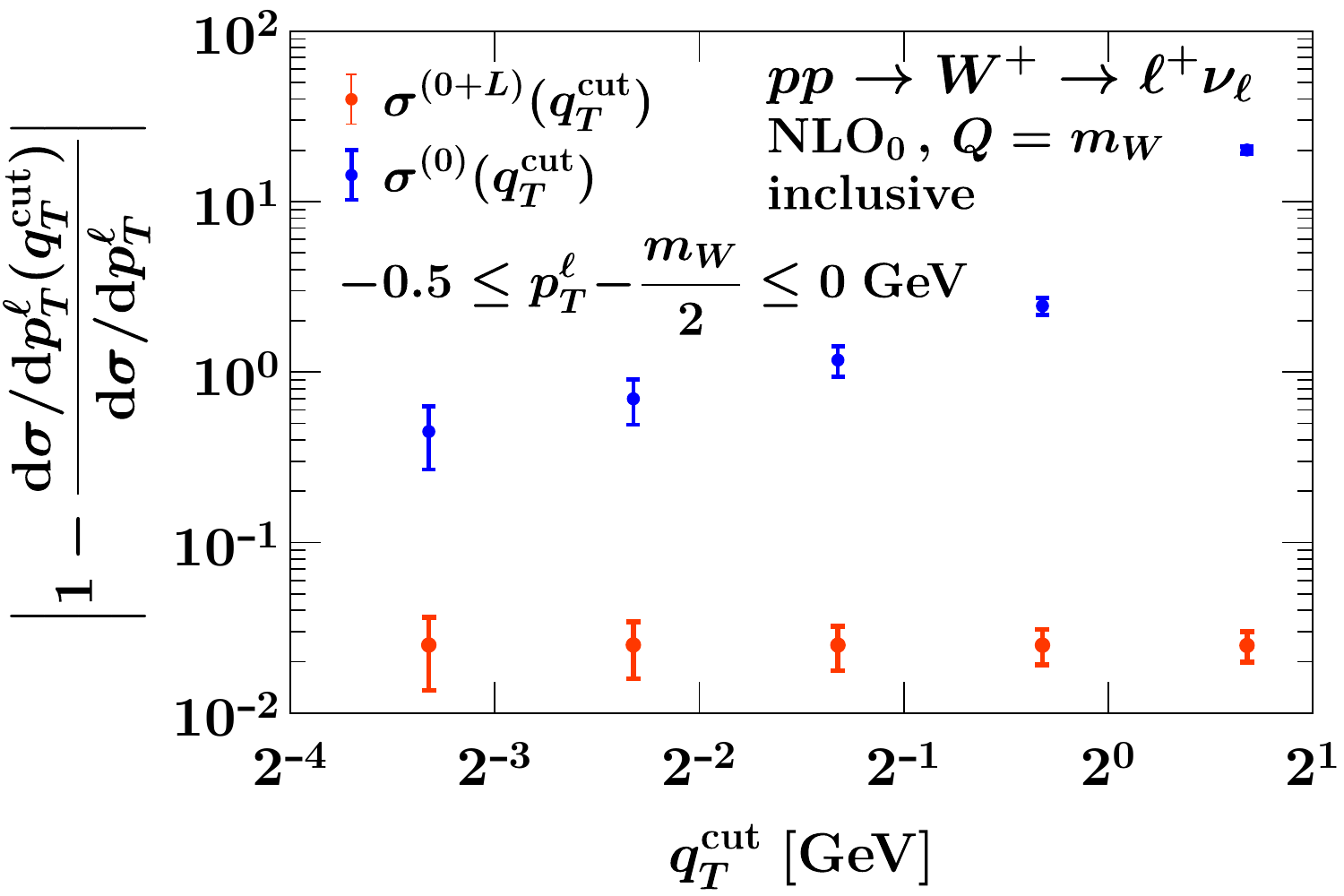}%
\caption{%
Relative difference of $\pTlep$ spectrum obtained with $q_T$ subtractions as a
function of $\qTcut$ compared to Catani-Seymour subtractions. The two panels
show two different bins in $\pTlep$ near the Jacobian peak, as indicated. The
$q_T$ subtraction are implement at strict LP (blue points) and including
fiducial power corrections (red points). The error bars indicate the MC
integration uncertainties.
}
\label{fig:pTlep_qTcut_dependence}
\end{figure*}

\Fig{pTlep_qTcut_dependence} shows the $\qTcut$ dependence of the lepton $p_T$
spectrum, for the bin $-1 \le \pTlep - \frac{m_W}{2} \le -\frac12~\GeV$ (left),
and the bin $-\frac12 \le \pTlep - \frac{m_W}{2} \le 0~\GeV$ (right), i.e.\ the
two bins left to the Jacobian peak. (Note the finer bin width of $0.5\GeV$,
compared to $1\GeV$ above). The result for the first bin right of the peak is
very similar to the right figure, while bins further away have negligible
$\qTcut$ dependence that is entirely driven by MC integration uncertainties, and
hence we do not explicitly show these. We also only show the inclusive case. In
\fig{pTlep_qTcut_dependence}, we show the relative difference of the result
obtained using $q_T$ subtractions to the \mcfm\ result using Catani-Seymour
subtractions. As before, the blue and red points correspond to implementing the
$q_T$ subtractions without and with fiducial power corrections, and the error
bars indicate MC integration uncertainties.

In the lower bin (left panel), the $\qTcut$ dependence of the $q_T$ subtraction
including fiducial power corrections (red points) is consistent with zero within
the MC uncertainties, illustrating the extremely good convergence of the subtractions.
In contrast, at strict LP (blue points) there is a sizable difference to the exact
result. While it is falling with $\qTcut$, one needs a very small value
$\qTcut \sim 0.2 - 0.4\GeV$ to achieve an accuracy of $\cO(10^{-2})$.
In the bin adjacent to the peak (right figure; note the different plot range),
the difference between Catani-Seymour and strict-LP $q_T$-subtraction results
is consistently larger by a factor $\sim 10^2$, indicating the much stronger
sensitivity to small $q_T \sim \pTlep - m_W / 2$. As before, at strict LP (blue points)
there is a strong dependence on $\qTcut$, but even at the extremely small value
of $\qTcut = 0.1\GeV$ the result has an $\ord{1}$ uncertainty due to missing
power corrections, showing the breakdown of the method.
After including the fiducial power corrections
(red points) the $\qTcut$ dependence is negligible within the MC integration uncertainties.

Note that there seems to be a systematic difference compared to the \mcfm\ result
obtained with Catani-Seymour subtractions. However, in this extreme region of
phase space it is not a priori clear which of the results is more reliable
and whether the reported MC uncertainties are still trustworthy.
Thus, we refrain from a definite statement which subtraction method performs
better in this scenario.

%-------------------------------------------------------------------------------
%-------------------------------------------------------------------------------

%-------------------------------------------------------------------------------
\begin{figure*}
\includegraphics[width=\WidthTwoSubfigs]{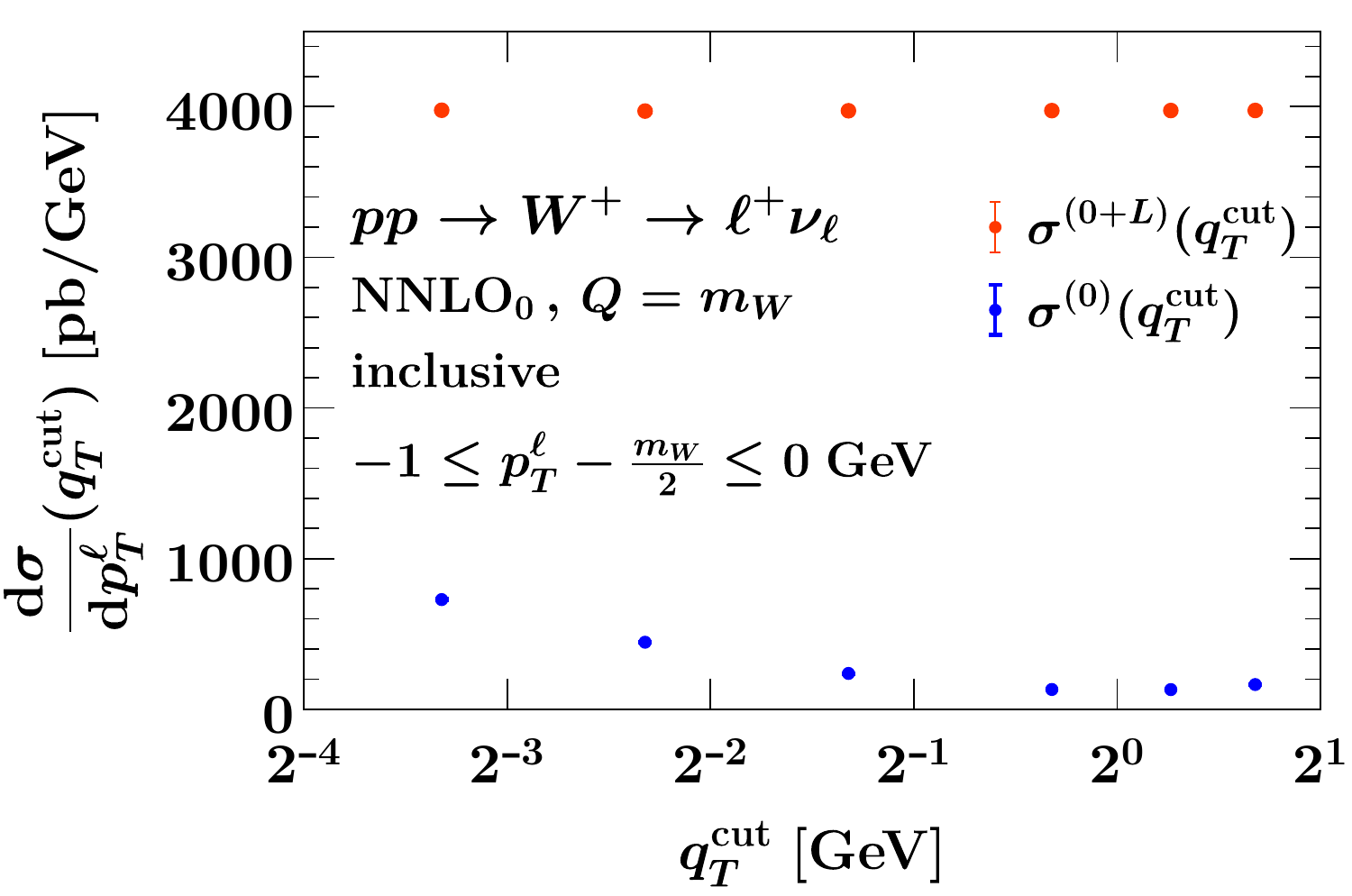}%
\hfill%
\includegraphics[width=\WidthTwoSubfigs]{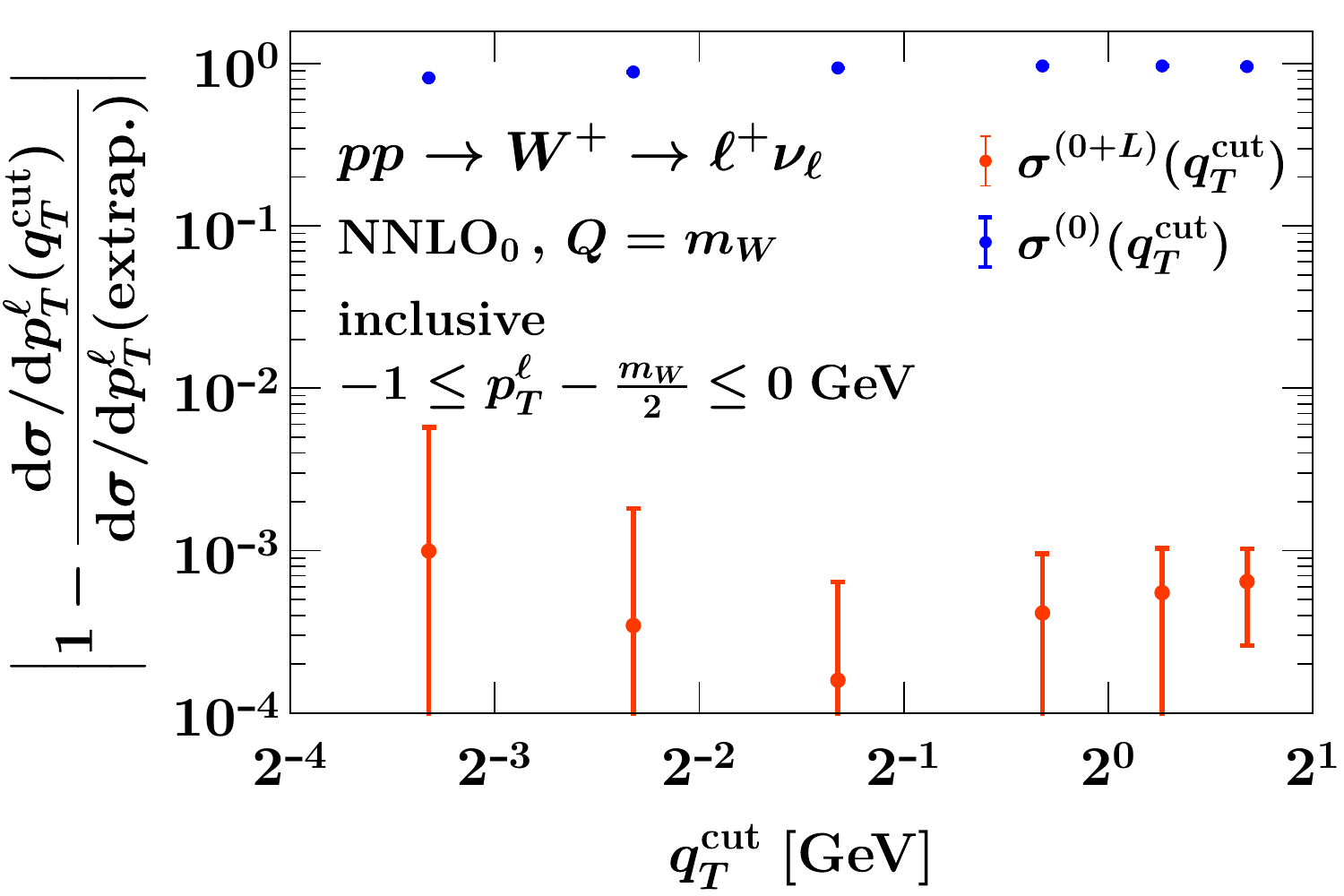}%
\caption{$q_T^\cut$ dependence of the $\pTlep$ cross section in the bin $-1 \le m_W/2 \le 0~\GeV$
with ($\sigma^{(0+L)}$, red points) and without ($\sigma^{(0)}$, blue points) fiducial power corrections included in the subtractions.
In the left, we show the corresponding  absolute cross sections in this bin,
in the right we show their relative difference to a quadratic extrapolation
of the $\qTcut$ dependence of $\sigma^{(0+L)}$.}
\label{fig:qTcut_NNLO}
\end{figure*}
%-------------------------------------------------------------------------------

Our numerical checks confirm the theoretical observation that close to the peak,
where one is sensitive to small $q_T$, reliable predictions can only be obtained
when taking the fiducial power corrections into account. In bins sufficiently far away from the peak,
this is not as crucial, but can still significantly improve convergence
of the method, i.e.\ it allows one to use much larger values of $\qTcut$ than
the strict LP subtraction whenever linear fiducial corrections are present.
This becomes particularly important at higher orders, where it becomes numerically
challenging to push $\qTcut$ to such small values.

In our analysis in \sec{numerics_pTlep}, we require the $\pTlep$ spectrum at NNLO$_0$,
which we obtain by combining the NLO$_1$ calculation of $pp \to W^+ + 1$ parton from
\mcfm\ with a $q_T$ subtraction including fiducial power corrections supplied by \scetlib,
choosing $\qTcut = 0.8\GeV$. We verify that this yields a sufficiently
precise result in the first bin left of the Jacobian edge, $-1 \le m_W/2 \le 0~\GeV$.
Compared to above we have increased the bin width from $0.5\GeV$ to $1\GeV$ to
reduce numerical uncertainties.
In the left panel in \fig{qTcut_NNLO}, we show the absolute cross section in this
bin as a function of $\qTcut$ for $\sigma^{(0+L)}$
(red points) and $\sigma^{(0)}$ (blue points). Clearly, only the result using
$\sigma^{(0+L)}$ shows a stable convergent behavior by giving a result
insensitive to $\qTcut$, whereas the result for $\sigma^{(0)}$ does not converge
and is quite far from the true result.
We then use a simple quadratic extrapolation of $\sigma^{(0+L)}(\qTcut)$
to $\qTcut\to0$ as a proxy for the exact result in this bin.
In the right figure, we show the relative difference of $\sigma^{(0+L)}$ (red points)
and $\sigma^{(0)}$ (blue points) to this extrapolation.
As expected, $\sigma^{(0)}$ shows $\cO(1)$ deviations due to the clear lack
of convergence towards small $\qTcut$ due to uncontrolled leptonic power corrections.
In contrast, $\sigma^{(0+L)}$ is in good agreement
with the extrapolation, with deviations of $\cO(10^{-3})$ that are largely driven
by numerical uncertainties. This justifies our choice of
$\qTcut=0.8\GeV$ for the results presented in \sec{numerics_pTlep}.
It also confirms the expectation that the systematic pattern of improvement from
including fiducial power corrections in the subtractions observed at NLO$_0$
will carry over to NNLO$_0$. We leave a more detailed investigation at NNLO$_0$
to future work.

%===============================================================================
\subsection{Application to Drell-Yan with symmetric cuts}
\label{sec:qT_subtraction_DY}
%===============================================================================

As a second application to illustrate the $q_T$ subtraction including fiducial
power corrections, we consider Drell-Yan $pp \to Z/\gamma^* \to \ell^+\ell^-$
with fiducial cuts on the final-state leptons.
In particular, we focus on enhanced power corrections that arise when one
applies \emph{symmetric} cuts, where both leptons are required to have
$p_T^\ell \ge \pTmin$. For this case, it was already remarked in \refcite{Grazzini:2017mhc}
that the $q_T$ subtractions become very sensitive to $\qTcut$ leading to instabilities
in the perturbative calculation. In fact, these instabilities have already been
pointed out long ago in \refcite{Frixione:1997ks}, and traced back to a sensitivity
to small transverse momenta, which may require resummation to obtain well-behaved results.
In addition, as explained in \refcite{Ebert:2019zkb} and in \sec{numerics_qT},
the lepton cuts induce linear power corrections. Here, we will illustrate in
more detail that the enhanced linear power corrections
can be avoided by employing $q_T$ subtraction with fiducial power corrections.
Since this allows one to reach the
precision of a LP subtraction at larger values of $\qTcut$, it also reduces the
region of phase space where fixed-order calculations need to be evaluated,
and thus should also reduce the effect of the aforementioned instabilities.

In the following we consider the total cross section for the Drell-Yan process,
studying the effect of different $p_T$ cuts on the two leptons, namely
\begin{align} \label{eq:DY_cuts}
 p_{T\,h} &\ge 25~\GeV \,, \quad 25.1~\GeV \,,\quad 25.5~\GeV \,,\quad 26~\GeV
\,,\nn\\
 p_{T\,s} &\ge 25~\GeV
\,.\end{align}
Here, $p_{T\,h}$ ($p_{T\,s}$) is the harder (softer) transverse momentum.
We only vary the cut on the harder lepton, while the cut on the softer lepton
is fixed to $p_{T\,s} \ge 25~\GeV$, and we always apply the same rapidity cut
$|\eta_\ell| \le 2.4$ for both leptons.
As in \sec{qT_subtraction_pTlep}, we consider the cross section at NLO,
combining a one-loop below-cut calculation of $pp \to Z/\gamma^* \to \ell^+\ell^-$
from \scetlib, with a tree level above-cut
calculation of $pp \to (Z/\gamma^* \to \ell^+\ell^-) + j$ from \mcfm,
always working in the narrow-width approximation.
We consider the following values of the $\qTcut$ subtraction cutoff:
\begin{align}
 \qTcut \in \{ 0.1,\, 0.2,\, 0.4,\, 0.8,\, 1.6,\, 3.2,\, 6.4\}~\GeV
\,.\end{align}

\begin{figure*}
\includegraphics[width=\WidthTwoSubfigs]{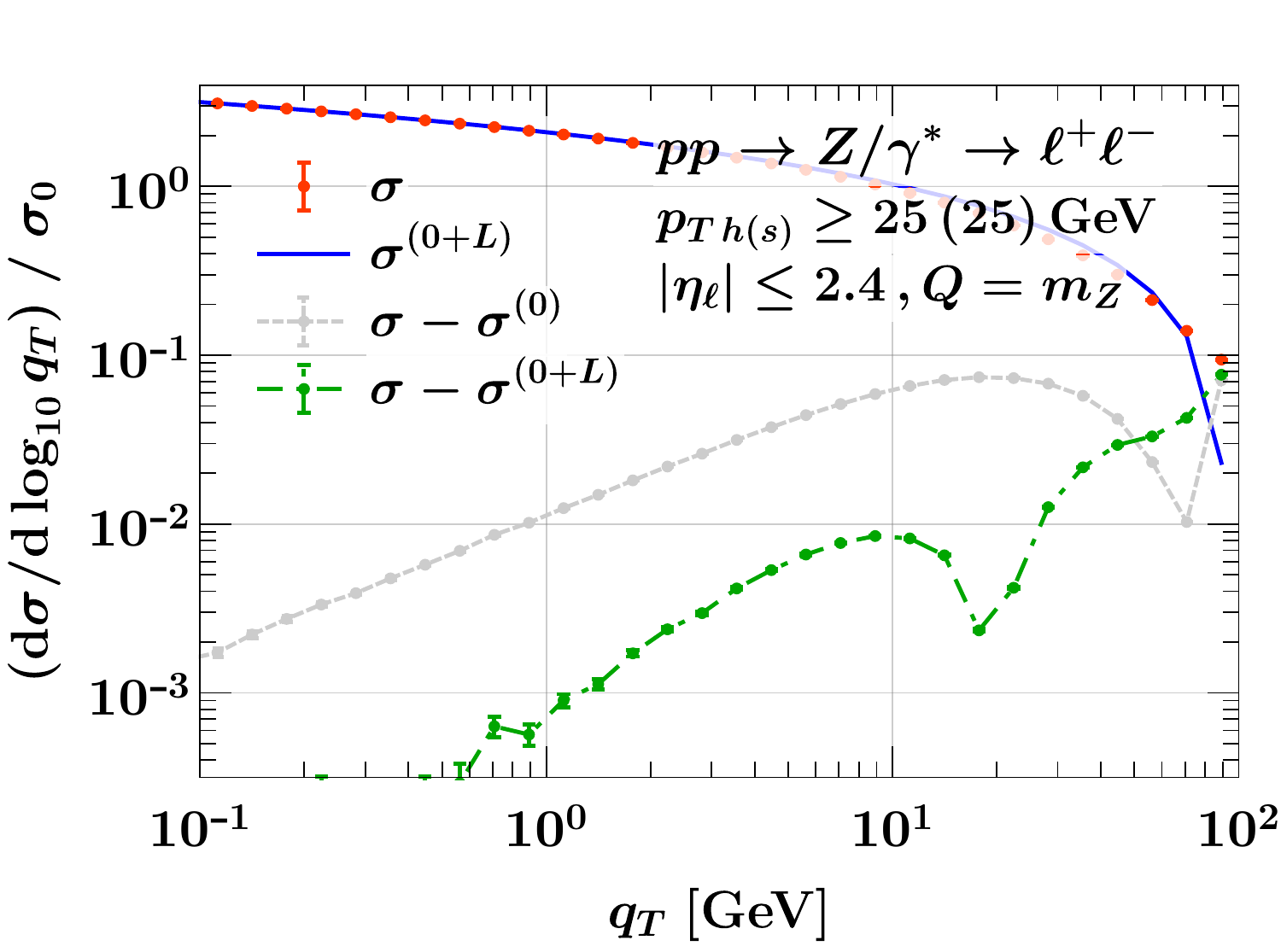}%
\hfill%
\includegraphics[width=\WidthTwoSubfigs]{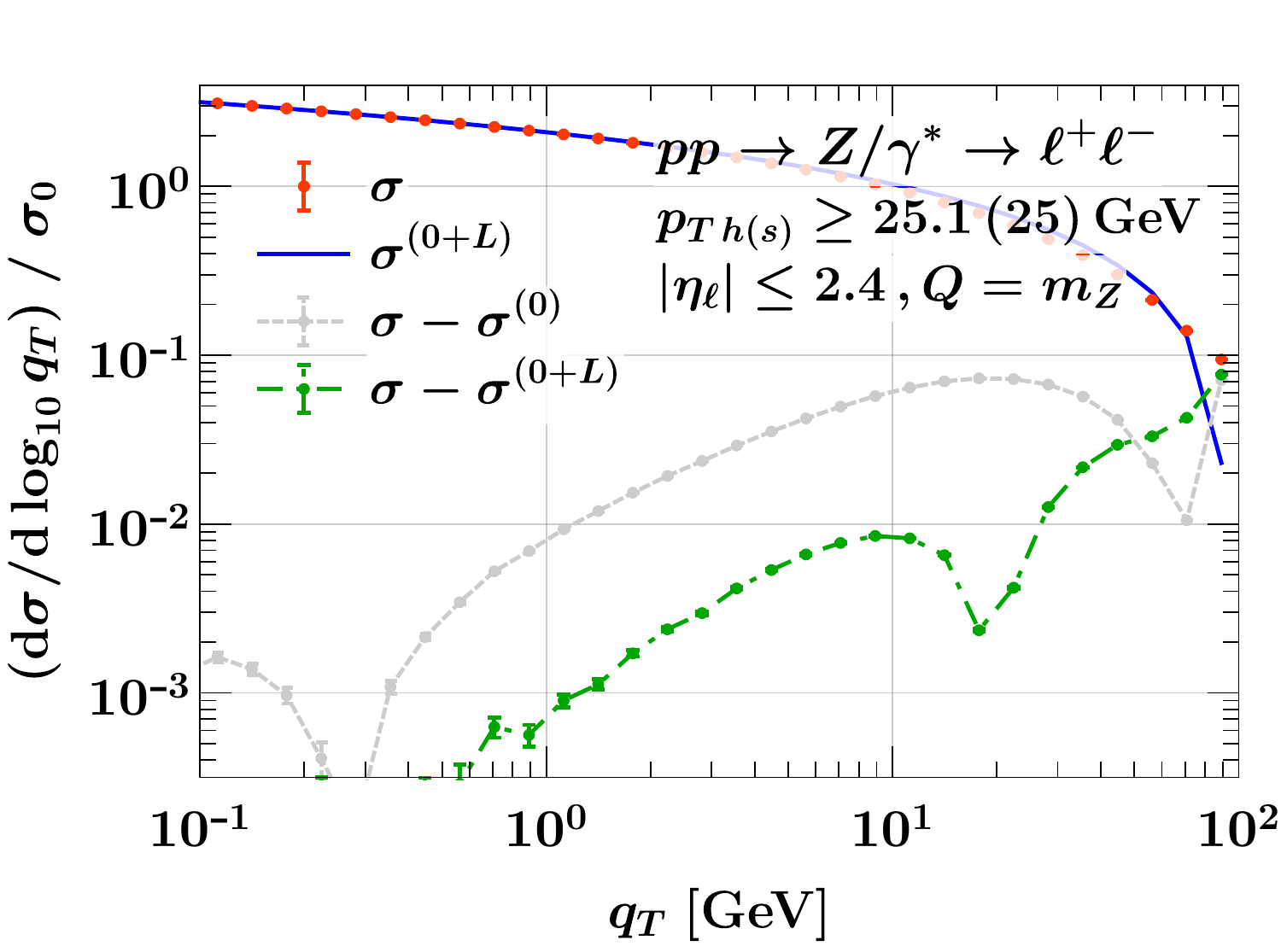}%
\\%
\includegraphics[width=\WidthTwoSubfigs]{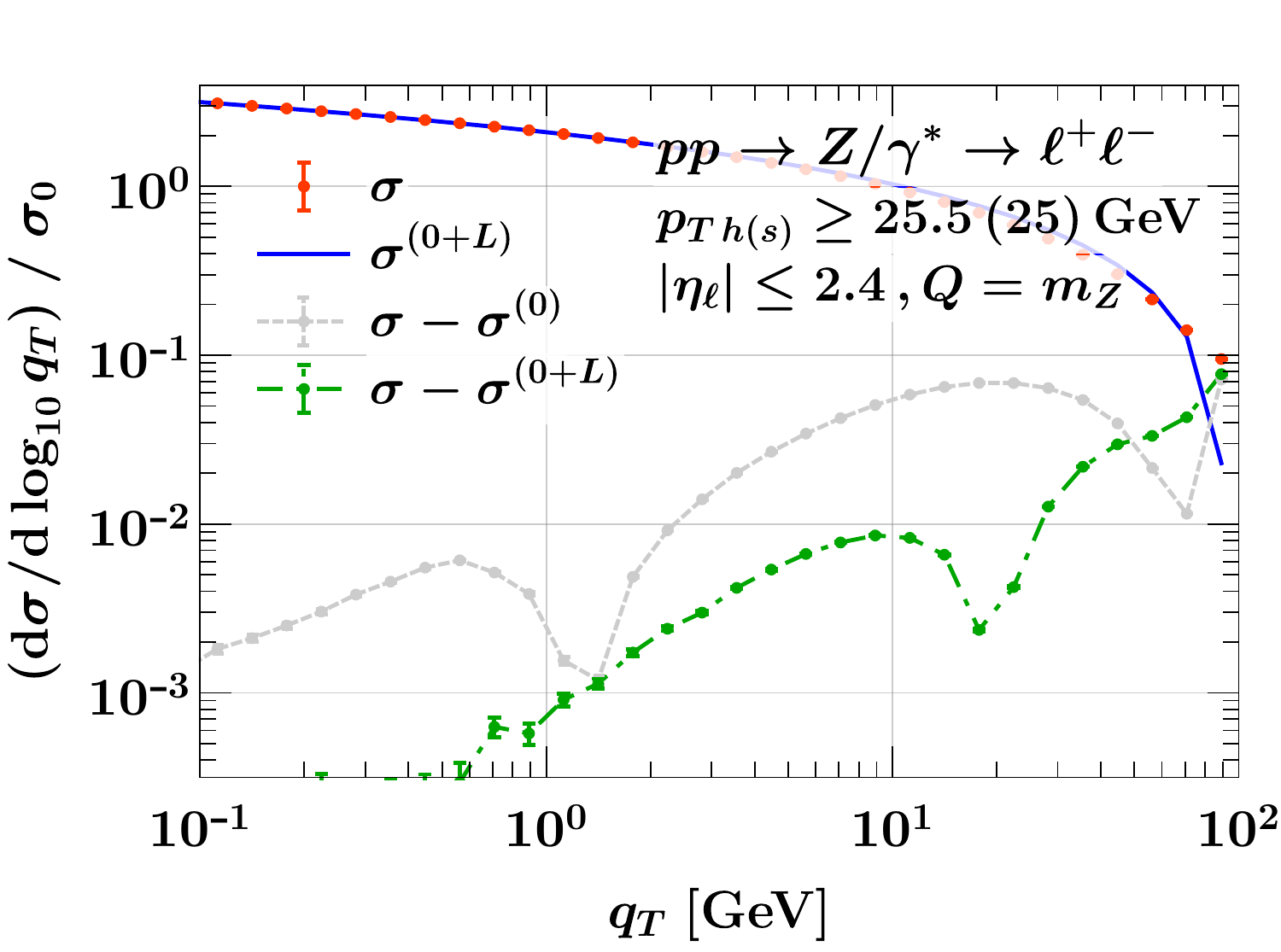}%
\hfill%
\includegraphics[width=\WidthTwoSubfigs]{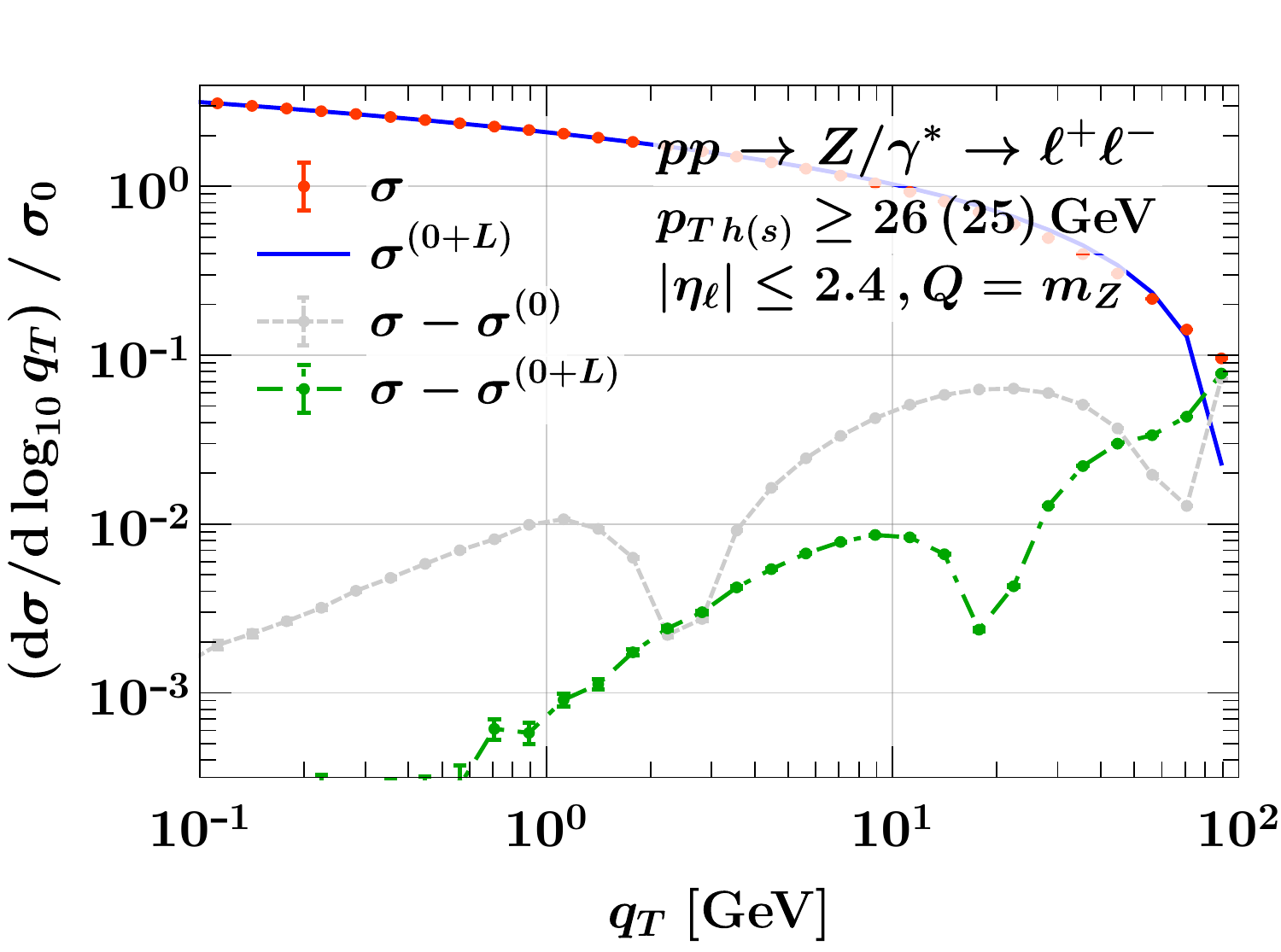}%
\caption{%
Drell-Yan $q_T$ spectrum at fixed $\ord{\alpha_s}$ for different symmetric and
slightly asymmetric lepton $p_T$ cuts, as indicated in the labels.
The full result $\sigma$ is shown by the red points and the LP limit
$\sigma^{(0+L)}$ including fiducial power corrections by the blue line.
The gray and green curves show the leftover power corrections $\sigma-\sigma^{(0)}$
and $\sigma-\sigma^{(0+L)}$. The error bars indicate the MC integration
uncertainties.
}
\label{fig:Z_qT_sing_nons_lepton_cuts}
\end{figure*}

First, we study the $q_T$ spectrum, comparing the exact result
with $\sigma^{(0)}$ and $\sigma^{(0+L)}$. This is shown in
\fig{Z_qT_sing_nons_lepton_cuts}, where the four panels correspond to
the four lepton cuts in \eq{DY_cuts}, and the corresponding plot without fiducial
cuts is shown in \fig{Z_qT_sing_nons_13TeV}.
In each panel, the red points show the full result $\sigma$ and the blue line
the LP prediction including fiducial power corrections $\sigma^{(0+L)}$.
In all cases, including the fiducial power corrections substantially improves the accuracy,
with the leftover power corrections $\sigma-\sigma^{(0+L)}$ (dashed green)
being quadratically suppressed and of similar size as in the inclusive case.
On the other hand, using the strict LP limit, the leftover power corrections
$\sigma-\sigma^{(0)}$ (dotted gray) are around an order of magnitude larger and only
linearly suppressed. Varying the cut on the harder lepton has almost
no effect on the green curves, but changes the gray curve, which has a dip
from changing its sign, whose position is very sensitive to the difference
of the two $p_T$ cuts.

%-------------------------------------------------------------------------------
%-------------------------------------------------------------------------------

%-------------------------------------------------------------------------------
\begin{figure*}
\includegraphics[width=\WidthTwoSubfigs]{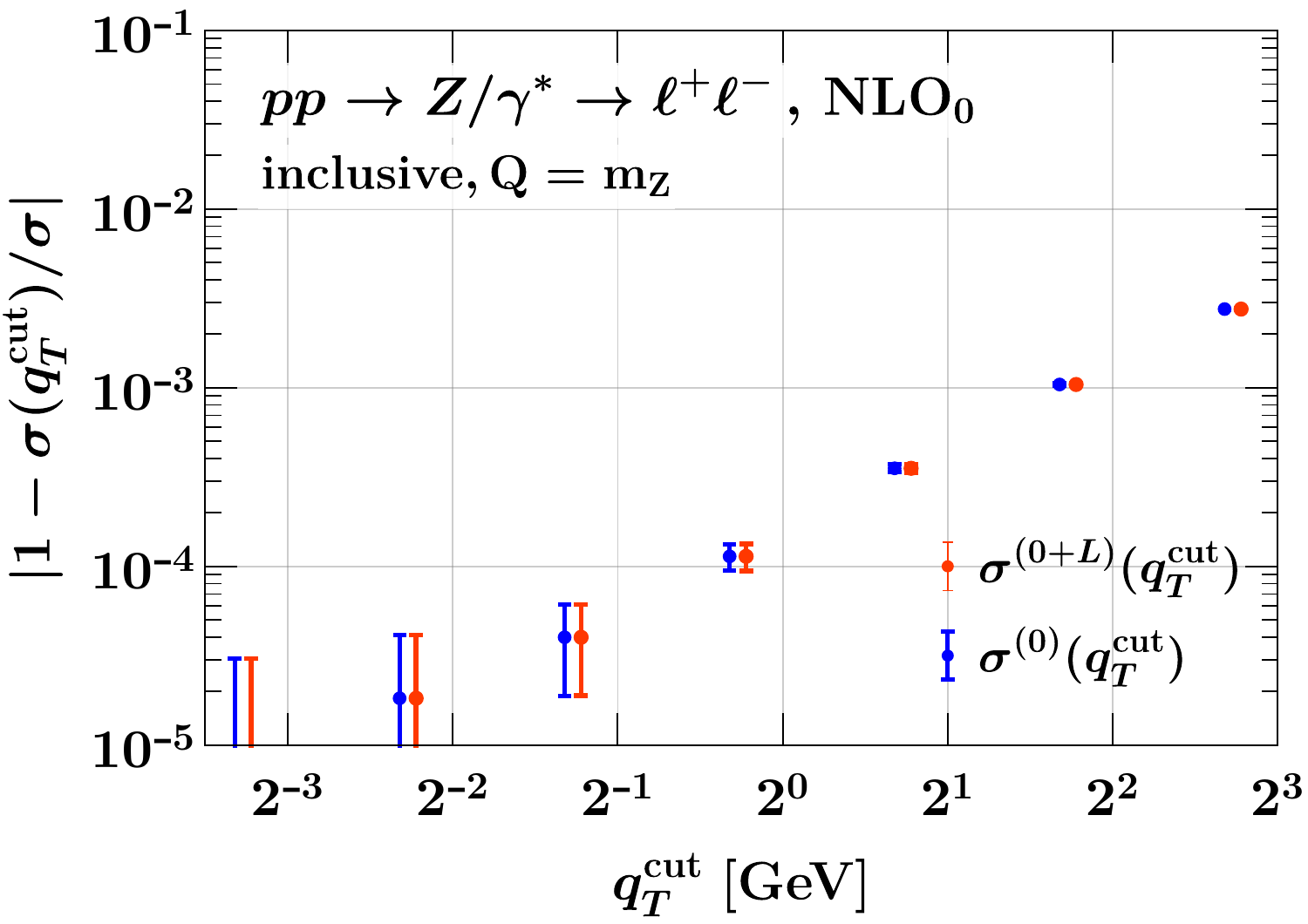}%
\hfill%
\includegraphics[width=\WidthTwoSubfigs]{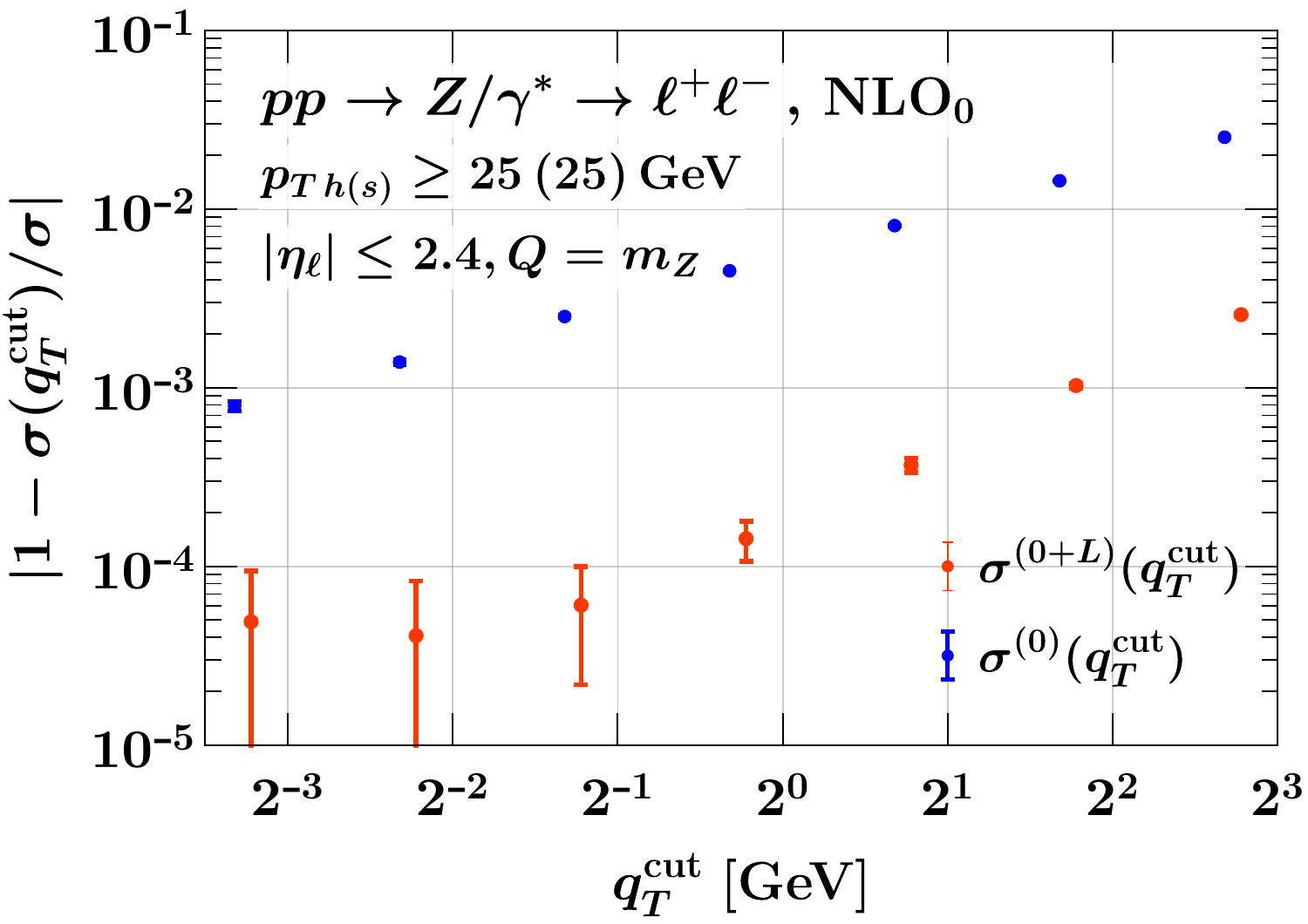}%
\\%
\includegraphics[width=\WidthTwoSubfigs]{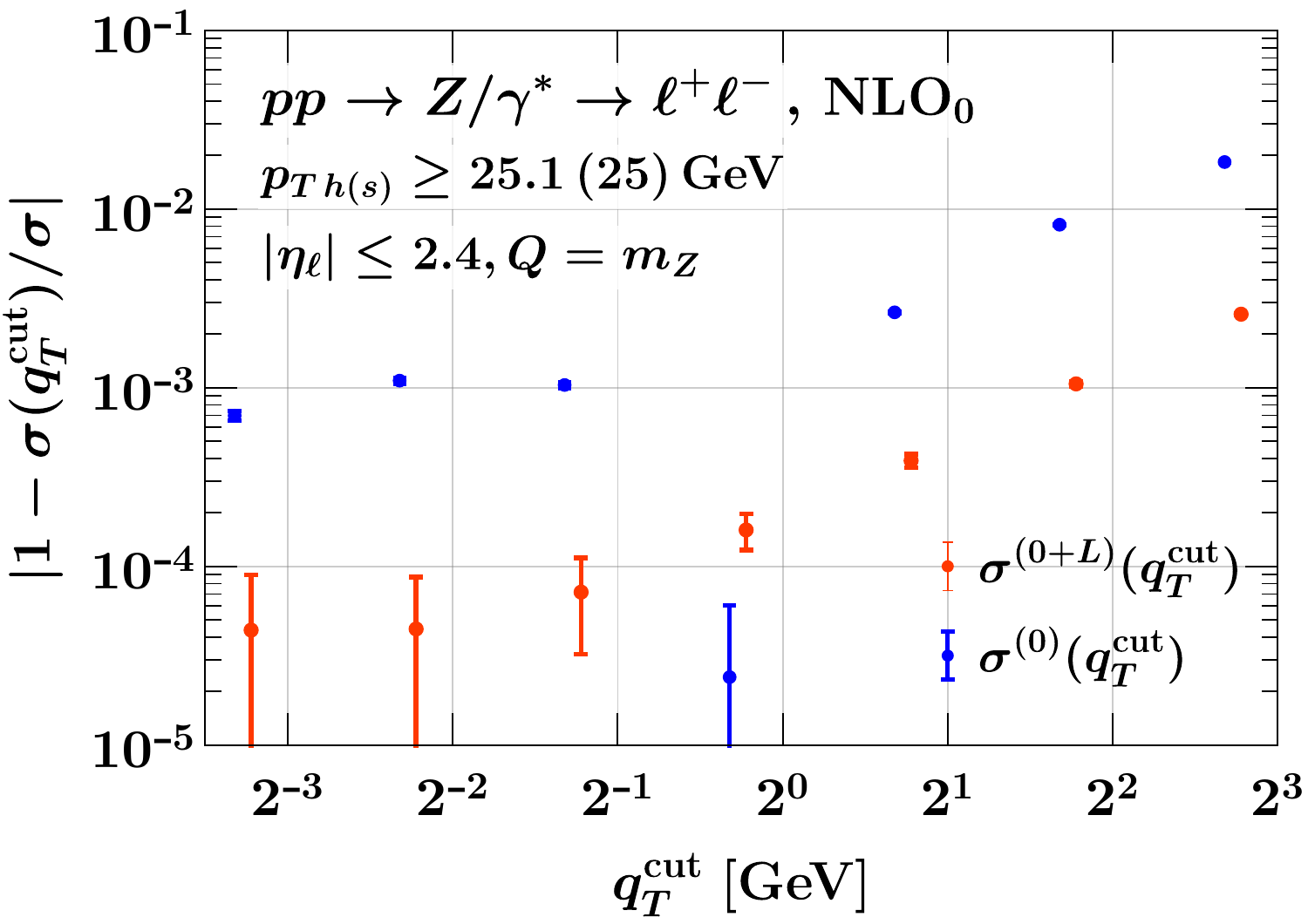}%
\hfill%
\includegraphics[width=\WidthTwoSubfigs]{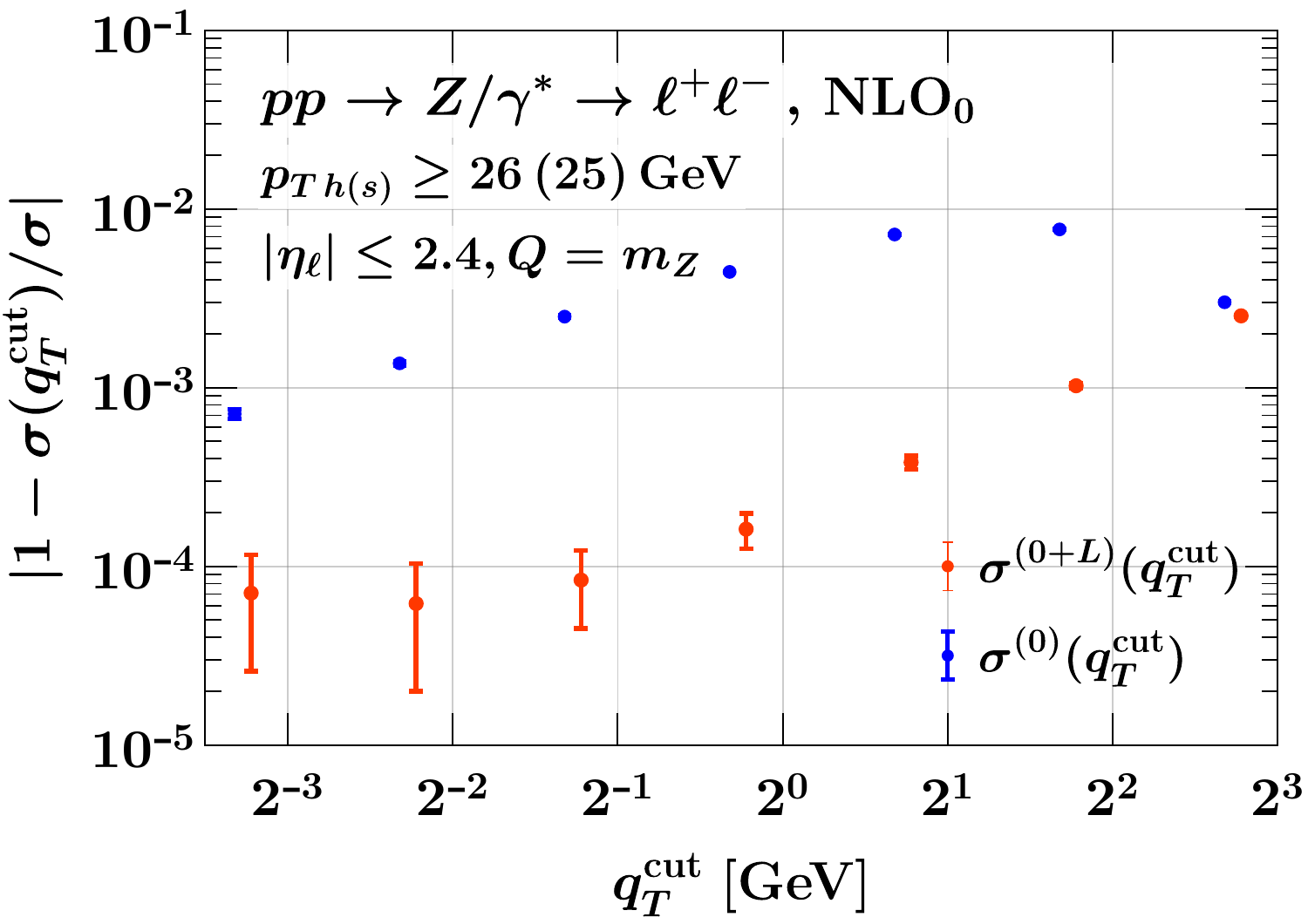}%
\caption{%
Relative difference between the exact NLO Drell-Yan cross section and its
approximation with $q_T$ subtractions as a function of $\qTcut$ for different
fiducial cuts, as indicated in the labels. The red and blue points correspond to
implementing the $q_T$ subtraction with $\sigma^\sub = \sigma^{(0)}$ and
$\sigma^\sub = \sigma^{(0+L)}$, respectively. The blue points are slightly
shifted for clarity, and the error bars indicate the MC integration
uncertainties.
}
\label{fig:Z_qTcut_dependence}
\end{figure*}
%-------------------------------------------------------------------------------

To show the precise effect on the $q_T$ subtraction, we show in
\fig{Z_qTcut_dependence} the relative difference between the total cross section
obtained via $q_T$ subtraction to the reference NLO result obtained from
\mcfm~using Catani-Seymour subtractions as a function of $\qTcut$. The blue
points show the result from $q_T$ subtractions at strict LP, $\sigma^\sub =
\sigma^{(0)}$, while the red points show the result including fiducial power
corrections in the subtractions, $\sigma^\sub = \sigma^{(0+L)}$. The blue points
are slightly shifted for better visibility. The uncertainties show the MC
integration uncertainties.

For reference, the top-left panel shows the result without any fiducial cuts,
in which case there is no difference between the two subtractions.
The remaining power corrections are quadratically suppressed, and thus one
observes a very good convergence, with deviations to the exact result
of only $0.1\%$ already for $\qTcut = 3.2 \GeV$.
In the other panels, we show different leptonic cuts. The top-right panel
shows the case of equal cuts, $p_{T\,h},  p_{T\,s} \ge 25\GeV$, while in the bottom we
increase the cut on $p_{T\,h}$ to $25.1\GeV$ (left) and $26\GeV$ (right).
(We do not show $p_{T\,h} \ge 25.5\GeV$, which does not show any new features.)
In all three cases, the power corrections to the subtraction with fiducial power
corrections (red points) are of similar size as in the inclusive case.
Even though the $\qTcut$ dependence becomes somewhat smaller in magnitude
with increasing difference $p_{T\,h}^{\rm min} - p_{T\,s}^{\rm min}$,
reflecting a sensitivity on $q_T \sim |p_{T\,h}^{\rm min} - p_{T\,s}^{\rm min}|$,
this is mostly accounted for by including the fiducial power corrections,
and as before choosing $q_T^\cut \sim 3.2\GeV$ suffices for permille accuracy.

In contrast, when implementing the $q_T$ subtraction at strict LP (blue points),
the size of missing power corrections increases by one to two orders of magnitude,
reflecting the enhanced power corrections from the fiducial cuts. To reach the
$1\permil$ precision goal requires a substantially smaller $\qTcut \approx 0.1\GeV$.
Furthermore, we observe again that the shape of the $\qTcut$ dependence
of $\sigma^{(0)}(\qTcut)$ strongly
depends on the precise difference $|p_{T\,h}^{\rm min} - p_{T\,s}^{\rm min}|$
between the lepton cuts. In particular, for
$p_{T\,h} \ge 25.1\GeV$~(bottom left), we see that the $\qTcut$ dependence
accidentally vanishes around $\qTcut = 0.8\GeV$, while it is quite flat
in the region $\qTcut \sim 0.1 - 0.8\GeV$. This region roughly corresponds to the
the sign change visible in the corresponding $q_T$ spectrum, see
\fig{Z_qT_sing_nons_lepton_cuts} (top right). This behavior is very dangerous,
as it can easily be confused with an apparent convergence if the $\qTcut$
dependence is not properly extrapolated.

In summary, we again find that including the fiducial power corrections in the
$q_T$ subtraction significantly improves the convergence in the $\qTcut\to0$ limit, and
thus makes the method much more reliable. In particular, for the case of
symmetric lepton $p_T$ cuts, it recovers the same level of power corrections as
for the inclusive case.

%%%%%%%%%%%%%%%%%%%%%%%%%%%%%%%%%%%%%%%%%%%%%%%%%%%%%%%%%%%%%%%%%%%%%%%%%%%%%%%%
\FloatBarrier
\section{Comparison to data}
\label{sec:data_comparison}
%%%%%%%%%%%%%%%%%%%%%%%%%%%%%%%%%%%%%%%%%%%%%%%%%%%%%%%%%%%%%%%%%%%%%%%%%%%%%%%%

In this section, we compare our N$^3$LL$+$NNLO$_0$ resummed predictions for
$q_T$ and $\phi^*$ for Drell-Yan, $pp \to Z/\gamma^* \to \ell^+ \ell^-$ ($\ell = e,\mu)$,
with the following precision LHC measurements:
%%%
\begin{itemize}
 \item The ATLAS measurement from \refcite{Aad:2015auj}
       using $20.3~\fb^{-1}$ of 8 TeV data. We consider the
       $m_{\ell^+ \ell^-} \in [66,116]~\GeV$ invariant mass bin.
       The fiducial lepton cuts are $p_T > 20~\GeV, |\eta| < 2.4$
       for both electrons and muons.
       Separate results for the electron and muon channels are reported, in
       both cases we compare to the measurements using Born leptons.
       (While the event selection in \refcite{Aad:2015auj} excludes electrons
       in the region $1.37 < |\eta| < 1.52$, this exclusion is removed during
       the unfolding and not part of the fiducial volume in which the measurements
       are reported.)
 \item The CMS measurement from \refcite{Sirunyan:2019bzr}
       using $35.9~\fb^{-1}$ of 13 TeV data in the $m_{\ell^+\ell^-} = m_Z \pm 15~\GeV$
       invariant mass bin. The fiducial cuts are given by $p_T > 25~\GeV$ and $|\eta| < 2.4$
       for both electrons and muons. We compare to the combined measurements
       of dressed electrons and muons.
\end{itemize}
%%%
We consider two sets of predictions: The strict LP resummation with fiducial
power corrections only included via the fixed-order matching is denoted as
N$^3$LL$^\zero+$NNLO$_0$, and analogously at lower orders.
The resummation including fiducial power corrections is denoted as
N$^3$LL$^{(0+L)}+$NNLO$_0$, and analogously at lower orders. In this case,
the fixed-order matching only adds the remaining genuine (non-fiducial) power
corrections.

By default we compare to the measured spectra that are normalized to
the total cross section. We correspondingly normalize our
predictions to the total fiducial cross section at the corresponding order
obtained by integrating the central value
for the spectrum to infinity. This effectively amounts to obtaining the total
cross section via $q_T$ or $\phi^*$ subtractions including fiducial power corrections. Since
the uncertainties for the total cross section are much smaller than for the
spectrum, they are practically irrelevant for this purpose. We have also checked
that treating the $\mu_\FO$ variation in a correlated fashion or treating it
fully uncorrelated between spectrum and normalization and adding it in
quadrature leads to essentially identical estimates of the total perturbative
uncertainty.
For completeness, we also provide analogous comparisons for the unnormalized
CMS measurements in \app{data_comparison}.

%===============================================================================
\subsection{\texorpdfstring{$q_T$}{qT} spectrum}
\label{sec:data_qt}
%===============================================================================

%-------------------------------------------------------------------------------
\begin{figure*}
\centering
\includegraphics[width=0.49\textwidth]{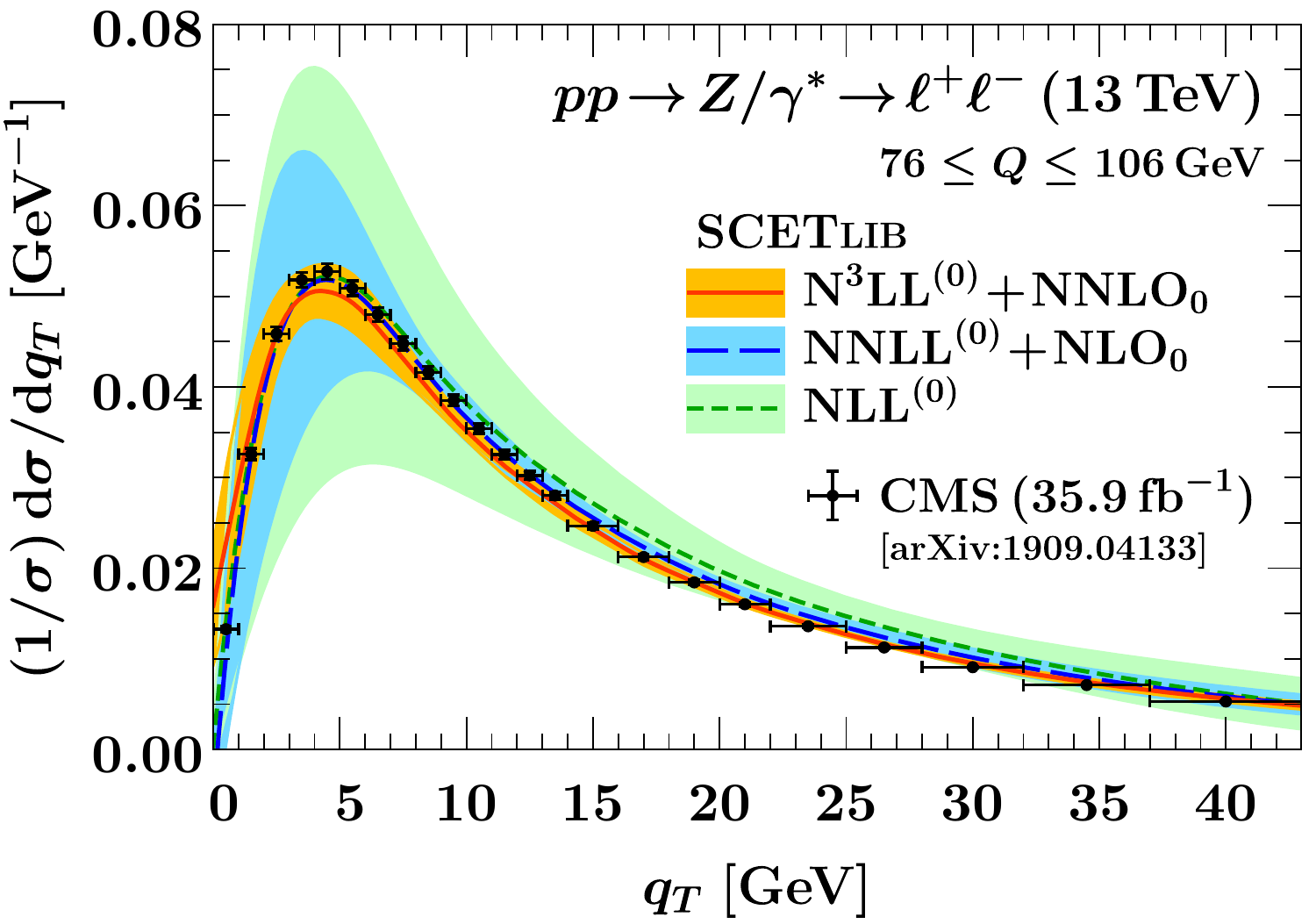}%
\hfill%
\includegraphics[width=0.49\textwidth]{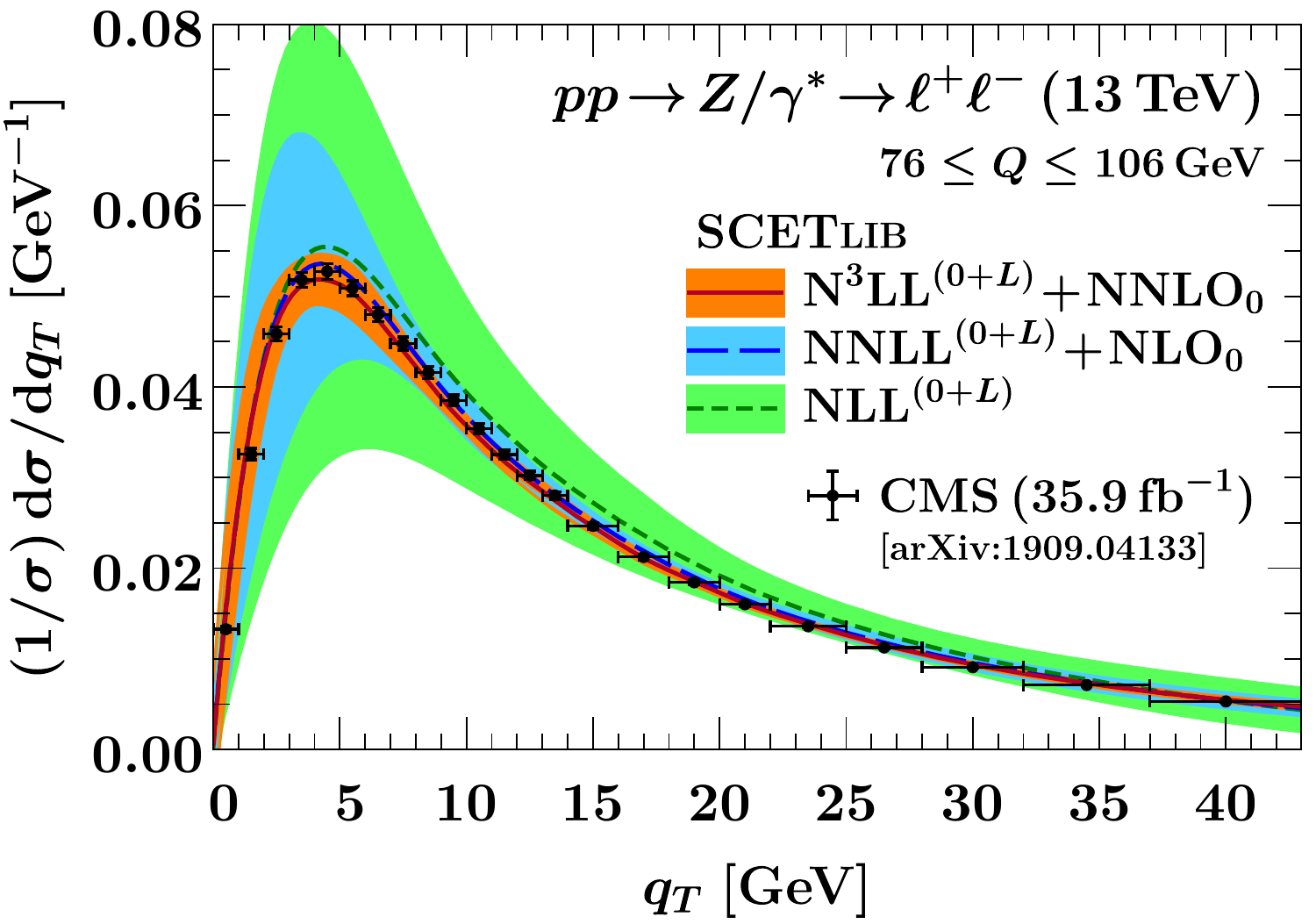}%
\\
\includegraphics[width=0.49\textwidth]{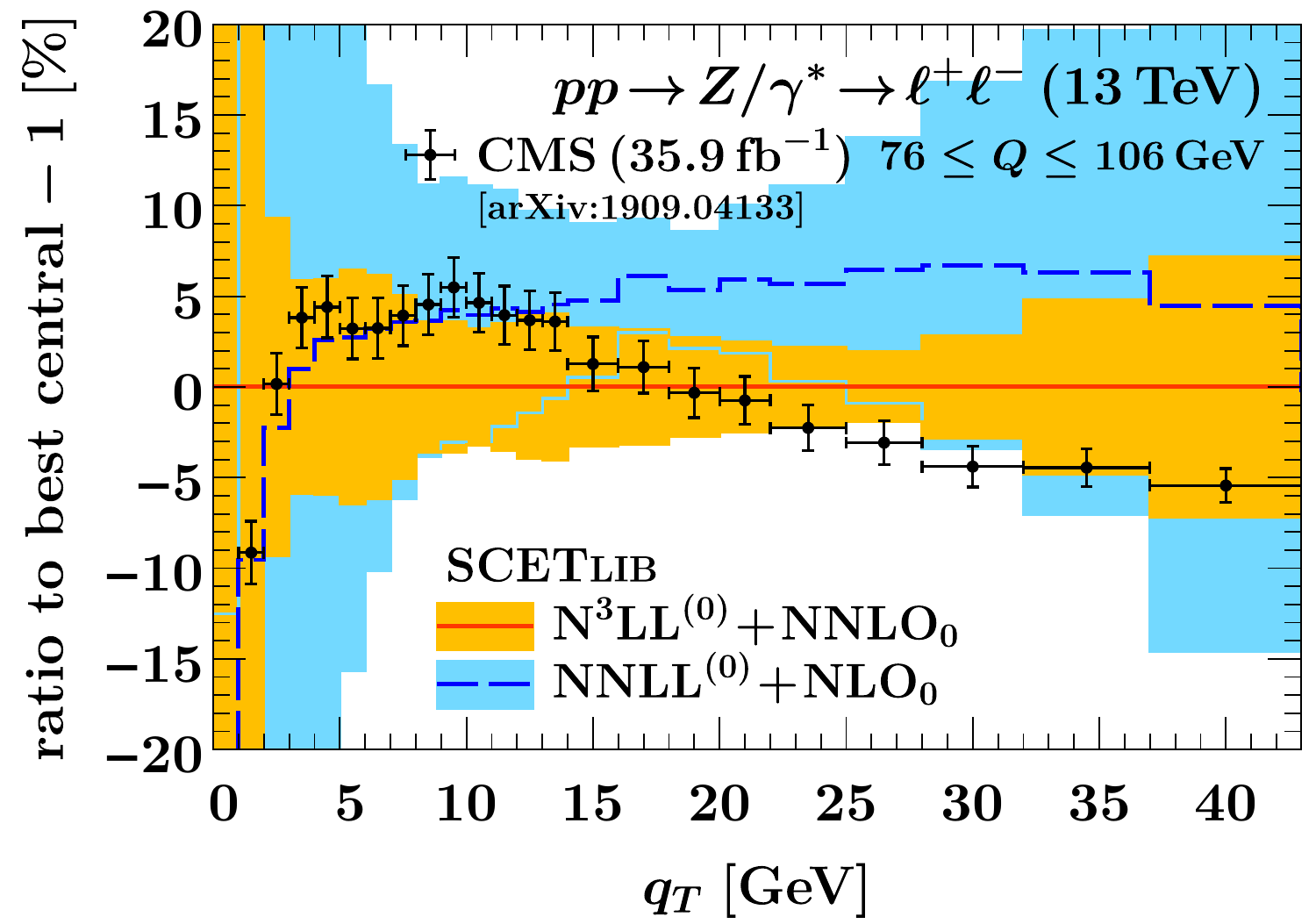}%
\hfill%
\includegraphics[width=0.49\textwidth]{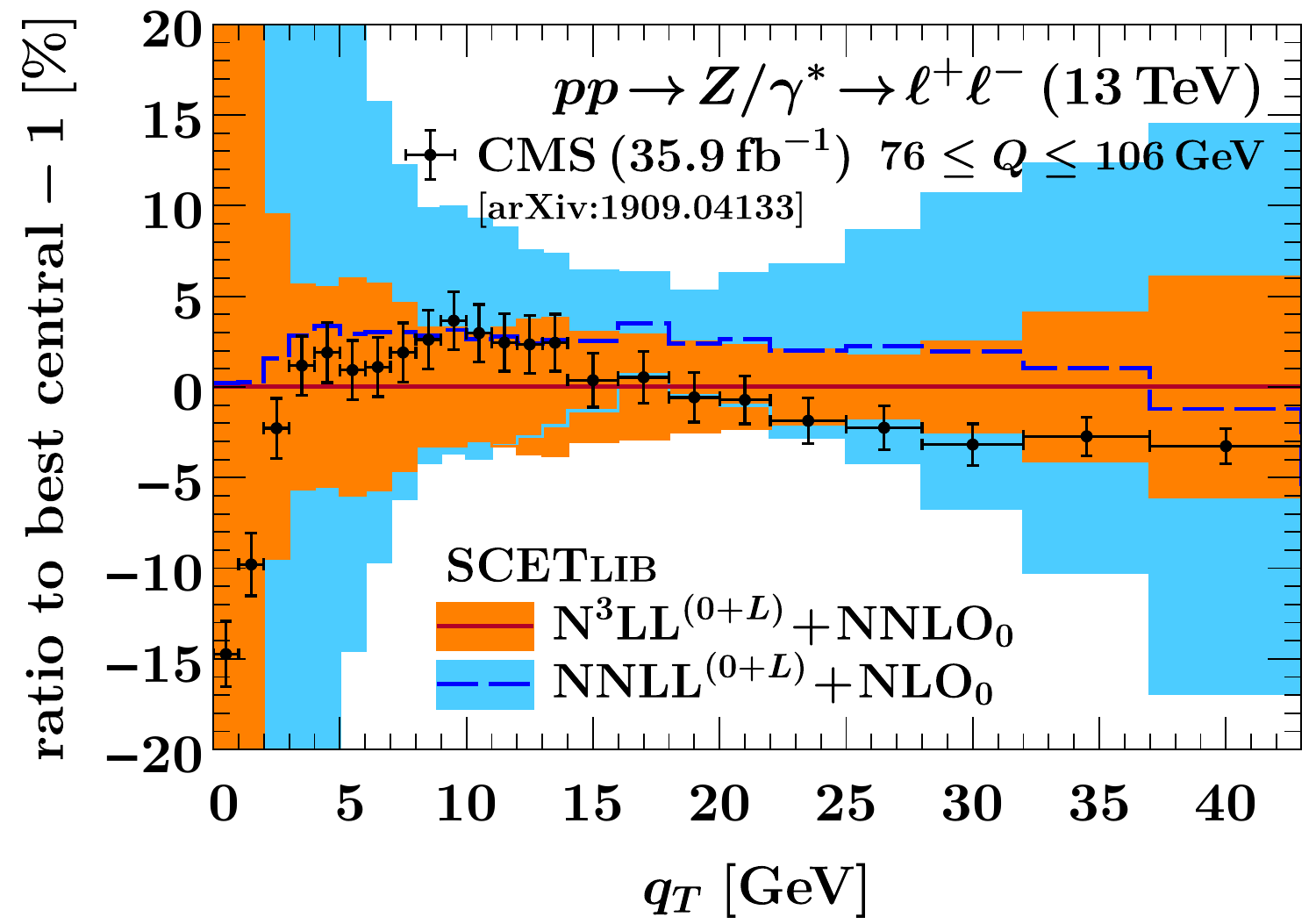}%
\caption{%
Predictions for the normalized Drell-Yan fiducial $q_T$ spectrum without (left)
and with (right) resummed fiducial power corrections compared to CMS 13 TeV
measurements~\cite{Sirunyan:2019bzr}. The top panels show the spectrum, with the
theory predictions drawn as smooth curves for better visibility. The bottom
panels show the percent differences to the respective highest-order prediction
central value.
}
\label{fig:Z_qT_13TeV_CMS}
\end{figure*}
%-------------------------------------------------------------------------------

%-------------------------------------------------------------------------------
\begin{figure*}
\centering
\includegraphics[width=0.49\textwidth]{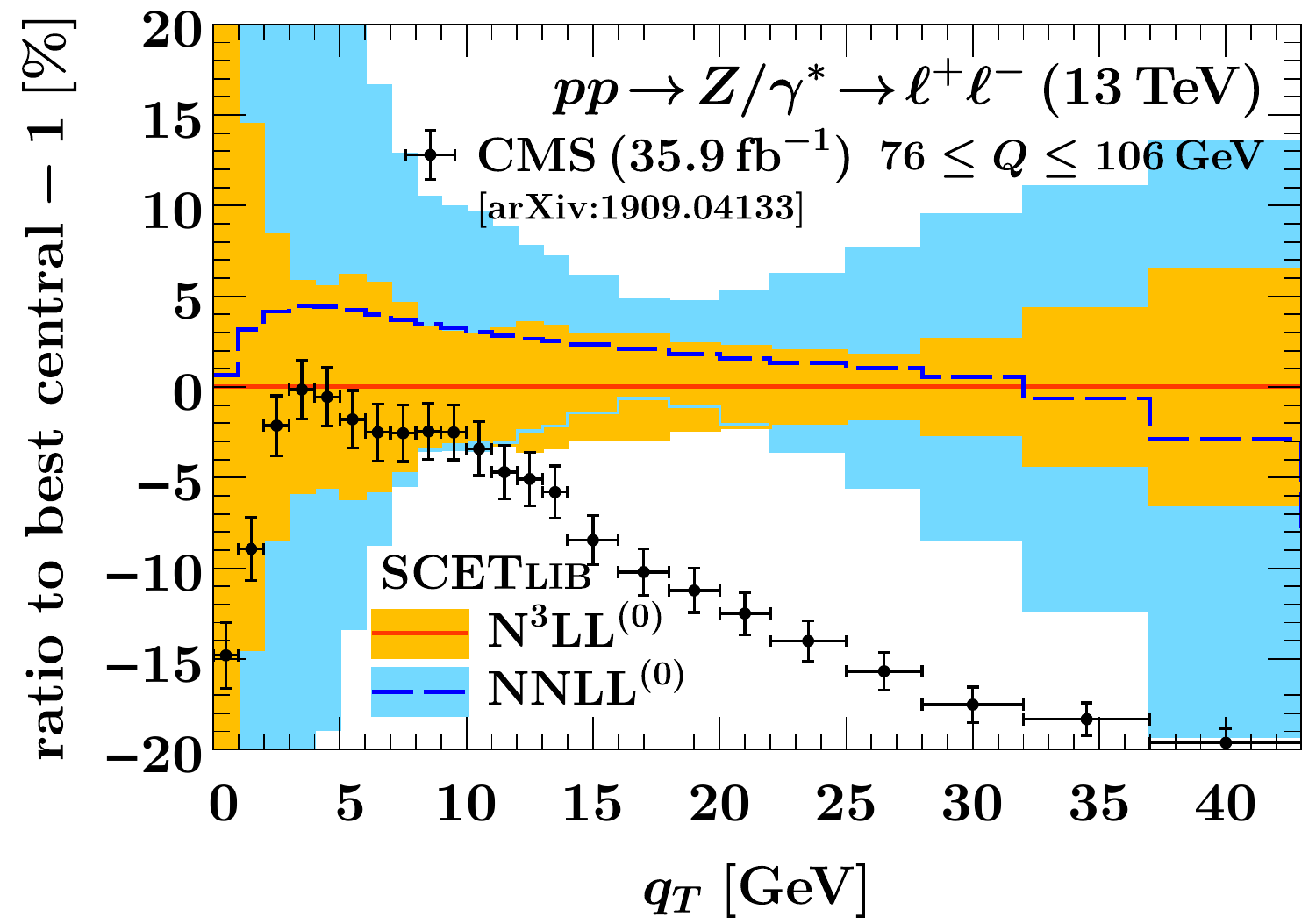}%
\hfill%
\includegraphics[width=0.49\textwidth]{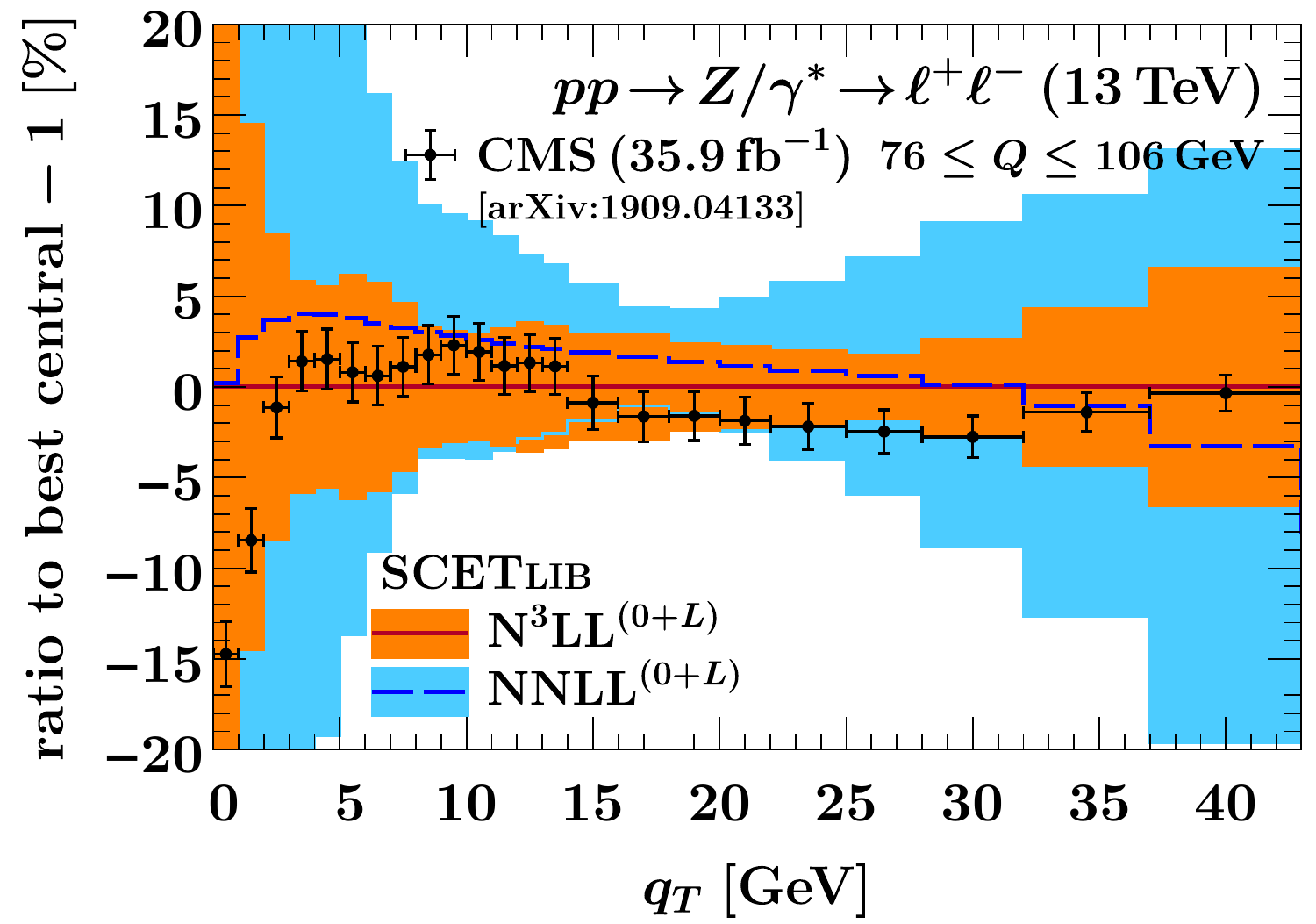}%
\caption{%
Same as the bottom row of \fig{Z_qT_13TeV_CMS}, but without including power
corrections from the fixed-order matching.
}
\label{fig:Z_qT_13TeV_CMS_no_matching}
\end{figure*}
%-------------------------------------------------------------------------------

%-------------------------------------------------------------------------------
\begin{figure*}
\centering
\includegraphics[width=0.49\textwidth]{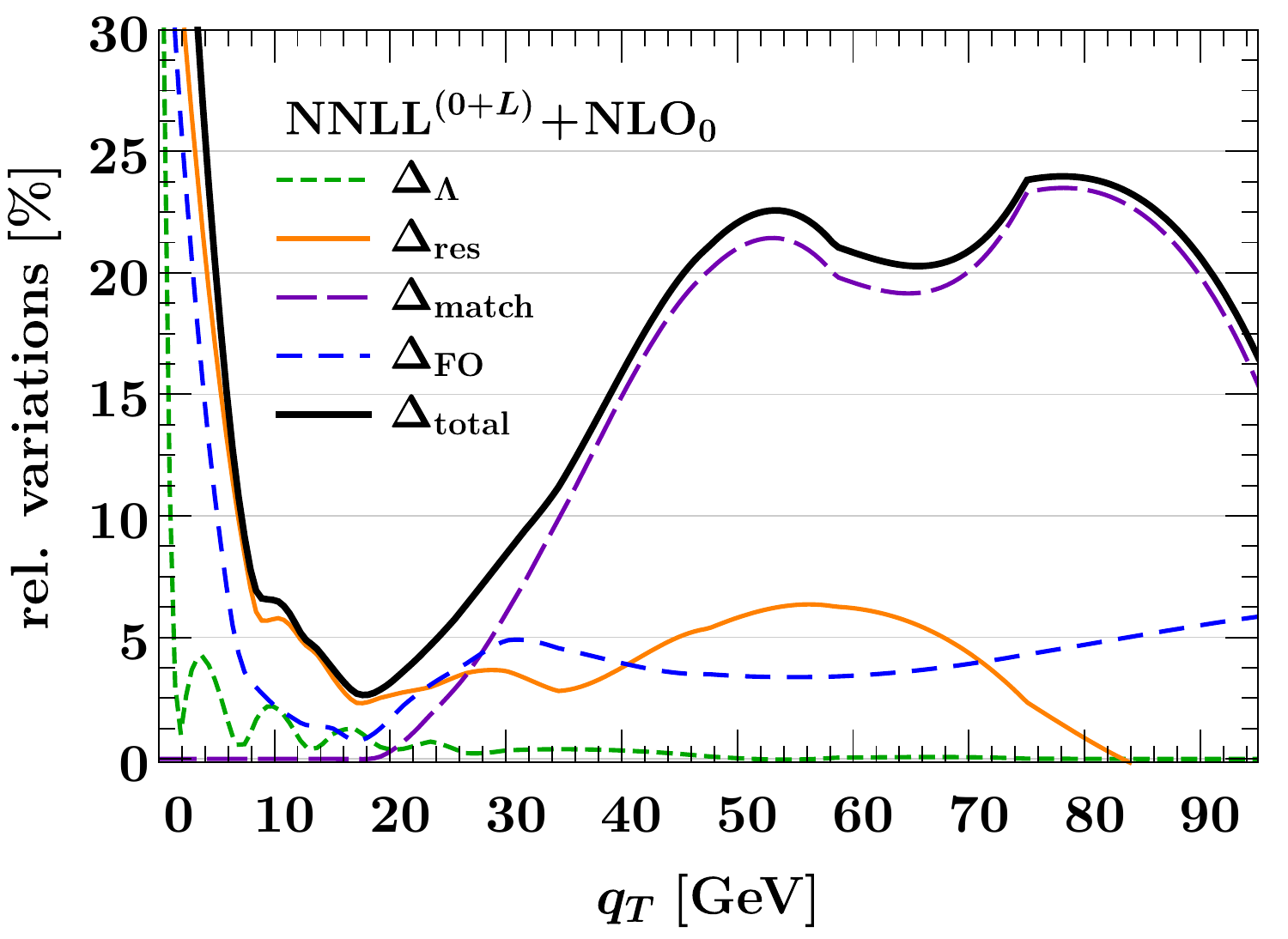}%
\hfill%
\includegraphics[width=0.49\textwidth]{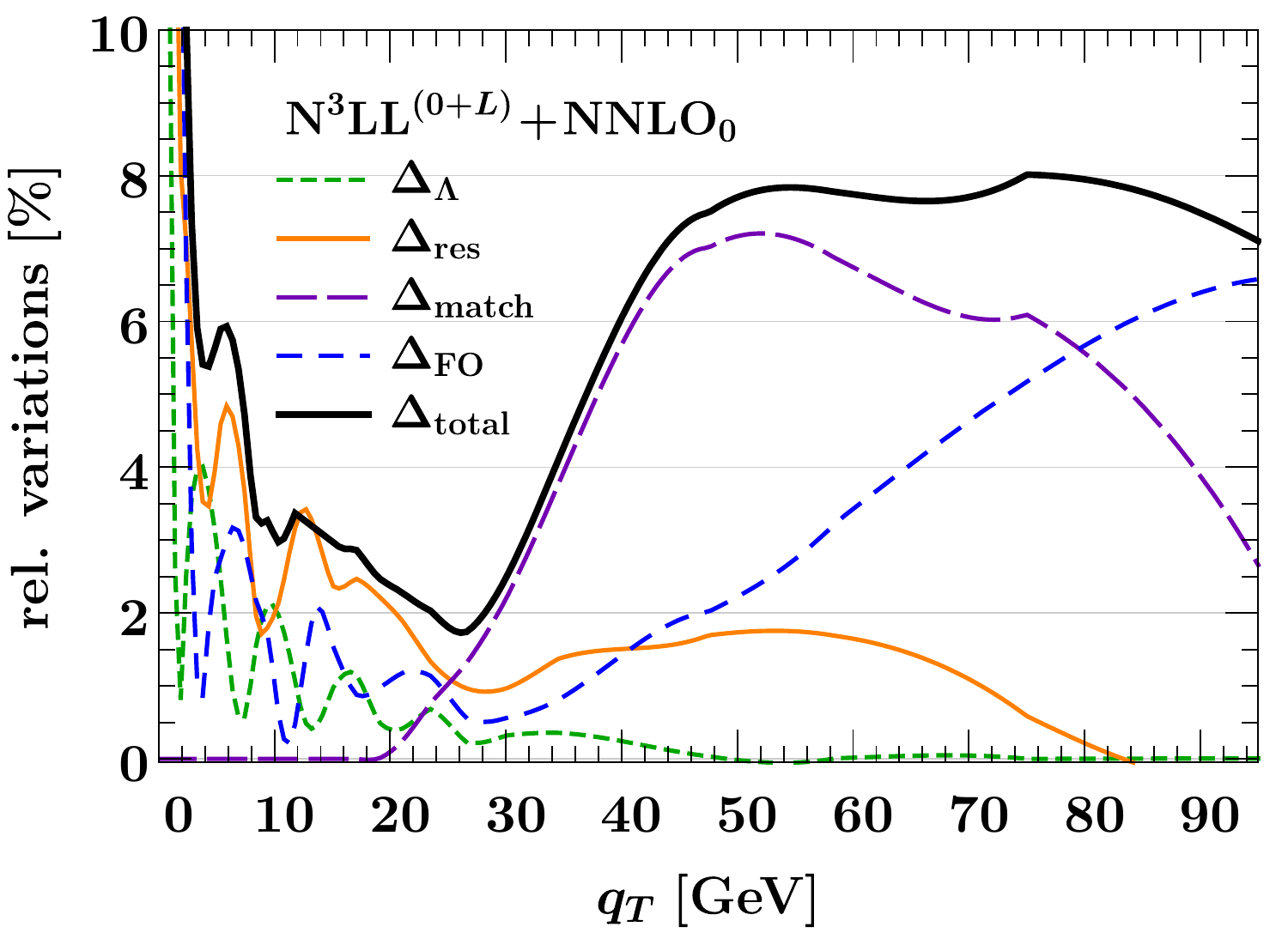}%
\caption{%
Breakdown of the total uncertainty estimate in the bottom-right panel of \fig{Z_qT_13TeV_CMS}
at NNLL$^{(0+L)}+$NLO$_0$ (left) and N$^3$LL$^{(0+L)}+$NNLO$_0$ (right).
We note the difference in vertical scale.
The various contributions are discussed in \sec{estimate_pert_uncerts}.
The total uncertainty estimate (solid black) is obtained
by adding the individual contributions in quadrature.
}
\label{fig:Z_qT_13TeV_CMS_uncertainty_breakdown}
\end{figure*}
%-------------------------------------------------------------------------------

In \fig{Z_qT_13TeV_CMS}, we compare our results to the CMS 13 TeV measurements
for $q_T$. The top panels show the $q_T$ spectrum at NLL (green), NNLL$+$NLO$_0$
(blue), and N$^3$LL$+$NNLO$_0$ (orange). The bands show the estimated perturbative
uncertainties as discussed in \sec{estimate_pert_uncerts}. Note that the theory
predictions are obtained for the measured binning, but are drawn as smooth
curves for clearer visibility. The lower panels show the same results normalized
to the respective highest-order prediction. The results using strict LP
resummation are shown on the left (lighter shading),
while those including the resummation of fiducial power corrections are
shown on the right (darker shading). In both cases, we observe
good convergence of the resummed predictions, with substantially reduced
perturbative uncertainties at subsequent higher orders, as well as good
agreement with the data. Nevertheless, resumming the fiducial power corrections
on the right further improves the perturbative convergence and also yields a
systematically better agreement with the data. The data agreement deteriorates
in the first two bins, which can be attributed to small-$q_T$ nonperturbative
effects. These are expected to become important for $q_T \lesssim 2\GeV$, but
the nonperturbative ingredients necessary to account for these effects are not
included in our predictions. This is also reflected in the substantially
increased perturbative uncertainties in this region.

A detailed breakdown of our uncertainty estimate
is shown in \fig{Z_qT_13TeV_CMS_uncertainty_breakdown}.
As expected, the nonperturbative uncertainty $\Delta_\Lambda$ (short-dashed green)
only has an impact at low values of $q_T \lesssim 10 \GeV$.
The resummation uncertainty $\Delta_\res$ (solid orange) dominates
up to $q_T \approx 30 \GeV$ at which point the matching uncertainty $\Delta_\match$
(long-dashed violet) takes over as the dominant component.
The fixed-order uncertainty $\Delta_\FO$ (dashed blue)
starts to dominate at very high values of $q_T \gtrsim 80 \GeV$ at N$^3$LL$+$NNLO$_0$,
but also has an appreciable impact at low $q_T$, where
it estimates the fixed-order uncertainties within the resummation.
We stress that in the intermediate range $30 \leq q_T \leq 70 \GeV$,
the $\Delta_\FO$ is known to underestimate the perturbative uncertainty in the
pure fixed-order results. Supplementing it by the matching uncertainty $\Delta_\match$ is
thus reasonable as the latter probes a viable set of all-order terms beyond the strict
fixed-order result.
Note that several of the uncertainty components begin to oscillate strongly for $q_T \lesssim 10 \GeV$.
These oscillations are due to the specific implementation
of quark flavor thresholds in the PDFs as provided by LHAPDF. The flavor thresholds
lead to sharp features in $b_T$ space in individual scale variations around $\mu_B \sim b_0/b_T$,
which then translate into oscillatory artifacts in $\df \sigma/\df q_T$.
They are of no concern for the total uncertainty for which they largely average out.
It will be interesting to properly address the effect of flavor thresholds
in this region along the lines of \refcite{Pietrulewicz:2017gxc} in the future.

To further illustrate the importance and impact of the fiducial power corrections,
in \fig{Z_qT_13TeV_CMS_no_matching} we show the analog of the bottom panel
of \fig{Z_qT_13TeV_CMS} but comparing to the pure resummed results, i.e., without
including the fixed-order matching corrections to the spectrum.
(We still normalize to the same total cross section as in \fig{Z_qT_13TeV_CMS}.)
The strict LP resummation (left)
completely fails to describe the data, showing that in this case the fixed-order
matching corrections that supply the fiducial power corrections at fixed order are
essential. On the other hand, upon resumming the fiducial power corrections (right),
the excellent data agreement is restored even without the fixed-order matching.
In other words, with the fiducial power corrections included in the resummation,
the fixed-order matching becomes essentially unimportant for $q_T \lesssim 40\GeV$,
both at NNLL and N$^3$LL.

%-------------------------------------------------------------------------------
\begin{figure*}
\centering
\includegraphics[width=0.49\textwidth]{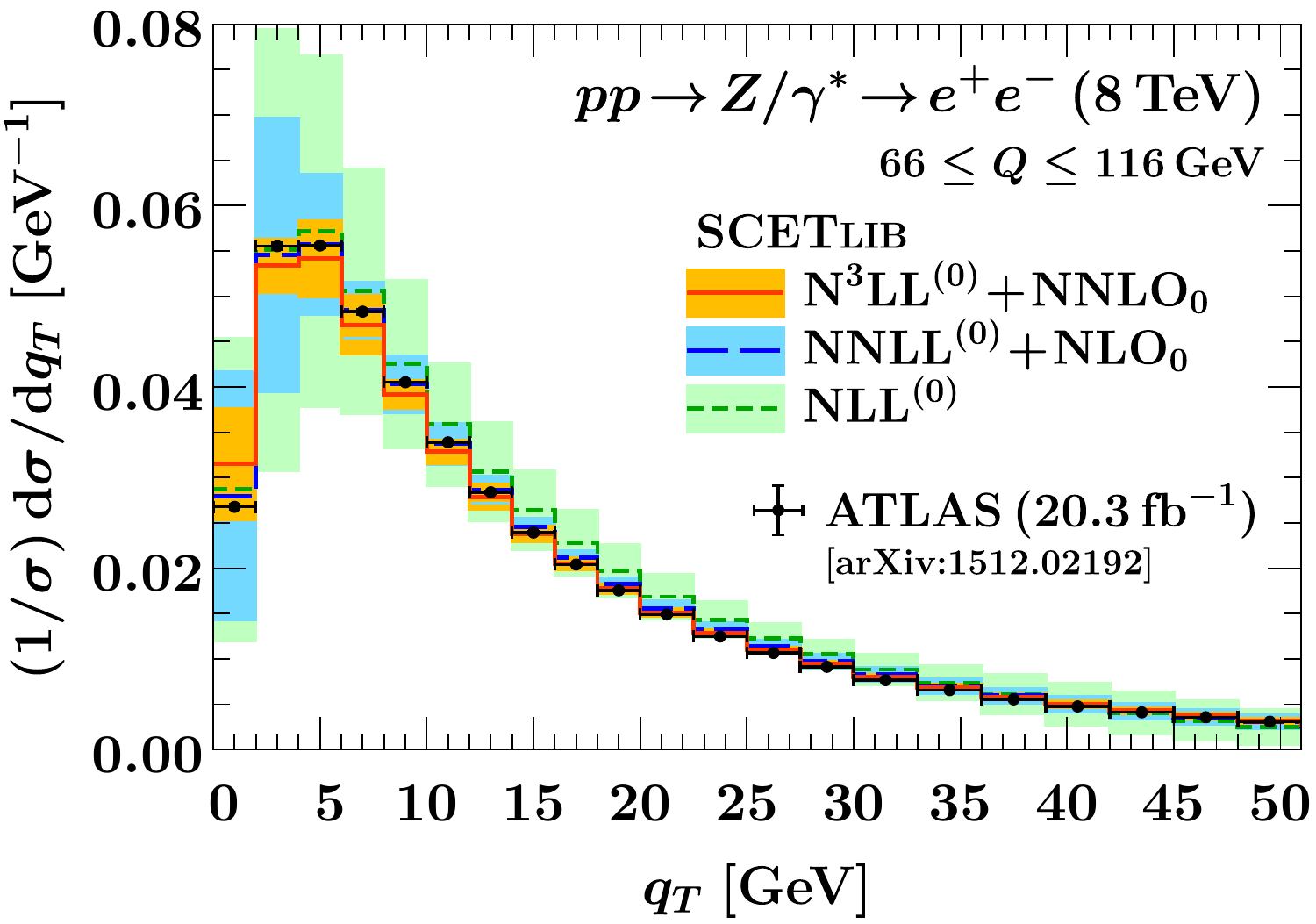}%
\hfill%
\includegraphics[width=0.49\textwidth]{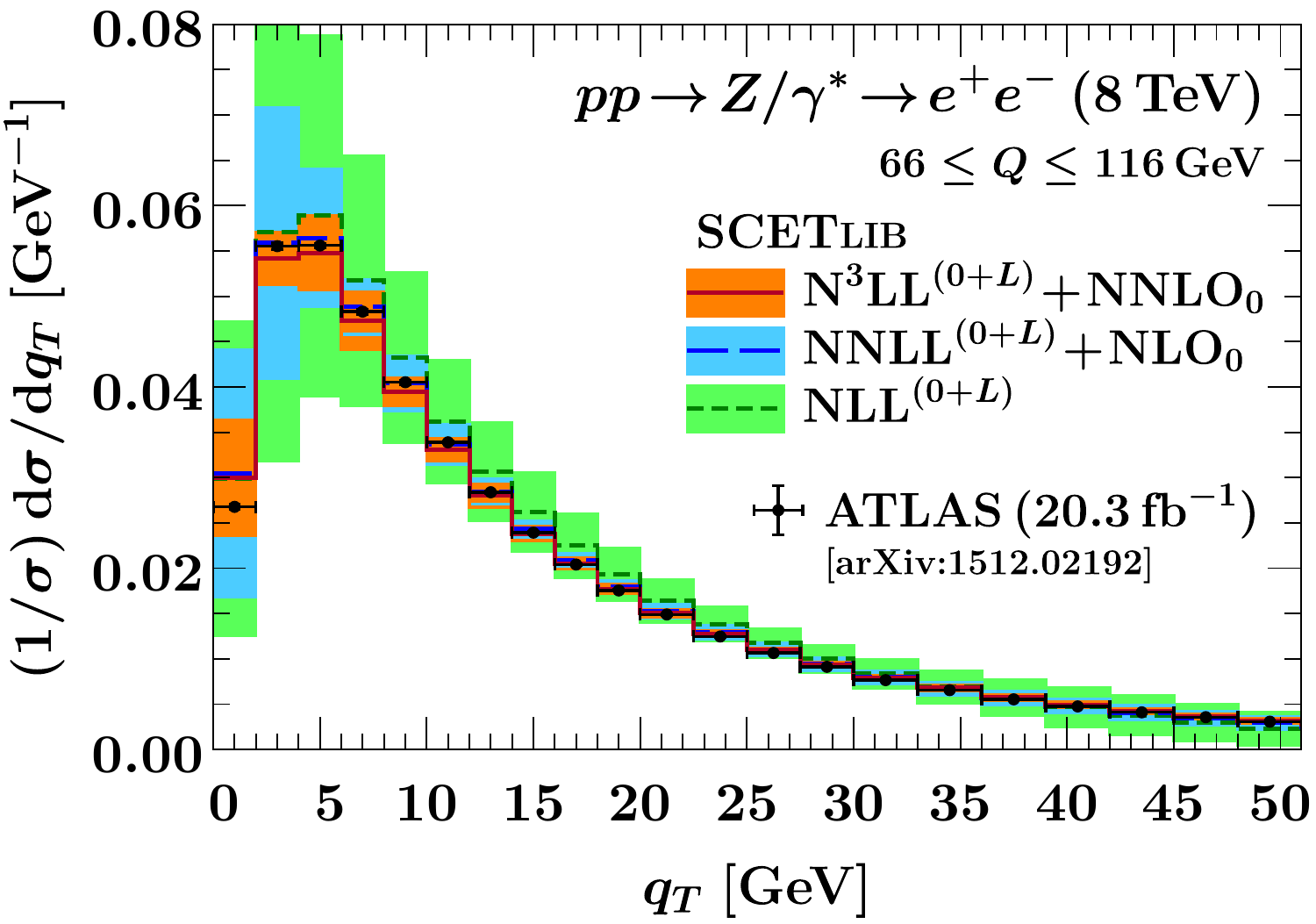}%
\\
\includegraphics[width=0.49\textwidth]{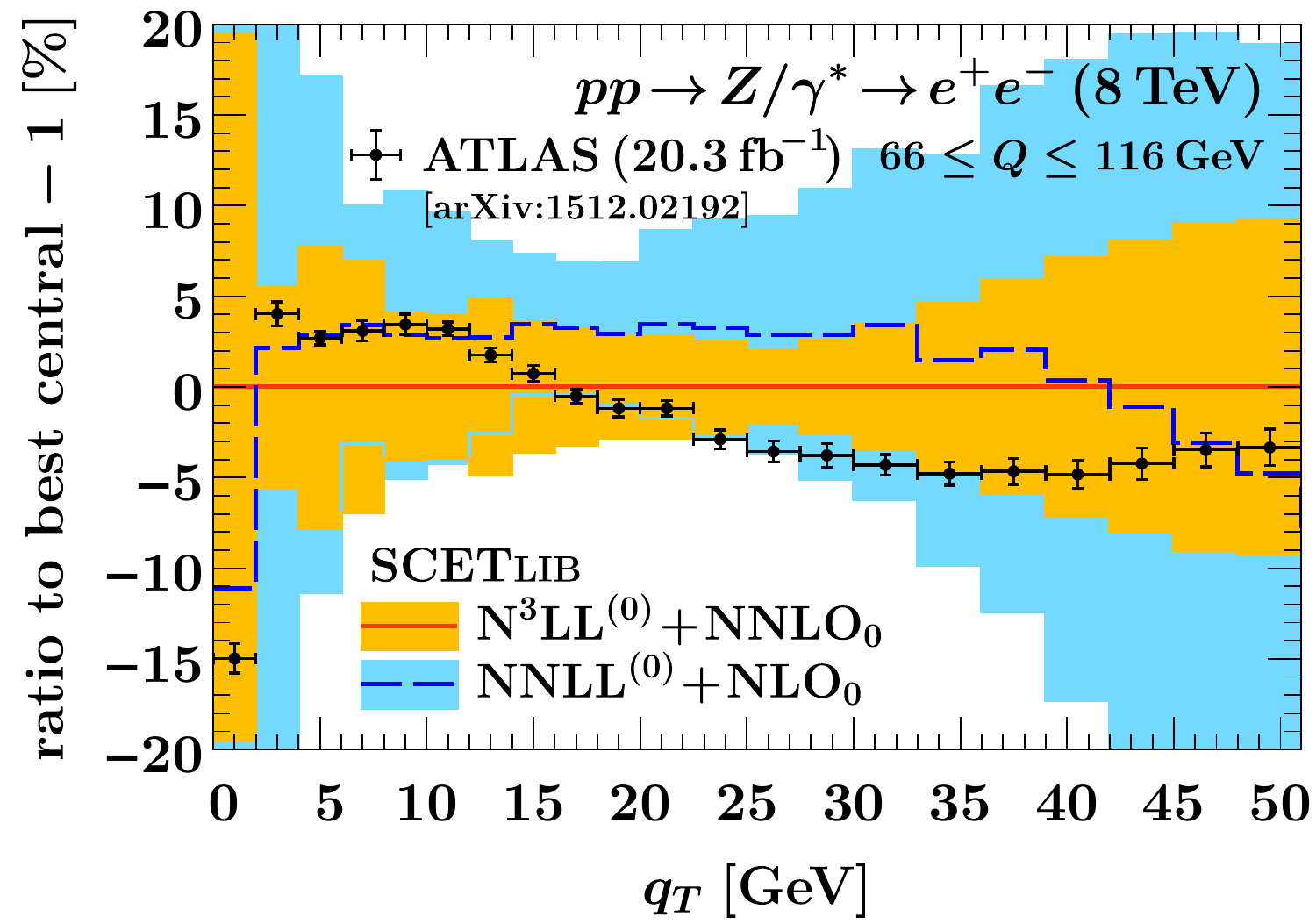}%
\hfill%
\includegraphics[width=0.49\textwidth]{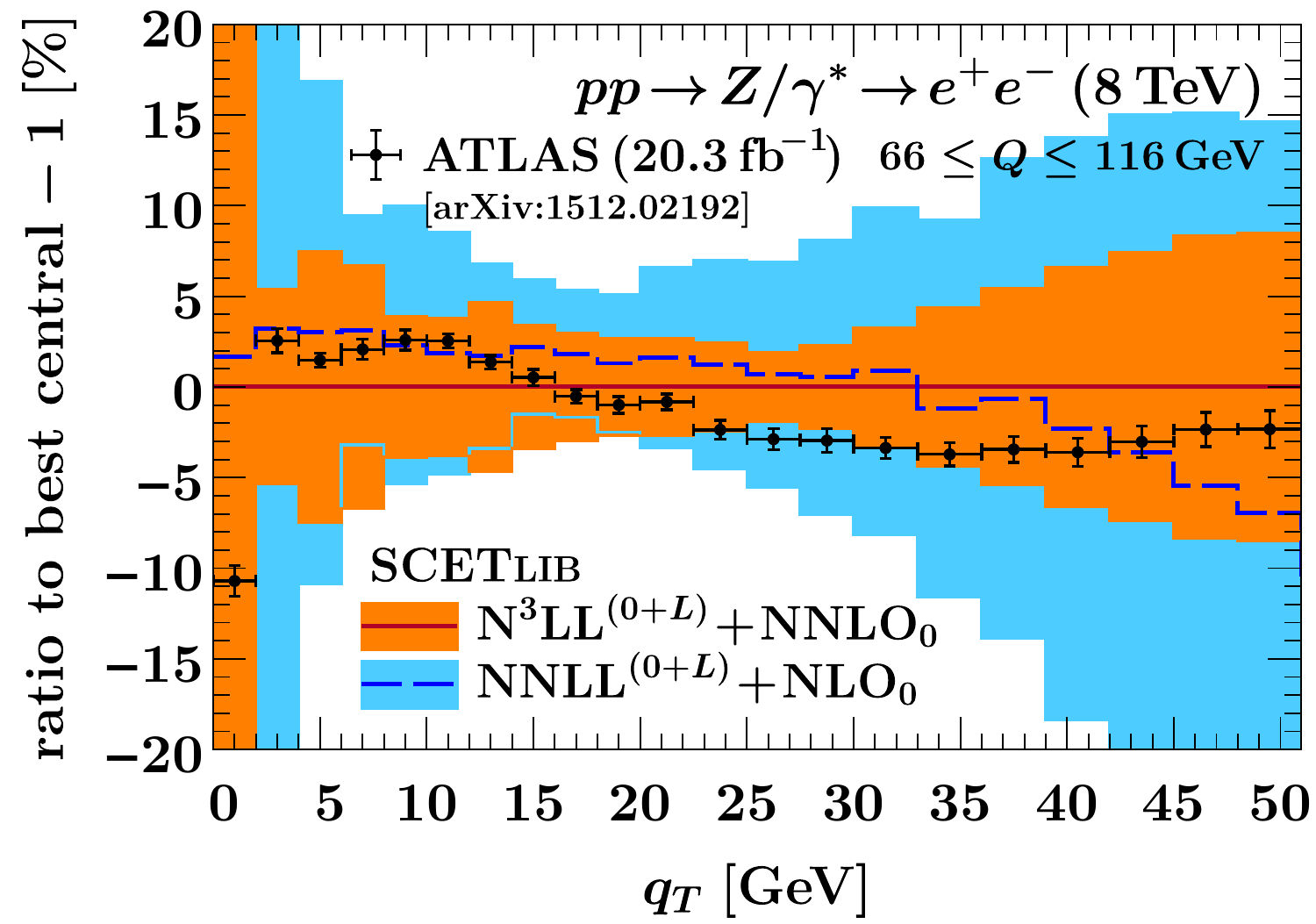}%
\caption{%
Predictions for the normalized Drell-Yan fiducial $q_T$ spectrum without (left)
and with (right) resummed fiducial power corrections compared to ATLAS 8 TeV
measurements~\cite{Aad:2015auj} in the $e^+e^-$ channel. The top panels
show the normalized spectrum. The bottom panels show the percent differences to
the respective highest-order prediction central value. The analogous results for
the $\mu^+\mu^-$ channel can be seen in \app{data_comparison} in
\fig{Z_qT_8TeV_ATLAS_mu}.
}
\label{fig:Z_qT_8TeV_ATLAS_el}
\end{figure*}
%-------------------------------------------------------------------------------

In \fig{Z_qT_8TeV_ATLAS_el}, we show the analogous comparison for the ATLAS 8
TeV measurements~\cite{Aad:2015auj} in the electron channel, with the top panel
showing the $q_T$ spectrum itself, while the bottom panel shows the relative
differences to the respective highest-order prediction.
As before, we see good perturbative convergence of the predictions, as well as
good agreement with the data. The data agreement again improves when resumming
the fiducial power corrections on the right, leading to an overall flatter shape
and reduced size in the difference between predictions and measurement. The
results in the muon channel are practically identical, and are provided for
completeness in \fig{Z_qT_8TeV_ATLAS_mu} in \app{data_comparison}.

%===============================================================================
\subsection{\texorpdfstring{$\phi^*$}{phistar} distribution}
\label{sec:data_phistar}
%===============================================================================

%-------------------------------------------------------------------------------
\begin{figure*}
\centering
\includegraphics[width=0.49\textwidth]{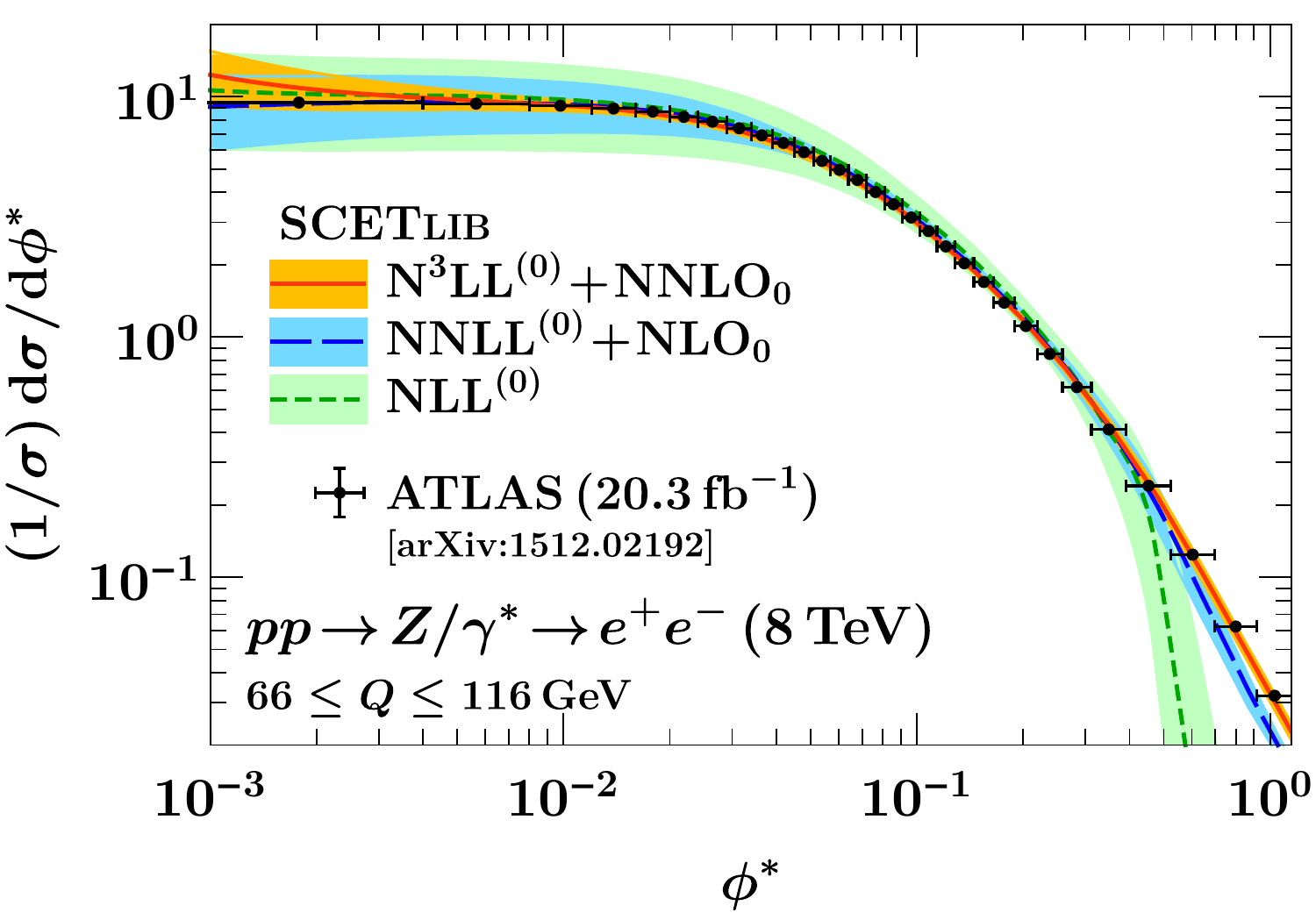}%
\hfill%
\includegraphics[width=0.49\textwidth]{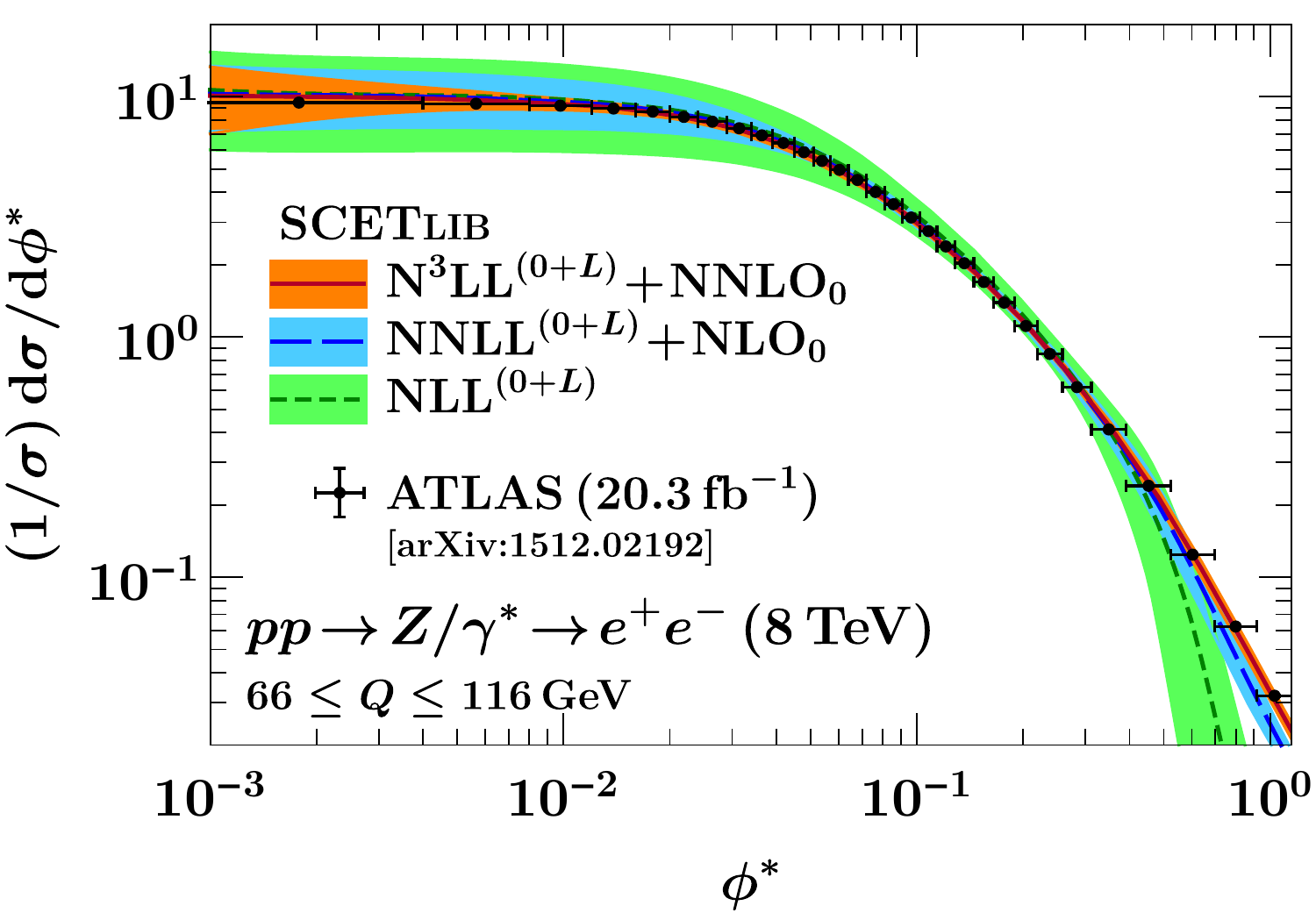}%
\\
\includegraphics[width=0.49\textwidth]{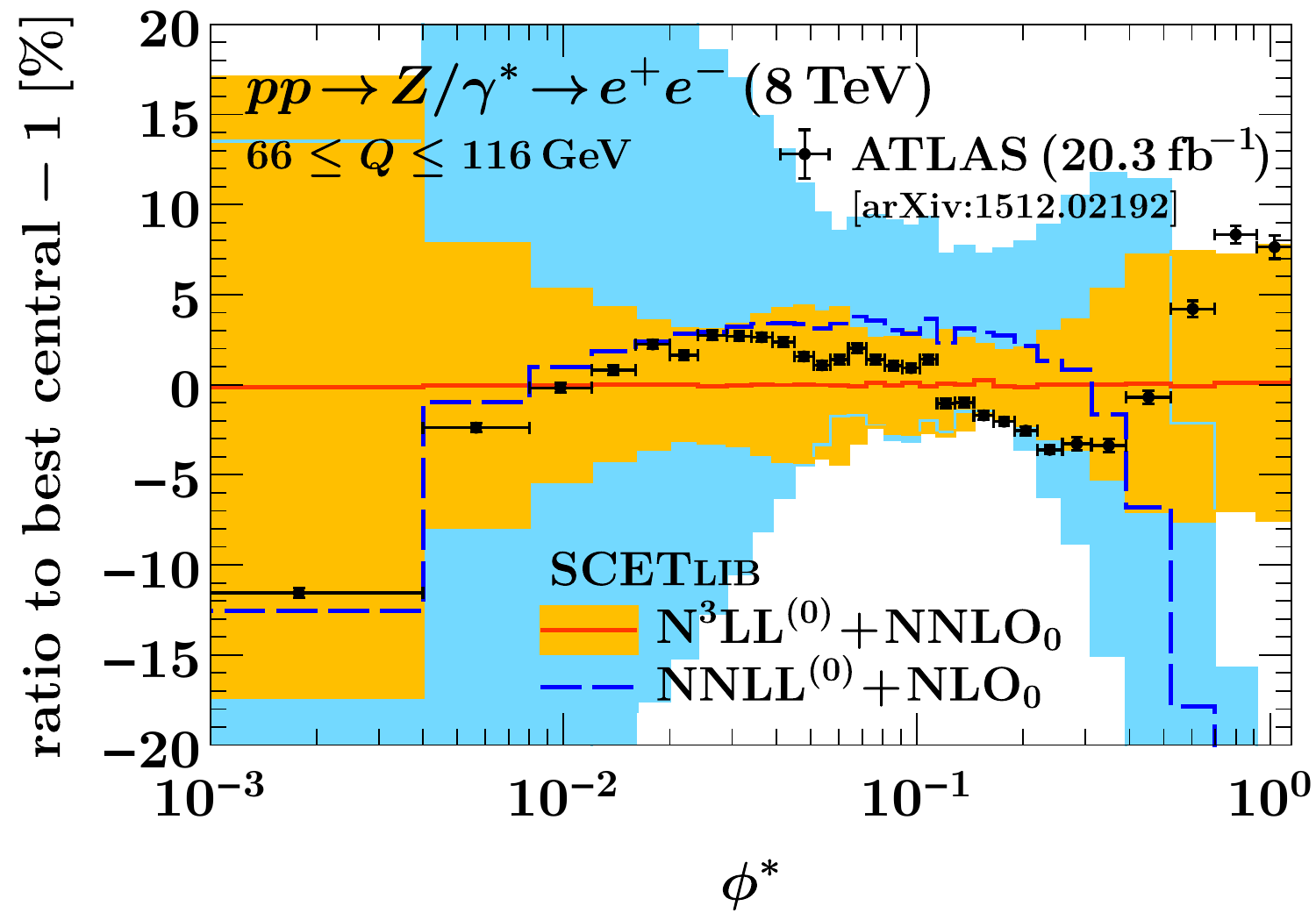}%
\hfill%
\includegraphics[width=0.49\textwidth]{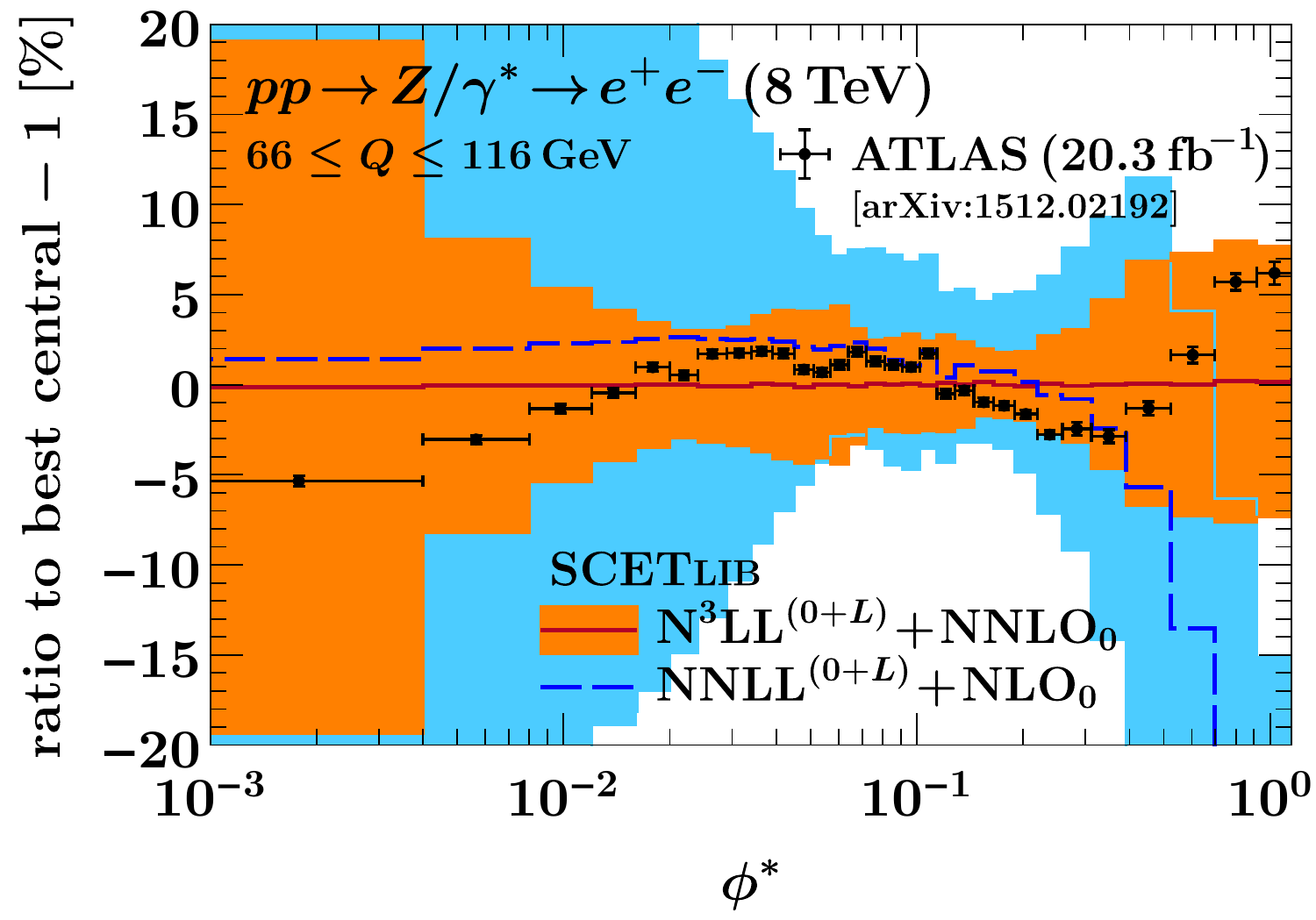}%
\caption{%
Predictions for the normalized Drell-Yan fiducial $\phi^*$ spectrum using the LP
resummation (left) and including the resummation of fiducial power corrections
(right) compared to ATLAS 8 TeV measurements~\cite{Aad:2015auj} in the $e^+e^-$
channel. The top panels show the spectrum, with the predictions drawn as smooth
curves for better visibility. The bottom panels show the percent differences to
the respective highest-order prediction central value. The analogous result for
the $\mu^+\mu^-$ channel can be seen in \app{data_comparison} in
\fig{Z_phi_star_8TeV_ATLAS_mu}.
}
\label{fig:Z_phi_star_8TeV_ATLAS_el}
\end{figure*}
%-------------------------------------------------------------------------------

%-------------------------------------------------------------------------------
\begin{figure*}
\centering
\includegraphics[width=0.49\textwidth]{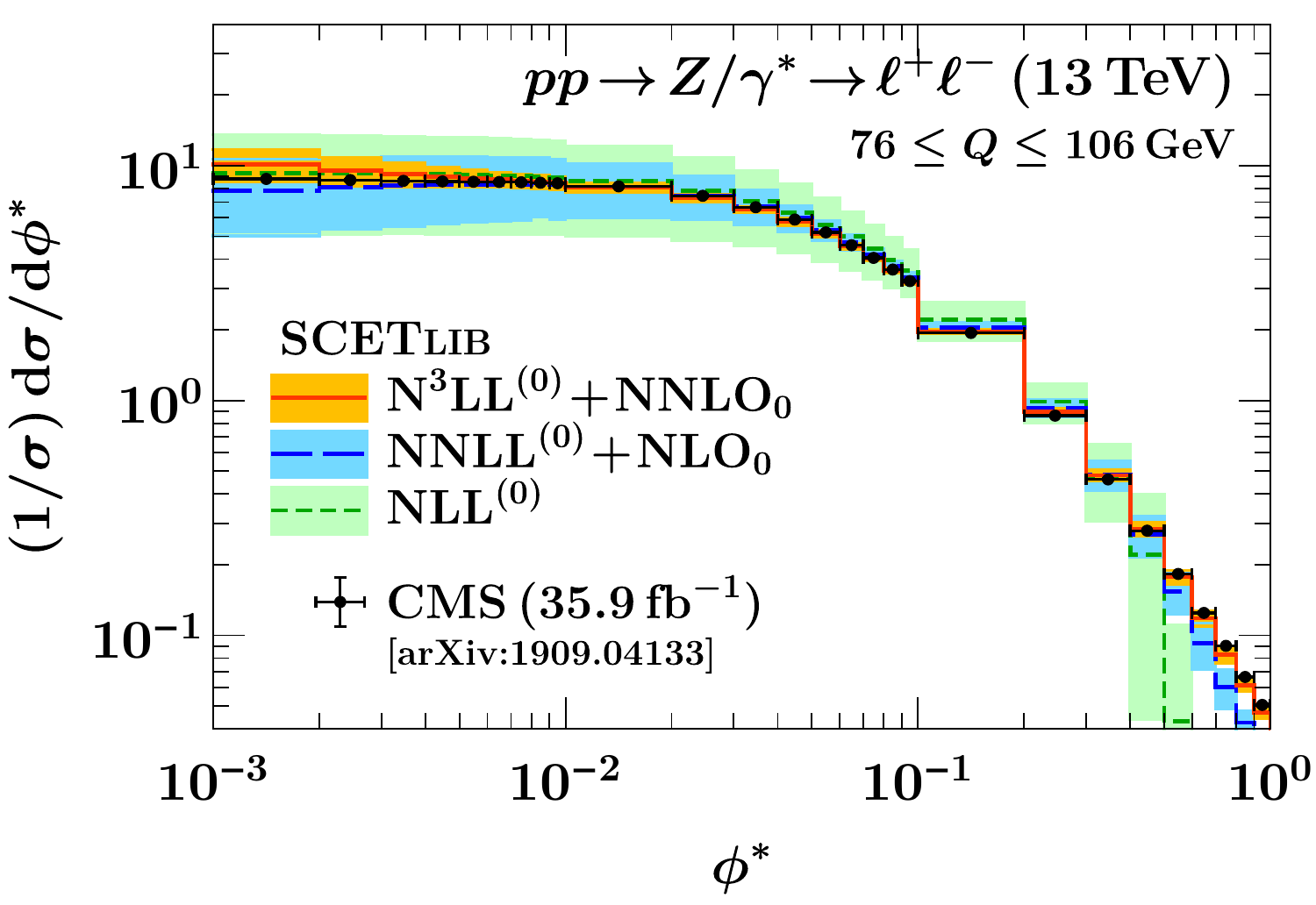}%
\hfill%
\includegraphics[width=0.49\textwidth]{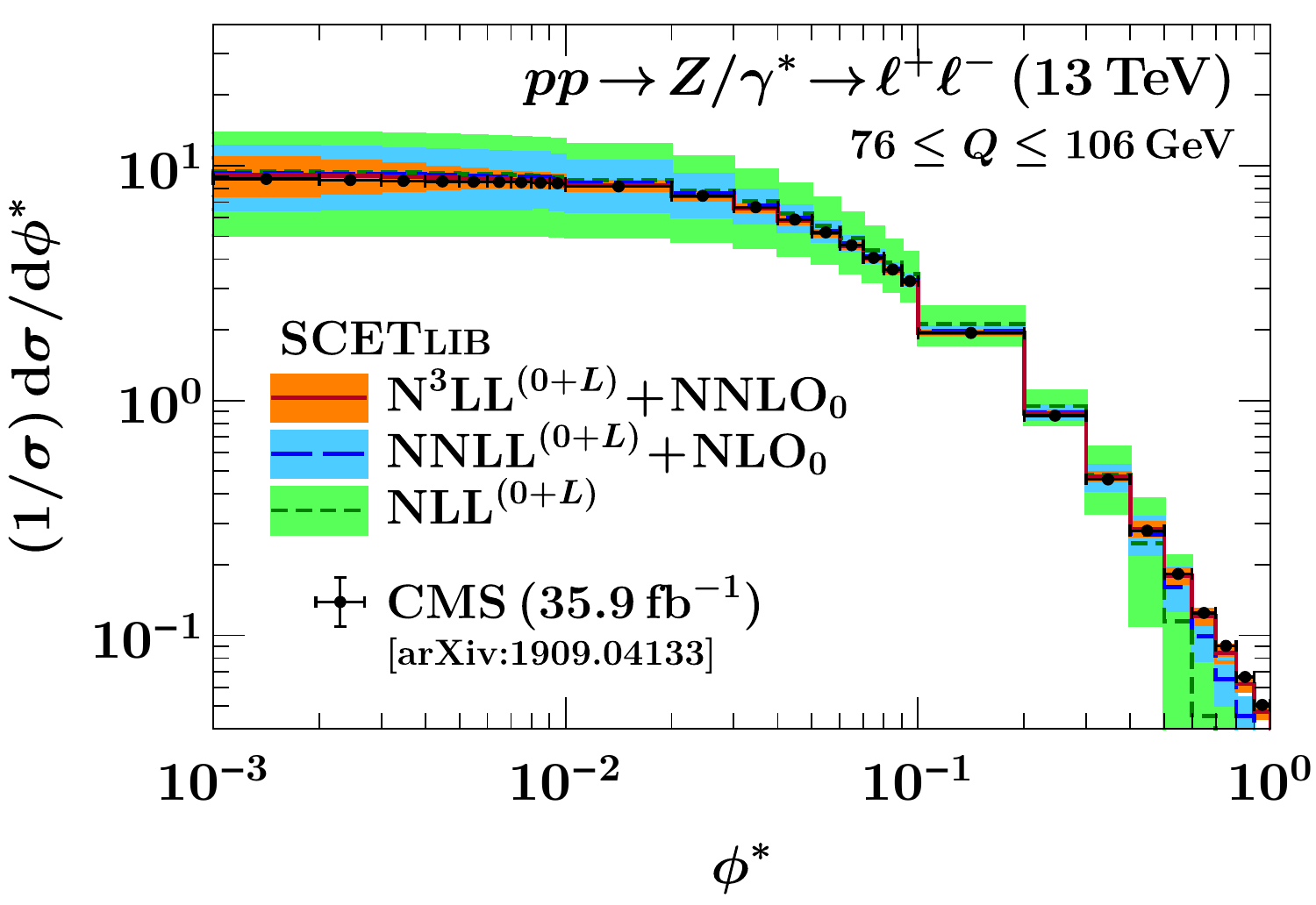}%
\\
\includegraphics[width=0.49\textwidth]{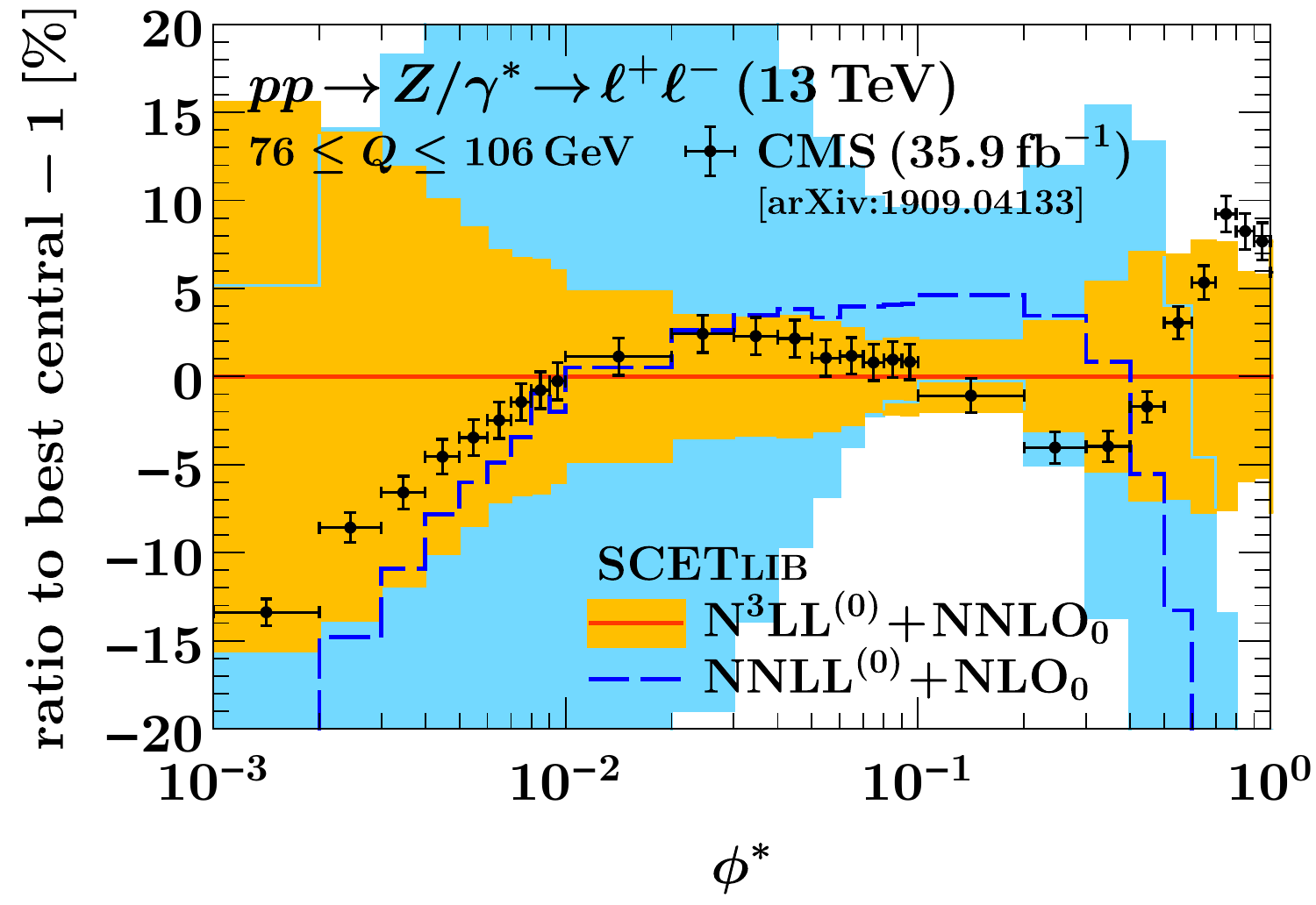}%
\hfill%
\includegraphics[width=0.49\textwidth]{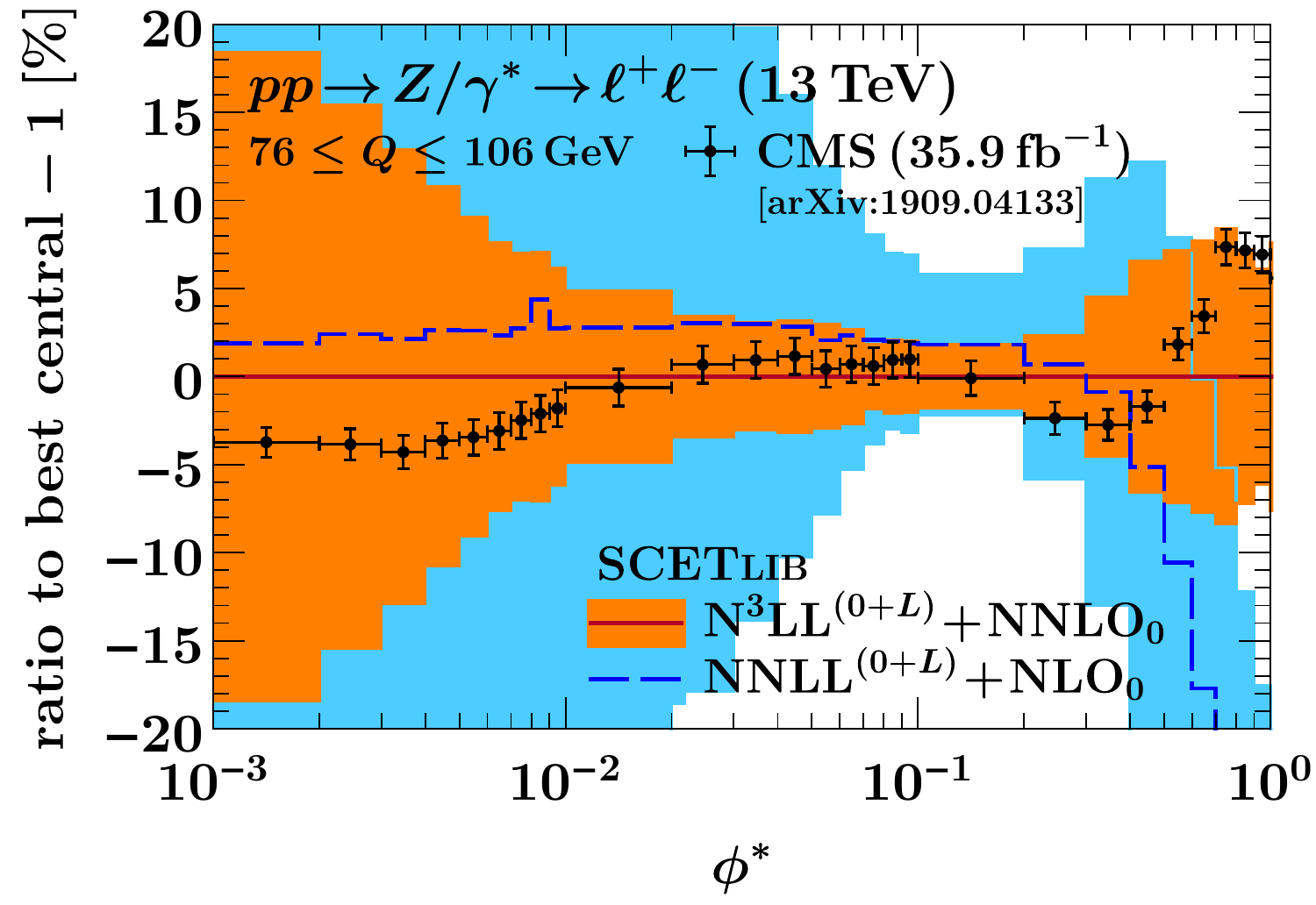}%
\caption{%
Predictions for the normalized Drell-Yan fiducial $\phi^*$ spectrum using the LP
resummation (left) and including the resummation of fiducial power corrections
(right) compared to CMS 13 TeV measurements~\cite{Sirunyan:2019bzr}.
The top panels show the spectrum, and the bottom panels show the percent differences to
the respective highest-order prediction central value.
}
\label{fig:Z_phi_star_13TeV_CMS}
\end{figure*}
%-------------------------------------------------------------------------------

%-------------------------------------------------------------------------------
\begin{figure*}
\centering
\includegraphics[width=0.49\textwidth]{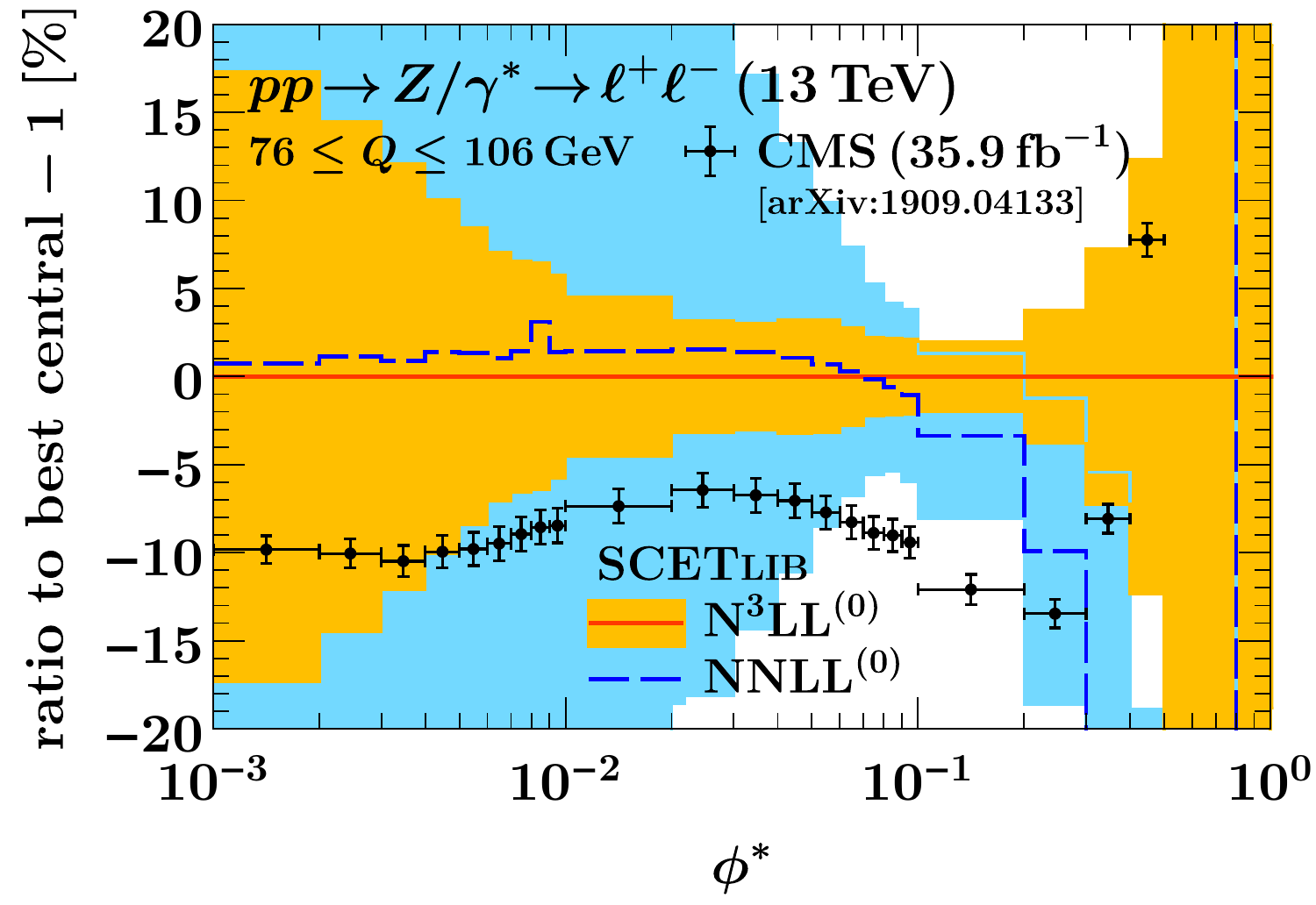}%
\hfill%
\includegraphics[width=0.49\textwidth]{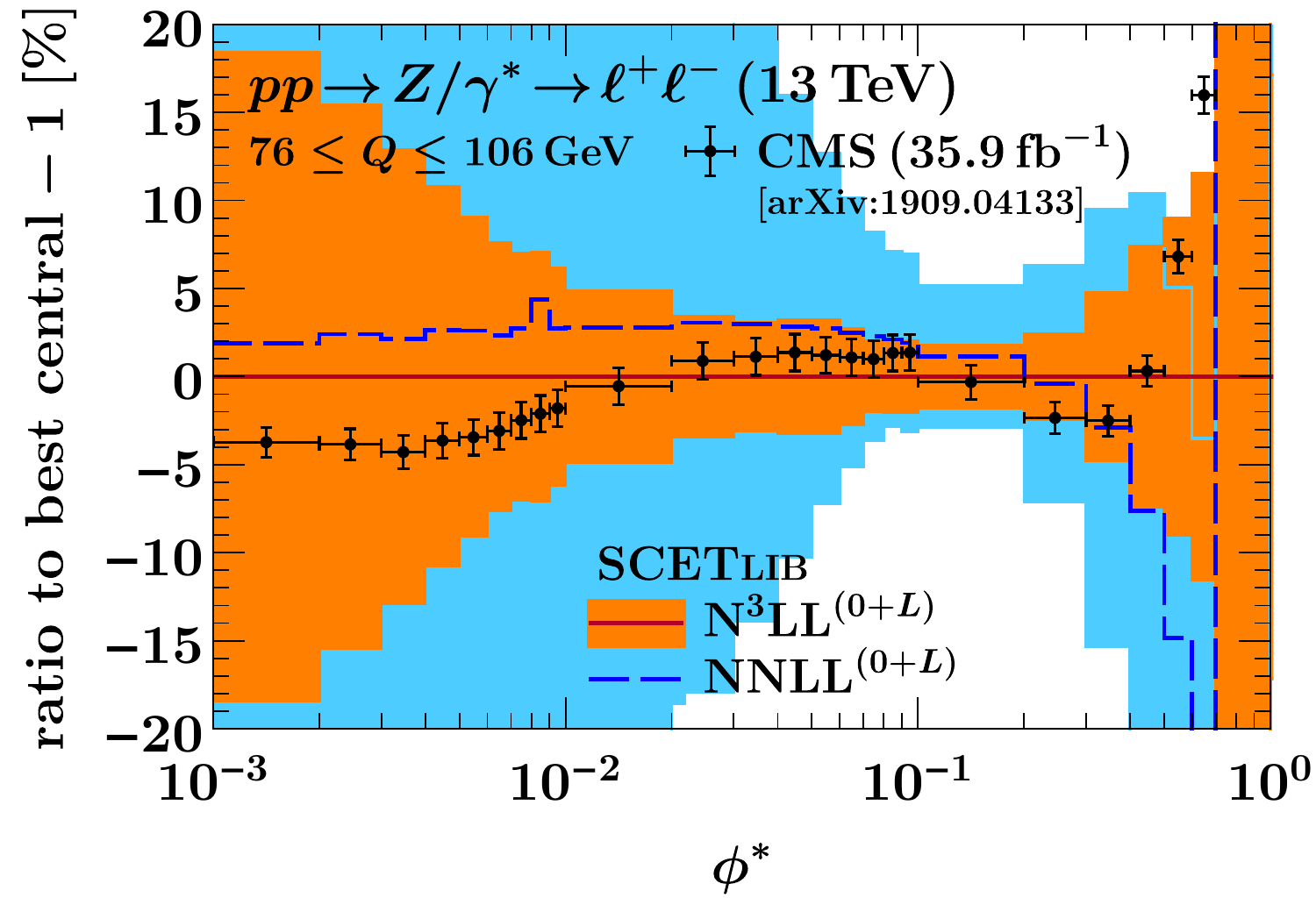}%
\caption{Same as the bottom row of \fig{Z_phi_star_13TeV_CMS},
but without including power corrections from the fixed-order matching.}
\label{fig:Z_phi_star_13TeV_CMS_no_matching}
\end{figure*}
%-------------------------------------------------------------------------------

In \fig{Z_phi_star_8TeV_ATLAS_el}, we compare our results for the
$\phi^*$ spectrum to the ATLAS 8 TeV measurements~\cite{Aad:2015auj} in the
electron channel. The analogous results in the muon channel are provided in
\app{data_comparison} in \fig{Z_phi_star_8TeV_ATLAS_mu}. The top panel shows the
predictions at NLL (green), NNLL$+$NLO$_0$ (blue), and N$^3$LL$+$NNLO$_0$ (orange),
with the bands showing the estimated perturbative uncertainties as discussed in
\sec{estimate_pert_uncerts}. The predictions are obtained with the experimental
binning but are drawn as smooth curves for better visibility. The lower panels
show the same results normalized to the respective highest-order predictions.
The results using strict LP resummation for both the observable itself and the
fiducial cuts are shown on the left (lighter shading), while those including the resummation of
fiducial power corrections for observable and cuts are shown on the right (darker shading). In
both cases we observe good convergence of the resummed predictions. For large
$\phi^* \gtrsim 0.5$, the spectrum enters the fixed-order region and
consequently the NLL (green) results start to deviate substantially, and to
lesser extent also the NNLL$+$NLO$_0$ (blue) results. The first one or two bins
are again sensitive to small-$q_T$ nonperturbative effects, which is reflected
in their increased perturbative uncertainties. As for the $q_T$ spectrum, we
find excellent agreement with the data, which is further improved on the right
by resumming the fiducial power corrections, especially at NNLL$+$NLO$_0$ where
the shape improves significantly.

In \fig{Z_phi_star_13TeV_CMS}, we show the analogous comparison for the CMS 13
TeV $\phi^*$ measurements~\cite{Sirunyan:2019bzr}. The top panels show the
spectrum itself, and the bottom panels the relative difference to the respective
highest-order prediction. The predictions show the same behaviour as at 8 TeV,
and we again find good agreement with the data. Here, the improvements from
resumming the fiducial power corrections are even more striking. While the
strict LP resummation on the left shows a clear trend of overshooting the data
at small $\phi^*$, we find nigh-perfect agreement across the spectrum with
resummed fiducial power corrections. To further highlight this, in
\fig{Z_phi_star_13TeV_CMS_no_matching} we show the analog of the bottom panel of
\fig{Z_phi_star_13TeV_CMS} but comparing to the pure resummed results only,
i.e., without including fixed-order matching corrections. As we already saw for
the $q_T$ spectrum, the LP resummation alone (left) basically fails to describe
the data, showing that in this case the fixed-order matching is necessary in
order to supply the fiducial power corrections at least at fixed order. In
contrast, when resumming the fiducial power corrections (right), we find the
same excellent data agreement as before up to $\phi^* \lesssim 0.5$. This shows
that the $\phi^*$ spectrum has rather large sensitivity to power corrections
throughout its spectrum and profits enormously from including them in the
resummation. At the same time, the remaining fixed-order power corrections
become almost negligible in this range. Beyond $\phi^* \gtrsim 0.5$, we enter the
fixed-order region and as expected, the pure resummed results quickly
deteriorate and matching to the fixed-order results becomes strictly necessary.

%%%%%%%%%%%%%%%%%%%%%%%%%%%%%%%%%%%%%%%%%%%%%%%%%%%%%%%%%%%%%%%%%%%%%%%%%%%%%%%%
\FloatBarrier
\section{Conclusions}
\label{sec:conclusions}
%%%%%%%%%%%%%%%%%%%%%%%%%%%%%%%%%%%%%%%%%%%%%%%%%%%%%%%%%%%%%%%%%%%%%%%%%%%%%%%%

We have studied the impact of fiducial cuts and other generic leptonic measurements on
the factorization of the Drell-Yan process at small transverse momentum $q_T \ll Q$.
They generically induce fiducial power corrections in $q_T/Q$ relative
to the well-studied leading-power terms predicted by $q_T$ (equivalently TMD)
factorization, which are significantly larger than the quadratic power
corrections arising for the inclusive $q_T$ spectrum.

Using a Lorentz-covariant tensor decomposition of the leptonic
and hadronic tensors combined with formal power-counting arguments in SCET,
we have shown that for a large class of observables (those that are azimuthally
symmetric at leading power), the fiducial power
corrections are the only source of linear power corrections.
Furthermore, by retaining the
exact leptonic structure functions, the fiducial power
corrections are unambiguously predicted from factorization and are
correctly resummed to the same order as the leading-power corrections.

We have also shown that the naive power expansion in $q_T/Q \ll 1$
can break down near the edge of Born phase space due to uncontrolled leptonic
power corrections $\sim q_T/p_L$, where $p_L$ is the distance from the edge of
Born phase space.
In such regions, it is strictly required to keep all leptonic power corrections
$\sim q_T/p_L$ to correctly describe the actual \emph{leading-power} limit.
An important example is the $\pTlep$ spectrum near the Jacobian peak $\pTlep = Q/2$
with $p_L = Q - 2\pTlep$. This provides another formal reason to keep
the exact leptonic structure functions, because doing so guarantees
that all required leptonic power corrections are automatically retained.
The kinematic recoil prescriptions used in practical implementations usually
yield an exact description of the leptonic decay and measurements. Our analysis
shows for the first time that this is not only justified, but even necessary to
obtain a description that is formally valid across the entire leptonic phase space.
These conclusions also immediately apply to scalar processes such as Higgs production.

The tensor decomposition can be interpreted as a specific choice of vector-boson
rest frame, which naturally emerged to be the Collins-Soper frame as defined by
boosting from the lab frame, even when keeping nonzero masses of the initial state hadrons.
The CS tensor decomposition yields nine Lorentz-scalar hadronic structure functions,
which are defined for an arbitrary leptonic final state, and
for $Z/\gamma^* \to \ell\ell$ or $W\to\ell\nu$ decays directly map onto
the commonly used angular coefficients for the cross section in the
CS angles. We also discussed that Born leptons
can be theoretically well defined in terms of an IR-safe Born projection of the
full leptonic final state, including QED final-state radiation. We have shown
that the cross section in the CS angles of the so-defined Born leptons admits
a LO-like complete angular decomposition in terms of spherical harmonics, with
the corresponding generalized angular coefficients modified by QED corrections.

We have presented resummed predictions with and without the resummation of fiducial
power corrections at N$^3$LL. The comparison of our predictions to precision
Drell-Yan $q_T$ and $\phi^*$ measurements from ATLAS
and CMS confirms the importance of fiducial power corrections and their
resummation in a striking way: While the strict LP resummation is able to
describe the data within (theory) uncertainties, it fails to do so without
including the sizeable fixed-order corrections. On the other hand, including the
fiducial power corrections in the resummation systematically improves the
agreement with the data, particularly at very small $q_T$ and $\phi^*$.
Furthermore, the fixed-order matching corrections now become very small and
do not play much of a role below $q_T \lesssim 40 \GeV$ and $\phi^*\lesssim 0.5$.
Since the importance of unresummed, fixed-order power corrections is
significantly reduced by resumming the fiducial power corrections, this has
important implications for all phenomenological applications at small $q_T$,
for example the extraction of nonperturbative inputs to TMD distributions from data.

Furthermore, it is also much cheaper computationally to predict or obtain the
fiducial power corrections via factorization instead of including (or
extracting) them numerically from the full fixed-order calculations, since the
latter become expensive quickly toward small $q_T$ or may not even be available.
This is also reflected in a significantly improved performance for the $q_T$
subtraction method when the fiducial power corrections are included in the
subtraction term predicted from factorization. In particular, this extends the
applicability of the subtraction method to phase-space regions where it would
otherwise break down due to uncontrolled leptonic power corrections. We have
explicitly demonstrated this numerically for the $\pTlep$ spectrum near the
Jacobian edge $\pTlep = m_W/2$. As a second example, we have studied the $q_T$
subtraction for Drell-Yan production with symmetric cuts on the final-state
leptons, where we observed that enhanced power corrections from choosing
symmetric cuts are fully accounted for by our formalism.

We have also considered the resummed $\pTlep$ spectrum, which plays a crucial
role in the precision $m_W$ measurement at the LHC,
and have obtained the first N$^3$LL predictions for it.
We have demonstrated that a reliable prediction of the physical $\pTlep$ spectrum
near the Jacobian peak is possible and that it relies in an essential way on the
interplay between small-$q_T$ resummation effects and the exact treatment of
leptonic power corrections that describe the recoil of the leptonic system.
The effective enhancement from leptonic power corrections also induces an
enhanced sensitivity to additional hadronic structure functions.
We stress that these effects are suppressed in the $q_T$ spectrum itself, where
they only enter indirectly through fiducial cuts. In contrast, they are directly
exposed in the $\pTlep$ spectrum near the Jacobian peak. We therefore encourage precise,
unfolded measurements of the $\pTlep$ spectrum for both $W$ and $Z$ in this region.
Other interesting measurements would be the double-differential spectrum in the lepton
momenta $p_{T1}, p_{T2}$,
or equivalently $p_{T1} \pm p_{T2}$, which can be expected to provide further
sensitivity to these effects.

%%%%%%%%%%%%%%%%%%%%%%%%%%%%%%%%%%%%%%%%%%%%%%%%%%%%%%%%%%%%%%%%%%%%%%%%%%%%%%%%
\acknowledgments
We thank Daniel Froidevaux, Ludovica Aperio Bella, Stefano Camarda, Maarten Boone\-kamp,
Aaron Armbruster, Kerstin Tackmann, Aram Apyan, and Josh Bendavid for many
stimulating discussions.
We also thank Markus Diehl and Gherardo Vita for helpful discussions.
J.M.\ and F.T.\ thank the MIT Center for Theoretical Physics for hospitality
and M.E.\ thanks the DESY theory group for hospitality and Gherardo Vita for providing
steaks while part of this work was completed.
This work was supported in part by the Office of Nuclear Physics of the U.S.\
Department of Energy under Contract No.\ DE-SC0011090,
the Deutsche Forschungsgemeinschaft (DFG) under Germany's Excellence
Strategy -- EXC~2121 ``Quantum Universe'' -- 390833306,
the Alexander von Humboldt Foundation through a Feodor Lynen Research Fellowship,
the PIER Hamburg Seed Project PHM-2019-01, 
and within the framework of the TMD Topical Collaboration.
%%%%%%%%%%%%%%%%%%%%%%%%%%%%%%%%%%%%%%%%%%%%%%%%%%%%%%%%%%%%%%%%%%%%%%%%%%%%%%%%

%%%%%%%%%%%%%%%%%%%%%%%%%%%%%%%%%%%%%%%%%%%%%%%%%%%%%%%%%%%%%%%%%%%%%%%%%%%%%%%%
\appendix
%%%%%%%%%%%%%%%%%%%%%%%%%%%%%%%%%%%%%%%%%%%%%%%%%%%%%%%%%%%%%%%%%%%%%%%%%%%%%%%%

%%%%%%%%%%%%%%%%%%%%%%%%%%%%%%%%%%%%%%%%%%%%%%%%%%%%%%%%%%%%%%%%%%%%%%%%%%%%%%%%
\section{Hard functions and leptonic tensors for Drell-Yan}
\label{app:DY_ingredients}
%%%%%%%%%%%%%%%%%%%%%%%%%%%%%%%%%%%%%%%%%%%%%%%%%%%%%%%%%%%%%%%%%%%%%%%%%%%%%%%%

%===============================================================================
\subsection{Hard functions}
\label{app:hard_functions}
%===============================================================================

At leading power in $\la$, the hard contribution to the hadronic tensor in the
limit $\la \ll 1$ is given by~\cite{Stewart:2009yx},%
\footnote{The additional factor of $2$ compared to \refcite{Stewart:2009yx}
is because the hadronic tensor there is defined for $\df\sigma/\df Q^2 \df Y \df^2 \qt$
whereas here it is defined for
$\df\sigma/\df^4 q = 2\,\df\sigma/\df Q^2 \df Y \df^2 \qt$.}
%%%
\begin{align} \label{eq:def_hard_function_tensor}
H_{VV'\, q\bq'}^{\mu\nu}(n_a, n_b; \w_a, \w_b)
&= 2\frac{1}{4N_c} \operatorname{tr} \Bigl[ \frac{\slashed{n}_a}{2} \bC^{(0)\mu}_{V \,q\bq'}(n_b, n_a; \w_b, \w_a) \frac{\slashed{n}_b}{2} C^{(0)\nu}_{V' \, q\bq'}(n_b, n_a; \w_b, \w_a) \Bigr]
\nn \\ & \quad
+ \Bigl(\text{terms}\, \propto \operatorname{tr}\Bigl[
   \slashed{n}_a \slashed{n}_{\perp a} \bC^{(0)\mu}_{V \,q\bq'}\,
   \slashed{n}_{\perp b} \slashed{n}_{b} \, C^{(0)\nu}_{V' \, q\bq'} \Bigr] \Bigr)
\,,\end{align}
%%%
where $C^{\zero\mu\,\alpha\beta}_{V, q\bq'}$
are the hard matching coefficients in \eq{wilson_coeff_hard_matching_lp},
$\bC^{\mu \beta\alpha} = [\gamma^0 C^{\dagger \mu} \gamma^0]^{\beta \alpha}$,
and the trace is over the Dirac indices.
The additional terms in parenthesis in the second line do not contribute to
$H_{-1,4}$ but only to $H_{2,5}$ relevant for the Boer-Mulders effect, see
\eq{tmd_factorization_m1_2_4_5}, where the $n_{\perp a,b}$ are transverse unit
vectors associated with the Boer-Mulders functions $h_{1 a,b}^\perp$.

We remind the reader that the $VV'$ indices were largely left implicit in the
main text. Using \eq{def_W_i}, the hard functions $H_{i\, VV'\,q\bq'}$ for
$i = -1,4$ are given by the projections onto
$x_\mu x_\nu + y_\mu y_\nu = -g_{\perp \mu\nu} + \ord{\la}$
and $2\img \bigl(x_\mu y_\nu - x_\nu y_\mu \bigr) = 2\img \eps_{\perp \mu\nu} + \ord{\la}$,
%%%
\begin{align} \label{eq:def_hard_function_m1_4}
H_{-1 \, VV' \, q \bq'}(q^2)
= -g_{\perp \mu\nu} H_{VV'\, q\bq'}^{\mu\nu}(n_a, n_b; \w_a, \w_b)
\,, \nn \\
H_{4 \, VV' \, q \bq'}(q^2)
= 2\img \eps_{\perp\mu\nu} H_{VV'\, q\bq'}^{\mu\nu}(n_a, n_b; \w_a, \w_b)
\,.\end{align}
%%%
Here we used that the projected hard function can only depend on the Lorentz scalar $q^2 = \w_a \w_b$.

The case of an incoming antiquark in the $a$ direction follows from $a \leftrightarrow b$,
%%%
\begin{align}
H_{VV'\, \bq' q}^{\mu\nu}(n_a, n_b; \w_a, \w_b)
= H_{VV'\, q\bq'}^{\mu\nu}(n_b, n_a; \w_b, \w_a)
\,.\end{align}
%%%
This implies
%%%
\begin{align}
H_{-1\,VV' \, \bq' q}(q^2) = + H_{-1 \, VV' \, q\bq'}(q^2)
\,, \qquad
H_{4 \, VV' \, \bq' q}(q^2) = - H_{4 \, VV' \, q\bq'}(q^2)
\,,\end{align}
%%%
as expected from parity.
In the inclusive case discussed in \sec{factorization_inclusive},
we used the shorthand
%%%
\begin{align}
H_{VV'\,ab}(q^2) \equiv H_{-1\,VV'\,ab}(q^2)
\,.\end{align}
%%%
Inserting the expression in \eq{wilson_coeff_hard_matching_lp} into \eq{def_hard_function_tensor}, and suppressing the $q^2$ argument for brevity,
we find
%%%
\begin{align}
H_{-1\,ZZ \, q\bq'}
 &= \frac{8\pi \aem}{N_c} \delta_{qq'} \Bigl\{ (v_q^2 + a_q^2) \abs{C_q}^2 + 2 \Re \sum_f \bigl( v_q v_f C_q^* C_{vf} + a_q a_f C_q^* C_{af} \bigr) + \ord{\as^4} \Bigr\}
\,, \nn \\
H_{-1\,\gamma \gamma \, q\bq'}
&= \frac{8\pi \aem}{N_c} \delta_{qq'} \Bigl\{ Q_q^2 \abs{C_q}^2 + 2\Re \sum_f Q_q Q_f C_q^* C_{vf} + \ord{\as^6} \Bigr\}
\,, \nn \\
H_{-1\,Z\gamma \, q\bq'}
&= \frac{8\pi \aem}{N_c} \delta_{qq'} \Bigl\{ - v_q Q_q \abs{C_q}^2 - \sum_f\bigl( v_f Q_q C_{vf}^* C_q + v_q Q_f C_q^* C_{vf} \bigr) + \ord{\as^6} \Bigr\}
\nn \\
&= H^*_{-1\,\gamma Z \, q\bq'}
\,,\end{align}
%%%
where $\Re$ denotes the real part,
the vector ($v_f$) and axial ($a_f$) couplings of a quark of flavor $f$ to the $Z$ boson were given in \eq{Z_vector_axial_coupling}. The terms denoted as $\ord{\as^{4,6}}$
are proportional to the square of the singlet matching coefficients,
which only start at $C_{af} = \ord{\as^2}$ and $C_{vf} = \ord{\as^3}$,
whereas the nonsinglet coefficient $C_q = 1 + \ord{\as}$.
In the parity-odd case, we have
%%%
\begin{align}
H_{4\,ZZ \, q\bq'}
&= \frac{16\pi\aem}{N_c} \delta_{qq'} \Bigl\{ 2v_q a_q \abs{C_q}^2 + 2\Re \sum_f \bigl( v_q a_f C_q^* C_{af} + a_q v_f C_q^* C_{vf} \bigr) + \ord{\as^5} \Bigr\}
\,, \nn \\
H_{4\,\gamma \gamma \, q\bq'}
&= 0
\,, \nn \\
H_{4\,Z\gamma \, q\bq'}
&= \frac{16\pi \aem}{N_c} \delta_{qq'} \Bigl\{ -a_q Q_q \abs{C_q}^2 - \sum_f \bigl(a_f Q_q C_{af}^* C_q + a_q Q_f C_q^* C_{vf} \bigr) + \ord{\as^5} \Bigr\}
\nn \\
&= H^*_{4\,\gamma Z \, q\bq'}
\,. \end{align}
%%%
For $W^\pm$ exchange, we have
%%%
\begin{align}
H_{-1 \, W^+ W^+ \, q\bq'} &= \frac{2\pi \aem}{N_c} \frac{\abs{V_{qq'}}^2}{\sin^2 \theta_w} \abs{C_q}^2
\,, \qquad
H_{4 \, W^+ W^+ \, q\bq'} = \frac{4\pi \aem}{N_c} \frac{\abs{V_{qq'}}^2}{\sin^2 \theta_w} \abs{C_q}^2
\,, \nn \\
H_{-1 \, W^- W^- \, q\bq'} &= \frac{2\pi \aem}{N_c} \frac{\abs{V_{q'q}}^2}{\sin^2 \theta_w} \abs{C_q}^2
\,, \qquad
H_{4 \, W^- W^- \, q\bq'} = \frac{4\pi \aem}{N_c} \frac{\abs{V_{q'q}}^2}{\sin^2 \theta_w} \abs{C_q}^2
\,,\end{align}
%%%
where $V_{qq'}$ denotes the CKM-matrix element for $q \in \{u, c, t\}$ and
$q' \in \{d, s, b\}$ (and we take it to vanish in all other cases).
The overall relative factor of $2$ between $H_{-1}$ and $H_4$ is due to the
conventional normalization of $g_4(\theta, \varphi) = \cos \theta$ rather
than $2\cos \theta$ in \eq{def_g_i}.

The renormalized matching coefficients $C_q(q^2, \mu)$ and $C_{vf,af}(q^2, \mu)$
can be extracted from the IR-finite parts of the $q\bar q$ vector and
axial-vector form factors, which
admit the same flavor decomposition as \eq{wilson_coeff_hard_matching_lp}.
Explicit expressions for $C_q(q^2, \mu)$ in our notation can be found in
\refcite{Ebert:2017uel}. The two-loop results which enter in our analysis
follow from the two-loop quark form
factors~\cite{Kramer:1986sg, Matsuura:1987wt, Matsuura:1988sm, Gehrmann:2005pd}.
In principle, there is an $\ord{\as^2}$ contribution to the
axial-vector singlet coefficient if the top quark is taken to be
massive at the hard scale~\cite{Stewart:2009yx}.
These contributions have however been found to be small at the level of the total cross section~\cite{Dicus:1985wx, Hamberg:1990np}, and we neglect them in this paper.

%===============================================================================
\subsection{Leptonic tensors}
\label{app:leptonic_tensors}
%===============================================================================

The leptonic scalar coefficients $L_{\pm\,VV'}(q^2)$ defined in \eq{Lmunu_1to2}
encode the squared electroweak decay matrix element
including the vector-boson propagator.
In the inclusive case discussed in \sec{factorization_inclusive},
we used the shorthand
%%%
\begin{align}
L(q^2) \equiv L_{+\,VV'}(q^2)
\,.\end{align}
%%%
For $Z/\gamma^* \to \ell^+\ell^-$, the parity-even leptonic coefficients read
%%%
\begin{alignat}{3}
L_{+\,ZZ}(q^2)
&= \frac{2}{3} \frac{\aem}{q^2}\, (v_\ell^2 + a_\ell^2)\, \Abs{P_Z(q^2)}^2
\,, \qquad &
L_{+\,\gamma \gamma}(q^2)
&= \frac{2}{3} \frac{\aem}{q^2} \, Q_\ell^2
\,, \nn \\
L_{+\,\gamma Z}(q^2)
&= \frac{2}{3} \frac{\aem}{q^2}\, (- v_\ell\, Q_\ell)\, P_Z(q^2)
\,, \qquad &
L_{+\,Z\gamma}(q^2)
&= L_{+\,\gamma Z}^*(q^2)
\,.\end{alignat}
%%%
The parity-odd coefficients arise from the interference of axial and vector current contributions,
%%%
\begin{alignat}{3}
L_{-\,ZZ}(q^2)
&= \frac{2}{3} \frac{\aem}{q^2}\, (2 v_\ell\, a_\ell)\, \Abs{P_Z(q^2)}^2
\,, \qquad &
L_{-\,\gamma \gamma}(q^2)
&= 0
\,, \nn \\
L_{-\,\gamma Z}(q^2)
&= \frac{2}{3} \frac{\aem}{q^2}\, (- a_\ell Q_\ell) \, P_Z(q^2)
\,, \qquad &
L_{-\,Z \gamma}(q^2)
&= L_{-\,\gamma Z}^*(q^2)
\,.\end{alignat}
%%%
Here, $Q_\ell = -1$, and the vector ($v_\ell$) and axial couplings ($a_\ell$) of the lepton $\ell$
to the $Z$ boson have the same form as \eq{Z_vector_axial_coupling} with $T_3^\ell = -1/2$.
The $Z\to \nu\bar\nu$ process is obtained by the replacement $\ell\to\nu$ with
$Q_\nu = 0$ and $T_3^\nu = +1/2$. The reduced propagator $P_V$ is given by
%%%
\begin{align}
P_V(q^2) = \frac{q^2}{q^2 - m_V^2 + \img \Gamma_V m_V}
\,.\end{align}
%%%
For $W \to \ell\nu$, the $L_{\pm}(q^2)$ are equal due to the current's $V-A$ structure,
and are given by
%%%
\begin{align}
L_{+\,W^\pm W^\pm}(q^2) = L_{-\,W^\pm W^\pm}(q^2)
= \frac{1}{6} \frac{\aem}{q^2}\, \frac{1}{\sin^2 \theta_w}\, \Abs{P_W(q^2)}^2
\,.\end{align}
%%%

%%%%%%%%%%%%%%%%%%%%%%%%%%%%%%%%%%%%%%%%%%%%%%%%%%%%%%%%%%%%%%%%%%%%%%%%%%%%%%%%
\section{Additional comparisons to data}
\label{app:data_comparison}
%%%%%%%%%%%%%%%%%%%%%%%%%%%%%%%%%%%%%%%%%%%%%%%%%%%%%%%%%%%%%%%%%%%%%%%%%%%%%%%%

%===============================================================================
\subsection{Unnormalized results}
%===============================================================================

%-------------------------------------------------------------------------------
\begin{figure*}
\includegraphics[width=0.49\textwidth]{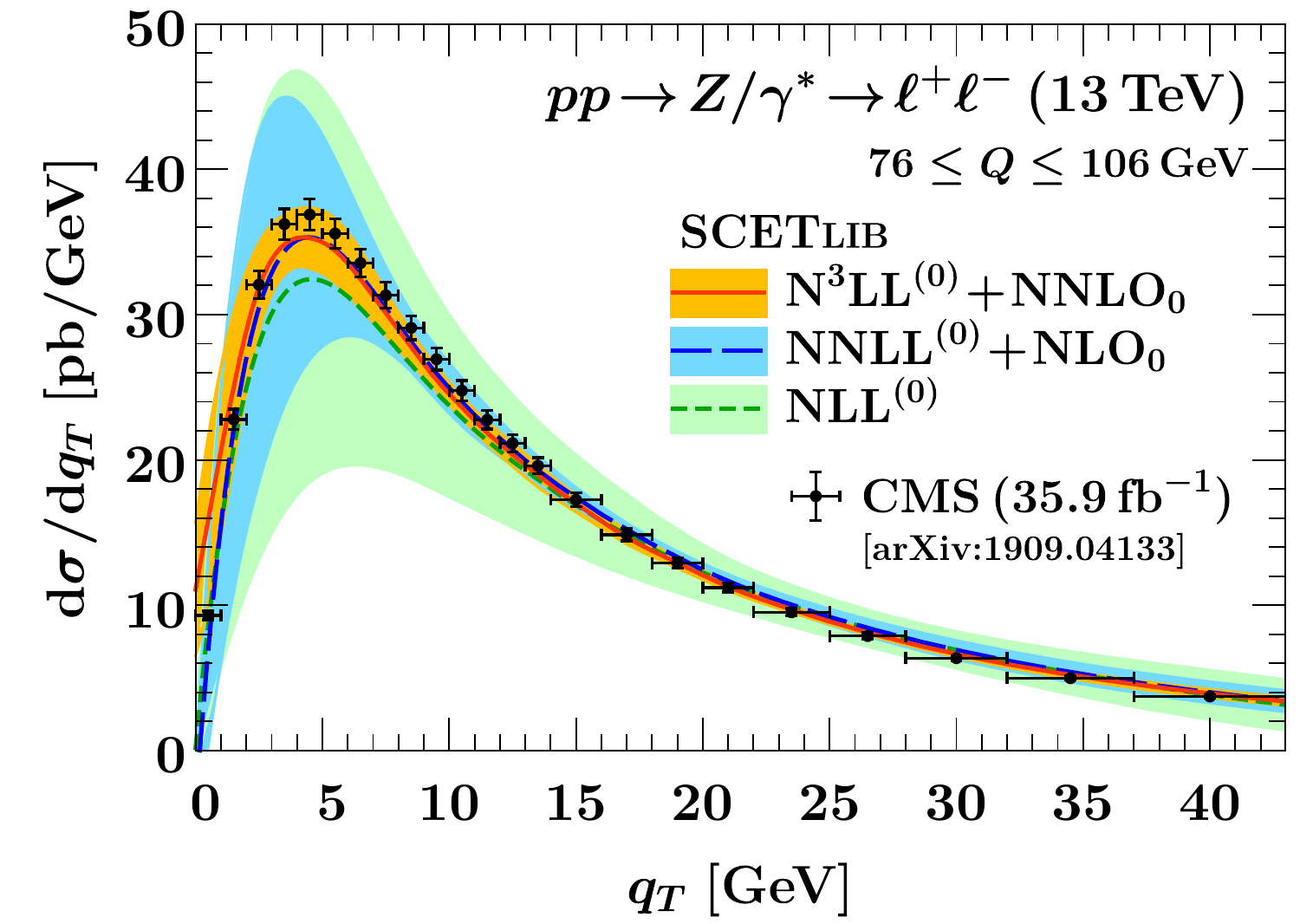}%
\hfill%
\includegraphics[width=0.49\textwidth]{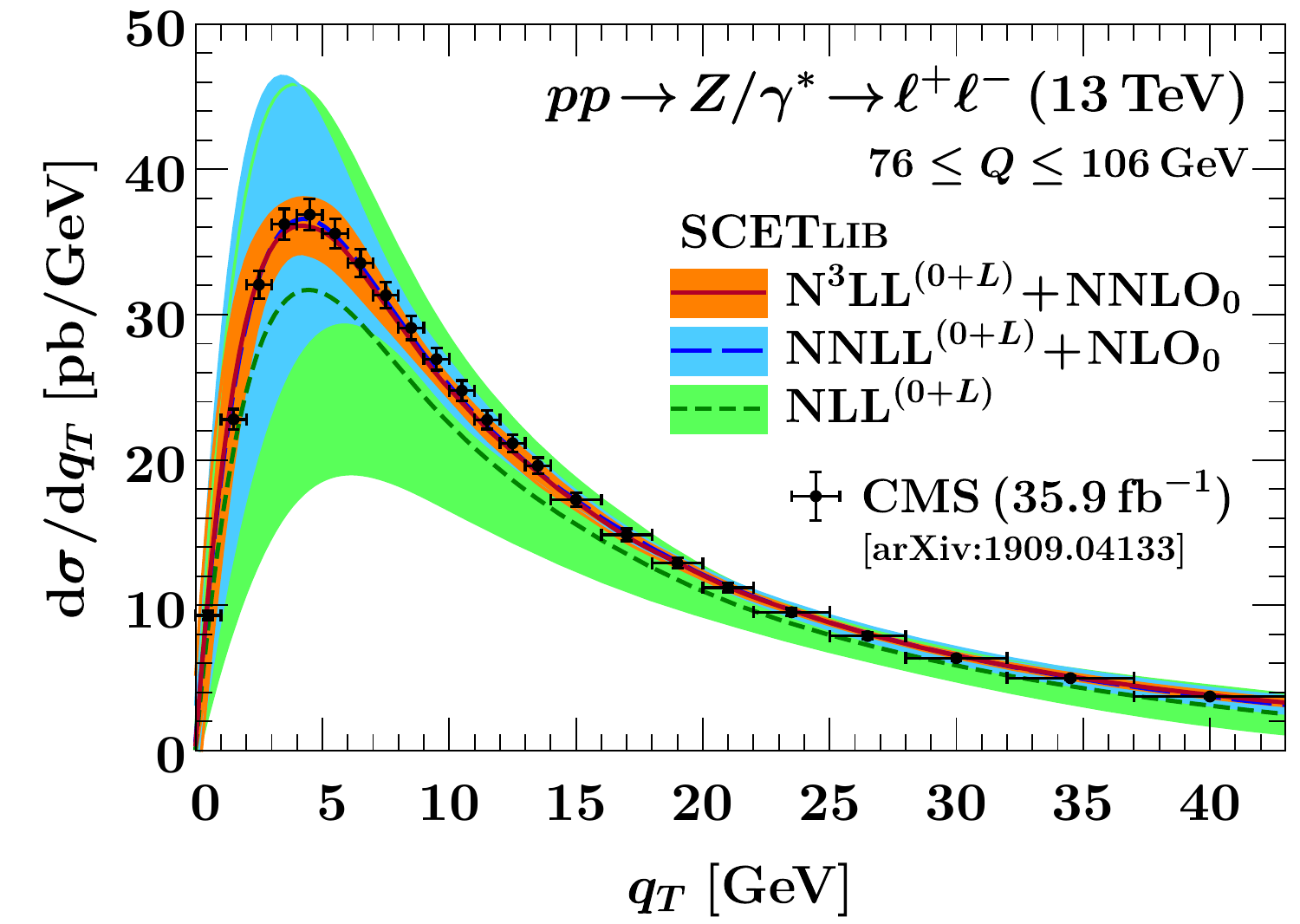}%
\\
\includegraphics[width=0.49\textwidth]{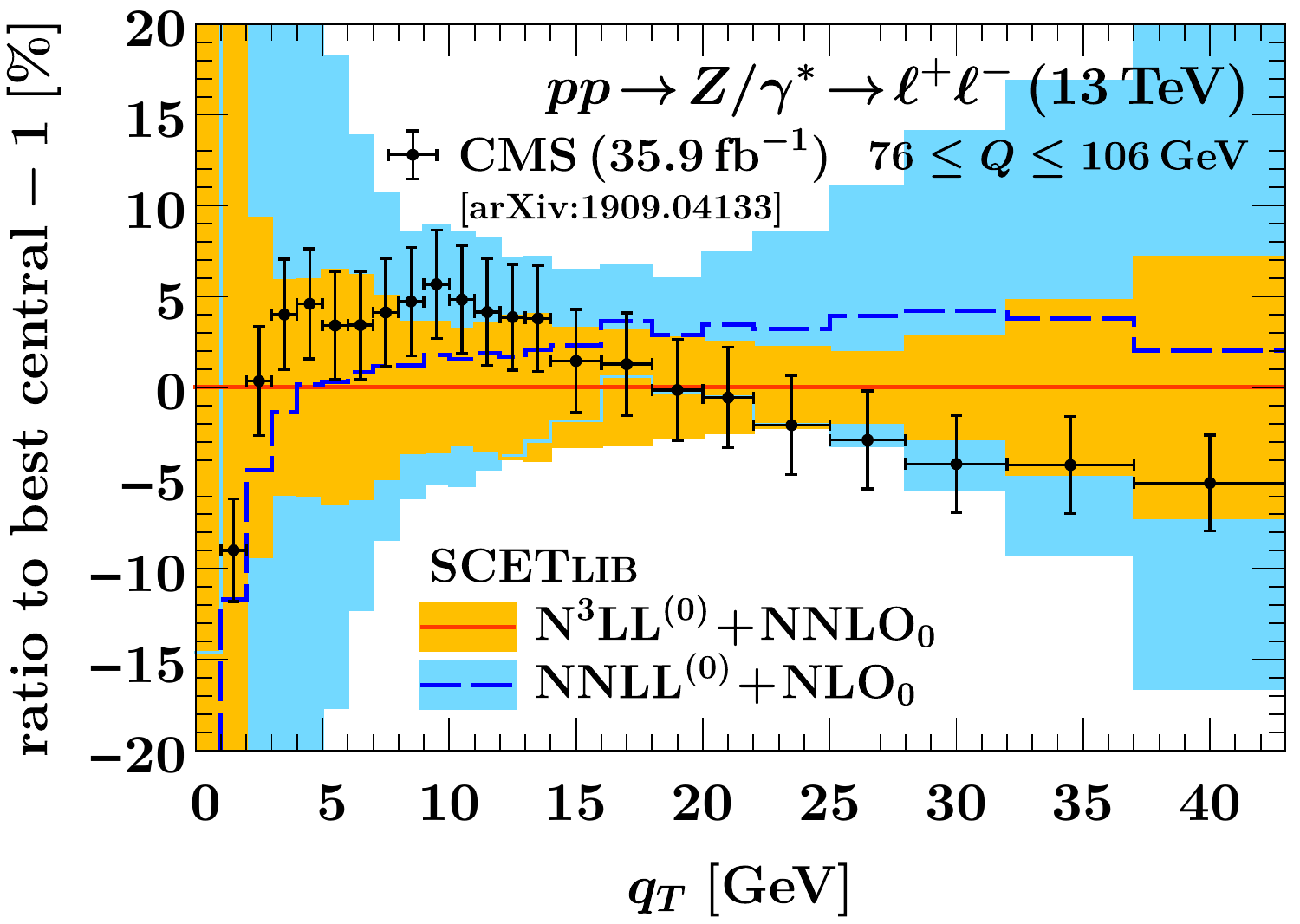}%
\hfill%
\includegraphics[width=0.49\textwidth]{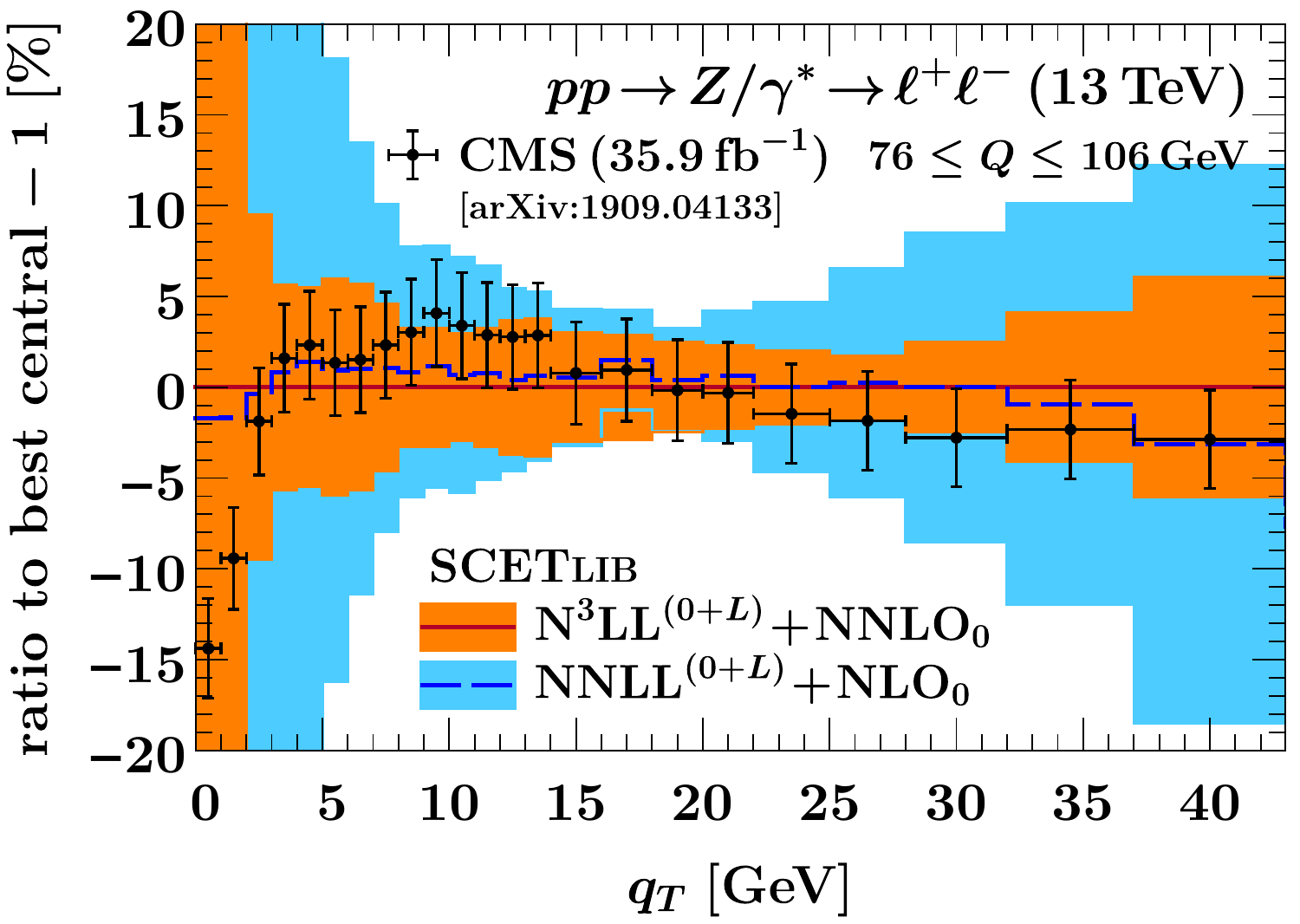}%
\vspace{-2ex}
\caption{Comparison to CMS $13\TeV$ measurements~\cite{Sirunyan:2019bzr}
analogous to \fig{Z_qT_13TeV_CMS} but for the unnormalized, absolute
$q_T$ spectrum.}
\label{fig:Z_qT_13TeV_CMS_absolute}
\vspace{-1ex}
\end{figure*}
%-------------------------------------------------------------------------------

%-------------------------------------------------------------------------------
\begin{figure*}
\includegraphics[width=0.49\textwidth]{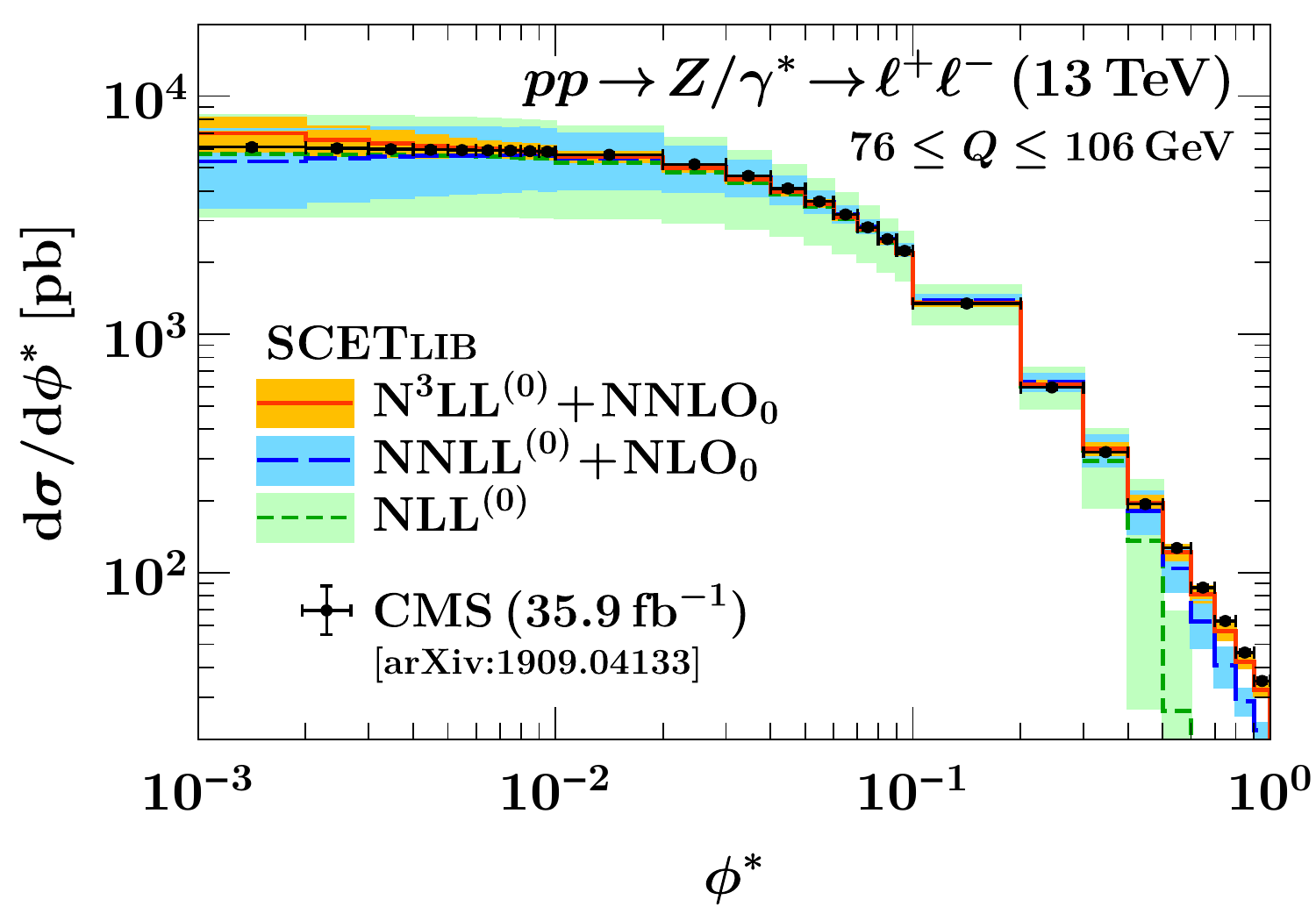}%
\hfill%
\includegraphics[width=0.49\textwidth]{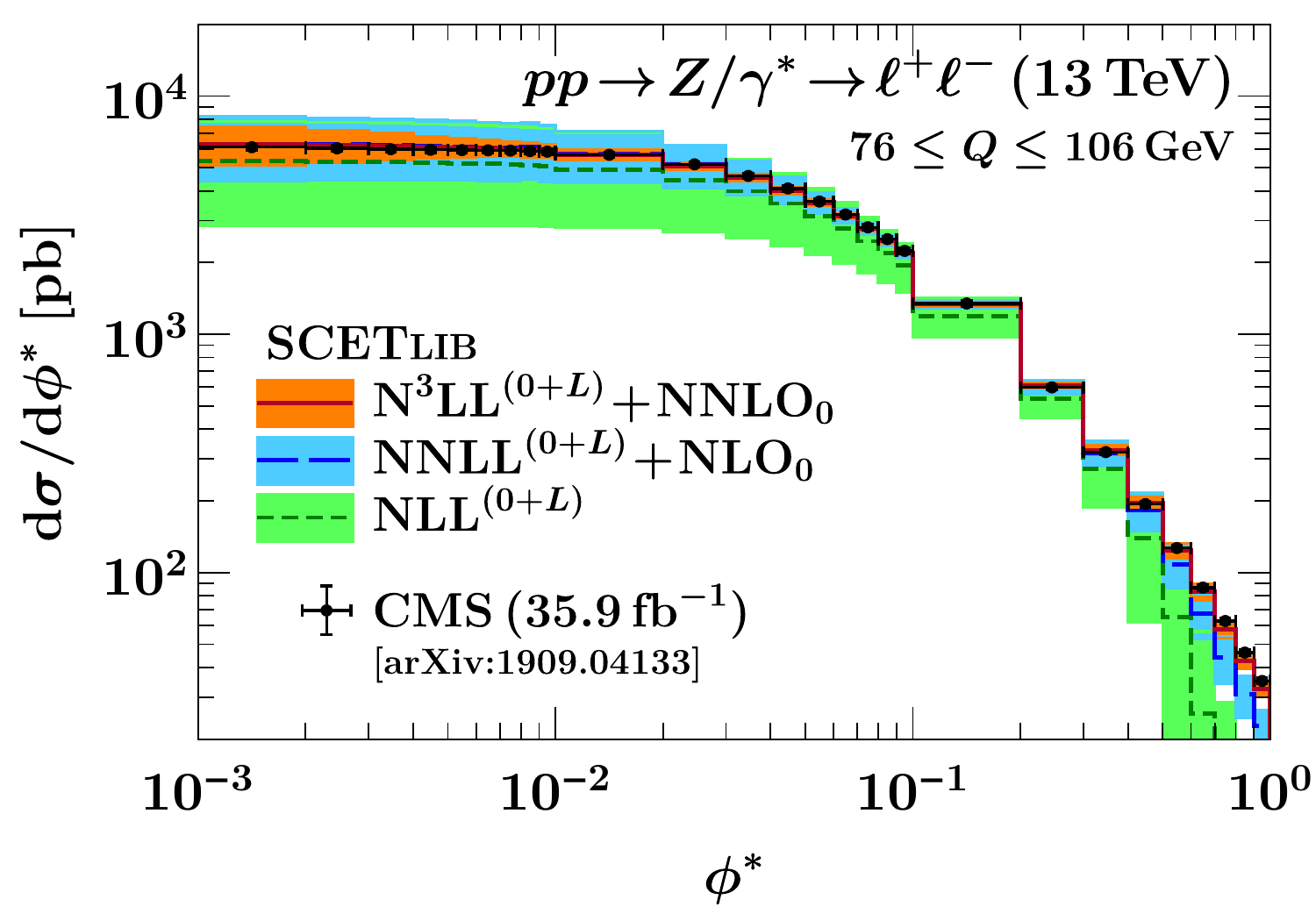}%
\\
\includegraphics[width=0.49\textwidth]{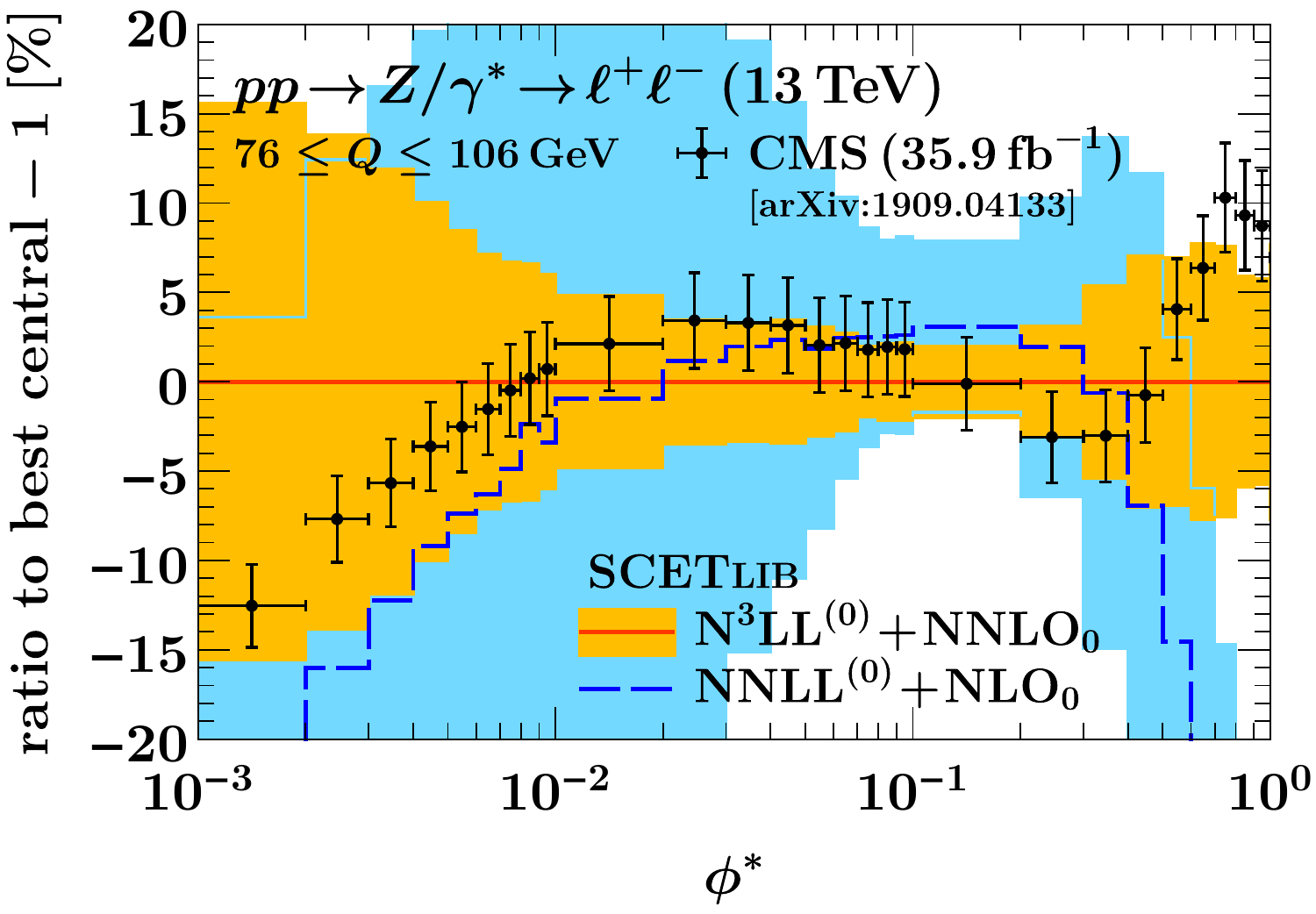}%
\hfill%
\includegraphics[width=0.49\textwidth]{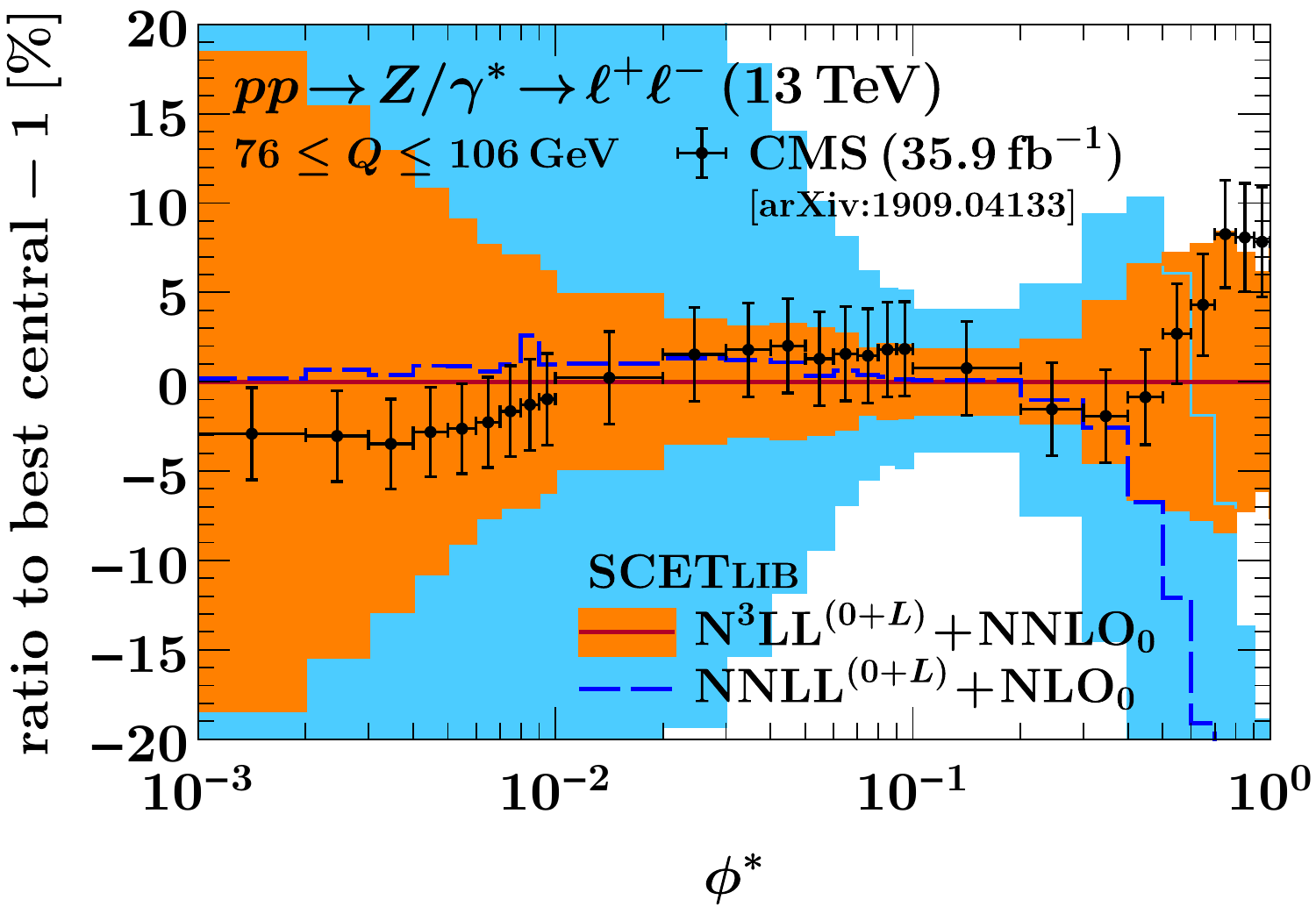}%
\vspace{-2ex}
\caption{Comparison to CMS $13\TeV$ measurements~\cite{Sirunyan:2019bzr}
analogous to \fig{Z_phi_star_13TeV_CMS} but for the unnormalized, absolute
$\phi^*$ spectrum.}
\label{fig:Z_phi_star_13TeV_CMS_absolute}
% \vspace{-2ex}
\end{figure*}
%-------------------------------------------------------------------------------

In \sec{data_comparison}, we compared our predictions to the CMS $13\TeV$
measurements from \refcite{Sirunyan:2019bzr} using the normalized $q_T$ and $\phi^*$ spectra. For completeness,
here we show the corresponding results for the unnormalized, absolute $q_T$
spectrum in \fig{Z_qT_13TeV_CMS_absolute} and $\phi^*$ spectrum in
\fig{Z_phi_star_13TeV_CMS_absolute}. The experimental uncertainties are larger
for the absolute spectra than the normalized ones but include a substantial
correlated component from the overall absolute normalization of the measurement.
The absolute spectra show the same systematic improvement in the predictions
from resumming the fiducial power corrections.

%===============================================================================
\subsection{Results for \texorpdfstring{$\mu^+\mu^-$}{mu+mu-} channel}
\label{sec:atlas_muon_data}
%===============================================================================

%-------------------------------------------------------------------------------
\begin{figure*}
\includegraphics[width=0.49\textwidth]{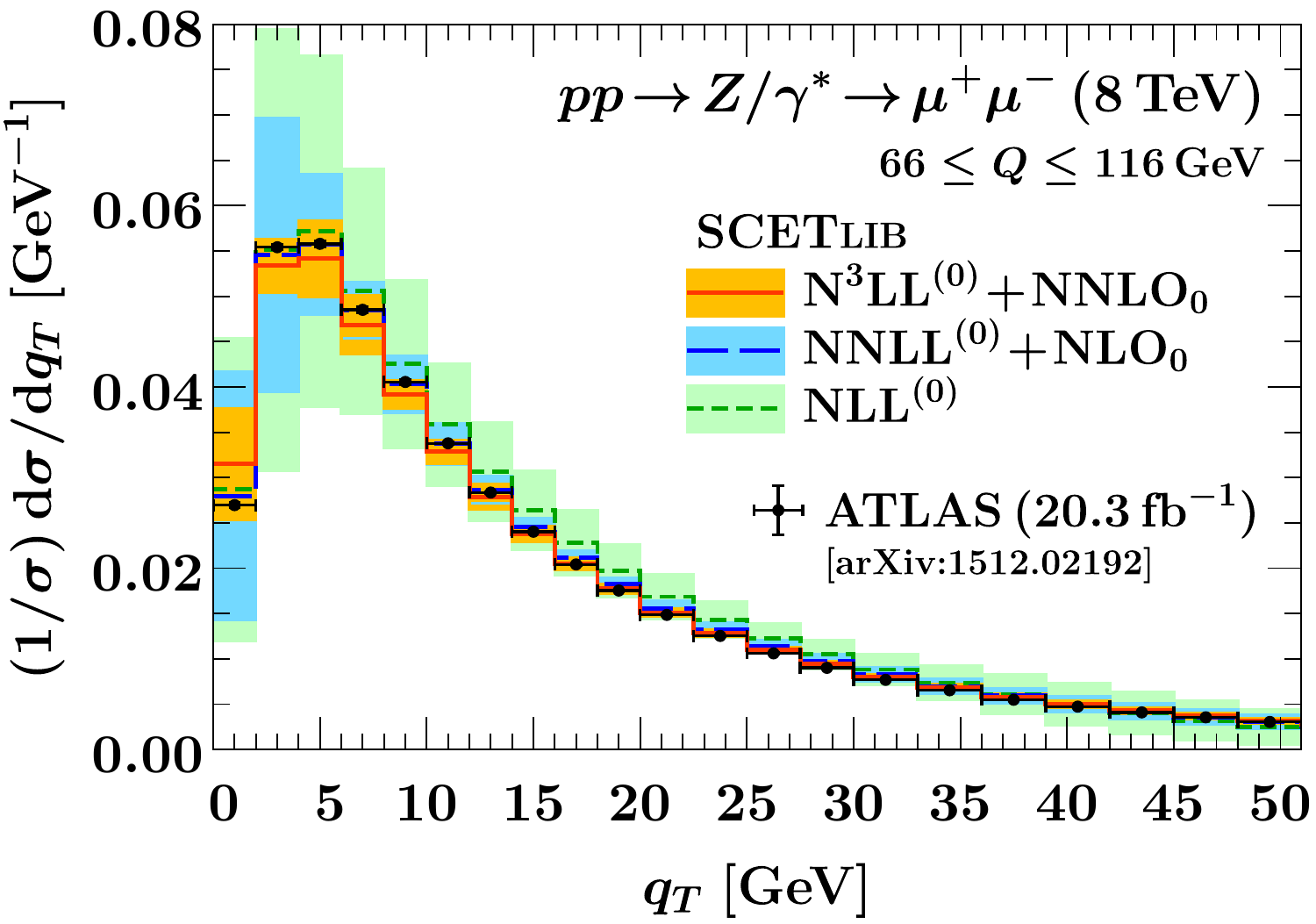}%
\hfill%
\includegraphics[width=0.49\textwidth]{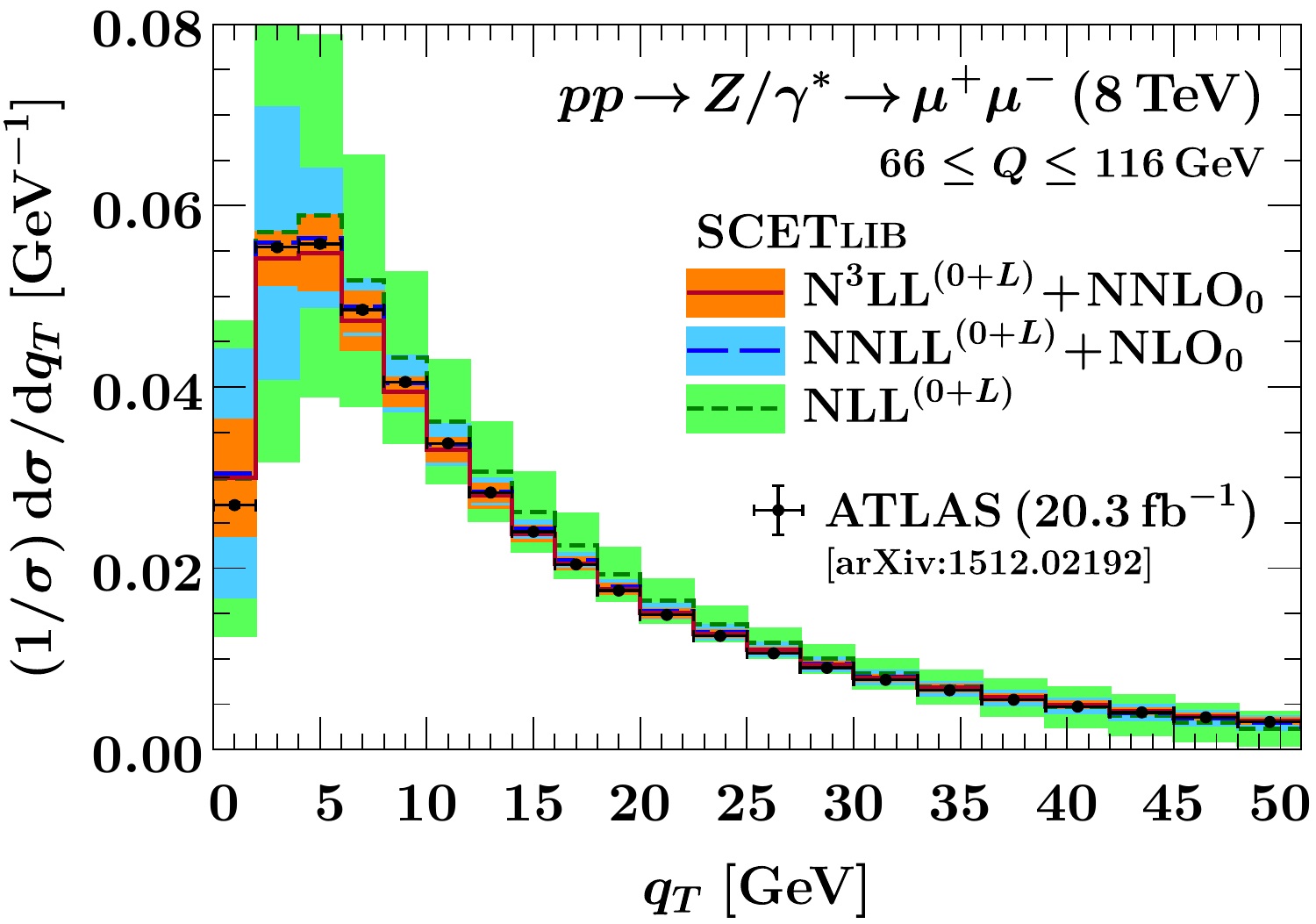}%
\\
\includegraphics[width=0.49\textwidth]{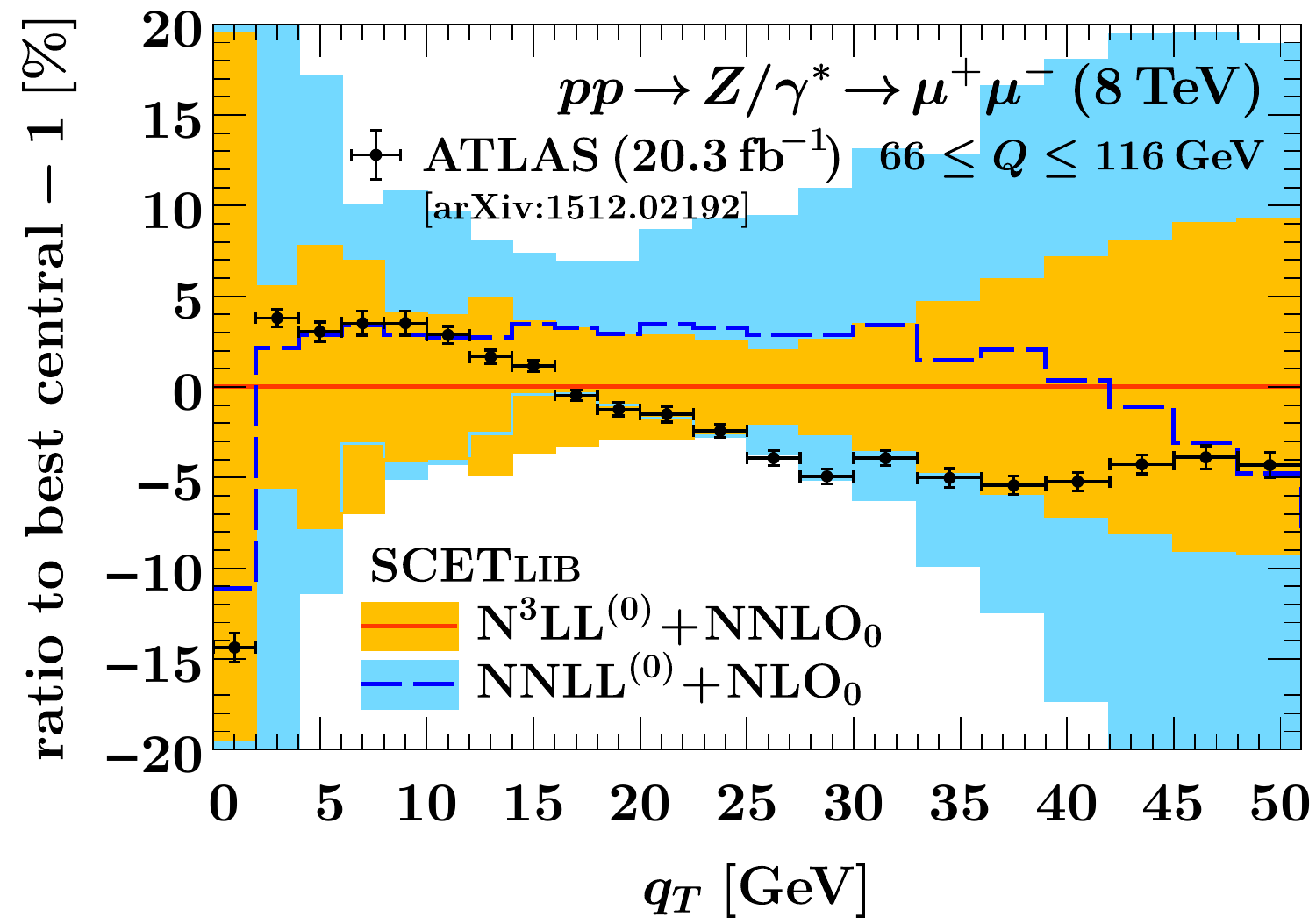}%
\hfill%
\includegraphics[width=0.49\textwidth]{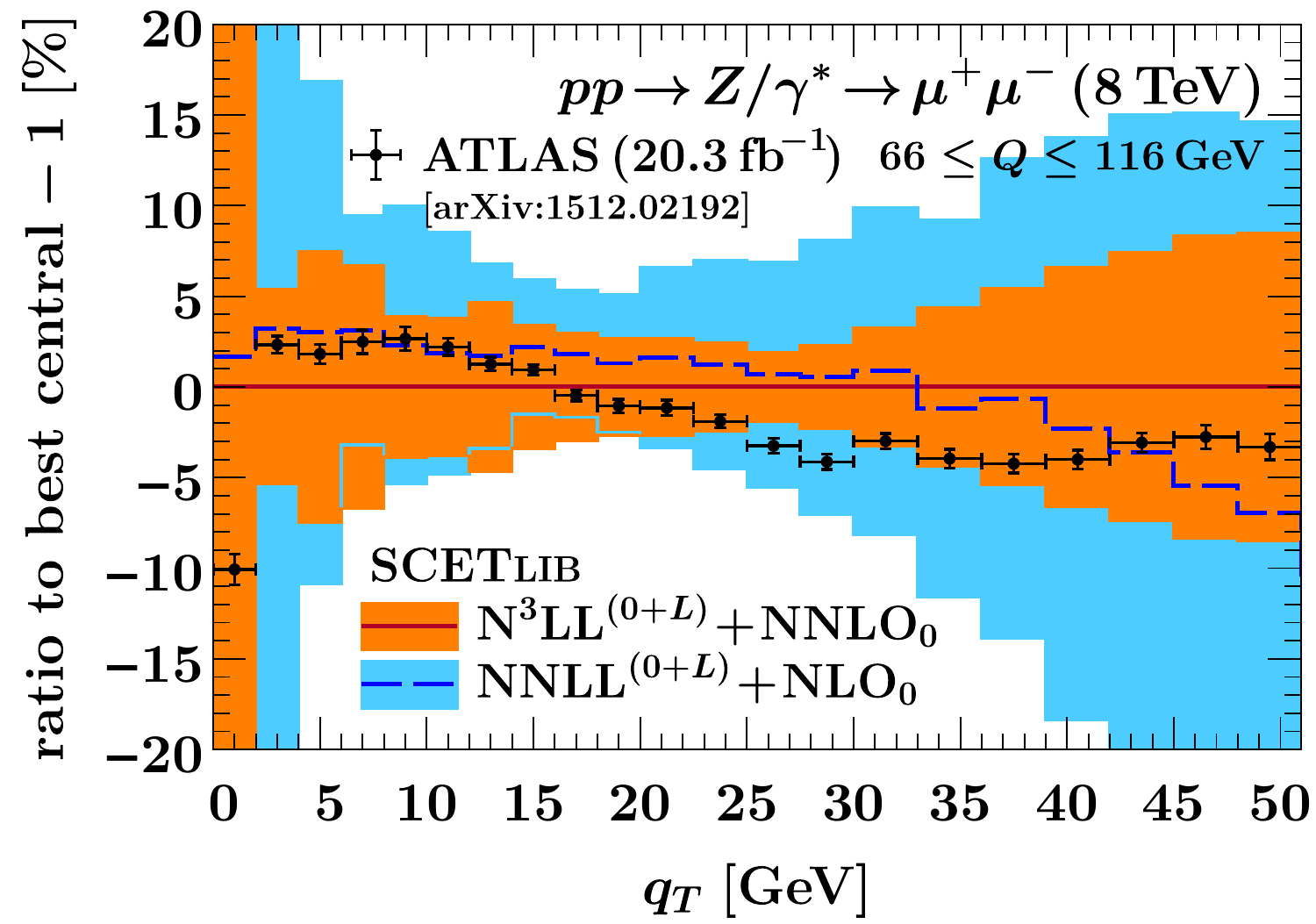}%
\vspace{-2ex}
\caption{Comparison to ATLAS $8\TeV$ $q_T$ measurements~\cite{Aad:2015auj}
in the $\mu^+\mu^-$ channel, analogous to the $e^+e^-$ channel in
\fig{Z_qT_8TeV_ATLAS_el}.}
\label{fig:Z_qT_8TeV_ATLAS_mu}
\vspace{-4ex}
\end{figure*}
%-------------------------------------------------------------------------------

%-------------------------------------------------------------------------------
\begin{figure*}
\includegraphics[width=0.49\textwidth]{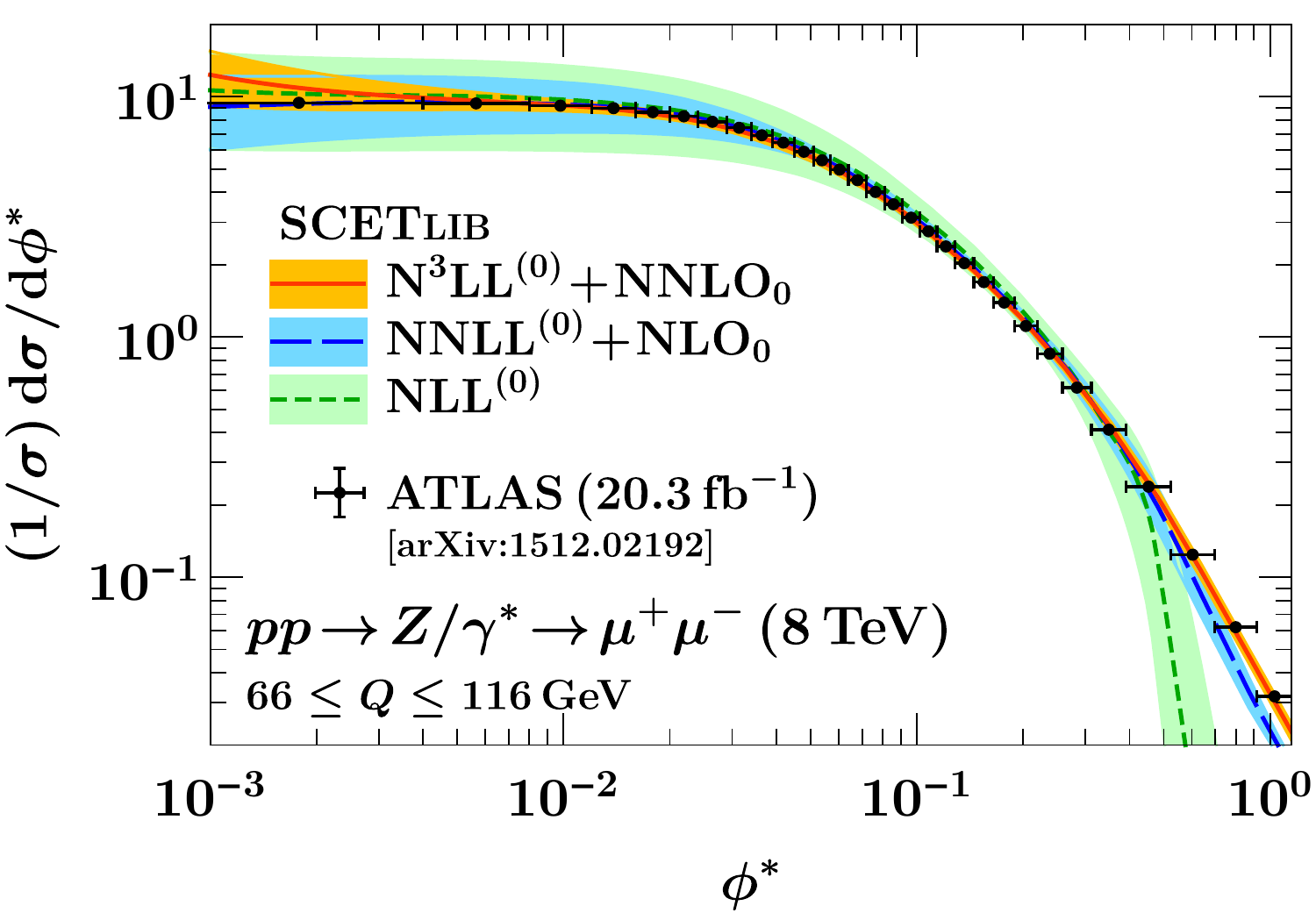}%
\hfill%
\includegraphics[width=0.49\textwidth]{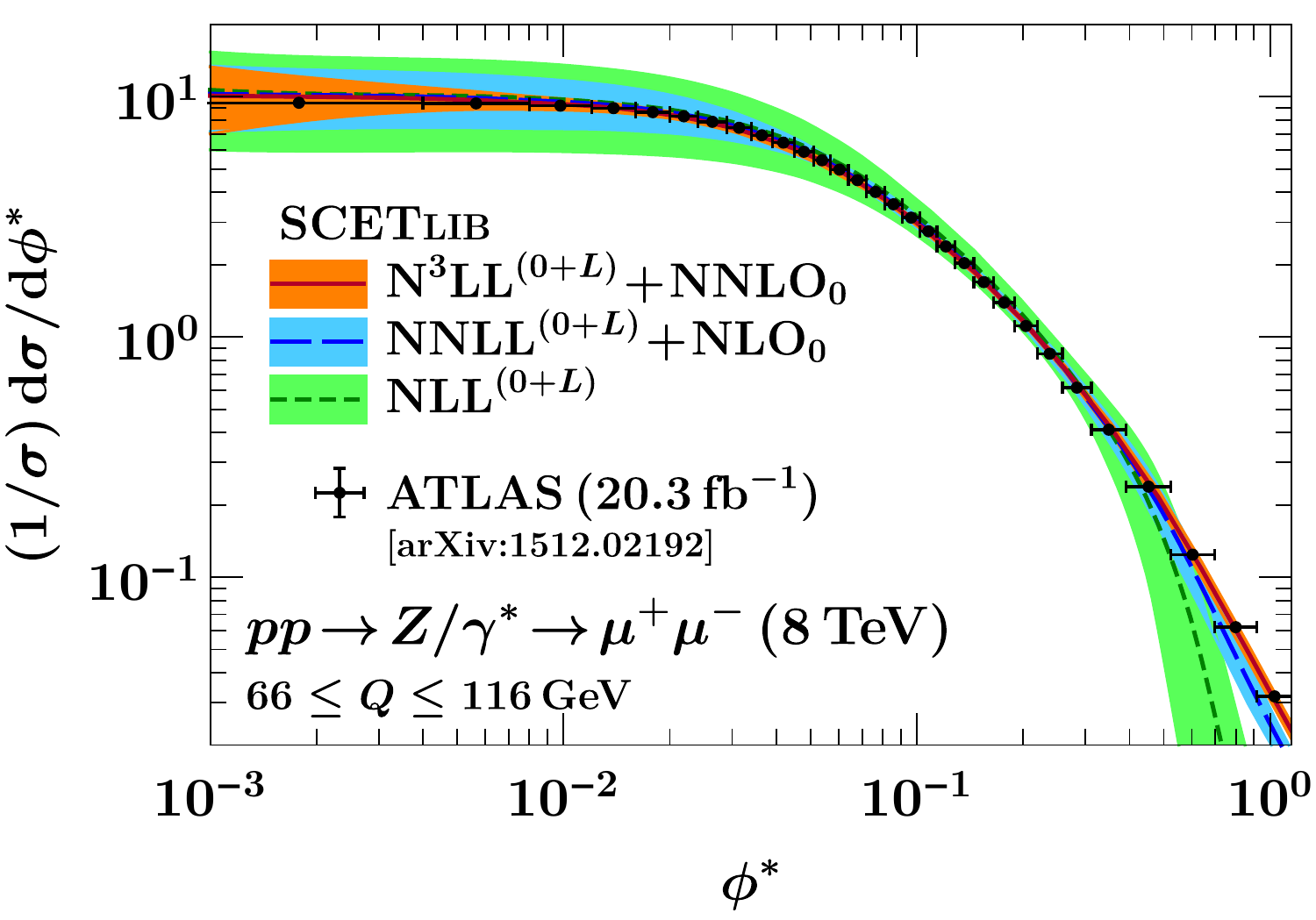}%
\\
\includegraphics[width=0.49\textwidth]{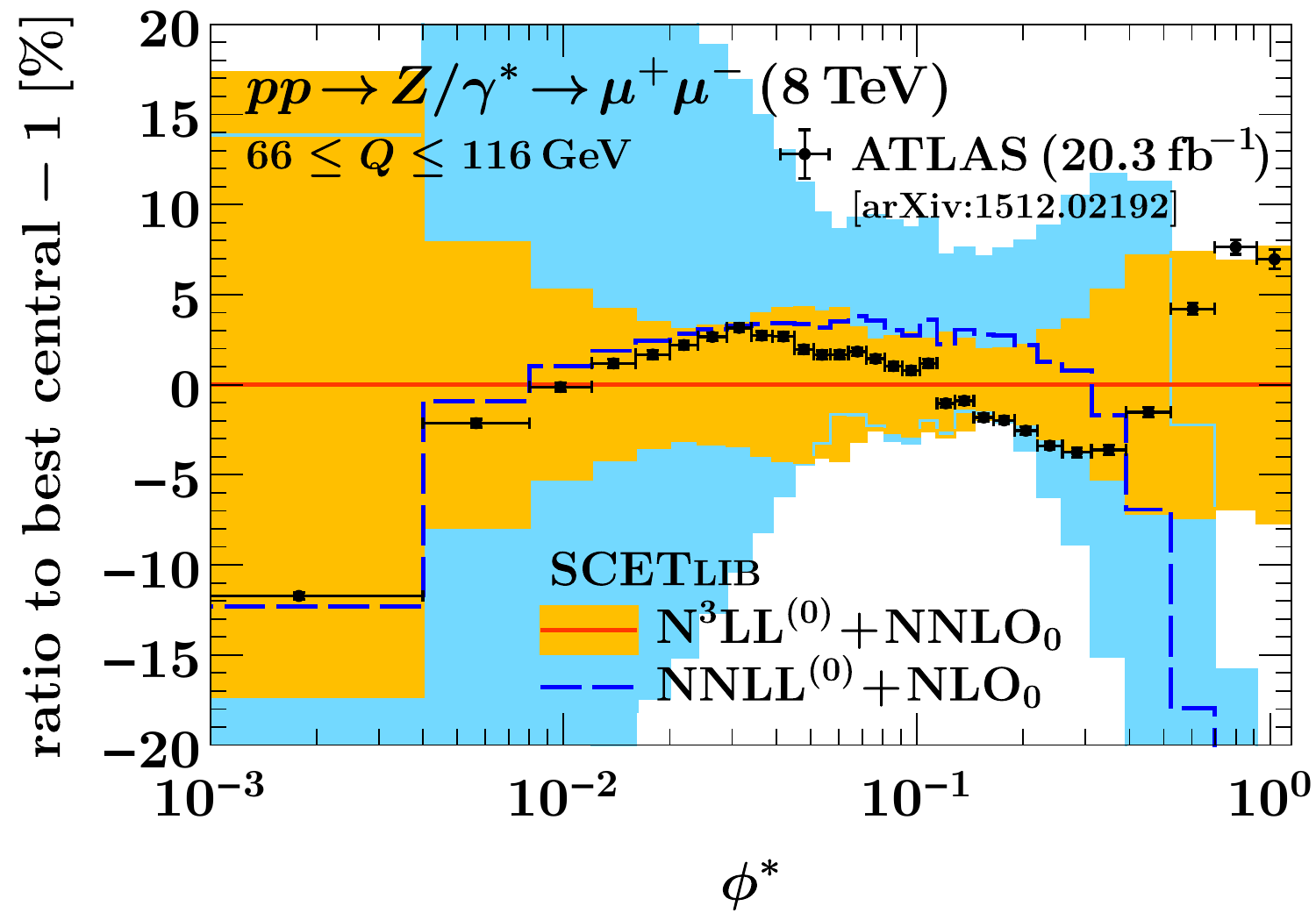}%
\hfill%
\includegraphics[width=0.49\textwidth]{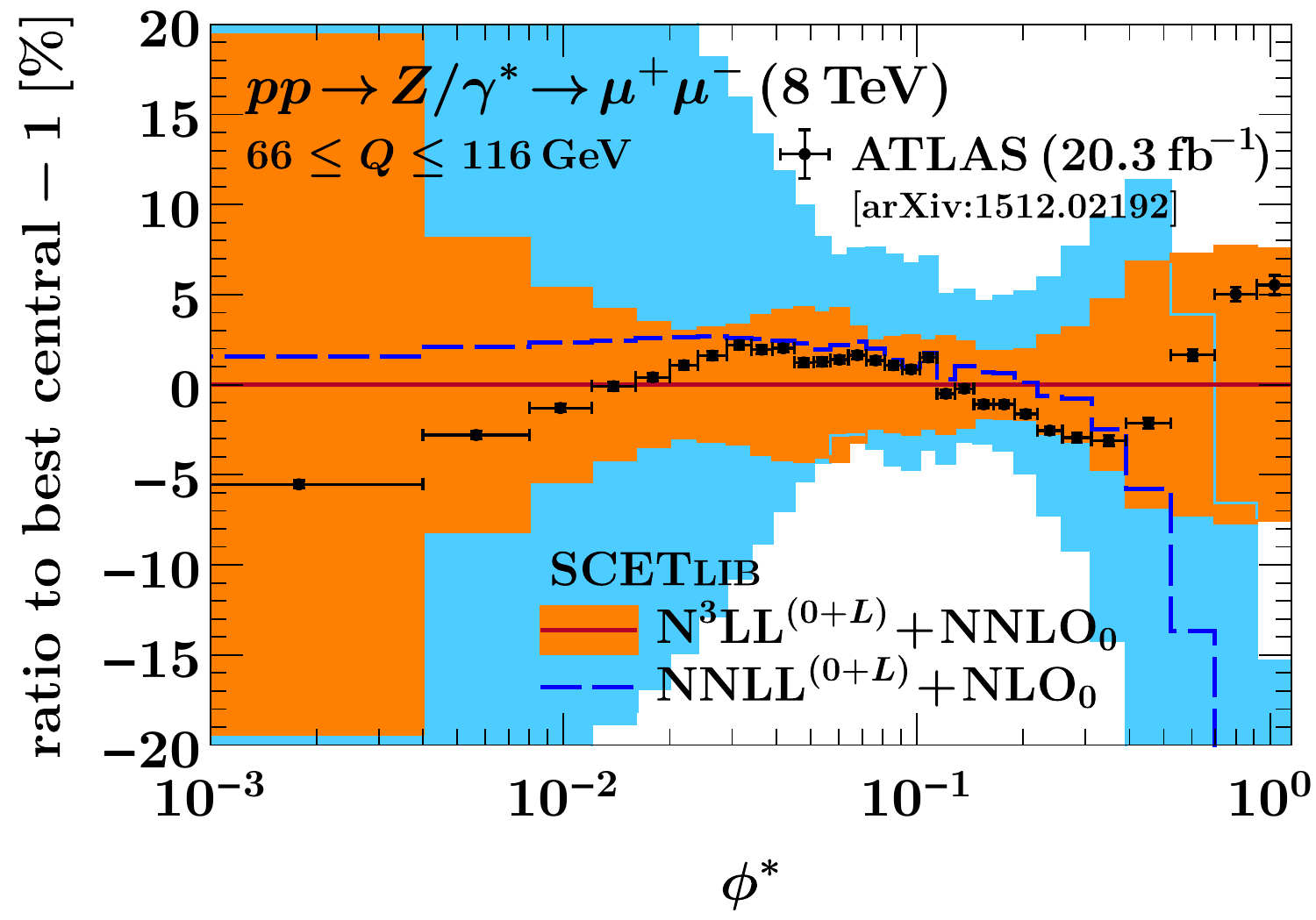}%
\vspace{-2ex}
\caption{Comparison to ATLAS $8\TeV$ $\phi^*$ measurements~\cite{Aad:2015auj}
in the $\mu^+\mu^-$ channel, analogous to the $e^+e^-$ channel in
\fig{Z_phi_star_8TeV_ATLAS_el}.}
% \vspace{-2ex}
\label{fig:Z_phi_star_8TeV_ATLAS_mu}
\end{figure*}
%-------------------------------------------------------------------------------

In \sec{data_comparison}, we compared our predictions to the ATLAS $8\TeV$
measurements~\cite{Aad:2015auj} in the $pp \to Z/\gamma^* \to e^+ e^-$ channel.
For completeness, here we compare the same predictions to the data taken in the
$pp \to Z/\gamma^* \to \mu^+ \mu^-$ channel.
In \fig{Z_qT_8TeV_ATLAS_mu}, we compare to the measurement of the $q_T$ spectrum in the muon channel,
which corresponds to \fig{Z_qT_8TeV_ATLAS_el} in the electron channel.
In \fig{Z_phi_star_8TeV_ATLAS_mu}, we show the results for the $\phi^*$ spectrum in the muon channel,
which corresponds to \fig{Z_phi_star_8TeV_ATLAS_el} in the electron channel.
The muon data follow the same trends and confirm the conclusions drawn in \sec{data_comparison}.

%%%%%%%%%%%%%%%%%%%%%%%%%%%%%%%%%%%%%%%%%%%%%%%%%%%%%%%%%%%%%%%%%%%%%%%%%%%%%%%%
\FloatBarrier
\addcontentsline{toc}{section}{References}
\bibliographystyle{jhep}
\bibliography{refs}

\end{document}